\documentclass[11pt]{article}
\usepackage{graphicx}
\usepackage{latexsym}
\usepackage{amssymb}
\usepackage{color}
\usepackage{amsfonts}
\usepackage{amsmath}
\usepackage{mathtools}
\usepackage{booktabs}
\usepackage{comment}
\usepackage[center]{subfigure}%Provides support for the manipulation and reference of small or 'sub' figures and tables within a single figure or table environment
\usepackage{url}
\newcommand{\BE}{\begin{equation}}
\newcommand{\EE}{\end{equation}}
\newcommand{\BEQ}{\begin{eqnarray}}
\newcommand{\EEQ}{\end{eqnarray}}
\newcommand{\BEQA}{\begin{eqnarray*}}
\newcommand{\EEQA}{\end{eqnarray*}}
\newcommand{\BA}{\begin{array}}
\newcommand{\EA}{\end{array}}

\newcommand{\CP}{\stackrel{p}{\longrightarrow}}

\newtheorem{theorem}{Theorem}
\newtheorem{lemma}{Lemma}

\newtheorem{corollary}{Corollary}
\newtheorem{remark}{Remark}
\newtheorem{assumption}{Assumption}

\newcommand{\argmax}{\operatornamewithlimits{argmax}}
\newcommand{\argmin}{\operatornamewithlimits{argmin}}

\usepackage[numbers,square,sort]{natbib} % package for bibliography style and citation; numbers and square enable squared brackets and numeric citations (The command \citet adds the name of the author to the citation mark, regardless of the citation style.), sort orders multiple citations according to the list of references
\bibpunct{(}{)}{,}{a}{,}{,} % this changes default

\begin{document}

\title{On the asymptotic behavior of bubble date estimators\footnote{\footnotesize{We thank the participants at the Center for Econometrics and Business Analytics (CEBA, St. Petersburg State University) seminar series and the transdisciplinary econometrics and data science seminar (TEDS) for helpful discussions and comments. Anton Skrobotov' research for this paper was supported in part by a grant from the Russian Science Foundation (Project No. 20-78-10113). %by Russian Science Foundation, Project No. 16-18-10432.
Eiji Kurozumi's research was supported by JSPS KAKENHI Grant Number 19K01585 and 22K01422.
Address correspondence to Eiji Kurozumi, Department of Economics, Hitotsubashi University, 2-1 Naka, Kunitachi, Tokyo 186-8601, Japan; e-mail: kurozumi@stat.hit-u.ac.jp
}}}
\author{
Eiji Kurozumi$^{a}$, Anton Skrobotov$^{b,c}$ \\
{\small {$^{a}$ Hitotsubashi University}}\\
{\small {$^{b}$ Russian Presidential Academy of National Economy and Public Administration}}\\
{\small {$^{c}$ Saint Petersburg University (Center for Econometrics and Business Analytics)}}
}
%\date{\today}
\date{February 22, 2022\\
Revised; September 1, 2022}
\maketitle

%\begin{center}
%CONFERENCE DRAFT PAPER -- PLEASE DO NOT CITE OR CIRCULATE WITHOUT THE %AUTHOR'S PERMISSION
%\end{center}

\begin{abstract}
In this study, we extend the three-regime bubble model of \citet{pang2021estimating} to allow the forth regime followed by the unit root process after recovery. We provide the asymptotic and finite sample justification of the consistency of the collapse date estimator in the two-regime AR(1) model. The consistency allows us to split the sample before and after the date of collapse and to consider the estimation of the date of exuberation and date of recovery separately. We have also found that the limiting behavior of the recovery date varies depending on the extent of explosiveness and recovering.
\medskip


\medskip


\end{abstract}

%------------------------------- 1.INTRODUCTION ----------------------------------

\newpage
\renewcommand{\baselinestretch}{1.4}

\section{Introduction}
\baselineskip= 6mm

The estimation of the dates of regime changes has a long history in the econometric literature (see, e.g., \citet{Casini2019} for a recent review). In this study, we address the issue of estimating the break dates in explosive bubble model. Such a bubble model is often  generated (see \citet{PWY2011}) as a unit root process followed by explosive process, which in turn followed by a unit root regime (with possible stationary recovering regime after the collapse of the bubble). For estimation of bubble dates, \citet{phillips2015a,phillips2015b} (PSY hereinafter) proposed the recursive algorithm based on the right-tailed ADF test. Generally speaking, the origination of the bubble is taken at the date for which the test statistic begins to exceed the critical value, and the date of collapse is taken at the date for which the test statistic subsequently falls below the critical value (see \citet{Skrobotov2021} for review).

However, as demonstrated in \citet{harvey2017improving}, the bubble dates estimates are more accurate if they are obtained by minimizing the sum of the squared residuals over all possible dates. For the model with only two regimes, \citet{chong2001structural} investigated consistency and the limiting distribution of the break date estimator for the AR(1) model with a break in the coefficient. The break date estimator is based on minimizing the sum of the squared residuals (SSR) over all possible break dates. The results depend on the direction of the coefficient change. \citet{chong2001structural} considered three cases, both regimes are stationary, the first regime is a unit root regime, and the second regime is stationary and \textit{vice versa}. \citet{pang2018structural} extended the result of \citet{chong2001structural}, allowing the AR(1) coefficient to be dependent on the sample size. They also considered the case with change in persistence from a unit root to moderately explosive and \textit{vice versa}. \citet{pang2021estimating} (PDC hereinafter) considered the model with three regimes that reflect the explosive financial bubble model: the first regime is a unit root regime, followed by an explosive regime and then by a stationary collapsing regime. PDC investigated the sample splitting strategy: The first break date is estimated for the two regime AR(1) model. Results indicate that the break date estimator is consistent for the date of collapse (i.e., the second break). Then, one could consider the sample before the date of collapse and investigate the asymptotic behavior of the date of origination of the bubble -- the results are the same as that in \citet{pang2018structural}. From a different perspective, the estimation procedure used by PDC is the same as that in \citet{chong1995} and \citet{bai1997multiple}, in which multiple breaks are estimated one at a time in the regime-wise stationary model with level shifts.

In this study, we extend the results of PDC by allowing the forth unit root regime. The advantage of this extension is that we can investigate not only the emerging and collapsing dates of the bubble but also the recovering date to the normal market. As a result, we can totally investigate the abnormal market behavior. In our procedure, we first prove the consistency of the collapse date estimator by minimizing the sum of the squared residuals using the two-regime model, allowing nonstationary volatility. Second, as the estimated break date is consistent with the collapsing date, one could split the sample at the estimated break date and consider the estimator of the market recovery date (i.e., the date of change from stationary collapsing regime to unit root regime) using the subsample after the break date detected in the first step.\footnote{The asymptotics for the date of the beginning of the explosive regime were considered in PDC.} Third, we allow a weak dependence in the innovation errors. Our approach closely resembles that of \citet{harvey2017improving} by minimizing the full SSR based on the four regimes model. However, \citet{harvey2017improving} estimated all the break dates simultaneously, while we use the sample-splitting approach. As demonstrated in PDC, this performs better in three-regime case  and computationally less involved (we perform the three SSR minimization with one break each with $O(T)$ computations, while \citet{harvey2017improving} requires minimizing the three break model over all possible combinations of these breaks). Interestingly, the limiting behaviour of the market recovery date estimator depends on the relationship between the extent of the explosive regime and the stationary collapsing (recovering) regime.

It should be noted that our approach can be easily extended to the multiple bubbles context. \citet{harvey2020date} proposed to initially identify the bubble regimes based on PSY approach and then obtain more precise bubble date estimates during each bubble episode. Subsequently, one can use our approach in the second step.

The remainder of this paper is organized as follows. Section 2 formulates the model and assumptions. In Section 3, we define the main procedure and provide the limiting behaviors of the break date estimators. The model with serially correlated shocks is considered in Section 4. The finite sample performance of the estimated break dates is demonstrated in Section 5, and the empirical example is given in Section 6. Section 7 concludes the paper. All proofs are collected in the  appendix.

\section{Model and Assumptions}

Let us consider the following bubble's emerging and collapsing model for $t=1,2,\ldots,T$:
\BE
y_t=\left\{\BA{lcl}
c_0T^{-\eta_0}+y_{t-1}+\varepsilon_t & : & 1 \leq t \leq k_e,\\
\phi_a y_{t-1}+\varepsilon_t & : & k_e+1\leq t \leq k_c, \\
\phi_b y_{t-1}+\varepsilon_t & : & k_c+1\leq t \leq k_r,\\
c_1T^{-\eta_1}+y_{t-1}+\varepsilon_t & : & k_r+1\leq t\leq T,
\EA\right.
\label{model:0}
\EE
where $y_0=o_p(T^{1/2})$, $c_0\geq 0$, $\eta_0 > 1/2$, $\phi_a\coloneqq 1+c_a/T^{a}$ with $c_a > 0$ and $0<a<1$, $\phi_b\coloneqq 1-c_b/T^b$ with $c_b > 0$ and $0 < b < 1$, $c_1\geq 0$, and $\eta_1 > 1/2$. In this model, the process evolves up to time $k_e$ as a unit root process with a possibly positive drift shrinking to zero, and then $y_t$ behaves mildly explosive and next starts collapsing at $t=k_c+1$. Thereafter, the adjustment (collapsing and recovering) period lasts up to $k_r$ and then returns to a unit root process. We denote the mildly explosive regime with $\phi_a\coloneqq 1+c_a/T^a$ and the mildly stationary regime with $\phi_b\coloneqq 1-c_b/T^b$ for ease of exposition. However, we can also consider these AR(1) parameters in more general forms, such as $\phi_a\coloneqq 1+c_a/h_T$ and $\phi_b\coloneqq 1-c_b/k_T$, where $h_T+T/h_T\to \infty$ and $k_T+T/k_T\to \infty$. The break points $k_e$, $k_c$, and $k_r$ are unknown, and the corresponding break fractions are defined as $\tau_e\coloneqq k_e/T$, $\tau_c\coloneqq k_c/T$, and $\tau_r\coloneqq k_r/T$, respectively. Note that PDC considered a model up to $k_r$. Thus, our model is an extension of theirs to include the normal market behavior. Note that the result on the estimator of $k_r$ is new to the existing literature.

For model \eqref{model:0}, we make the following assumption.

\begin{assumption}
\label{assumption:tau}
$0 < \tau_e < \tau_c < \tau_r < 1$.
\end{assumption}

\begin{assumption}
\label{assumption:base}
$\varepsilon_t\coloneqq\sigma_t e_t$, where $\{e_t\}\sim i.i.d.(0,1)$ with $E[e_t^4]<\infty$ and $\sigma_t\coloneqq \omega(t/T)$ where $\omega(\cdot)$ is a nonstochastic and strictly positive function on $[0,1]$ satisfying $\underline{\omega} < \omega(\cdot) <\overline{\omega} < \infty$.
\end{assumption}

Assumption \ref{assumption:tau} implies that break dates are distinct and not quite close each other, which is typically assumed in the existing literature. Assumption \ref{assumption:base} is general enough to allow for nonstationary unconditional volatility. 

Following \citet{CavaliereTaylor2007a,CavaliereTaylor2007b}, the functional central limit theorem (FCLT) holds for the partial sum process of $\{\varepsilon_t\}$ under Assumption \ref{assumption:base}:
\BE
\frac{1}{\sqrt{T}}\sum_{t=1}^{\lfloor \tau T \rfloor}\varepsilon_t\Rightarrow \tilde{\omega}W(\kappa(\tau))\eqqcolon W^{\kappa}(\tau),
\label{FCLT}
\EE
for $0\leq \tau \leq 1$, where $\Rightarrow$ signifies weak convergence of associated probability measures, $W(\cdot)$ is a standard Brownian motion, $\kappa(\tau)\coloneqq \int_0^{\tau}\omega(s)ds/\int_0^1\omega^2(s)ds$ is called the variance profile, and $\tilde{\omega}^2\coloneqq \int_0^1\omega^2(s)ds$. Note that, when $\sigma_t$ is constant, $W^{\kappa}(\tau)$ reduces to a constant volatility times a standard Brownian motion.

Assumption \ref{assumption:base} is used mainly for the estimation of the collapsing date $k_c$ and the market recovering date $k_r$ for $a < b$. For the estimation of the emerging date of a bubble $k_e$ and the last break point $k_r$ for $a > b$, we impose the following stronger assumption:

\begin{assumption}
\label{assumption:iid}
$\{\varepsilon_t\}\sim i.i.d.(0,\sigma^2)$ with $E[\varepsilon_t^4]<\infty$.
\end{assumption}

Under Assumption \ref{assumption:iid}, the FCLT in \eqref{FCLT} becomes simpler and is given by
\BE
\frac{1}{\sqrt{T}}\sum_{t=1}^{\lfloor \tau T \rfloor}\varepsilon_t\Rightarrow \sigma W(\tau).
\label{FCLT2}
\EE
This result is used for deriving the limiting distributions of the estimators of $k_e$ and $k_r$ for $a > b$ in Theorems \ref{theorem:ke}(iii) and \ref{theorem:kr}(ii), respectively.

\section{Individual Estimation of Break Dates}

Following PDC, we estimate the model with a one-time break. Note that model \eqref{model:0} can be expressed as
\BE
y_t
=
\left\{\BA{l}
\phi_0y_{t-1}+u_t \\
\phi_ay_{t-1}+u_t \\
\phi_by_{t-1}+u_t \\
\phi_1y_{t-1}+u_t 
\EA\right.
\quad \mbox{where}\quad
\BA{l}
\phi_0\coloneqq 1 \\
\phi_a\coloneqq 1+c_a/T^a \\
\phi_b\coloneqq 1-c_b/T^b \\
\phi_1\coloneqq 1 
\EA,\quad
u_t
\coloneqq
\left\{\BA{l}
c_0/T^{\eta_0}+\varepsilon_t \\
\varepsilon_t \\
\varepsilon_t \\
c_1/T^{\eta_1}+\varepsilon_t.
\EA\right.
\label{model:est1}
\EE

For a given $1 \leq k \leq T-1$, we denote the sum of the squared residuals as
\BE
SSR(k/T)\coloneqq \sum_{1}^{k}\left(y_t-\hat{\phi}_a(k/T)y_{t-1}\right)^2+\sum_{k+1}^{T}\left(y_t-\hat{\phi}_b(k/T)y_{t-1}\right)^2
\label{ssr:k}
\EE
where $\sum_{t=\ell}^m$ is abbreviated just as $\sum_{\ell}^m$ and
\[
\hat{\phi}_a(k/T)=\frac{\sum_1^k y_{t-1}y_t}{\sum_1^k y_{t-1}^2}
\quad\mbox{and}\quad
\hat{\phi}_b(k/T)=\frac{\sum_{k+1}^T y_{t-1}y_t}{\sum_{k+1}^T y_{t-1}^2}.
\]
Defining the break point estimator as $\hat{k}\coloneqq \arg\min_{k} SSR(k/T)$, we have the following theorem.

\begin{theorem}
\label{theorem:k}
Suppose that Assumptions \ref{assumption:tau} and  \ref{assumption:base} hold. Then, we have $\lim_{T\to\infty} P(\hat{k}=k_c)=1$.
\end{theorem}

Theorem \ref{theorem:k} is the same as Theorem 3(a) in PDC and implies that, even if the last regime with a unit root process is additionally included in the sample period, we can consistently estimate the bubble collapsing date. This may not be a surprising result because the explosive behavior in the second regime is quite different from those in the other regimes. Therefore, even if we estimate only a one time break date, the estimator is consistent with the collapsing date of the bubble. We also note that we allow nonstationary volatility in $\{\varepsilon_t\}$, general enough to apply our method to practical analyses. It might be possible to estimate the break dates by minimizing the weighted sum of the squared residuals using the volatility estimated by \citet{HLZ2022volatility}, but this is beyond the scope of this paper.

The consistency of the collapsing date is intuitively explained as follows. As given in the proof of Lemma 2, during the emergence of the bubble in the second regime, the process can be expressed as
\[
y_k
=
\phi_a^{k-k_e}y_{k_e}+\sum_{j=k_e+1}^{k}\phi_a^{(k-j)}u_j 
\sim_a
\phi_a^{k-k_e}y_{k_e}
\]
for $k_e+1\leq k \leq k_c$, where $y_{k_e}/\sqrt{T}$ weakly converges to a (time-transformed) Brownian motion. Therefore, the process evolves monotonically at a geometric rate (noting that $\phi_a > 1$) and attains its peak at $k=k_c$. Conversely, the process collapses at a geometrically decaying rate (noting that $\phi_b < 1$) because it can be expressed as
\BE
y_k=\phi_b^{(k-k_c)}y_{k_c}+\sum_{j=k_c+1}^{k}\phi_b^{(k-j)}u_j,
\label{yk:3rd}
\EE
for $k_c+1\leq k \leq k_r$, where the first term on the right hand side can be shown to dominate the second term, at least, when $k$ is relatively close to $k_c$. These geometric explosive and collapsing behaviors help us identify the collapsing date $k_r$ consistently.

After we obtained the consistent estimator $\hat{k}$ of $k_c$, we decompose the whole sample into two sub-samples for $t=1,\ldots,\hat{k}$ and $\hat{k}+1,\ldots,T$ and estimate $k_e$ and $k_r$ from the first and second sub-samples, respectively. Because $\hat{k}$ is a consistent estimator of $k_c$, we can treat $k_c$ as a known break point. Thus, we use $k_c$, instead of $\hat{k}$, in the following.

Estimation of $k_e$ is based on the minimization of the sum of the squared residuals using the first sub-sample, and the estimator is defined as
\[
\hat{k}_e\coloneqq \arg\min_{\underline{\tau}_e\leq k/T\leq \bar{\tau}_e}  SSR_1(k/T)
\]
where $0 <\underline{\tau}_e < \tau_e < \bar{\tau}_e <{\tau}_c$ and
\[
SSR_1(k/T) \coloneqq \sum_{1}^k\left(y_t-\hat{\phi}_c(k/T)y_{t-1}\right)^2+\sum_{k+1}^{k_c}\left(y_t-\hat{\phi}_d(k/T)y_{t-1}\right)^2
\]
\[
\mbox{with}\quad \hat{\phi}_c(k/T)=\frac{\sum_{1}^k y_{t-1}y_t}{\sum_{1}^k y_{t-1}^2}\quad\mbox{and}\quad
\hat{\phi}_d(k/T)=\frac{\sum_{k+1}^{k_c} y_{t-1}y_t}{\sum_{k+1}^{k_c} y_{t-1}^2}.
\]
The corresponding break fraction estimator is defined by $\hat{\tau}_e\coloneqq \hat{k}_e/T$.

The consistency of the $\hat{\tau}_e$ has been already established in Theorem 1.3 %and Lemmas B.1 and C.2 
of \citet{pang2018structural} and stated in the next theorem.

\begin{theorem}
\label{theorem:ke}
(\citet{pang2018structural}) 
Suppose that Assumptions \ref{assumption:tau} and  \ref{assumption:iid} hold. Then,\\
(i) when $a < 1/2$, $\lim_{T\to\infty} P(\hat{k}_e=k_e)=1$ .\\
(ii) when $a=1/2$, $|\hat{k}_e-k_e|=O_p(1)$.\\
(iii) when $a > 1/2$, 
\[
(1-\phi_a)^2T^2(\hat{\tau}_e-\tau_e)\Rightarrow \arg\max_{v\in R}\left\{\frac{W^*(v)}{W_1(\tau_e)}-\frac{|v|}{2}\right\},
\]
where $W^*(v)$ is a two-sided Brownian motion on $R$ defined to be $W^*(v)=W_1(-v)$ for $v\leq 0$ and $W^*(v)=W_2(v)$ for $v > 0$ with $W_1(\cdot)$ and $W_2(\cdot)$ being two independent Brownian motion on $R^+$.
\end{theorem}

On the other hand, for the estimation of $k_r$, we minimize the sum of the squared residuals using the second sub-sample, and the estimator is defined as
\[
\hat{k}_r\coloneqq \arg\min_{\underline{\tau}_r\leq k/T\leq \bar{\tau}_r}  SSR_2(k/T)
\]
where $\tau_c <\underline{\tau}_r <\tau_r < \bar{\tau}_r <1$ and
\[
SSR_2(k/T) \coloneqq \sum_{k_c+1}^k\left(y_t-\hat{\phi}_e(k/T)y_{t-1}\right)^2+\sum_{k+1}^T\left(y_t-\hat{\phi}_f(k/T)y_{t-1}\right)^2
\]
\[
\mbox{with}\quad \hat{\phi}_e(k/T)=\frac{\sum_{k_c+1}^k y_{t-1}y_t}{\sum_{k_c+1}^k y_{t-1}^2}\quad\mbox{and}\quad
\hat{\phi}_f(k/T)=\frac{\sum_{k+1}^T y_{t-1}y_t}{\sum_{k+1}^T y_{t-1}^2}.
\]
The corresponding break fraction estimator is defined by $\hat{\tau}_r\coloneqq \hat{k}_r/T$.

To investigate the asymptotic property of $\hat{k}_r$, we must distinguish the two cases: $a < b$ and $a >b$. In the case of $a < b$, the explosive behavior is faster than the collapsing (recovering) speed, and we continue to make Assumption \ref{assumption:base}. On the other hand, to derive the limiting distribution, we need to impose stronger restrictions on the shocks given by Assumption \ref{assumption:iid} in the case of $a > b$, in which the process recovers relatively faster than the evolution of the bubble.

The following theorem reveals that the convergence order of $\hat{\tau}_r$ depends on whether $a < b$ or $a > b$.
\begin{theorem}
\label{theorem:kr}
Suppose that Assumptions \ref{assumption:tau} and  \ref{assumption:iid} hold.
(i) When $a < b$,
\[
\lim_{T\to\infty} P(\hat{k}_r=k_r)=1.
\]
(ii) When $a > b$,
\[
 (1-\phi_b)T(\hat{\tau}_r-\tau_r)\Rightarrow \arg\max_{v\in R}\left\{C_{c_b}^{*}(v)-\frac{|v|}{2}\right\},
\]
where $C_{c_b}^*(v)\coloneqq 2\int_0^{|v|}\tilde{B}_{c_b}(s)dB_1(s)-c_b\int_0^{|v|}\left(\tilde{B}_{c_b}^2(s)-1/(2c_b)\right)ds$ for $v < 0$ and $C_{c_b}^*(v)\coloneqq -\left\{B_2(v)+\int_0^{v}B_2(s)dB_2(s)/\tilde{B}_{c_b}(0)+c_b\int_0^v\left(B_2(s)/(2\tilde{B}_{c_b}(0))+1\right)B_2(s)ds\right\}/(c_b\tilde{B}_{c_b}(0))$ for $v \geq 0$ with $\tilde{B}_{c_b}(s)\coloneqq \int_{s}^{\infty} \exp\left(-c_b(t-s)\right)dB_1(t)$ for $s\geq 0$ and  $B_1(\cdot)$ and $B_2(\cdot)$ being two independent standard Brownian motions on $R^+$.
\end{theorem}

\begin{remark}
\label{remark:1}
Evidently, Theorems \ref{theorem:ke} and \ref{theorem:kr} hold under Assumptions \ref{assumption:tau} and \ref{assumption:base}, except for the derivation of the limiting distributions. That is, the same convergence orders of the break date estimators will be obtained with independent and heteroskedastic errors. We made Assumption \ref{assumption:iid} to derive the limiting distributions in Theorems \ref{theorem:ke}(iii) and \ref{theorem:kr}(ii), which are the same as those obtained in the existing literature.
\end{remark}

From Theorem \ref{theorem:kr}, we can observe that the break date from the collapsing (recovering) regime to the normal market can be consistently estimated if $a<b$; that is, the explosive speed of the process is faster than the collapsing speed. In this case, as given in \eqref{yk:3rd}, the process decays at geometric rate for $k_c+1\leq k \leq k_r$ (because the first term of \eqref{yk:3rd} dominates the second one) and reaches at
\BEQA
y_{k_r}
& = &
\phi_b^{(k_r-k_c)}y_{k_c}+\sum_{j=k_c+1}^{k_r}\phi_b^{(k-j)}u_j \\
& \sim_a &
\phi_a^{(k_c-k_e)}\phi_b^{(k_r-k_e)}y_{k_e}.
\EEQA
Note that $\phi_a^{(k_c-k_e)}\phi_b^{(k_r-k_e)}\to \infty$ when the explosive speed is faster ($a<b$) because
\BEQ
\log\left(\phi_a^{(k_c-k_e)}\phi_b^{(k_r-k_e)}\right)
& = &
(\tau_c-\tau_e)T\log\left(1+\frac{c_a}{T^a}\right)+(\tau_r-\tau_c)T\log\left(1-\frac{c_b}{T_b}\right)
\nonumber \\
& = &
(\tau_c-\tau_e)c_aT^{1-a}\left(1+o(1)\right)-(\tau_r-\tau_c)c_bT^{1-b}\left(1+o(1))\right)
\label{logphiphi} \\
& \to &
\infty
\nonumber
\EEQ
because $a < b$. That is, the explosive effect induced by $\phi_a$ remains in the process during the whole collapsing (recovering) regime, after which the process evolves as
\[
y_k
=
y_{k_r}+c_1\frac{k-k_r}{T^{\eta_1}}+\sum_{j=k_r+1}^{k}\varepsilon_j
\sim_a
y_{k_r}
\]
because we can observe from \eqref{logphiphi} that $T^{-\alpha}\phi_a^{(k_c-k_e)}\phi_b^{(k_r-k_c)}\to \infty$ for any $\alpha > 0$. This monotonically geometrical decay during the collapsing regime followed by the asymptotically flat behavior makes $\hat{k}_r$ consistent.

On the other hand, when the collapsing speed is faster ($a>b$), the effect of the explosive component in $y_k$ in the recovering regime diminishes to zero as $k$ approaches $k_r$ because $\phi_a^{(k_c-k_e)}\phi_b^{(k_r-k_c)}\to 0$ when $a > b$, as can be observed in \eqref{logphiphi}. That is, the initial value effect by $y_{k_c}$ in the recovering regime gradually disappears and the process behaves as if it started from the small initial value. Therefore, we obtain the same result as Theorem 1 of \citet{pang2018structural}.

The asymptotically different convergence orders in Theorem \ref{theorem:kr} reflect the finite sample performance of $\hat{k}_r$, as presented in Section 5.

\begin{remark}
\label{remark:2}
It is possible to consider the case where $a=b$. As is seen from the expansion given by \eqref{logphiphi}, the case where $(\tau_c-\tau_e)c_a > (\tau_r-\tau_c)c_b$ will have the same result as the case of $a<b$ and the reverse relation corresponds to the case of $a > b$. We can also consider the knife edge case in which $(\tau_c-\tau_e)c_a =(\tau_r-\tau_c)c_b$. In this case, the asymptotic behavior of $\phi_a^{(k_c-k_e)}\phi_b^{(k_r-k_e)}$ depends on the higher order expansion of the logarithms as well as the values of $a=b$. Because this knife edge case will make our analysis complicated, we do not pursue this case and concentrate on the two cases where $a<b$ or $a>b$ to clearly state our main results.
\end{remark}

\section{Serially Correlated Case}
In this section, we extend model \eqref{model:0} with $\{\varepsilon_t\}$ being serially correlated. We consider the case where $\{\varepsilon_t\}$ is a linear process given by
\BE
\varepsilon_t=\sum_{j=0}^{\infty}\psi_jv_{t-j}
\quad
\mbox{where}
\quad
\sum_{j=0}^{\infty}j^{3/2}|\psi_j| < \infty
\label{model:0:correl}
\EE
For the innovations $\{v_t\}$, we make the following assumption.

\begin{assumption}
\label{assumption:correl}
$\{v_t\}\sim i.i.d.(0,\sigma^2)$ with $E|v_t|^{\gamma} < \infty$ for some $\gamma$ where $\gamma > \max(4,2/a, 2/b)$.
\end{assumption}

Linear process \eqref{model:0:correl} can be expressed as, by the BN decomposition,
\BE
\varepsilon_t=\psi v_t+\tilde{v}_{t-1}-\tilde{v}_{t}
\quad
\mbox{where}
\quad
\tilde{v}_t\coloneqq \sum_{\ell=0}^{\infty}\tilde{\psi}_{\ell}v_{t-\ell}
\quad
\mbox{with}
\quad
\tilde{\psi}_{\ell}\coloneqq \sum_{k=\ell+1}^{\infty}\psi_k
\label{BNdecomposition}
\EE
and $\psi=\sum_{j=0}^{\infty}\psi_j$. It is also well known that the summability condition in \eqref{model:0:correl} implies $\sum_{\ell=0}^{\infty}|\tilde{\psi}_{\ell}|<\infty$. In this case, we have the following corollary.

\begin{corollary}
\label{corollary:correl}
Suppose that we estimate $k_c$, $k_e$, and $k_r$ as in the previous section and that Assumptions \ref{assumption:tau} and \ref{assumption:correl} holds. Then, we obtain the same results as given in Theorems \ref{theorem:k}, \ref{theorem:ke}, and \ref{theorem:kr}(i), while Theorem \ref{theorem:kr}(ii) holds with the limiting distribution replaced by.
\[
 (1-\phi_b)T(\hat{\tau}_r-\tau_r)\Rightarrow \arg\max_{v\in R}\left\{C_{c_b}^{*}(v)-\frac{|v|}{2}\psi^*\right\},
\]
where $\psi^*$ depends on whether $v<0$ or $v\geq 0$ and is defined in Appendix C. 
\end{corollary}

By Corollary \ref{corollary:correl}, we can estimate the three break dates by using the misspecified AR(1) model with the same convergence orders of the estimators as before, although we may expect that the finite sample accuracy of the estimators would be improved by correctly specifying the correlation structure by considering the ADF type regression with an additional lag structure.

\section{Monte-Carlo Simulations}

In this section, we examine the performance of the estimates of the bubble regimes dates in finite samples. Our purpose is to demonstrate how the finite sample properties of the estimated break dates reflect the asymptotic results obtained in the previous section.

The Monte-Carlo simulations reported in this section are based on the series generated by \eqref{model:0} with $y_0=0$ and $\{\varepsilon_t\}\sim IIDN(0,1)$. Data were generated from this DGP for samples of $T=(400,800)$ with $50,000$ replications.\footnote{All simulations were programmed in R with rnorm random number generator.} We set the drift terms in the first and fourth regimes to $c_0T^{-\eta_0}=1/800$ and $c_1T^{-\eta_0}=1/800$, respectively, following PDC. In this experiment, we focus on two cases. In the first case, the explosive coefficient $\phi_a$ changes with a fixed $\phi_b$; $\phi_b$ is set to $0.96$ while $\phi_a$ takes values among $(1.01, 1.05, 1.09)$. In the second case, the collapsing coefficient $\phi_b$ varies with a fixed $\phi_a$; $\phi_a$ is set to 1.05 while $\phi_b$ varies among $(0.98, 0.96, 0.94)$. We set the localising parameters $c_a=c_b=1$; then, the values of $a$ and $b$ are uniquely determined based on the definitions given by $\phi_a=1+1/T^a$ and $\phi_b=1-1/T^b$. For the dates of bubble regimes, we use $(\tau_e,\tau_c,\tau_r)$ to be equal to (0.4,0.6,0.7).\footnote{To compare the estimator of $k_c$ with that of PDC, we also set $(\tau_e,\tau_c,\tau_r)$ to be equal to (0.4,0.6,1),  that is, without the last unit root regime (the normal market behavior after the collapse and recovering). The results are entirely similar and omitted for brevity.}

Further, in the minimization of $SSR(k/T)$, $SSR_1(k/T)$, and $SSR_2(k/T)$, we excluded the first and last 5\% observations from the permissible break date $k$. For example, when estimating $k_r$ based on $SSR_2(k/T)$, the permissible break date $k$ ranges from $\hat{k}_c+0.05T+1$ to $0.95T$. If the break date estimate $\hat{k}_c$ exceeds $0.95T$, then we cannot estimate $k_r$; we do not include such a case in any bins of the histogram and thus the sum of the heights of the bins does not necessarily equal one for $\hat{k}_r$ in some cases. Similarly, we cannot estimate $\hat{k}_e$ when $\hat{k}_c < 0.05T$. In the following, we investigate only $\hat{k}_c$ and $\hat{k}_r$ because, once $\hat{k}_c$ is obtained, the estimate of $k_e$ in the four regime model is the same as that in the three regime model, which has been already investigated by PDC.

Figure 1 presents the histograms of $\hat{k}_c$ when $\phi_b=0.96$ and $\phi_a=(1.01, 1.05, 1.09)$. As expected, when $\phi_a$ becomes larger and/or the sample size increases, the $\hat{k}_c$ becomes more accurate. When $a>b$ ($|\phi_a-1| < |\phi_b-1|$ in our setting) as in Figures 1(a) and 1(b), the frequency of selecting the true break date is not very high (it is around 30\% for $T=400$ and 65\% for $T=800$), but it increases to almost 100\% as in Figures 1(c)--1(f).

In the second case (Figure 2) where $\phi_a=1.05$ and $\phi_b$ varies among $(0.98, 0.96, 0.94)$, the estimate $\hat{k}_c$ is quite accurate for all the three cases with the almost 100\% frequency of selecting the true break date. 

Figure 3 demonstrates the histograms of $\hat{k}_r$ for the fixed value of $\phi_b=0.96$ and the different values of $\phi_a$. For the small value of $\phi_a=1.01$, as shown in Figures 3(a) and 3(b), which corresponds to the case of $a >b$, we can observe that the accuracy of $\hat{k}_r$ deteriorates, while for $a<b$ ($\phi_a=1.05$ and $\phi_a=1.09$) the estimate becomes more accurate. In particular, for the large sample size of $T=800$, the frequency of selecting the true break date is very close to 100\%.

In the case where $\phi_a$ is fixed at $1.05$ and $\phi_b$ varies, we can observe the good performance of the estimate $\hat{k}_r$ even for the case of $a>b$ ($\phi_a=1.05$, $\phi_b=0.94$) as shown in Figures 4(e) and 4(f). Reflecting the result in Theorem \ref{theorem:kr}, the estimate $\hat{k}_r$ is most accurate when $\phi_a=1.05$ and $\phi_b=0.98$; the frequency of selecting the true break date is approximately $75$\% and 100\% for $T=400$ and $800$, respectively.

We next consider the case where the explosive and collapsing regimes are shorter such as $(\tau_e,\tau_c,\tau_r)=(0.5,0.55,0.6)$, which is relatively close to NASDAQ data in empirical application. Figures are collected in the appendix and, as expected, the finite sample properties deteriorate in most case, although they improve as $T$ gets larger and/or the explosive speed gets relatively faster (Figures D.1--D.4).

We also experiment with shorter minimum separation periods;  we exclude the first and last 1\% samples from the permissible break date $k$. The results provided in the appendix demonstrate that there is virtually no significant difference between two trimming parameters by comparing Figures 1--4 with D.5--D.8.

Finally, we investigate the finite sample behavior of the break date estimates when the volatility process $\sigma_t$ follows the single shift model, $\omega(s) = \sigma_{0} + (\sigma_{1} - \sigma_{0})1(s > \tau_{\sigma})$, where $\tau_{\sigma}=0.5$ and $\sigma_1/\sigma_0\in\{1/3,3\}$. Note that $\sigma_1/\sigma_0=1$ corresponds to the case of constant unconditional volatility.

For $\sigma_1/\sigma_0=1/3$, with which the volatility in the first half of the sample is higher than in the later sample, $\hat{k}_c$ and $\hat{k}_r$ perform similarly to the case of the constant unconditional volatility; the frequency of correctly selecting the true collapsing date $k_c$ is relatively high, whereas the performance of $\hat{k}_r$ is not satisfactory when $\phi_a=1.01$ and $\phi_b=0.96$ (Figures D.9--D.12).

In the case where the low volatility regime shifts to the high volatility one at the middle of the sample ($\sigma_1/\sigma_0=3$), it becomes more difficult to select the true collapsing and recovering dates than in the case of $\sigma_1/\sigma_0=1/3$, in particular, when $\phi_a=1.01$ and $\phi_b=0.96$. The possible reason is that both $k_c$ and $k_r$ are in the volatile regime when $\sigma_1/\sigma_0=3$ and thus it would be difficult to distinguish between the shifts in the parameters and the large shocks. However, the performance of the estimates improves for the larger values of $\phi_a$ and/or as $T$ gets larger (Figures D.13--D.16).

\section{Empirical Application}

In this section, we demonstrate the application of the bubble dates estimation method for the four regime model investigated in the previous sections to two time series data sets. It is worth noticing that even though we may detect emergence and collapse of a bubble by some tests, we are not sure whether the final observation available for estimation is included in the collapsing regime or the unit root regime. Therefore, before estimating the break dates, we need to determine whether the three or four regime model is appropriate. Following \citet{harvey2017improving}, we implement the BIC with the penalty term given by the number of estimated parameters plus the number of breaks to determine the model. In the following, we define by $BIC_2$ the BIC obtained from the model without the collapsing regime; by $BIC_3$ the BIC obtained from the model without the recovering regime; by $BIC_4$ the BIC obtained from the full four-regime model.

The first example is the close prices of monthly data from January 1985 to August 2013 of NASDAQ Composite Index.\footnote{The data was downloaded from https://finance.yahoo.com} First, we select the model using the BICs; $BIC_{2}$, $BIC_{3}$, and $BIC_4$ equals 3409.296, 3404.268, and 3387.727, respectively. Thus, the four regime model has the minimum BIC and is selected. As explained in the previous section, we fit the AR(1) model with two regimes and the estimated break date corresponds to the collapsing date, which is February 2000, as depicted in Figure 5. %\ref{fig_emp}. 
Next, by splitting the whole sample at the estimated collapsing date, the date of origination of the bubble is detected at August 1998 from the first subsample, while the date of recovery is estimated at September 2001 from the seconod subsample. As is observed in Figure 5, %\ref{fig_emp} 
our method can detect the explosive and collapsing behavior very well. 

The second example is an application of our method to the logarithm of the U.S.\ house price index from January 1991 to December 2012, provided by the Federal Housing Finance Agency, adjusted by the consumer price index.\footnote{The house price index and the CPI are available from https://www.fhfa.gov/DataTools/Downloads/pages/house-price-index.aspx and https://fred.stlouisfed.org/} Again, because the BIC selects the four regime model ($BIC_{2}$, $BIC_{3}$, and $BIC_4$ equal $-2809.591$, $-2895.905$, and $-2918.223$, respectively), we first estimate the collapsing date, which is estimated at November 2006. As is seen in Figure 6, the explosive behavior becomes mild around early in 2006 but the collapsing date is estimated just before the series staring crashing. By splitting the whole sample at the estimated collapsing date, the emergence date of the explosive behavior is estimated at September 1997, while the recovering date is at May 2011. These estimated dates are consistent with the visual inspection and the proposed method works very well.

\section{Concluding Remarks}
In this paper, we considered the four regime bubble model and investigated the break date estimators using the sample splitting approach; the break dates are estimated one at a time. We showed that the break date estimated initially is consistent with the collapsing date of the bubble. We used the second subsample after the first estimated break date to estimate the break date returning to the normal market. The results revealed that this estimator is consistent with the recovering date if the explosive speed of the process during the bubble period is faster than the collapsing (recovering) speed, whereas in the case where the explosive speed is relatively slower, only the corresponding break fraction estimator is consistent. We need further investigation when the two speeds are the same.

Although we considered the case where the bubble occurred only once, we can extend our approach to the multiple bubble model in conjunction with the approach used by \citet{harvey2020date}. In this case, we first identify the bubble regimes based on the PSY approach and then implement our method to each estimated regime.

%--------------------------------- BIBLIOGRAPY ------------------------------------
\bibliographystyle{apalike} % or try abbrvnat or unsrtnat
\bibliography{ref_joe}

\newpage
\appendix% {\centerline{\textbf{\LARGE{Appendices}}}} 
\renewcommand{\thesection}{\Alph{section}.}
\renewcommand{\theequation}{\Alph{section}.\arabic{equation}}
\setcounter{section}{0}
\setcounter{equation}{0}
\begin{center}
{\Large\bf Appendix}
\end{center}

In Appendix A, we provide the proof of theorems. All the lemmas related to these proofs are given in Appendix B, while Appendix C gives the proof in the case of the serially correlated errors. The additional figures are collected in Appendix D. In the appendices, we denote $C$ a generic constant that may differ all over and $a_T \sim _a b_T$ implies $a_T/b_T\CP 1$ as $T\to \infty$.

\section{Proofs of Theorems}

Let $M_T$ be some positive and increasing sequence such that $M_T\to \infty$ and $M_T/T^{\min(a,b)}\to 0$. We use the following relations repeatedly:
\BE
O_p(T^{\alpha_1}M_T^{\alpha_2})=o_p(\phi_a^{\alpha_3 T})
\quad\mbox{and}\quad
O_p(T^{\alpha_1}M_T^{\alpha_2}\phi_b^{\alpha_3 T})=o_p(1)
\label{app:phiaphib1}
\EE
for any values of $\alpha_1$, $\alpha_2$ and $\alpha_3>0$. These relations hold because
\BEQA
\log(T^{\alpha_1}M_T^{\alpha_2}/\phi_a^{\alpha_3 T})
& = &
\alpha_1\log(T)+\alpha_2\log(M_T)-\alpha_3 T\log\left(1+\frac{c_a}{T^a}\right) \\
& = &
-\alpha_3c_a T^{1-a}(1+o(1))\to -\infty.  \\
\log(T^{\alpha_1}M_T^{\alpha_2}\phi_b^{\alpha_3T})
& = &
\alpha_1\log(T)+\alpha_2\log(M_T)+\alpha_3T\log\left(1-\frac{c_b}{T^b}\right) \\
& = &
-\alpha_3c_bT^{1-b}(1+o(1))\to -\infty.
\EEQA
We also note that, in the proofs, the orders of the terms of our interest may depend on $\phi_a^{(k_c-k_e)}\phi_b^{(k_r-k_c)}$ in some places, which diverges to infinity when $a<b$, because
\begin{eqnarray*}
\log(\phi_a^{(k_c-k_e)}\phi_b^{(k_r-k_e)})
& = &
(k_c-k_e)\log\left(1+\frac{c_a}{T^a}\right)+(k_r-k_e)\log\left(1-\frac{c_b}{T^b}\right) \\
& = &
(\tau_c-\tau_e)T\left(\frac{c_a}{T^a}+O_p\left(\frac{1}{T^{2a}}\right)\right)-(\tau_r-\tau_c)T\left(\frac{c_b}{T^b}+O_p\left(\frac{1}{T^{2b}}\right)\right) \\
& \to &
\infty
\end{eqnarray*}
when $a<b$, but it goes to $-\infty$ when $a > b$. This implies that the following relations similar to \eqref{app:phiaphib1} hold:
\[
O_p(T^{\alpha_1}M_T^{\alpha_2})=o_p(\phi_a^{\alpha_3T}\phi_b^{\alpha_4T})
\quad\mbox{when $a < b$ and}\quad
O_p(T^{\alpha_1}M_T^{\alpha_2}\phi_a^{\alpha_3T}\phi_b^{\alpha_4T})=o_p(1)\quad\mbox{when $a > b$}
\]
for any values of $\alpha_1$, $\alpha_2$, $\alpha_3$, and $\alpha_4 > 0$.

\noindent
\textbf{Proof of Theorem \ref{theorem:k}}: 
To prove the consistency of $\hat{k}$, we first show that $\hat{k}-k_c=O_p(1)$. Similarly to \citet{chong2001structural}, we prove this by the contradiction argument. Suppose that $\hat{k}-k_c$ is not $O_p(1)$, which implies that, for some increasing sequence $M_T$ such that $M_T\to \infty$ and $M_T/T \to 0$,
\BE
\lim_{T\to \infty}\inf P\left(|\hat{k}-k_c| \geq M_T \right) \geq \alpha
\label{consistency:fraction}
\EE
holds for some $0 < \alpha\leq 1$. We shall prove that \eqref{consistency:fraction} does not occur as $T\to \infty$. For this purpose, we decompose the whole sample into five subsamples such that $D_{0T}=\{k:\;k_c-M_T+1\leq k \leq k_c+M_T\}$, $D_{1T}=\{k:\;1\leq k \leq k_e\}$, $D_{2T}=\{k:\;k_e+1\leq k \leq k_c-M_T\}$, $D_{3T}=\{k:\;k_c+M_T+1\leq k \leq k_r\}$, $D_{4T}=\{k:\;k_r+1\leq k \leq T\}$. Because 
\BEQA
P\left(|\hat{k}-k_c| \geq M_T\right)
& = &
P\left(\inf_{k\in \cup_{i=1}^4 D_{iT}}SSR(k/T) < \inf_{k\in D_{0T}} SSR(k/T) \right) \\
& \leq &
P\left(\inf_{k\in \cup_{i=1}^4 D_{iT}}SSR(k/T) < SSR(k_c/T)  \right) \\
& \leq &
\sum _{i=1}^4P\left(\inf_{k\in D_{iT}}SSR(k/T) < SSR(\tau_c)  \right),
\EEQA
we show that
\BE
P\left(\inf_{k\in D_{iT}}SSR(k/T) < SSR(\tau_c)  \right)\to 0
\label{infssr}
\EE
for $i=1,\ldots,4$, which contradict \eqref{consistency:fraction}.

To derive the asymptotic order of $SSR(\tau_c)$, we decompose it into
\BEQA
SSR(\tau_c)
& = &
\sum_{1}^{k_e}\left(y_t-\hat{\phi}_a(k_c/T)y_{t-1}\right)^2
+
\sum_{k_e+1}^{k_c}\left(y_t-\hat{\phi}_a(k_c/T)y_{t-1}\right)^2
\\
& &
+
\sum_{k_c+1}^{k_r}\left(y_t-\hat{\phi}_b(k_c/T)y_{t-1}\right)^2
+
\sum_{k_r+1}^{T}\left(y_t-\hat{\phi}_b(k_c/T)y_{t-1}\right)^2 \\
& = &
S_{1T}+S_{2T}+S_{3T}+S_{4T},\mbox{ say}.
\EEQA
Since
\BE
\hat{\phi}_a(k_c/T)
=
\frac{\sum_{1}^{k_c}y_{t-1}u_t+\sum_{1}^{k_e}y_{t-1}^2+\phi_a\sum_{k_e+1}^{k_c}y_{t-1}^2}{\sum_{1}^{k_c}y_{t-1}^2},
\label{phia:kc}
\EE
we have
\BEQA
S_{1T}
& = &
\sum_1^{k_e}\left(u_t+y_{t-1}-\hat{\phi}_a(k_c/T)y_{t-1}\right)^2 \\
& = &
\sum_{1}^{k_e}\left(u_t+\frac{\sum_1^{k_e}y_{t-1}^2+\sum_{k_e+1}^{k_c}y_{t-1}^2}{\sum_1^{k_c}y_{t-1}^2}y_{t-1}-\frac{\sum_{1}^{k_c}y_{t-1}u_t+\sum_1^{k_e}y_{t-1}^2+\phi_a\sum_{k_e+1}^{k_c}y_{t-1}^2}{\sum_1^{k_c}y_{t-1}^2}y_{t-1}\right)^2 \\
& = &
\sum_{1}^{k_e}\left(u_t-\frac{\sum_{1}^{k_c}y_{t-1}u_t+(\phi_a-1)\sum_{k_e+1}^{k_c}y_{t-1}^2}{\sum_1^{k_c}y_{t-1}^2}y_{t-1}\right)^2 \\
& = &
\sum_{1}^{k_e}\left(u_t-\varphi_{1T} y_{t-1}\right)^2,\mbox{say}.
\EEQA
To investigate the order of $\varphi_{1T}$, we can observe from \eqref{lemma:1:yu} and \eqref{lemma:2:yu} that
\BEQA
\sum_{1}^{k_c}y_{t-1}u_t
& = &
\sum_{1}^{k_e}y_{t-1}u_t+\sum_{k_e+1}^{k_c}y_{t-1}u_t \\
& = &
O_p(T)+O_p(T^{(a+1)/2}\phi_a^{(k_c-k_e)})=O_p(T^{(a+1)/2}\phi_a^{(k_c-k_e)}),
\EEQA
while from \eqref{lemma:2:y2},
\[
(\phi_a-1)\sum_{k_e+1}^{k_c}y_{t-1}^2=O_p(T\phi_a^{2(k_c-k_e)})
\]
 and thus, the second term in the numerator of $\varphi_{1T}$ dominates the first term. Because
\[
\sum_1^{k_c}y_{t-1}^2=\sum_1^{k_e}y_{t-1}^2+\sum_{k_e+1}^{k_c}y_{t-1}^2=O_p(T^2)+O_p(T^{a+1}\phi_a^{2(k_c-k_e)})=O_p(T^{a+1}\phi_a^{2(k_c-k_e)})
\]
from \eqref{lemma:1:y2} and \eqref{lemma:2:y2}, the second term in the denominator of $\varphi_{1T}$ dominates the first term. Therefore, we can observe that
\[
\varphi_{1T}\sim_a \frac{(\phi_a-1)\sum_{k_e+1}^{k_c}y_{t-1}^2}{\sum_{k_e+1}^{k_c}y_{t-1}^2}=O_p\left(\frac{1}{T^a}\right).
\]
Using this result, we have
\BEQ
S_{1T}
& = &
\sum_{1}^{k_e}u_t^2-2\varphi_{1T}\sum_{1}^{k_e}y_{t-1}u_t+\varphi_{1T}^2\sum_{1}^{k_e}y_{t-1}^2 \nonumber \\
& = &
\sum_{1}^{k_e}u_t^2+O_p(T^{-a})O_p(T)+O_p(T^{-2a})O_p(T^2) \nonumber \\
& = &
\sum_{1}^{k_e}u_t^2+O_p(T^{2(1-a)})
\label{s1:kc}
\EEQ
from \eqref{lemma:1:y2} and \eqref{lemma:1:yu}.

Similarly, for $S_{2T}$, we can observe that
\BEQA
S_{2T}
& = &
\sum_{k_e+1}^{k_c}\left(u_t-\frac{\sum_{1}^{k_c}y_{t-1}u_t+(1-\phi_a)\sum_{1}^{k_e}y_{t-1}^2}{\sum_1^{k_c}y_{t-1}^2}y_{t-1}\right)^2 \\
& = &
\sum_{k_e+1}^{k_c}\left(u_t-\varphi_{2T} y_{t-1}\right)^2,\mbox{say}.
\EEQA
From \eqref{lemma:1:y2}, \eqref{lemma:1:yu},\eqref{lemma:2:y2} and \eqref{lemma:2:yu}, we can observe that
\[
\varphi_{2T}
=
\frac{O_p(T^{(a+1)/2}\phi_a^{(k_c-k_e)})+O_p(T^{2-a})}{O_p(T^{a+1}\phi_a^{2(k_c-k_e)})} 
=
O_p\left(\frac{1}{T^{(a+1)/2}\phi_a^{(k_c-k_e)}}\right).
\]
Then, we have
\BEQ
S_{2T}
& = &
\sum_{k_e+1}^{k_c}u_t^2-2\varphi_{2T}\sum_{k_e+1}^{k_c}y_{t-1}u_t+\varphi_{2T}^2\sum_{k_e+1}^{k_c}y_{t-1}^2 \nonumber \\
& = &
\sum_{k_e+1}^{k_c}u_t^2+O_p(1)
\label{s2:kc}
\EEQ
from \eqref{lemma:2:y2} and \eqref{lemma:2:yu}.

For $S_{3T}$ and $S_{4T}$, we first note that
\BE
\hat{\phi}_b(k_c/T)
=
\frac{\sum_{k_c+1}^{T}y_{t-1}u_t+\phi_b\sum_{k_c+1}^{k_r}y_{t-1}^2+\sum_{k_r+1}^Ty_{t-1}^2}{\sum_{k_c+1}^{T}y_{t-1}^2}.
\label{phib:kc}
\EE
Then, we have
\BEQA
S_{3T}
& = &
\sum_{k_c+1}^{k_r}\left(u_t-\frac{\sum_{k_c+1}^{T}y_{t-1}u_t+(1-\phi_b)\sum_{k_r+1}^{T}y_{t-1}^2}{\sum_{k_c+1}^{T}y_{t-1}^2}y_{t-1}\right)^2 \\
& = &
\sum_{k_c+1}^{k_r}\left(u_t-\varphi_{3T} y_{t-1}\right)^2,\mbox{ say}.
\EEQA
We investigate the order of $\varphi_{3T}$.  From \eqref{lemma:3:yu}, \eqref{lemma:4:y2-a}, \eqref{lemma:4:yu-a}, \eqref{lemma:4:y2-b}, and \eqref{lemma:4:yu-b}, we can observe that
\BEQA
\sum_{k_c+1}^{T}y_{t-1}u_t+(1-\phi_b)\sum_{k_r+1}^{T}y_{t-1}^2
& = &
O_p(T^{(b+1)/2}\phi_a^{(k_c-k_e)}) \\
& &
+\left\{\BA{l}
O_p(T\phi_a^{(k_c-k_e)}\phi_b^{(k_r-k_c)})+O_p(T^{2-b}\phi_a^{2(k_c-k_e)}\phi_b^{2(k_r-k_c)}) \\
O_p(T)+O_p(T^{2-b}) \EA\right. \\
& = &
\left\{\BA{ll}
O_p(T^{2-b}\phi_a^{2(k_c-k_e)}\phi_b^{2(k_r-k_c)}) & (a < b) \\
O_p(T^{(b+1)/2}\phi_a^{(k_c-k_e)}) & (a > b), \EA\right.
\EEQA
while from \eqref{lemma:3:y2}, \eqref{lemma:4:y2-a}, and \eqref{lemma:4:y2-b},
\BEQA
\sum_{k_c+1}^{T}y_{t-1}^2
& = &
\sum_{k_c+1}^{k_r}y_{t-1}^2
+\sum_{k_r+1}^{T}y_{t-1}^2 \\
& = & 
O_p(T^{b+1}\phi_a^{2(k_c-k_e)}) 
+\left\{\BA{ll} O_p(T^2\phi_a^{2(k_c-k_e)}\phi_b^{2(k_r-k_c)}) & (a < b) \\ O_p(T^2) & (a > b)\EA\right. \\
& = &
O_p(T^{b+1}\phi_a^{2(k_c-k_e)})
\EEQA
because $T^\alpha \phi_b^{(k_r-k_c)}\to 0$ for any value of $\alpha$.
Then, we have
\[
\varphi_{3T}=\left\{\BA{ll}
O_p(T^{1-2b}\phi_b^{2(k_r-k_c)}) & (a<b) \\
O_p(T^{-(b+1)/2}\phi_a^{-(k_c-k_e)}) & (a > b)
\EA\right. .
\]
Using this result, we have
\BEQ
S_{3T}
& = &
\sum_{k_c+1}^{k_r}u_t^2-2\varphi_{3T}\sum_{k_c+1}^{k_r}y_{t-1}u_t+\varphi_{3T}^2\sum_{k_c+1}^{k_r}y_{t-1}^2 \nonumber \\
& = &
\sum_{k_c+1}^{k_r}u_t^2
+\left\{\BA{ll}
O_p(T^{3(1-b)}\phi_a^{2(k_c-k_e)}\phi_b^{4(k_r-k_c)}) & (a< b) \\
O_p(1) & (a > b)
\EA\right.
\label{s3:kc}
\EEQ
from \eqref{lemma:3:y2} and \eqref{lemma:3:yu}.

Similarly, we have
\BEQA
S_{4T}
& = &
\sum_{k_r+1}^{T}\left(u_t-\frac{\sum_{k_c+1}^{T}y_{t-1}u_t+(\phi_b-1)\sum_{k_c+1}^{k_r}y_{t-1}^2}{\sum_{k_c+1}^{T}y_{t-1}^2}y_{t-1}\right)^2 \\
& = &
\sum_{k_r+1}^{T}\left(u_t-\varphi_{4T} y_{t-1}\right)^2,\mbox{say}.
\EEQA
From \eqref{lemma:3:y2}, \eqref{lemma:3:yu}, \eqref{lemma:4:yu-a}, and \eqref{lemma:4:yu-b}, we can observe that the second term of the numerator of $\varphi_{4T}$ dominates the first term, and it is $O_p(T\phi_{a}^{2(k_c-k_e)})$, while the denominator is $O_p(T^{b+1}\phi_a^{2(k_c-k_e)})$ as shown previously. As a result, we have $\varphi_{4T}=O_p(T^{-b})$. Using this result, we have
\BEQ
S_{4T}
& = &
\sum_{k_r+1}^{T}u_t^2-2\varphi_{4T}\sum_{k_r+1}^{T}y_{t-1}u_t+\varphi_{4T}^2\sum_{k_r+1}^{T}y_{t-1}^2 \nonumber \\
& = &
\sum_{k_r+1}^{T}u_t^2
+\left\{\BA{ll}
O_p(T^{2(1-b)}\phi_a^{2(k_c-k_e)}\phi_b^{2(k_r-k_c)}) & (a< b) \\
O_p(T^{2(1-b)}) & (a > b)
\EA\right. .
\label{s4:kc}
\EEQ

Combining \eqref{s1:kc}, \eqref{s2:kc}, \eqref{s3:kc}, and \eqref{s4:kc}, we can observe that
\BE
SSR(\tau_c)=\sum_1^Tu_t^2+\left\{\BA{ll}
O_p(T^{2(1-b)}\phi_a^{2(k_c-k_e)}\phi_b^{2(k_r-k_c)}) & (a< b) \\
O_p(T^{2(1-b)}) & (a > b)
\EA\right. .
\label{ssr:kc}
\EE

We next show that \eqref{infssr} holds for $i=1,\ldots,4$. Let us first consider the case where $k\in D_{1T}$. Because
\BEQA
SSR(k/T) 
& = &
\sum_{1}^{k}(y_t-\hat{\phi}_a(k/T) y_{t-1})^2
+\sum_{k+1}^{k_e}(y_t-\hat{\phi}_b(k/T) y_{t-1})^2 \\
& &
+\sum_{k_e+1}^{k_c}(y_t-\hat{\phi}_b(k/T) y_{t-1})^2
+\sum_{k_c+1}^{k_r}(y_t-\hat{\phi}_b(k/T) y_{t-1})^2
+\sum_{k_r+1}^{T}(y_t-\hat{\phi}_b(k/T) y_{t-1})^2 \\
& \geq &
\sum_{k_e+1}^{k_c}(y_t-\hat{\phi}_b(k/T)y_{t-1})^2,
\EEQA
it is sufficient to show that the last term diverges to infinity faster than $SSR(\tau_c)$ uniformly over $k\in D_{1T}$. From the definition of $\hat{\phi}_b$, we can observe that
\BEQA
\sum_{k_e+1}^{k_c}(y_t-\hat{\phi}_b(k/T)y_{t-1})^2
& = &
\sum_{k_e+1}^{k_c}\left(u_t-\varphi_{T} y_{t-1}\right)^2,\mbox{say},
\EEQA
where 
\[
\varphi_T\coloneqq
\frac{\sum_{k+1}^{T}y_{t-1}u_t+(1-\phi_a)\sum_{k+1}^{k_e}y_{t-1}^2+(\phi_b-\phi_a)\sum_{k_c+1}^{k_r}y_{t-1}^2+(1-\phi_a)\sum_{k_r+1}^Ty_{t-1}^2}{\sum_{k+1}^{T}y_{t-1}^2}.
\]
The order of each term of the denominator becomes
\BEQA
\sum_{k+1}^{T}y_{t-1}u_t
& = &
O_p(T)+O_p(T^{(a+1)/2}\phi_a^{(k_c-k_e)})+O_p(T^{(b+1)/2}\phi_a^{(k_c-k_e)})+
\left\{\BA{ll} O_p(T\phi_a^{(k_c-k_e)}\phi_b^{(k_r-k_c)}) & (a < b) \\ O_p(T) & (a > b) \EA\right. \\
& = &
O_p(T^{(\max(a,b)+1)/2}\phi_a^{(k_c-k_e)})
\EEQA
uniformly over $k\in D_{1T}$ from \eqref{lemma:1:yub}, \eqref{lemma:2:yu}, \eqref{lemma:3:yu}, \eqref{lemma:4:yu-a}, and \eqref{lemma:4:yu-b}, 
\[
|1-\phi_a|\sum_{k+1}^{k_e}y_{t-1}^2
\leq
(\phi_a-1)\sum_{1}^{k_e}y_{t-1}^2 \\
=
O_p(T^{2-a})
\]
from \eqref{lemma:1:y2},
\[
|\phi_b-\phi_a|\sum_{k_c+1}^{k_r}y_{t-1}^2
=\left(\frac{c_b}{T^b}+\frac{c_a}{T^a}\right)O_p(T^{b+1}\phi_a^{2(k_c-k_e)})=O_p(T^{b+1-\min(a,b)}\phi_a^{2(k_c-k_e)})
\]
from \eqref{lemma:3:y2}, and
\[
|1-\phi_a|\sum_{k_r+1}^Ty_{t-1}^2
=\frac{c_a}{T^a}\times
\left\{\BA{ll} O_p(T^2\phi_a^{2(k_c-k_e)}\phi_b^{2(k_r-k_c)})=O_p(T^{2-a}\phi_a^{2(k_c-k_e)}\phi_b^{2(k_r-k_c)}) & (a < b) \\
O_p(T^2)=O_p(T^{2-a}) & (a > b)
\EA\right.
\]
from \eqref{lemma:4:y2-a} and \eqref{lemma:4:y2-b}.
Then, in the numerator of $\varphi_T$, the third term dominates the others and it is $O_p(T^{b+1-\min(a,b)}\phi_a^{2(k_c-k_e)})$. Similarly, the order of the denominator of $\varphi_T$ is
\BEQA
\sum_{k+1}^Ty_{t-1}^2
& = &
\sum_{k+1}^{k_e}y_{t-1}^2
+
\sum_{k_e+1}^{k_c}y_{t-1}^2
+
\sum_{k_c+1}^{k_r}y_{t-1}^2
+
\sum_{k_r+1}^{k_e}y_{t-1}^2
 \\
 & = &
 O_p(T^2)+O_p(T^{a+1}\phi_a^{2(k_c-k_e)})+O_p(T^{b+1}\phi_a^{2(k_c-k_e)})
 +\left\{\BA{ll} O_p(T^2\phi_a^{2(k_c-k_e)}\phi_b^{2(k_r-k_c)}) & (a < b) \\ O_p(T^2) & (a > b) \EA\right. \\
 & = &
 O_p(T^{\max(a,b)+1}\phi_a^{2(k_c-k_e)})
 \EEQA
 uniformly over $k\in D_{1T}$ from \eqref{lemma:1:y2}, \eqref{lemma:2:y2}, \eqref{lemma:3:y2}, \eqref{lemma:4:y2-a}, and \eqref{lemma:4:y2-b}. Therefore, we have
 \[
 \varphi_T=O_p(T^{b-\min(a,b)-\max(a,b)})=O_p(T^{-a}),
 \]
uniformly over $k\in D_{1T}$.  Then, we obtain 
\BEQ
\sum_{k_e+1}^{k_c}\left(u_t-\varphi_{T} y_{t-1}\right)^2
& = &
\sum_{k_e+1}^{k_c}u_t^2-2\varphi_T\sum_{k_e+1}^{k_c} y_{t-1}u_t+\varphi_T^2\sum_{k_e+1}^{k_c}y_{t-1}^2 \nonumber  \\
& = &
O_p(T^{1-a}\phi_a^{2(k_c-k_e)}),
\label{ssr:1}
\EEQ
uniformly over $k\in D_{1T}$, where we note that the dominant term is positive and thus \eqref{ssr:1} diverges to infinity. This implies that
\[
\inf_{k\in D_{1T}}SSR(k/T) \geq O_p(T^{1-a}\phi_a^{2(k_c-k_e)}),
\]
which is strictly faster than \eqref{ssr:kc}. Therefore, \eqref{infssr} is proved for $i=1$.

For $k\in D_{2T}$, we have
\BEQA
SSR(k/T) 
& \geq &
\sum_{k_c+1}^{k_r}(y_t-\hat{\phi}_b(k/T) y_{t-1})^2.
\EEQA
Allowing for an abuse of notation, we repeatedly redefine $\varphi_T$ in the following. Similarly, we can observe that
\BEQA
\sum_{k_c+1}^{k_r}(y_t-\hat{\phi}_b(k/T) y_{t-1})^2
& = &
\sum_{k_e+1}^{k_c}\left(u_t-\varphi_{T} y_{t-1}\right)^2,\mbox{say},
\EEQA
where 
\[
\varphi_T\coloneqq
\frac{\sum_{k+1}^{T}y_{t-1}u_t+(\phi_a-\phi_b)\sum_{k+1}^{k_c}y_{t-1}^2+(1-\phi_b)\sum_{k_r+1}^{T}y_{t-1}^2}{\sum_{k+1}^{T}y_{t-1}^2}.
\]
The order of the first term of the numerator becomes
\BEQA
\sum_{k+1}^{T}y_{t-1}u_t
& = &
O_p(T^{(a+1)/2}\phi_a^{(k_c-k_e)})+O_p(T^{(b+1)/2}\phi_a^{(k_c-k_e)})+
\left\{\BA{ll} O_p(T\phi_a^{(k_c-k_e)}\phi_b^{(k_r-k_c)}) & (a < b) \\ O_p(T) & (a > b) \EA\right. \\
& = &
O_p(T^{(\max(a,b)+1)/2}\phi_a^{(k_c-k_e)})
\EEQA
uniformly over $k\in D_{2T}$ from \eqref{lemma:2:yub}, \eqref{lemma:3:yu}, \eqref{lemma:4:yu-a}, and \eqref{lemma:4:yu-b}, while the third term is
\[
(1-\phi_b)\sum_{k_r+1}^{T}y_{t-1}^2
=\frac{c_b}{T^b}\times
\left\{\BA{ll} O_p(T^2\phi_a^{2(k_c-k_e)}\phi_b^{2(k_r-k_c)})=O_p(T^{2-b}\phi_a^{2(k_c-k_e)}\phi_b^{2(k_r-k_c)}) & (a < b) \\
O_p(T^2)=O_p(T^{2-b}) & (a > b)
\EA\right.
\]
from \eqref{lemma:4:y2-a} and \eqref{lemma:4:y2-b}. For the second term, note that
\[
\sum_{k_c-M_T+1}^{k_c}y_{t-1}^2 \leq\sum_{k+1}^{k_c}y_{t-1}^2\leq \sum_{k_e+1}^{k_c}y_{t-1}^2
\]
over $k\in D_{2T}$; thus, using \eqref{lemma:2:y2} and \eqref{lemma:2:y2c}, the following relation holds.
\BE
O_p(T^{1-\min(a,b)}M_T\phi_a^{2(k_c-k_e)})
\leq
(\phi_a-\phi_b)\sum_{k+1}^{k_c}y_{t-1}^2
\leq
O_p(T^{a+1-\min(a,b)}\phi_a^{2(k_c-k_e)}).
\label{varphi:num:d2}
\EE
Thus, in the numerator of $\varphi_T$, the second term, which is always positive, dominates the others, and its order is bounded from below and above, as in \eqref{varphi:num:d2}. On the other hand, the order of the denominator of $\varphi_T$ is bounded below by
\BE
\sum_{k+1}^Ty_{t-1}^2
\geq
\sum_{k_c+1}^{k_r}y_{t-1}^2
=O_p(T^{b+1}\phi_a^{2(k_c-k_e)}),
\label{varphi:den:d2a}
\EE
from \eqref{lemma:3:y2} and bounded above by
\BEQ
\sum_{k+1}^Ty_{t-1}^2
& \leq &
\sum_{k_e+1}^{k_c}y_{t-1}^2
+
\sum_{k_c+1}^{k_r}y_{t-1}^2
+
\sum_{k_r+1}^{k_e}y_{t-1}^2
\nonumber \\
 & = &
O_p(T^{a+1}\phi_a^{2(k_c-k_e)})+O_p(T^{b+1}\phi_a^{2(k_c-k_e)})
 +\left\{\BA{ll} O_p(T^2\phi_a^{2(k_c-k_e)}\phi_b^{2(k_r-k_c)}) & (a < b) \\ O_p(T^2) & (a\geq b) \EA\right. \nonumber \\
 & = &
 O_p(T^{\max(a,b)+1}\phi_a^{2(k_c-k_e)})
\label{varphi:den:d2b}
\EEQ
from \eqref{lemma:2:y2}, \eqref{lemma:3:y2}, \eqref{lemma:4:y2-a}, and \eqref{lemma:4:y2-b}. By combining \eqref{varphi:num:d2}, \eqref{varphi:den:d2a}, and \eqref{varphi:den:d2b}, we can observe that $\varphi_T$ is asymptotically positive and
\BEQ
\lefteqn{
O_p(T^{-(a+b)}M_T)=\frac{O_p(T^{1-\min(a,b)}M_T\phi_a^{2(k_c-k_e)})}{O_p(T^{\max(a,b)+1}\phi_a^{2(k_c-k_e)})}
} \nonumber  \\
& \leq & 
\varphi_T
\leq 
\frac{O_p(T^{a+1-\min(a,b)}\phi_a^{2(k_c-k_e)})}{O_p(T^{b+1}\phi_a^{2(k_c-k_e)})}
=
O_p(T^{a-b-\min(a,b)}).
\label{varphi:bound2}
\EEQ
Using this result, we obtain 
\BEQ
\sum_{k_c+1}^{k_r}\left(u_t-\varphi_{T} y_{t-1}\right)^2
& = &
\sum_{k_c+1}^{k_r}u_t^2-2\varphi_T\sum_{k_c+1}^{k_r} y_{t-1}u_t+\varphi_T^2\sum_{k_c+1}^{k_r}y_{t-1}^2 \nonumber  \\
& \geq &
O_p(T^{-2(a+b)}M_T^2)O_p(T^{b+1}\phi_a^{2(k_c-k_e)}) \nonumber \\
& = &
O_p(T^{1-2a-b}M_T^2\phi_a^{2(k_c-k_e)}),
\label{ssr:2}
\EEQ
uniformly over $k\in D_{2T}$ from \eqref{lemma:3:y2}, \eqref{lemma:3:yu}, and \eqref{varphi:bound2}, where the inequality holds because the third term dominates the others in the first equality. This implies that
\[
\inf_{k\in D_{2T}}SSR(k/T) \geq O_p(T^{1-2a-b}M_T^2\phi_a^{2(k_c-k_e)}),
\]
which is strictly faster than \eqref{ssr:kc}. Therefore, \eqref{infssr} is proved for $i=2$.

For $k\in D_{3T}$, we have
\BEQA
SSR(k/T) 
& \geq &
\sum_{k_e+1}^{k_c}(y_t-\hat{\phi}_a(k/T) y_{t-1})^2.
\EEQA
Then, we can observe that
\BEQA
\sum_{k_e+1}^{k_c}(y_t-\hat{\phi}_a(k/T) y_{t-1})^2
& = &
\sum_{k_e+1}^{k_c}\left(u_t-\varphi_{T} y_{t-1}\right)^2,\mbox{ say},
\EEQA
where 
\[
\varphi_T\coloneqq
\frac{\sum_{1}^{k}y_{t-1}u_t+(1-\phi_a)\sum_{1}^{k_e}y_{t-1}^2+(\phi_b-\phi_a)\sum_{k_c+1}^{k}y_{t-1}^2}{\sum_{1}^{k}y_{t-1}^2}.
\]
The order of the first term of the numerator becomes
\BEQA
\sum_{1}^{k}y_{t-1}u_t
& = &
O_p(T)+O_p(T^{(a+1)/2}\phi_a^{(k_c-k_e)})+O_p(T^{(b+1)/2}\phi_a^{(k_c-k_e)}) \\
& = &
O_p(T^{(\max(a,b)+1)/2}\phi_a^{(k_c-k_e)})
\EEQA
uniformly over $k\in D_{3T}$ from \eqref{lemma:1:yu}, \eqref{lemma:2:yu}, and \eqref{lemma:3:yua}, while the second term is
\[
(1-\phi_a)\sum_{1}^{k_e}y_{t-1}^2
=O_p(T^{2-a})
\]
from \eqref{lemma:1:y2}. For the third term, note that
\[
\sum_{k_c+1}^{k_c+M_T}y_{t-1}^2 \leq\sum_{k_c+1}^{k}y_{t-1}^2\leq \sum_{k_c+1}^{k_r}y_{t-1}^2
\]
over $k\in D_{3T}$; thus, using \eqref{lemma:3:y2} and \eqref{lemma:3:y2c}, the following relation holds.
\BE
O_p(T^{1-\min(a,b)}M_T\phi_a^{2(k_c-k_e)})
\leq
|\phi_b-\phi_a|\sum_{k_c+1}^{k}y_{t-1}^2
\leq
O_p(T^{b+1-\min(a,b)}\phi_a^{2(k_c-k_e)}).
\label{varphi:num:d3}
\EE
Thus, in the numerator of $\varphi_T$, the third term, which is always negative, dominates the others, and its order is bounded from below and above, as in \eqref{varphi:num:d3}. On the other hand, the order of the denominator of $\varphi_T$ is bounded below by
\BE
\sum_{1}^{k}y_{t-1}^2
\geq
\sum_{k_e+1}^{k_c}y_{t-1}^2
=O_p(T^{a+1}\phi_a^{2(k_c-k_e)}),
\label{varphi:den:d3a}
\EE
from \eqref{lemma:2:y2} and bounded above by
\BEQ
\sum_{1}^{k}y_{t-1}^2
& \leq &
\sum_{1}^{k_e}y_{t-1}^2
+
\sum_{k_e+1}^{k_c}y_{t-1}^2
+
\sum_{k_c+1}^{k_r}y_{t-1}^2
\nonumber \\
 & = &
 O_p(T^{\max(a,b)+1}\phi_a^{2(k_c-k_e)})
\label{varphi:den:d3b}
\EEQ
from \eqref{lemma:1:y2}, \eqref{lemma:2:y2}, and \eqref{lemma:3:y2}. By combining \eqref{varphi:num:d3}, \eqref{varphi:den:d3a}, and \eqref{varphi:den:d3b}, we can observe that $\varphi_T$ is asymptotically negative and
\BEQ
\lefteqn{
O_p(T^{-(a+b)}M_T)=\frac{O_p(T^{1-\min(a,b)}M_T\phi_a^{2(k_c-k_e)})}{O_p(T^{\max(a,b)+1}\phi_a^{2(k_c-k_e)})}
} \nonumber  \\
& \leq & 
|\varphi_T|
\leq 
\frac{O_p(T^{b+1-\min(a,b)}\phi_a^{2(k_c-k_e)})}{O_p(T^{a+1}\phi_a^{2(k_c-k_e)})}
=
O_p(T^{b-a-\min(a,b)}).
\label{varphi:bound3}
\EEQ
Using this result, we obtain 
\BEQ
\sum_{k_e+1}^{k_c}\left(u_t-\varphi_{T} y_{t-1}\right)^2
& = &
\sum_{k_e+1}^{k_c}u_t^2-2\varphi_T\sum_{k_e+1}^{k_c} y_{t-1}u_t+\varphi_T^2\sum_{k_e+1}^{k_c}y_{t-1}^2 \nonumber  \\
& \geq &
O_p(T^{-2(a+b)}M_T^2)O_p(T^{a+1}\phi_a^{2(k_c-k_e)}) \nonumber \\
& = &
O_p(T^{1-a-2b}M_T^2\phi_a^{2(k_c-k_e)}),
\label{ssr:3}
\EEQ
uniformly over $k\in D_{3T}$ from \eqref{lemma:2:y2}, \eqref{lemma:2:yu}, and \eqref{varphi:bound3}, where the inequality holds because the third term asymptotically  dominates the others. This implies that
\[
\inf_{k\in D_{3T}}SSR(k/T) \geq O_p(T^{1-a-2b}M_T^2\phi_a^{2(k_c-k_e)}),
\]
which is strictly faster than \eqref{ssr:kc}. Therefore, \eqref{infssr} is proved for $i=3$.

For $k\in D_{4T}$, we have
\BEQA
SSR(k/T) 
& \geq &
\sum_{k_e+1}^{k_c}(y_t-\hat{\phi}_a(k/T) y_{t-1})^2.
\EEQA
We can see that
\BEQA
\sum_{k_e+1}^{k_c}(y_t-\hat{\phi}_a(k/T) y_{t-1})^2
& = &
\sum_{k_e+1}^{k_c}\left(u_t-\varphi_{T} y_{t-1}\right)^2,\mbox{ say},
\EEQA
where 
\[
\varphi_T\coloneqq
\frac{\sum_{1}^{k}y_{t-1}u_t+(1-\phi_a)\sum_{1}^{k_e}y_{t-1}^2+(\phi_b-\phi_a)\sum_{k_c+1}^{k_r}y_{t-1}^2+(1-\phi_a)\sum_{k_r+1}^{k}y_{t-1}^2}{\sum_{1}^{k}y_{t-1}^2}.
\]
The order of each term of the numerator becomes
\BEQA
\sum_{1}^{k}y_{t-1}u_t
& = &
O_p(T)+O_p(T^{(a+1)/2}\phi_a^{(k_c-k_e)})+O_p(T^{(b+1)/2}\phi_a^{(k_c-k_e)}) \\
&  &
+\left\{\BA{ll} O_p(T\phi_a^{(k_r-k_c)}\phi_b^{(k_r-k_c)}) & (a<b)\\ O_p(T) & (a > b) \EA\right\}=O_p(T^{(\max(a,b)+1)/2}\phi_a^{k_c-k_e})
\EEQA
uniformly over $k\in D_{4T}$ from \eqref{lemma:1:yu}, \eqref{lemma:2:yu}, \eqref{lemma:3:yu}, \eqref{lemma:4:yua-a}, and \eqref{lemma:4:yua-b},
\[
(1-\phi_a)\sum_{1}^{k_e}y_{t-1}^2
=O_p(T^{2-a})
\]
from \eqref{lemma:1:y2},
\[
(\phi_b-\phi_a)\sum_{k_c+1}^{k_r}y_{t-1}^2=O_p(T^{b+1-\min(a,b)}\phi_a^{2(k_c-k_e)})
\]
from \eqref{lemma:3:y2}, and
\[
|1-\phi_a|\sum_{k_r+1}^{k}y_{t-1}^2\leq |1-\phi_a|\sum_{k_r+1}^{T}y_{t-1}^2=\left\{\BA{ll} O_p(T^{2-a}\phi_a^{2(k_c-k_e)}\phi_b^{2(k_r-k_c)}) & (a < b) \\ O_p(T^{2-a}) & (a > b) \EA\right.
\]
from \eqref{lemma:4:y2-a} and \eqref{lemma:4:y2-b}; thus, the third term of the numerator of $\varphi_T$, which takes negative values, dominates the others, and it is $O_p(T^{b+1-\min(a,b)}\phi_a^{2(k_c-k_e)})$. On the other hand, the order of the denominator is given by
\BEQA
\sum_1^ky_{t-1}^2
& = &
O_p(T^2)+O_p(T^{a+1}\phi_a^{2(k_c-k_e)})+O_p(T^{b+1}\phi_a^{2(k_c-k_e)}) \\
&  &
+\left\{\BA{ll} O_p(T^2\phi_a^{2(k_c-k_r)}\phi_b^{2(k_r-k_c)}) & (a<b) \\ O_p(T^2) & (a > b)\EA\right\}
=O_p(T^{\max(a,b)+1}\phi_a^{2(k_c-k_e)})
\EEQA
from \eqref{lemma:1:y2}, \eqref{lemma:2:y2}, \eqref{lemma:3:y2}, \eqref{lemma:4:y2-a}, and \eqref{lemma:4:y2-b}. Then, we have
\[
\varphi_T=O_p(T^{b-\min(a,b)-\max(a,b)})=O_p(T^{-a})
\]
uniformly over $k\in D_{4T}$. Therefore, we obtain
\BEQ
\sum_{k_e+1}^{k_c}\left(u_t-\varphi_{T} y_{t-1}\right)^2
& = &
\sum_{k_e+1}^{k_c}u_t^2-2\varphi_T\sum_{k_e+1}^{k_c} y_{t-1}u_t+\varphi_T^2\sum_{k_e+1}^{k_c}y_{t-1}^2 \nonumber  \\
& = &
O_p(T)+O_p(T^{-a})O_p(T^{(a+1)/2}\phi_a^{(k_c-k_e)})+O_p(T^{-2a})O_p(T^{a+1}\phi_a^{2(k_c-k_e})) \nonumber \\
& = &
O_p(T^{1-a}\phi_a^{2(k_c-k_e)}),
\label{ssr:4}
\EEQ
uniformly over $k\in D_{4T}$ from \eqref{lemma:2:y2} and \eqref{lemma:2:yu}. This implies that
\[
\inf_{k\in D_{4T}}SSR(k/T) \geq O_p(T^{1-a}\phi_a^{2(k_c-k_e)}),
\]
which is strictly faster than \eqref{ssr:kc}. Therefore, \eqref{infssr} is proved for $i=4$.

As we established $\hat{k}-k_c=O_p(1)$, there exists a positive integer $M$ for any given $\epsilon$ such that $P(|\hat{k}-k_c| > M) < \epsilon$ for all sufficiently large $T$. In the same manner as \citet{pang2021estimating}, we can observe that
\BEQA
P(\hat{k}\neq k_c)
& = &
P(|\hat{k}-k_c|>M)+P(|\hat{k}-k_c|\leq M,\;\hat{k}\neq k_c) \\
&\leq &
\epsilon+\sum_{m=1}^MP(SSR(\tau_c-m/T)-SSR(\tau_c)<0)+\sum_{m=1}^MP(SSR(\tau_c+m/T)-SSR(\tau_c)<0) \\
& = &
\epsilon+
\sum_{m=1}^MP\left(\frac{1}{T(\phi_a-\phi_b)^2\phi_a^{2(k_c-k_e)}}\left\{SSR(\tau_c-m/T)-SSR(\tau_c)\right\}<0\right) \\
& &
+\sum_{m=1}^MP\left(\frac{1}{T(\phi_a-\phi_b)^2\phi_a^{2(k_c-k_e)}}\left\{SSR(\tau_c+m/T)-SSR(\tau_c)\right\}<0\right).
\EEQA
 We show that the probabilities on the right-hand side go to 0. Note that the diverging order of $SSR(k_c)$ has been already obtained in \eqref{ssr:kc}, which is strictly smaller than $T(\phi_a-\phi_b)^2\phi_a^{2(k_c-k_r)}$; thus, we can ignore $SSR(\tau_c)$ divided by the scaling term.
 
 We first focus on $SSR(\tau_c-m/T)$. Let $k=(\tau_c-m/T)T=k_c-m$. Then, allowing an abuse of notation, $SSR(\tau_c-m/T)$ is decomposed into
 \BEQA
 SSR(\tau_c-m/T)
 & = & 
 \sum_{1}^{k_e}(y_t-\hat{\phi}_ay_{t-1})^2+ \sum_{k_e+1}^{k}(y_t-\hat{\phi}_ay_{t-1})^2 \\
 & &
  +\sum_{k+1}^{k_c}(y_t-\hat{\phi}_b y_{t-1})^2+ \sum_{k_c+1}^{k_r}(y_t-\hat{\phi}_b y_{t-1})^2+ \sum_{k_r+1}^{T}(y_t-\hat{\phi}_by_{t-1})^2 \\
  & = & 
  S_{1T}+S_{2T}+S_{3T}+S_{4T}+S_{5T},\mbox{ say}.
  \EEQA
  We investigate each of $S_{it}$ for $i=1,\ldots,5$.
  
  For $S_{1T}$, we can observe that
  \[
  S_{1T}
  =
  \sum_{1}^{k_e}(u_t+ y_{t-1}-\hat{\phi}_a y_{t-1})^2=
   \sum_{1}^{k_e}(u_t-\varphi_{1T} y_{t-1})^2,
  \]
  \[
  \mbox{where}\quad
  \varphi_{1T}=\frac{\sum_{1}^{k}y_{t-1}u_t+(\phi_a-1)\sum_{k_e+1}^{k}y_{t-1}^2}{\sum_{1}^k y_{t-1}^2}=O_p(T^{-a}),
  \]
which is obtained by using \eqref{lemma:1:y2}, \eqref{lemma:1:yu}, \eqref{lemma:2:y2}, and \eqref{lemma:2:yu} because $m$ is a fixed integer value ($k=k_c-m$). Then, we obtain
 \[
    S_{1T}=\sum_{1}^{k_e}(u_t^2-2\varphi_{1T}y_{t-1}u_t+\varphi_{1T}^2y_{t-1}^2)=O_p(T)+O_p(T^{1-a})+O_p(T^{2(1-a)})
 \]
 from \eqref{lemma:1:y2} and \eqref{lemma:1:yu}, which is smaller order than $T(\phi_a-\phi_b)^2\phi_a^{2(k_c-k_r)}$; thus, $S_{1T}$ divided by the scaling term is negligible.
  
  For $S_{2T}$, we can observe that
  \[
  S_{2T}
  =
  \sum_{k_e+1}^{k}(u_t+\phi_a y_{t-1}-\hat{\phi}_a y_{t-1})^2=
   \sum_{k_e+1}^{k}(u_t-\varphi_{2T} y_{t-1})^2,
  \]
  \[
  \mbox{where}\quad
  \varphi_{2T}=\frac{\sum_{1}^{k}y_{t-1}u_t+(1-\phi_a)\sum_{1}^{k_e}y_{t-1}^2}{\sum_{1}^k y_{t-1}^2}=O_p(T^{-(a+1)/2}\phi_a^{-(k_c-k_e)}),
  \]
which is obtained by using \eqref{lemma:1:y2}, \eqref{lemma:1:yu}, \eqref{lemma:2:y2}, and \eqref{lemma:2:yu}. Then, we obtain
 \[
    S_{2T}=\sum_{k_e+1}^{k}(u_t^2-2\varphi_{2T}y_{t-1}u_t+\varphi_{2T}^2y_{t-1}^2)=O_p(T)+O_p(1)+O_p(1)
 \]
 from \eqref{lemma:2:y2} and \eqref{lemma:2:yu}, which is smaller order than $T(\phi_a-\phi_b)^2\phi_a^{2(k_c-k_e)}$; thus, $S_{2T}$ divided by the scaling term is negligible.

For $S_{3T}$, 
  \[
  S_{3T}
  =
  \sum_{k+1}^{k_c}(u_t+\phi_ay_{t-1}-\hat{\phi}_by_{t-1})^2=
   \sum_{k+1}^{k_c}(u_t-\varphi_{3T} y_{t-1})^2,
  \]
  \[
  \mbox{where}\quad
  \varphi_{3T}=\frac{\sum_{k+1}^{T}y_{t-1}u_t+(\phi_b-\phi_a)\sum_{k_c+1}^{k_r}y_{t-1}^2+(1-\phi_a)\sum_{k_r+1}^{T}y_{t-1}^2}{\sum_{k+1}^T y_{t-1}^2}.
  \]
As $k=k_c-m$ and $m$ is a fixed integer, we can observe from \eqref{lemma:2:y}, \eqref{lemma:2:yub},  \eqref{lemma:3:y2}, \eqref{lemma:3:yu},  \eqref{lemma:4:y2-a}, \eqref{lemma:4:yu-a}, \eqref{lemma:4:y2-b}, and \eqref{lemma:4:yu-b} that the second term in the numerator dominates the others, whereas
\[
\sum_{k+1}^Ty_{t-1}^2\sim_a \sum_{k_c+1}^{k_r}y_{t-1}^2.
\]
Therefore, we can observe that
\[
\varphi_{3T}\sim_a \frac{(\phi_b-\phi_a)\sum_{k_c+1}^{k_r}y_{t-1}^2}{\sum_{k_c+1}^{k_r}y_{t-1}^2}=(\phi_b-\phi_a).
\]
Using this result, we have
\BEQA
\frac{1}{T(\phi_a-\phi_b)^2\phi_a^{2(k_c-k_e)}}S_{3T}
& = &
\frac{1}{T(\phi_a-\phi_b)^2\phi_a^{2(k_c-k_e)}}\left(\sum_{k+1}^{k_c}u_{t}^2-2\varphi_{3T}\sum_{k+1}^{k_c}y_{t-1}u_t+\varphi_{3T}^2\sum_{k+1}^{k_c}y_{t-1}^2\right) \\
& \sim_a&
\frac{1}{T(\phi_a-\phi_b)^2\phi_a^{2(k_c-k_e)}}\varphi_{3T}^2\sum_{k_c-m+1}^{k_c}y_{t-1}^2 \\
& \sim_a &
\frac{1}{T(\phi_a-\phi_b)^2\phi_a^{2(k_c-k_e)}}(\phi_b-\phi_a)^2 m \phi_a^{2(k_c-k_e)}y_{k_e}^2 \\
& \Rightarrow &
m (W^{\kappa}(\tau_e))^2
\EEQA
from \eqref{lemma:1:y} and \eqref{lemma:2:y}.

For $S_{4T}$, 
  \[
  S_{4T}
  =
  \sum_{k_c+1}^{k_r}(u_t+\phi_by_{t-1}-\hat{\phi}_by_{t-1})^2=
   \sum_{1}^{k_e}(u_t-\varphi_{4T} y_{t-1})^2,
  \]
  \[
  \mbox{where}\quad
  \varphi_{3T}=\frac{\sum_{k+1}^{T}y_{t-1}u_t+(\phi_a-\phi_b)\sum_{k+1}^{k_c}y_{t-1}^2+(1-\phi_b)\sum_{k_r+1}^{T}y_{t-1}^2}{\sum_{k+1}^T y_{t-1}^2}.
  \]
From \eqref{lemma:2:y}, \eqref{lemma:2:yub}, \eqref{lemma:3:yu}, \eqref{lemma:4:y2-a}, \eqref{lemma:4:yu-a}, \eqref{lemma:4:y2-b}, and \eqref{lemma:4:yu-b}, we can see that the second term of the numerator is the dominant term, which is $O_p(T^{1-\min(a,b)}\phi_a^{2(k_c-k_e)})$, while the denominator is $O_p(T^{b+1}\phi_a^{2(k_c-k_e)})$. Then,  we have
\[
\varphi_{4T}=O_p(T^{-b-\min(a,b)}).
\]
Thus,
 \BEQA
 S_{4T}
 & = &
 \sum_{k_c+1}^{k_r}(u_t^2-2\varphi_{4T}y_{t-1}u_t+\varphi_{4T}^2y_{t-1}^2) \\
 & = &
 O_p(T)+O_p(T^{(1-b)/2-\min(a,b)}\phi_a^{k_c-k_e})+O_p(T^{1-b-2\min(a,b)}\phi_a^{2(k_c-k_e)})
 \EEQA
 from \eqref{lemma:3:y2} and \eqref{lemma:3:yu}, which is smaller order than $T(\phi_a-\phi_b)^2\phi_a^{2(k_c-k_e)}$; thus, $S_{4T}$ divided by the scaling term is negligible.

For $S_{5T}$, 
  \[
  S_{5T}
  =
  \sum_{k_r+1}^{T}(u_t+y_{t-1}-\hat{\phi}_by_{t-1})^2=
   \sum_{1}^{k_e}(u_t-\varphi_{5T} y_{t-1})^2,
  \]
  \[
  \mbox{where}\quad
  \varphi_{5T}=\frac{\sum_{k+1}^{T}y_{t-1}u_t+(\phi_a-1)\sum_{k+1}^{k_c}y_{t-1}^2+(\phi_b-1)\sum_{k_c+1}^{k_r}y_{t-1}^2}{\sum_{k+1}^T y_{t-1}^2}.
  \]
From \eqref{lemma:2:y}, \eqref{lemma:2:yub}, \eqref{lemma:3:y2}, \eqref{lemma:3:yu}, \eqref{lemma:4:y2-a}, \eqref{lemma:4:yu-a}, \eqref{lemma:4:y2-b}, and \eqref{lemma:4:yu-b}, we can observe that the third term of the numerator is the dominant term, which is $O_p(T\phi_a^{2(k_c-k_e)})$, while the denominator is $O_p(T^{b+1}\phi_a^{2(k_c-k_e)})$. Then,  we have $\varphi_{5T}=O_p(T^{-b})$ and thus,
 \BEQA
 S_{5T}
 & = &
 \sum_{k_r+1}^{T}(u_t^2-2\varphi_{5T}y_{t-1}u_t+\varphi_{5T}^2y_{t-1}^2) \\
 & = &
 O_p(T)+\left\{\BA{ll}
 O_p(T^{1-b}\phi_a^{(k_c-k_e)}\phi_b^{(k_r-k_c)})+O_p(T^{2(1-b)}\phi_a^{2(k_c-k_e)}\phi_b^{2(k_r-k_c)})  & (a<b) \\
 O_p(T^{1-b})+O_p(T^{2(1-b)}) & (a > b) \EA\right.
 \EEQA
 from \eqref{lemma:4:y2-a}, \eqref{lemma:4:yu-a}, \eqref{lemma:4:y2-b}, and \eqref{lemma:4:yu-b}, which is smaller order than $T(\phi_a-\phi_b)^2\phi_a^{2(k_c-k_e)}$; thus, $S_{5T}$ divided by the scaling term is negligible.

To summarize the result, we have, for a given positive integer value of $m$,
\[
\frac{1}{T(\phi_a-\phi_b)^2\phi_a^{2(k_c-k_e)}}\left(SSR(\tau_c-m/T)-SSR(\tau_c)\right)
\Rightarrow
m(W^{\kappa}(\tau_e))^2,
\]
which implies that 
\[
P\left(\frac{1}{T(\phi_a-\phi_b)^2\phi_a^{2(k_c-k_e)}}\left\{SSR(\tau_c-m/T)-SSR(\tau_c)\right\}<0\right) \to 0.
\]

A similar result is obtained for $SSR(\tau_c+m/T)$. Let $k=(\tau_c+m/T)T=k_c+m$. Then, in exactly the same manner, we can show that
\[
\frac{1}{T(\phi_a-\phi_b)^2\phi_a^{2(k_c-k_e)}}\left(SSR(\tau_c+m/T)-SSR(\tau_c)\right)
\Rightarrow
m(W^{\kappa}(\tau_e))^2.
\]
Thus, for a given positive integer value of $m$, 
\[
P\left(\frac{1}{T(\phi_a-\phi_b)^2\phi_a^{2(k_c-k_e)}}\left\{SSR(\tau_c+m/T)-SSR(\tau_c)\right\}<0\right) \to 0.\blacksquare
\]

\noindent
{\bf Proof of Theorem \ref{theorem:kr}}: (i) We first show that $\hat{k}_r-k_r=O_p(1)$. That is, 
\BE
\lim_{T\to \infty}\inf P\left(|\hat{k}_r-k_r| \geq M_T \right) \geq \alpha
\label{consistency:fraction:r}
\EE
does not hold for any $0 < \alpha\leq 1$ where $M_T\to\infty$ and $M_T/T\to 0$ For this purpose, we decompose the range of the permissible break dates into three sub-samples such that $E_{0T}=\{k:\;k_r-M_T+1\leq k \leq k_r+M_T\}$, $E_{1T}=\{k:\;\underline{k}_r+1\leq k \leq k_r-M_T\}$, and  $E_{2T}=\{k:\;k_r+M_T+1\leq k \leq \overline{k}_r\}$, where $\underline{k}_r\coloneqq \underline{\tau}_rT$ and $\overline{k}_r\coloneqq \overline{\tau}_rT$.  Because 
\BEQA
P\left(|\hat{k}_r-k_r| \geq M_T\right)
& = &
P\left(\inf_{k\in E_{1T}\cup E_{2T}}SSR_2(k/T) < \inf_{k\in E_{0T}} SSR_2(k/T) \right) \\
& \leq &
P\left(\inf_{k\in E_{1T}\cup E_{2T}}SSR_2(k/T) < SSR_2(k_r/T) \right) \\
& \leq &
\sum _{i=1}^2P\left(\inf_{k\in E_{iT}}SSR_2(k/T) < SSR_2(\tau_r)  \right),
\EEQA
we will show that, for $i=1$ and $2$,
\BE
P\left(\inf_{k\in E_{iT}}SSR_2(k/T) < SSR_2(\tau_r)  \right)\to 0.
\label{infssr:r}
\EE

We first consider the case where $k\in E_{1T}$. Following (B.2) in \citet{chong2001structural}, we have
\BEQ
SSR_2(k/T)-SSR_2(\tau_r)
& = &
2(1-\phi_b)\left(\frac{\sum_{k+1}^{k_r}{y_{t-1}^2}\sum_{k_r+1}^{T}{y_{t-1}u_t}}{\sum_{k+1}^{T}{y_{t-1}^2}}-\frac{\sum_{k_r+1}^{T}{y_{t-1}^2}\sum_{k+1}^{k_r}{y_{t-1}u_t}}{\sum_{k+1}^{T}{y_{t-1}^2}}\right) \nonumber \\
& &
+(1-\phi_b)^2\frac{\sum_{k_r+1}^{T}{y_{t-1}^2}\sum_{k+1}^{k_r}{y_{t-1}^2}}{\sum_{k+1}^{T}{y_{t-1}^2}}+\Lambda_T(k/T),
\label{Chong:B2}
\EEQ
where
\begin{equation}
    \Lambda_T(k/T)=\frac{\left(\sum_{k_c+1}^{k_r}{y_{t-1}u_t}\right)^2}{\sum_{k_c+1}^{k_r}{y_{t-1}^2}}-\frac{\left(\sum_{k_c+1}^{k}{y_{t-1}u_t}\right)^2}{\sum_{k_c+1}^{k}{y_{t-1}^2}}+\frac{\left(\sum_{k_r+1}^{T}{y_{t-1}u_t}\right)^2}{\sum_{k_r+1}^{T}{y_{t-1}^2}}-\frac{\left(\sum_{k+1}^{T}{y_{t-1}u_t}\right)^2}{\sum_{k+1}^{T}{y_{t-1}^2}}.
\label{Chong:Lambda}
\end{equation}
Thus, as explained in Appendix C on page 122 of \citet{chong2001structural}, it is sufficient to show that $A_1=o_p(\lambda)$, $A_2=o_p(\lambda)$, and $A_3=o_p(\lambda^2)$, where $\lambda=1-\phi_b=c_b/T^b$,
\[
A_1=\frac{\sum_{k_r+1}^{T}{y_{t-1}u_t}}{\sum_{k_r+1}^{T}{y_{t-1}^2}},\quad
A_2=\max_{k\in E_{1T}}\frac{\sum_{k+1}^{k_r}{y_{t-1}u_t}}{\sum_{k+1}^{k_r}{y_{t-1}^2}},\quad
A_3=\max_{k\in E_{1T}}\left|\frac{\sum_{k+1}^{T}{y_{t-1}^2}}{\sum_{k_r+1}^{T}{y_{t-1}^2}\sum_{k+1}^{k_r}{y_{t-1}^2}}\Lambda_T (k/T)\right|.
\]
Using \eqref{lemma:4:y2-a} and \eqref{lemma:4:yu-a}, we have, for $a < b$,
\[
A_1=O_p\left(\frac{1}{T\phi_a^{(k_c-k_e)}\phi_b^{(k_r-k_c)}}\right)=o_p\left(\frac{1}{T^b}\right),
\]
while using \eqref{lemma:3:y2d} and \eqref{lemma:3:yub},
\[
|A_2|\leq \frac{\max_{k\in E_{1T}}|\sum_{k+1}^{k_r}y_{t-1}u_t|}{\sum_{k_r-M_T+1}^{k_r}y_{t-1}^2}=O_p\left(\frac{1}{T^{(1-b)/2}M_T\phi_a^{(k_c-k_e)}\phi_b^{2(k_r-k_c)}}\right)=o_p\left(\frac{1}{T^b}\right).
\]
To derive the order of $A_3$, we expand it, as given in Appendix C of \citet{chong2001structural},
\begin{eqnarray}
A_3
&\leq&
\left(\frac{1}{\sum_{k_r+1}^{T}{y_{t-1}^2}}+\frac{1}{\sum_{k_r-M_T+1}^{kr}{y_{t-1}^2}}\right) \nonumber \\
&&
\times\left(\max_{k\in E_{1T}}\left|\frac{\left(\sum_{k_c+1}^{k_r}{y_{t-1}u_t}\right)^2}{\sum_{k_c+1}^{k_r}{y_{t-1}^2}}-\frac{\left(\sum_{k_c+1}^{k}{y_{t-1}u_t}\right)^2}{\sum_{k_c+1}^{k}{y_{t-1}^2}}\right|\right. \nonumber \\
&&
\left.+\max_{k\in E_{1T}}\left|\frac{\left(\sum_{k_r+1}^{T}{y_{t-1}u_t}\right)^2}{\sum_{k_r+1}^{T}{y_{t-1}^2}}-\frac{\left(\sum_{k+1}^{T}{y_{t-1}u_t}\right)^2}{\sum_{k+1}^{T}{y_{t-1}^2}}\right|\right) \label{a3:ineq} \\
&=&
\left(O_p\left(\frac{1}{T^2\phi_a^{2(k_c-k_e)}\phi_b^{2(k_r-k_c)}}\right)+O_p\left(\frac{1}{TM_T\phi_a^{2(k_c-k_e)}\phi_b^{2(k_r-k_c)}}\right)\right)  \nonumber \\
&&
\times
\left(O_p(1)+O_p(1)+O_p(1)+O_p\left(\frac{1}{T^{1-b}\phi_b^{2(k_r-k_c)}}\right)\right) \nonumber  \\
& = &
O_p\left(\frac{1}{T^{2-b}M_T\phi_a^{2(k_c-k_e)}\phi_b^{4(k_r-k_c)}}\right)=o_p\left(\frac{1}{T^{2b}}\right) \nonumber 
\end{eqnarray}
using Lemmas \ref{lemma:3} and \ref{lemma:4}, where the second terms in the two maxima are evaluated as
\[
\max_{k\in E_{1T}}\frac{\left(\sum_{k_c+1}^{k}{y_{t-1}u_t}\right)^2}{\sum_{k_c+1}^{k}{y_{t-1}^2}}
\leq 
\frac{\max_{k\in E_{1T}}\left(\sum_{k_c+1}^{k}{y_{t-1}u_t}\right)^2}{\sum_{k_c+1}^{\underline{k}_r}y_{t-1}^2}=\frac{O_p(T^{b+1}\phi_a^{2(k_c-k_e)})}{O_p(T^{b+1}\phi_a^{2(k_c-k_e)})}=O_p(1)
\]
by \eqref{lemma:3:y-1} and \eqref{lemma:3:yua}, and
\BEQA
\max_{k\in E_{1T}}\frac{\left(\sum_{k+1}^{T}{y_{t-1}u_t}\right)^2}{\sum_{k+1}^{T}{y_{t-1}^2}}
& \leq &
\frac{\left(\max_{k\in E_{1T}}|\sum_{k+1}^{k_r}{y_{t-1}u_t}|+|\sum_{k_r+1}^{T}{y_{t-1}u_t}|\right)^2}{\sum_{k_r+1}^{T}{y_{t-1}^2}} \\
& = &
\frac{O_p(T^{b+1}\phi_a^{2(k_c-k_e)})}{O_p(T^2\phi_a^{2(k_c-k_e)}\phi_b^{2(k_r-k_c)})}=O_p\left(\frac{1}{T^{1-b}\phi_b^{2(k_r-k_c)}}\right)
\EEQA
by \eqref{lemma:3:yub}, \eqref{lemma:4:y2-a}, and \eqref{lemma:4:yu-a}, respectively. Then, we obtained \eqref{infssr:r} for $i=1$.

In the case of $k\in E_{2T}$, we have, following (B.4) in \citet{chong2001structural},
\BEQ
SSR_2(k/T)-SSR_2(\tau_r)
& = &
2(1-\phi_b)\left(\frac{\sum_{k_r+1}^{k}{y_{t-1}u_t}\sum_{k_c+1}^{k_r}{y_{t-1}^2}}{\sum_{k_c+1}^{k}{y_{t-1}^2}}-\frac{\sum_{k_c+1}^{k_r}{y_{t-1}u_t}\sum_{k_r+1}^{k}{y_{t-1}^2}}{\sum_{k_c+1}^{k}{y_{t-1}^2}}\right) \nonumber \\
& &
    +(1-\phi_b)^2\frac{\sum_{k_r+1}^{k}{y_{t-1}^2}\sum_{k_c+1}^{k_r}{y_{t-1}^2}}{\sum_{k_c+1}^{k}{y_{t-1}^2}}+\Lambda_T(k/T). \label{Chong:B4}
\EEQ
Thus, as explained in Appendix C on page 124 of Chong (2001), it is sufficient to show that $A_4=o_p(\lambda)$, $A_5=o_p(\lambda)$, and $A_6=o_p(\lambda^2)$, where $\lambda=1-\phi_b=c_b/T^b$,
\[
A_4=\frac{\sum_{k_c+1}^{k_r}{y_{t-1}u_t}}{\sum_{k_c+1}^{k_r}{y_{t-1}^2}},\quad
A_5=\max_{k\in E_{2T}}\frac{\sum_{k_r+1}^{k}{y_{t-1}u_t}}{\sum_{k_r+1}^{k}{y_{t-1}^2}},\quad
A_6=\max_{k\in E_{2T}}\left|\frac{\sum_{k_c+1}^{k}{y_{t-1}^2}}{\sum_{k_r+1}^{k}{y_{t-1}^2}\sum_{k_c+1}^{k_r}{y_{t-1}^2}}\Lambda_T(k/T)\right|.
\]
Using \eqref{lemma:3:y2} and \eqref{lemma:3:yu}, we have, for $a < b$,
\[
A_4=O_p\left(\frac{1}{T^{(b+1)/2}\phi_a^{(k_c-k_e)}}\right)=o_p\left(\frac{1}{T^b}\right),
\]
while using \eqref{lemma:4:y2c-a} and \eqref{lemma:4:yua-a},
\[
|A_5|\leq \frac{\max_{k\in E_{2T}}|\sum_{k_r+1}^{k}y_{t-1}u_t|}{\sum_{k_r+1}^{k_r+M_T}y_{t-1}^2}=O_p\left(\frac{1}{M_T\phi_a^{(k_c-k_e)}\phi_b^{(k_r-k_c)}}\right)=o_p\left(\frac{1}{T^b}\right).
\]
To derive the order of $A_6$, we expand it, as given in Appendix C of \citet{chong2001structural},
\begin{eqnarray}
A_6
&\leq&
\left(\frac{1}{\sum_{k_c+1}^{k_r}{y_{t-1}^2}}+\frac{1}{\sum_{k_r+1}^{k_r+M_T}{y_{t-1}^2}}\right) \nonumber \\
&\times&\left(\max_{k\in E_{2T}}\left|\frac{\left(\sum_{k_c+1}^{k_r}{y_{t-1}u_t}\right)^2}{\sum_{k_c+1}^{k_r}{y_{t-1}^2}}-\frac{\left(\sum_{k_c+1}^{k}{y_{t-1}u_t}\right)^2}{\sum_{k_c+1}^{k}{y_{t-1}^2}}\right|\right. \nonumber \\
&&\left.+\max_{k\in E_{2T}}\left|\frac{\left(\sum_{k_r+1}^{T}{y_{t-1}u_t}\right)^2}{\sum_{k_r+1}^{T}{y_{t-1}^2}}-\frac{\left(\sum_{k+1}^{T}{y_{t-1}u_t}\right)^2}{\sum_{k+1}^{T}{y_{t-1}^2}}\right|\right) \label{a6:ineq} \\
%&=&
%\left(O_p\left(\frac{1}{T^{b+1}\phi_a^{2(k_c-k_e)}}\right)+O_p\left(\frac{1}{TM_T\phi_a^{2(k_c-k_e)}\phi_b^{2(k_r-k_e)}}\right)\right) \nonumber \\
%&&
%\times
%\left(O_p(1)+O_p(1)+O_p(1)+O_p(1)\right) \nonumber \\
& = &
O_p\left(\frac{1}{TM_T\phi_a^{2(k_c-k_e)}\phi_b^{2(k_r-k_e)}}\right)=o_p\left(\frac{1}{T^{2b}}\right) \nonumber
\end{eqnarray}
using Lemmas \ref{lemma:3} and \ref{lemma:4}, where the second terms in the two maxima are evaluated as
\BEQA
\max_{k\in E_{2T}}\frac{\left(\sum_{k_c+1}^{k}{y_{t-1}u_t}\right)^2}{\sum_{k_c+1}^{k}{y_{t-1}^2}}
& \leq & 
\frac{\left(|\sum_{k_c+1}^{k_r}{y_{t-1}u_t}|+\max_{k\in E_{2T}} \sum_{k_r+1}^{k}y_{t-1}u_t|\right)^2}{\sum_{k_c+1}^{k_r}y_{t-1}^2} \\
& = &
\frac{O_p(T^{b+1}\phi_a^{2(k_c-k_e)})}{O_p(T^{b+1}\phi_a^{2(k_c-k_e)})}=O_p(1)
\EEQA
by \eqref{lemma:3:y2}, \eqref{lemma:3:yu}, and \eqref{lemma:4:yua-a}, and
\BEQA
\max_{k\in E_{2T}}\frac{\left(\sum_{k+1}^{T}{y_{t-1}u_t}\right)^2}{\sum_{k+1}^{T}{y_{t-1}^2}}
& \leq &
\frac{\max_{k\in E_{2T}}\left(\sum_{k+1}^{T}{y_{t-1}u_t}\right)^2}{\sum_{\overline{k}_r+1}^{T}y_{t-1}^2} \\
& = &
\frac{O_p(T^{2}\phi_a^{2(k_c-k_e)}\phi_b^{2(k_r-k_c)})}{O_p(T^2\phi_a^{2(k_c-k_e)}\phi_b^{2(k_r-k_c)})}=O_p(1),
\EEQA
by \eqref{lemma:4:y-1} and \eqref{lemma:4:yub-a}, respectively. Then, we obtained \eqref{infssr:r} for $i=2$.

As we established $\hat{k}_r-k_r=O_p(1)$, there exists a positive integer $M$ for any given $\epsilon$ such that $P(|\hat{k}_r-k_r| > M) < \epsilon$ for all sufficiently large $T$. In the same manner as PDC, we have
\begin{eqnarray}
\lefteqn{
    P(\hat{k}_r\neq k_r)
    } \nonumber \\
    &=&P(|\hat{k}_r- k_r|>M)+P(|\hat{k}_r- k_r|\leq M,\hat{k}_r\neq k_r) \notag \\
    &\leq&\epsilon+\sum_{m=1}^M{P\left(\frac{1}{T(1-\phi_b)^2\phi_a^{2(k_c-k_e)}\phi_b^{2(k_r-k_c)}}\left(SSR_2\left(\tau_r-m/T\right)-SSR(\tau_r)\right)<0\right)}\notag\\
    &&+\sum_{m=1}^M{P\left(\frac{1}{T(1-\phi_b)^2\phi_a^{2(k_c-k_e)}\phi_b^{2(k_r-k_c)}}\left(SSR_2\left(\tau_r+m/T\right)-SSR(\tau_r)\right)<0\right)}.\label{PR:cons}
\end{eqnarray}

From expression \eqref{Chong:B2} with $k$ replaced by $k_r-m$, we can show that the second term of \eqref{Chong:B2} dominates the first and third terms using Lemmas \ref{lemma:3} and \ref{lemma:4}. By focusing the second term, we can observe that, for a given fixed $m$,
\BEQA
\frac{1}{T\phi_a^{2(k_c-k_e)}\phi_b^{2(k_r-k_c)}}\frac{\sum_{k_r+1}^{T}y_{t-1}^2\sum_{k_r-m+1}^{k_r}y_{t-1}^2}{\sum_{k_r-m+1}^{T}y_{t-1}^2} 
& = & 
\frac{1}{T\phi_a^{2(k_c-k_e)}\phi_b^{2(k_r-k_c)}}\frac{(\sum_{k_r+1}^{T}y_{t-1}^2)my_{k_r}^2(1+o_p(1))}{\sum_{k_r+1}^{T}y_{t-1}^2(1+o_p(1))} \\
& \sim_a&
\frac{m y_{k_r}^2}{T\phi_a^{2(k_c-k_e)}\phi_b^{2(k_r-k_c)}} \\
& \Rightarrow &
m(W^{\kappa}(\tau_e))^2
\EEQA
using \eqref{lemma:1:y} and \eqref{lemma:3:y-1}, which implies that
\[
P\left(\frac{1}{T(1-\phi_b)^2\phi_a^{2(k_c-k_e)}\phi_b^{2(k_r-k_c)}}\left(SSR_2\left(\tau_r-m/T\right)-SSR(\tau_r)\right)<0\right)\to 0.
\]

Similarly, from expression \eqref{Chong:B4} with $k$ replaced by $k_r+m$, we can show that the second term of \eqref{Chong:B4} dominates the first and third terms using Lemmas \ref{lemma:3} and \ref{lemma:4}. By focusing on the second term, we can observe that, for a given fixed $m$,
\BEQA
\frac{1}{T\phi_a^{2(k_c-k_e)}\phi_b^{2(k_r-k_c)}}\frac{\sum_{k_c+1}^{k_r}y_{t-1}^2\sum_{k_r+1}^{k_r+m}y_{t-1}^2}{\sum_{k_c+1}^{k_r+m}y_{t-1}^2}  
& = & 
\frac{1}{T\phi_a^{2(k_c-k_e)}\phi_b^{2(k_r-k_c)}}\frac{(\sum_{k_c+1}^{k_r}y_{t-1})^2my_{k_r}^2(1+o_p(1))}{\sum_{k_c+1}^{k_r}y_{t-1}^2(1+o_p(1))} \\
& \sim_a&
\frac{m y_{k_r}^2}{T\phi_a^{2(k_c-k_e)}\phi_b^{2(k_r-k_c)}} \\
& \Rightarrow &
m(W^{\kappa}(\tau_e))^2
\EEQA
using \eqref{lemma:1:y} and \eqref{lemma:3:y-1} and thus, for a given positive integer value of $m$,
\[
P\left(\frac{1}{T(1-\phi_b)^2\phi_a^{2(k_c-k_e)}\phi_b^{2(k_r-k_c)}}\left(SSR_2\left(\tau_r+m/T\right)-SSR(\tau_r)\right)<0\right)\to 0.
\]

In case (ii), we first prove \eqref{consistency:fraction:r} with $M_T$ replaced by $M_{bT}$ does not hold, where $M_{bT}\to\infty$ and $T^{b+\epsilon}/M_{bT}\to 0$ for a given $\epsilon >0$. That is, for any given $\epsilon > 0$, the objective function cannot be minimized outside the $T^{b+\epsilon}$ neighborhood of $k_r$. Let, for a given $\epsilon$, $E_{b0T}=\{k:\;k_r-M_{bT}+1\leq k \leq k_r+M_{bT}\}$, $E_{b1T}=\{k:\;\underline{k}_r+1\leq k \leq k_r-M_{bT}\}$, and  $E_{b2T}=\{k:\;k_r+M_{bT}+1\leq k \leq \overline{k}_r\}$, where we suppress the dependence on $\epsilon$ for notational simplicity. Then, as in case (i), it is sufficient to prove that
\BE
P\left(\inf_{k\in E_{biT}}SSR_2(k/T) < SSR_2(\tau_r)  \right)\to 0
\label{infssr:r:b}
\EE
for $i=1$ and $2$ when $a > b$.

In exactly the same manner as in case (i), the difference in the sums of the squared residuals is expressed as \eqref{Chong:B2} for $k\in E_{b1T}$ and \eqref{Chong:B4} for $k\in E_{b2T}$. Thus, it is sufficient to prove that $A_1=o_p(1/T^b)$, $A_{b2}=o_p(1/T^b)$, and $A_{b3}=o_p(1/T^{2b})$, $A_4=o_p(1/T^b)$, $A_{b5}=o_p(1/T^b)$, and $A_{b6}=o_p(1/T^{2b})$, where $A_{b2}$, $A_{b3}$, $A_{b5}$, and $A_{b6}$ are defined as $A_2$, $A_3$, $A_5$, and $A_6$ with $E_{iT}$ replaced by $E_{biT}$ for $i=1$ and $2$, respectively.

For $k\in E_{b1T}$, $A_1=O_p(1/T)=o_p(1/T^b)$ by \eqref{lemma:4:y2-b} and \eqref{lemma:4:yu-b}, while $A_{b2}=o_p(1/T^b)$ is obtained by \eqref{lemma:uniform:a2}. As in the proof in case (i), $A_{b3}$ is bounded above, as given in \eqref{a3:ineq} with $M_T$ and $E_{1T}$ replaced by $M_{bT}$ and $E_{b1T}$, respectively. Using \eqref{lemma:3:y-1}, we have
\[
\sum_{k_r-M_{bT}+1}^{k_r}y_{t-1}^2\sim_a \sum_{k_r-M_{bT}+1}^{k_r}\left(\sum_{j=k_c+1}^{t-1}\phi_b^{t-j-1} u_j\right)^2=O_p(T^bM_{bT})
\]
because $E[\left(\sum_{j=k_c+1}^{t}\phi_b^{t-j} u_j\right)^2]=O(T^b)$ for $k\in E_{b1T}$. Thus, with \eqref{lemma:4:y2-b}, we can observe that
\[
\left(\frac{1}{\sum_{k_r+1}^{T}{y_{t-1}^2}}+\frac{1}{\sum_{k_r-M_{bT}+1}^{k_r}{y_{t-1}^2}}\right)
=O_p\left(\frac{1}{T^2}\right)+O_p\left(\frac{1}{T^bM_{bT}}\right)=
O_p\left(\frac{1}{T^bM_{bT}}\right).
\]
For the first maximum in \eqref{a3:ineq}, the first term can be observed $O_p(1)$ by \eqref{lemma:3:y2} and \eqref{lemma:3:yu}, while for the second term,
\[
\sum_{k_c+1}^{k}{y_{t-1}^2}=O_p(T^{b+1}\phi_a^{2(k_c-k_e)})
\]
uniformly over $k\in E_{b1T}$ by using \eqref{lemma:3:y2a}. Therefore, we can observe with \eqref{lemma:3:yua} that the second term is $O_p(1)$ uniformly over $k\in E_{b1T}$. Similarly, for the second maximum in \eqref{a3:ineq}, the first term can be observed $O_p(1)$ by \eqref{lemma:4:y2-b} and \eqref{lemma:4:yu-b}, while for the second term, we can observe that
\BEQA
\sum_{k+1}^Ty_{t-1}^2
& = &
T^b(k_r-k)\left(\frac{1}{T^b(k_r-k)}\sum_{k+1}^{k_r}y_{t-1}^2\right)+\sum_{k_r+1}^Ty_{t-1}^2 \\
& = &
T^b(k_r-k)O_p(1)+O_p(T^2)=O_p(T^2)
\EEQA
uniformly over $k\in E_{b1T}$, where the second equality holds by \eqref{lemma:4:y2-b} and because the term in the parentheses uniformly converges in probability to $\sigma^2/(2c_b)$, as shown in the proof of Lemma \ref{lemma:uniform:b}, while the last equality is obtained by noting that $T^b(k_r-k)\leq CT^{b+1}$ for $k\in E_{b1T}$.  For the numerator of the second term, we have
\BEQA
\sum_{k+1}^Ty_{t-1}u_t
 & = &
(k_r-k)\left(\frac{1}{k_r-k}\sum_{k+1}^{k_r}y_{t-1}u_t\right)+\sum_{k_r+1}^Ty_{t-1}u_t \\
& = &
o_p(T)+O_p(T)=O_p(T)
\EEQA 
uniformly over $k\in E_{b1T}$, where the second equality holds by \eqref{lemma:4:yu-b} and because the term in the parentheses is uniformly $o_p(1)$, as shown in the proof of Lemma \ref{lemma:uniform:b}. Therefore, we can observe that the second term in the second maximum in \eqref{a3:ineq} is $O_p(1)$ uniformly over $E_{b1T}$. Summarizing these results, we can observe that $A_{b3}=O_p(1/(T_bM_{bT}))=o_p(1/T^{2b})$.

For $k\in E_{b2T}$, $A_4=O_p(1/(T^{(b+1)/2}\phi_a^{(k_c-k_e)}))=o_p(1/T^b)$ by \eqref{lemma:3:y2} and \eqref{lemma:3:yu}, while $A_{b5}=o_p(1/T^b)$ is obtained by \eqref{lemma:uniform:a5}. As in the proof in case (i), $A_{b6}$ is bounded above, as given in \eqref{a6:ineq} with $M_T$ and $E_{2T}$ replaced by $M_{bT}$ and $E_{b2T}$, respectively. Using \eqref{lemma:4:y-1}, we have
\[
\sum_{k_r+1}^{k_r+M_{bT}}y_{t-1}^2 \sim_a \sum_{k_r+1}^{k_r+M_{bT}}\left(\sum_{j=k_r+1}^{t-1} \varepsilon_j\right)^2=O_p(M_{bT}^2).
\]
Thus, with \eqref{lemma:3:y2}, we can observe that
\[
\left(\frac{1}{\sum_{k_c+1}^{k_r}{y_{t-1}^2}}+\frac{1}{\sum_{k_r+1}^{kr+M_{bT}}{y_{t-1}^2}}\right)
=O_p\left(\frac{1}{T^{b+1}\phi_a^{2(k_c-k_e)}}\right)+O_p\left(\frac{1}{M_{bT}^2}\right)=
O_p\left(\frac{1}{M_{bT}^2}\right).
\]
For the first maximum in \eqref{a6:ineq}, the first term is $O_p(1)$ as shown above for $A_{b3}$, while for the second term,
\[
\sum_{k_c+1}^{k}y_{t-1}^2\geq \sum_{k_c+1}^{k_r}y_{t-1}^2=O_p(T^{b+1}\phi_a^{2(k_c-k_e)})
\]
by \eqref{lemma:3:y2} and
\[
\left|\sum_{k_c+1}^{k}y_{t-1}u_t\right|\leq \left|\sum_{k_c+1}^{k_r}y_{t-1}u_t\right|+\left|\sum_{k_r+1}^{k}y_{t-1}u_t\right|
=
O_p\left(T^{(b+1)/2}\phi_a^{(k_c-k_e)}\right)+O_p(T)
\]
by \eqref{lemma:3:yu} and \eqref{lemma:4:yua-b}. Thus, the second term is $O_p(1)$ uniformly over $k\in E_{b2T}$. Similarly, for the second maximum in \eqref{a6:ineq}, the first term $O_p(1)$ as shown above for $A_{b3}$, while for the second term, we have
\[
\sum_{k+1}^Ty_{t-1}^2
\geq
\sum_{\overline{k}_r+1}^Ty_{t-1}^2=O_p(T^2)
\]
and the numerator is uniformly $O_p(T^2)$ by \eqref{lemma:4:yua-b}. Thus, the second term is $O_p(1)$ uniformly over $k\in E_{b2T}$. Summarizing these results, we can observe that $A_{b6}=O_p(1/(M_{bT}^2))=o_p(1/T^{2b})$.

To derive the limiting distribution of $(1-\phi_b)T(\hat{\tau}_r-\tau_r)$, let $k=k_r+vT^b$ where $v$ is a finite constant, and consider the $O(T^b)$ order neighborhood of $k_r$. We first derive the limiting distribution when $v < 0$ ($k=k_r-|v|T^b$) using \eqref{Chong:B2}, in which $\Lambda_T(k/T)$ can be shown to be $o_p(1)$ using the same technique as before.  Because time is measured in $1/T^b$ units rather than $1/T$ in this case, we note that
\BE
\frac{1}{T^{b/2}}\sum_{k+1}^{k_r}\varepsilon_t
=
\frac{1}{T^{b/2}}\sum_{i=0}^{|v|T^b-1}\varepsilon_{k_r-i}
\Rightarrow 
\sigma B_1(|v|),
\label{iid:FCLT1}
\EE
by the FCLT, where $B_1(\cdot)$ is a standard Brownian motion on $R^{+}$. Then, we can observe that
\BEQ
\frac{1}{T^{b/2}}y_{k-1}
& = &
\frac{1}{T^{b/2}}\phi_b^{(k-k_c-1)}y_{k_c}+\frac{1}{T^{b/2}}\sum_{j=k_c+1}^{k-1}\phi_b^{(k-j-1)}\varepsilon_j \\
& = &
O_p(T^{(1-b)/2}\phi_a^{(k_c-k_e)}\phi_b^{(k_r-|v|T^b-k_c)})+\frac{1}{T^{b/2}}\sum_{i=|v|T^b+1}^{k_r-k_c-1}\phi_b^{i-|v|T^b-1}\varepsilon_{k_r-i}
\nonumber \\
& = &
o_p(1)+\frac{1}{T^{b/2}}\sum_{i=|v|T^b+1}^{k_r-k_c-1}\left(1-\frac{c_b}{T^b}\right)^{T^b(i/T^b-|v|-1/T^b)}\varepsilon_{k_r-i} \\
& \Rightarrow &
\sigma \int_{|v|}^{\infty}\exp\left(-c_b(s-|v|)\right)dB_1(s)\eqqcolon \sigma \tilde{B}_{c_b}(|v|).
\label{iid:limit:y}
\EEQ
Using these results, each term in \eqref{Chong:B2} becomes, by noting that $\sum_{k+1}^{T}y_{t-1}^2=\sum_{k_r+1}^Ty_{t-1}^2(1+o_p(1))$ in this case,
\BE
2(1-\phi_b)\frac{\sum_{k+1}^{k_r}{y_{t-1}^2}\sum_{k_r+1}^{T}{y_{t-1}u_t}}{\sum_{k+1}^{T}{y_{t-1}^2}}
= 
\frac{2c_b}{T^b}\frac{O_p(T^{2b})O_p(T)}{O_p(T^2)}=O_p(T^{b-1})=o_p(1),
\label{limit:kr:1a}
\EE
\BEQ
-2(1-\phi_b)\frac{\sum_{k_r+1}^{T}{y_{t-1}^2}\sum_{k+1}^{k_r}{y_{t-1}u_t}}{\sum_{k+1}^{T}{y_{t-1}^2}}
& = &
-\frac{2c_b}{T^b}\sum_{k_r-|v|T^b+1}^{k_r}y_{t-1}\varepsilon_t(1+o_p(1)) \nonumber \\
& = &
-2c_b\sum_{i=0}^{|v|T^b-1}\frac{y_{k_r-i-1}}{T^{b/2}}\frac{\varepsilon_{k_r-i}}{T^{b/2}} \nonumber \\
& \Rightarrow &
-2\sigma^2c_b\int_0^{|v|}\tilde{B}_{c_b}(s)dB_1(s)
\label{limit:kr:2a}
\EEQ
by \eqref{iid:limit:y}, and
\BEQ
(1-\phi_b)^2\frac{\sum_{k_r+1}^{T}{y_{t-1}^2}\sum_{k+1}^{k_r}{y_{t-1}^2}}{\sum_{k+1}^{T}{y_{t-1}^2}}
& = &
\frac{c_b^2}{T^{2b}}\sum_{k_r-|v|T^{b}+1}^{k_r}y_{t-1}^2(1+o_p(1)) \nonumber \\
& = &
\frac{c_b^2}{T^{2b}}\sum_{i=0}^{|v|T^b-1}y_{k_r-i-1}^2(1+o_p(1)) \nonumber \\
& \Rightarrow &
\sigma^2c_b^2\int_0^{|v|} \tilde{B}_{c_b}^2(s)ds \nonumber \\
& = &
\sigma^2c_b^2\int_0^{|v|} \left(\tilde{B}_{c_b}^2(s)-\frac{1}{2c_b}\right)ds+\frac{\sigma^2c_b|v|}{2},
\label{limit:kr:3a}
\EEQ
where we note that $E[\tilde{B}_{c_b}^2(s)]=1/(2c_b)$ for any $s\geq 0$ in the last equality. Then, we have
\[
SSR_2(k/T)-SSR_2(\tau_r)
\Rightarrow
-\sigma^2c_b\left(2\int_0^{|v|}\tilde{B}_{c_b}(s)dB_1(s)-c_b\int_0^{|v|}\left(\tilde{B}_{c_b}^2(s)-\frac{1}{2c_b}\right)ds-\frac{|v|}{2}\right)
\]
for $v < 0$. Because $\argmin\{SSR_2(k/T)-SSR_2(\tau_r)\}=\argmax\{-(SSR_2(k/T)-SSR_2(\tau_r))\}$, we obtain the limiting distribution given in the theorem.

Similarly, for $k=k_r+vT^b$ with $v \geq 0$, it can be shown that $\Lambda_{T}(k/T)=o_p(1)$ in \eqref{Chong:B4}. Note that
\BEQ
\frac{1}{T^{b/2}}\sum_{k_r+1}^{k}u_t
& = &
\frac{1}{T^{b/2}}\sum_{k_r+1}^{k_r+vT^{b}}\varepsilon_t+o_p(1) \\
& \Rightarrow &
\sigma B_2(v)
\label{iid:FCLT2}
\EEQ
where $B_2(\cdot)$ is a standard Brownian motion on $R^+$ independent of $B_1(\cdot)$. Then, each term in \eqref{Chong:B4} becomes, noting that $\sum_{k_c+1}^{k}y_{t-1}^2=\sum_{k_c+1}^{k_r}y_{t-1}^2(1+o_p(1))$ in this case,
\BEQ
\lefteqn{
2(1-\phi_b)\frac{\sum_{k_r+1}^{k}{y_{t-1}u_t}\sum_{k_c+1}^{k_r}{y_{t-1}^2}}{\sum_{k_c+1}^{k}{y_{t-1}^2}}
} \nonumber \\
& = &
\frac{2c_b}{T^b}\sum_{k_r+1}^{k_r+vT^b}\left(y_{k_r}+\frac{c_1(t-k_r-1)}{T^{\eta_1}}+\sum_{j=k_r+1}^{t-1}\varepsilon_j\right)\left(\frac{c_1}{T^{\eta_1}}+\varepsilon_t\right)(1+o_p(1)) \nonumber \\
& = &
2c_b\left\{\frac{1}{T^b}y_{k_r}\sum_{k_r+1}^{k_r+vT^b}\varepsilon_t+\frac{1}{T^b}\sum_{k_r+1}^{k_r+vT^b}\left(\sum_{j=k_r+1}^{t-1}\varepsilon_j\right)\varepsilon_{t}+o_p(1)\right\}(1+o_p(1)) \nonumber \\
& \Rightarrow &
2\sigma^2 c_b\int_0^v\left(\tilde{B}_{c_b}(0)+B_2(s)\right)dB_2(s),
\label{limit:kr:1b}
\EEQ
where we used $y_{k_r}/T^{b/2}\Rightarrow \sigma \tilde{B}_{c_b}(0)$ by \eqref{iid:limit:y},
\BE
2(1-\phi_b)\frac{\sum_{k_c+1}^{k_r}{y_{t-1}u_t}\sum_{k_r+1}^{k}{y_{t-1}^2}}{\sum_{k_c+1}^{k}{y_{t-1}^2}}=o_p(1),
\label{limit:kr:2b}
\EE
and
\BEQ
(1-\phi_b)^2\frac{\sum_{k_r+1}^{k}{y_{t-1}^2}\sum_{k_c+1}^{k_r}{y_{t-1}^2}}{\sum_{k_c+1}^{k}{y_{t-1}^2}}
& = &
\frac{c_b^2}{T^{2b}}\sum_{k_r+1}^{k_r+vT^b}y_{t-1}^2(1+o_p(1)) \nonumber \\
& = &
\frac{c_b^2}{T^{2b}}\sum_{k_r+1}^{k_r+vT^b}\left(y_{k_r}+\frac{c_1(t-k_r-1)}{T^{\eta_1}}+\sum_{j=k_r+1}^{t-1}\varepsilon_j\right)^2(1+o_p(1)) \nonumber \\
& \Rightarrow &
\sigma^2c_b^2\int_0^v\left(\tilde{B}_{c_b}(0)+B_2(s)\right)^2ds.
\label{limit:kr:3b}
\EEQ
Then, we have
\BEQA
\lefteqn{
SSR_2(k/T)-SSR_2(\tau_r)
} \\
&\Rightarrow&
\sigma^2\left[2c_b\int_0^v\left(\tilde{B}_{c_b}(0)+B_2(s)\right)dB_2(s)+c_b^2\int_0^v\left(\tilde{B}_{c_b}(0)+B_2(s)\right)^2ds\right] \\
& = &
-2\sigma^2c_b^2\tilde{B}_{c_b}^2(0)
\left[-\left\{\frac{B_2(v)}{c_b\tilde{B}_{c_b}(0)}+\frac{\int_0^vB_2(s)dB_2(s)}{c_b\tilde{B}_{c_b}^2(0)}\right\} 
-\left\{\frac{v}{2} + \frac{\int_0^v B_2^2(s)ds}{2\tilde{B}_{c_b}^2(0)}+\frac{2\int_0^v \tilde{B}_{c_b}(0)B_2(s)ds}{2\tilde{B}_{c_b}^2(0)}\right\}\right].
\EEQA
Because the last two terms in the second braces can be expressed as
\BEQA
\frac{\int_0^v B_2^2(s)ds}{2\tilde{B}_{c_b}^2(0)}+\frac{2\int_0^v \tilde{B}_{c_b}(0)B_2(s)ds}{2\tilde{B}_{c_b}^2(0)}
& = &
\frac{1}{\tilde{B}_{c_b}(0)}\int_0^v\left(\frac{B_2(s)B_2(s)}{2\tilde{B}_{c_b}(0)}+B_2(s)\right)ds \\
& = &
\frac{1}{\tilde{B}_{c_b}(0)}\int_0^v\left(\frac{B_2(s)}{2\tilde{B}_{c_b}(0)}+1\right)B_2(s)ds,
\EEQA
the limiting distribution can be expressed as
\BEQA
\lefteqn{
SSR_2(k/T)-SSR_2(\tau_r)
} \\
&\Rightarrow&
-2\sigma^2c_b^2\tilde{B}_{c_b}^2(0)\left\{-\frac{1}{c_b\tilde{B}_{c_b}(0)}\left[B_2(v)+\int_0^v\frac{B_2(s)}{\tilde{B}_{c_b}(0)}dB_2(s)+c_b\int_0^v\left(\frac{B_2(s)}{2\tilde{B}_{c_b}(0)}+1\right)B_2(s)ds\right]-\frac{v}{2}\right\},
\EEQA
for $v \geq 0$. Then, we obtain the limiting distribution for $v \geq 0$ given in the theorem.$\blacksquare$

\newpage
\renewcommand{\theequation}{\Alph{section}.\arabic{equation}}
\setcounter{equation}{0}

\section{Lemmas}
 
 \begin{lemma}
\label{lemma:1}
Under Assumption \ref{assumption:base}, for $1\leq k \leq k_e$,
\BEQ
\frac{1}{\sqrt{T}}y_{k_e} &  \Rightarrow & W^{\kappa}(\tau_e). \label{lemma:1:y} \\
\sum_{1}^{k_e}y_{t-1}^2 & = & O_p(T^2).  \label{lemma:1:y2}\\
\sum_{1}^{k_e}y_{t-1}u_t & = & O_p(T). \label{lemma:1:yu}\\
\sum_{1}^{k}y_{t-1}u_t & = & O_p(T) \;\mbox{uniformly over $1\leq k \leq k_e$.}\label{lemma:1:yua}\\
\sum_{k+1}^{k_e}y_{t-1}u_t & = & O_p(T)\;\mbox{uniformly over $1\leq k \leq k_e-1$}. \label{lemma:1:yub}
\EEQ
\end{lemma}
{\bf Proof of Lemma \ref{lemma:1}}: 
From \eqref{model:est1}, we have
\[
y_t=y_0+c_0\frac{t}{T^{\eta_0}}+\sum_{1}^t\varepsilon_j.
\]
Then, by the FCLT \eqref{FCLT} and the continuous mapping theorem (CMT), we obtain \eqref{lemma:1:y} and \eqref{lemma:1:y2}. For \eqref{lemma:1:yu}--\eqref{lemma:1:yub}, we note that
\[
\frac{1}{T}\sum_{1}^ky_{t-1}u_t=\frac{1}{2T}\left(y_k^2-y_{0}^2-\sum_{1}^ku_t^2\right)
\]
and
\[
\sum_{1}^k u_t^2=\frac{c_0^2}{T^{2\eta_0}}+\frac{2c_0}{T^{\eta_0}}\sum_{1}^k\varepsilon_t+\sum_{1}^k\varepsilon_t^2.
\]
Thus, we can observe that \eqref{lemma:1:yu} and \eqref{lemma:1:yua} hold. \eqref{lemma:1:yub} holds from \eqref{lemma:1:yua} because $\sum_{k+1}^{k_e}y_{t-1}u_t=\sum_{1}^{k_e}y_{t-1}u_t-\sum_{1}^{k}y_{t-1}u_t$.$\blacksquare$

\begin{lemma}
\label{lemma:2}
Under Assumption \ref{assumption:base}, for $k_e+1\leq k \leq k_c$,
\BEQ
y_{k} &  \sim_a & \phi_a^{k-k_e}y_{k_e} \;\mbox{uniformly over $k_e+1\leq k \leq k_c$.} \label{lemma:2:y} \\
\sum_{k_e+1}^{k_c}y_{t-1}^2 & = & O_p(T^{a+1}\phi_a^{2(k_c-k_e)}).  \label{lemma:2:y2}\\
\sum_{k_c-M_T+1}^{k_c}y_{t-1}^2 & = & O_p(TM_T\phi_a^{2(k_c-k_e)}).  \label{lemma:2:y2c}
\EEQ
\BEQ
\sum_{k_e+1}^{k_c}y_{t-1}u_t & = & O_p(T^{(a+1)/2}\phi_a^{(k_c-k_e)}).  \label{lemma:2:yu}\\
\sum_{k_e+1}^{k}y_{t-1}u_t & = & O_p(T^{(a+1)/2}\phi_a^{(k_c-k_e)}) \;\mbox{uniformly over $k_e+1\leq k\leq k_c$.} \label{lemma:2:yua}\\
\sum_{k+1}^{k_c}y_{t-1}u_t & = & O_p(T^{(a+1)/2}\phi_a^{(k_c-k_e)}) \;\mbox{uniformly over $k_e+1\leq k\leq k_c-1$.}  \label{lemma:2:yub} \\
\sum_{k_c-M_T+1}^{k_c}y_{t-1}u_t & = & O_p(T^{1/2}M_T^{1/2}\phi_a^{(k_c-k_e)}). \label{lemma:2:yuc}
\EEQ
\end{lemma}
{\bf Proof of Lemma \ref{lemma:2}}:
For $k_e+1\leq k \leq k_c$, $y_k$ is expressed as
\BEQ
y_k
& = &
\phi_a^{(k-k_e)}y_{k_e}+\sum_{j=k_e+1}^{k}\phi_a^{(k-j)}u_j \nonumber \\
& = &
\phi_a^{(k-k_e)}y_{k_e}+\phi_a^{(k-k_e)}\sum_{j=k_e+1}^{k}\phi_a^{(k_e-j)}u_{j}  \nonumber \\
& = &
\phi_a^{(k-k_e)}\left(y_{k_e}+\sum_{j'=1}^{k-k_e}\phi_a^{-j'}u_{k_e+j'}\right).
\label{pf:lemma2:y}
\EEQ
Because $\{\phi_a^{-j'}u_{k_e+j'}\}_{j'=1}^{k_c-k_e}$ is a martingale difference sequence (MDS), we have, from the Haj\'eck-R\'enyi inequality (HRI) for martingale differences (6.6b in \citet{LinBai2010}),
\BEQA
P\left(\max_{1\leq k-k_e\leq k_c-k_e}\left|\sum_{j'=1}^{k-k_e}\phi_a^{-j'}u_{k_e+j'}\right|\geq N\right)
& \leq &
\frac{\bar{\omega}^2}{N^2}\sum_{j'=1}^{k_c-k_e}\phi_a^{-2j'} \\
& = &
\frac{\bar{\omega}^2}{N^2}\frac{1-\phi_a^{-2(k_c-k_e)}}{\phi_a^2-1} \\
& \leq &
\frac{C}{N^2}T^a
\EEQA
for any given $N > 0$. This implies that $\max\left|\sum_{j'=1}^{k-k_e}\phi_a^{-j'}u_{k_e+j'}\right|=O_p(T^{a/2})$. As $y_{k_e}=O_p(T^{1/2})$ by \eqref{lemma:1:y} in Lemma \ref{lemma:1}, we have \eqref{lemma:2:y}.

Using \eqref{lemma:2:y}, we can observe that
\BEQA
\sum_{k_e+1}^{k_c}y_{t-1}^2
& \sim_a &
y_{k_e}^2\sum_{k_e+1}^{k_c}\phi_a^{2(t-k_e-1)} \\
& = &
y_{k_e}^2\frac{1-\phi_a^{2(k_c-k_e)}}{1-\phi_a^2} = O_p(T^{a+1}\phi_a^{2(k_c-k_e)})
\EEQA
and thus we obtain \eqref{lemma:2:y2}.

Similarly, for \eqref{lemma:2:y2c}, we have
\BEQA
\sum_{k_c-M_T+1}^{k_c}y_{t-1}^2
& \sim_a &
y_{k_e}^2\frac{\phi_a^{2(k_c-k_e-M_T)}-\phi_a^{2(k_c-k_e)}}{1-\phi_a^2} \\
& \sim_a &
y_{k_e}^2\frac{T^a}{2c_a}\phi_a^{2(k_c-k_e)}\left(1-\phi_a^{-2M_T}\right).
\EEQA
In general, when $x\to 0$, $y\to \infty$ (or $y\to -\infty$), and $xy\to 0$, $(1+x)^y$ can be expanded as
\BEQ
(1+x)^y
& = &
\exp\left(y\log (1+x)\right) \nonumber \\
& = &
\exp\left(y(x+O(x^2))\right) \nonumber \\
& = &
1+xy+O(x^2y^2). \label{expand:a}
\EEQ
By applying \eqref{expand:a} to $\phi_a^{-2M_T}$, we have
\BEQ
1-\phi_a^{-2M_T}
& = &
1-\left(1+\frac{c_a}{T^a}\right)^{-2M_T} \nonumber \\
& = &
1-\left(1-2M_T\frac{c_a}{T^a}+O\left(\frac{M_T^2}{T^{2a}}\right)\right) \nonumber \\
& = &
2c_a\frac{M_T}{T^a}(1+o(1)).
\label{phiMT}
\EEQ
We thus obtain
\[
\sum_{k_c-M_T+1}^{k_c}y_{t-1}^2
\sim_a
y_{k_e}^2\frac{T^a}{2c_a}\phi_a^{2(k_c-k_e)}\times 2c_a\frac{M_T}{T^a}=O_p(TM_T\phi_a^{2(k_c-k_e}).
\]

\eqref{lemma:2:yu} is obtained by noting that
\[
\sum_{k_e+1}^{k_c}y_{t-1}u_t
\sim_a
y_{k_e}\sum_{k_e+1}^{k_c}\phi_a^{(t-k_e-1)}u_t
\]
from \eqref{lemma:2:y} and that
\[
Var\left[\sum_{k_e+1}^{k_c}\phi_a^{(t-k_e-1)}u_t\right]=O(T^a\phi_a^{2(k_c-k_e)}).
\]

Similarly, by using \eqref{lemma:2:y}, we can observe that
\BE
\sum_{k_e+1}^{k}y_{t-1}u_t\sim_a y_{k_e}\sum_{k_e+1}^{k}\phi_a^{t-k_e-1}u_t.
\label{lemma:2:yua-1}
\EE
Again, by the HRI for any fixed value of $N>0$,
\BEQA
P\left(\max_{k_e+1\leq k \leq k_c}\left|\sum_{k_e+1}^k\phi_a^{(t-k_e-1)}u_t\right| \geq N\right)
& = &
P\left(\max_{1\leq k-k_e\leq  k_c-k_e}\left|\sum_{j'=1}^{k-k_e}\phi_a^{j'-1}u_{k_e+j'}\right| \geq N\right) \\
& \leq & 
\frac{\bar{\omega}^2}{N^2}\sum_{j'=1}^{k_c-k_e}\phi_a^{2(j'-1)} \\
& \leq &
\frac{C}{N^2}T^a\phi_a^{2(k_c-k_e)},
\EEQA
which implies $\sum_{k_e+1}^{k}\phi_a^{(t-k_e-1)}u_{t}=O_p(T^{a/2}\phi_a^{(k_c-k_e)})$ uniformly over $k_e+1\leq k \leq k_c$. Combining this with \eqref{lemma:2:yua-1}, we have \eqref{lemma:2:yua}.

\eqref{lemma:2:yub} is obtained using \eqref{lemma:2:yu} and \eqref{lemma:2:yua} by noting that $\sum_{k+1}^{k_e}y_{t-1}u_t=\sum_{k_e+1}^{k_c}y_{t-1}u_t-\sum_{k_e+1}^k y_{t-1}u_t$.

Similarly to \eqref{lemma:2:yu}, we have, for \eqref{lemma:2:yuc},
\[
\sum_{k_c-M_T+1}^{k_c}y_{t-1}u_t
\sim_a
y_{k_e}\sum_{k_c-M_T+1}^{k_c}\phi_a^{(t-k_e-1)}u_t,
\]
and
\BEQA
Var\left[\sum_{k_c-M_T+1}^{k_c}\phi_a^{(t-k_e-1)}u_t\right]
& \leq &
\bar{\omega}^2\frac{\phi_a^{2(k_c-k_e-M_T)}-\phi_a^{2(k_c-k_e)}}{1-\phi_a^2} \\
& \leq &
C T^a\phi_a^{2(k_c-k_e)}(1-\phi_a^{-2M_T}) \\
& \leq &
CM_T\phi_a^{2(k_c-k_e)},
\EEQA
where the last inequality is obtained by using \eqref{phiMT}. Thus, we obtain  \eqref{lemma:2:yuc}.$\blacksquare$

\begin{lemma}
\label{lemma:3}
Under Assumption \ref{assumption:base}, for $k_c+1\leq k \leq k_r$,
\BEQ
y_{k_r} & = & O_p(T^{1/2}\phi_a^{(k_c-k_e)}\phi_b^{(k_r-k_c)})+O_p(T^{b/2}).
\label{lemma:3:y}\\
\sum_{k_c+1}^{k}y_{t-1}^2 & = & y_{k_c}^2\sum_{k_c+1}^k\phi_b^{2(t-k_c-1)}(1+o_p(1))\;\mbox{uniformly over $k_c+1\leq k \leq k_r$}.  \label{lemma:3:y2a}\\
\sum_{k_c+1}^{k_r}y_{t-1}^2 & = & O_p(T^{b+1}\phi_a^{2(k_c-k_e)}).  \label{lemma:3:y2}\\
\sum_{k_c+1}^{k_c+M_T}y_{t-1}^2 & = & O_p(TM_T\phi_a^{2(k_c-k_e)}).  \label{lemma:3:y2c}\\
\sum_{k_r-M_T+1}^{k_r}y_{t-1}^2 & = & O_p(TM_T\phi_a^{2(k_c-k_e)}\phi_b^{2(k_r-k_c)})\quad\mbox{when $a < b$.}  \label{lemma:3:y2d}
\EEQ
\BEQ
\sum_{k_c+1}^{k_r}y_{t-1}u_t & = & O_p(T^{(b+1)/2}\phi_a^{(k_c-k_e)}).  \label{lemma:3:yu}\\
\sum_{k_c+1}^{k}y_{t-1}u_t & = & O_p(T^{(b+1)/2}\phi_a^{(k_c-k_e)}) \;\mbox{uniformly over $k_c+1\leq k\leq k_r$.} \label{lemma:3:yua}\\
\sum_{k+1}^{k_r}y_{t-1}u_t & = & O_p(T^{(b+1)/2}\phi_a^{(k_c-k_e)}) \;\mbox{uniformly over $k_c+1\leq k\leq k_r-1$.}  \label{lemma:3:yub}\\
\sum_{k_c+1}^{k_c+M_T}y_{t-1}u_t & = & O_p(T^{1/2}M_T^{1/2}\phi_a^{(k_c-k_e)}).  \label{lemma:3:yuc}
\EEQ
\end{lemma}
{\bf Proof of Lemma \ref{lemma:3}}:
For $k_c+1\leq k \leq k_r$, $y_k$ is expressed as
\BE
y_k
=
\phi_b^{(k-k_c)}y_{k_c}+\sum_{j=k_c+1}^{k}\phi_b^{(k-j)}u_j.
\label{lemma:3:y-1}
\EE
In particular, when $k=k_r$, \eqref{lemma:3:y-1} becomes
\[
y_{k_r}\sim_a\phi_b^{(k_r-k_c)}\phi_a^{(k_c-k_e)}y_{k_e}+\sum_{j=k_c+1}^{k_r}\phi_b^{(k_r-j)}u_j=O_p(T^{1/2}\phi_a^{(k_c-k_r)}\phi_b^{(k_r-k_c)})+O_p(T^{b/2}).
\]
Note that the dominant order depends on the behavior of $\phi_a^{k_c-k_e}\phi_b^{k_r-k_c}$.

For \eqref{lemma:3:y2a},
\BE
\sum_{k_c+1}^k y_{t-1}^2
=
 y_{k_c}^2\sum_{k_c+1}^k\phi_b^{2(t-k_c-1)}
 +2y_{k_c}\sum_{k_c+1}^k\sum_{j=k_c+1}^{t-1}\phi_b^{(2t-k_c-j-2)}u_j
 + \sum_{k_c+1}^k\left(\sum_{j=k_c+1}^{t-1}\phi_b^{(t-j-1)}u_j\right)^2.
 \label{lemma:3:y2-1}
 \EE
 From the definition of $\phi_b$, we can observe that the first term on the right-hand side of \eqref{lemma:3:y2-1} is at least $O_p(y_{k_c}^2)=O_p(T\phi_a^{2(k_c-k_e)})$ (when $k$ is finite) and at most $O_p(T^{b+1}\phi_a^{2(k_c-k_e)})$ (when $k=k_r$). For the third term, we can observe that
\[
\sum_{k_c+1}^k\left(\sum_{j=k_c+1}^{t-1}\phi_b^{(t-j-1)}u_j\right)^2
\leq
\sum_{k_c+1}^{k_r}\left(\sum_{j=k_c+1}^{t-1}\phi_b^{(t-j-1)}u_j\right)^2
\]
and
\[
E\left[\sum_{k_c+1}^{k_r}\left(\sum_{j=k_c+1}^{t-1}\phi_b^{(t-j-1)}u_j\right)^2\right]
\leq \bar{\omega}^2\sum_{k_c+1}^{k_r}\sum_{j=k_c+1}^{t-1}\phi_b^{2(t-j-1)}=O(T^{b+1}).
\]
Therefore, the first term on the right hand side of \eqref{lemma:3:y2-1} dominates the third term for all $k_c+1\leq k \leq k_r$. We can also see that the former dominates the second term by Cauchy-Schwarz inequality; thus, we obtain \eqref{lemma:3:y2a}.

\eqref{lemma:3:y2} and \eqref{lemma:3:y2c} are easily obtained by \eqref{lemma:3:y2a} with $k=k_r$ and $k=k_c+M_T$, respectively, using \eqref{expand:a}.

For \eqref{lemma:3:y2d}, we have, similarly to \eqref{lemma:3:y2-1},
\BEQ
\sum_{k_r-M_T+1}^{k_r} y_{t-1}^2
& = &
 y_{k_c}^2\sum_{k_r-M_T+1}^{k_r}\phi_b^{2(t-k_c-1)}
 +2y_{k_c}\sum_{k_r-M_T+1}^{k_r}\sum_{j=k_c+1}^{t-1}\phi_b^{(2t-k_c-j-2)}u_j \nonumber \\
& &
  + \sum_{k_r-M_T+1}^{k_r}\left(\sum_{j=k_c+1}^{t-1}\phi_b^{(t-j-1)}u_j\right)^2.
 \label{lemma:3:y2-1-d}
\EEQ
The order of the first term of \eqref{lemma:3:y2-1-d} is given by
\BEQA
 y_{k_c}^2\sum_{k_r-M_T+1}^{k_r}\phi_b^{2(t-k_c-1)}
 & = &
 O_p(T\phi_a^{2(k_c-k_e)})\frac{\phi_b^{2(k_r-k_c-M_T)}-\phi_b^{2(k_r-k_c)}}{1-\phi_b^2} \\
 &\sim_a & 
 O_p(T\phi_a^{2(k_c-k_e)})\frac{T^b}{2c_b}\phi_b^{2(k_r-k_c)}(\phi_b^{-2M_T}-1) \\
 & \sim_a &
 O_p(T\phi_a^{2(k_c-k_e)})\frac{T^b}{2c_b}\phi_b^{2(k_r-k_c)}2c_b\frac{M_T}{T^b} \\
 & = &
 O_p(TM_T\phi_a^{2(k_c-k_e)}\phi_b^{2(k_r-k_c)}),
 \EEQA
 where the second last relation holds because
\BEQ
\phi_b^{-2M_T}-1
& = &
\left(1-\frac{c_b}{T^b}\right)^{-2M_T}-1 \nonumber \\
& = &
\left(1-2M_T\frac{-c_b}{T^b}+O\left(\frac{M_T^2}{T^{2b}}\right)\right)-1 \nonumber \\
& = &
2c_b\frac{M_T}{T^b}(1+o(1))
\label{phiMTb}
\EEQ
by applying \eqref{expand:a}. For the third term of \eqref{lemma:3:y2-1-d}, we have
\[
E\left[\sum_{k_r-M_T+1}^{k_r}\left(\sum_{j=k_c+1}^{t-1}\phi_b^{(t-j-1)}u_j\right)^2\right]
\leq \bar{\omega}^2 \sum_{k_r-M_T+1}^{k_r}\sum_{j=k_c+1}^{t-1} \phi_b^{2(t-j-1)} =O_p(T^bM_T),
\]
which is the smaller order than the first term when $a < b$. Because the second term of \eqref{lemma:3:y2-1-d} is shown to be dominated by the first term using Cauchy-Schwarz inequality, we have \eqref{lemma:3:y2d}.

For \eqref{lemma:3:yu}, we have, by \eqref{lemma:3:y-1},
\BE
\sum_{k_c+1}^{k_r}y_{t-1}u_t
=
y_{k_c}\sum_{k_c+1}^{k_r}\phi_b^{(t-k_c-1)}u_t
+\sum_{k_c+1}^{k_r}\left(\sum_{j=k_c+1}^{t-1}\phi_b^{(t-j-1)}u_j\right)u_t. \label{lemma:3:yu-1}
\EE
From direct calculation, we obtain
\[
Var\left[\sum_{k_c+1}^{k_r}\phi_b^{(t-k_c-1)}u_t\right]
\leq \bar{\omega}^2\sum_{k_c+1}^{k_r}\phi_b^{2(t-k_c-1)}=O(T^b),
\]
which implies that the first term on the right-hand side of \eqref{lemma:3:yu-1} is $(O_p(T^{(b+1)/2}\phi_a^{(k_c-k_e)})$. On the other hand, the variance of the second term becomes
\BEQA
Var\left[\sum_{k_c+1}^{k_r}\left(\sum_{j=k_c+1}^{t-1}\phi_b^{(t-j-1)}u_j\right)u_t\right]
& \leq &
\bar{\omega}^4\sum_{k_c+1}^{k_r}\sum_{j=k_c+1}^{t-1}\phi_b^{2(t-j-1)} \\
& = &
\bar{\omega}^4\sum_{k_c+1}^{k_r}\frac{1-\phi_b^{2(t-k_c-1)}}{1-\phi_b^2}\leq CT^{b+1}.
\EEQA
Then, the second term on the right-hand side of \eqref{lemma:3:yu-1} is $O_p(T^{(b+1)/2})$ and dominated by the first term. We then have \eqref{lemma:3:yu}.

Similarly to \eqref{lemma:3:yu}, we have, for \eqref{lemma:3:yua},
\[
\sum_{k_c+1}^{k}y_{t-1}u_t
=
y_{k_c}\sum_{k_c+1}^{k}\phi_b^{(t-k_c-1)}u_t
+\sum_{k_c+1}^{k}\left(\sum_{j=k_c+1}^{t-1}\phi_b^{(t-j-1)}u_j\right)u_t.
\]
The first term is at least $O_p(y_{kc})=O_p(T^{1/2}\phi_a^{(k_c-k_r)})$ whereas, by the HRI,
\BEQA
P\left(\max_{k_c+1\leq k\leq k_r}\left|\sum_{k_c+1}^k\left(\sum_{j=k_c+1}^{t-1}\phi_b^{(t-j-1)}u_j\right)u_{t}\right|\geq N\right)
& \leq &
\frac{\bar{\omega}^4}{N^2}\sum_{k_c+1}^{k_r}\sum_{j=k_c+1}^{t-1}\phi_b^{2(t-j-1)} \\
& \leq &
\frac{C}{N^2}T^{b+1}
\EEQA
and thus, the first term is dominant for all $k_c+1\leq k\leq k_r$. In addition, we have, again using the HRI,
\BEQA
P\left(\max_{k_c+1\leq k\leq k_r}\left|\sum_{k_c+1}^{k}\phi_b^{(t-k_c-1)}u_t\right|\geq N\right)
& \leq &
\frac{\bar{\omega}^2}{N^2}\sum_{k_c+1}^{k_r}\phi_b^{2(t-k_c-1)} \\
& \leq &
\frac{C}{N^2}T^{b}.
\EEQA
Then, because $y_{k_c}=O_p(T^{1/2}\phi_a^{(k_c-k_e)})$, we obtain \eqref{lemma:3:yua}.

\eqref{lemma:3:yub} is proved in the same manner as \eqref{lemma:2:yub}.

\eqref{lemma:3:yuc} is obtained by noting that
\[
\sum_{k_c+1}^{k_c+M_T}y_{t-1}u_t\sim_a y_{k_c}\sum_{k_c+1}^{k_c+M_T}\phi_b^{t-k_c-1}u_t
\]
and by evaluating the variance of the second term.$\blacksquare$

\begin{lemma}
\label{lemma:4}
For $k_r+1\leq k \leq T$, 
\BE
y_T = O_p(T^{1/2}\phi_a^{(k_c-k_e)}\phi_b^{(k_r-k_c)})+O_p(T^{1/2}).
\label{lemma:4:y}
\EE
When $a < b$,
\BEQ
\sum_{k_r+1}^{T}y_{t-1}^2 & = & O_p(T^2\phi_a^{2(k_c-k_e)}\phi_b^{2(k_r-k_c)}). \label{lemma:4:y2-a}\\
\sum_{k_r+1}^{k_r+M_T}y_{t-1}^2 & = & O_p(TM_T\phi_a^{2(k_c-k_e)}\phi_b^{2(k_r-k_c)}). \label{lemma:4:y2c-a}\\
\sum_{k_r+1}^{T}y_{t-1}u_t & = & O_p(T\phi_a^{(k_c-k_e)}\phi_b^{(k_r-k_c)}). \label{lemma:4:yu-a} \\
\sum_{k_r+1}^{k}y_{t-1}u_t & = & O_p(T\phi_a^{(k_c-k_e)}\phi_b^{(k_r-k_c)}) \;\mbox{uniformly over $k_r+1\leq k\leq T$}. \label{lemma:4:yua-a} \\
\sum_{k+1}^{T}y_{t-1}u_t & = & O_p(T\phi_a^{(k_c-k_e)}\phi_b^{(k_r-k_c)}) \;\mbox{uniformly over $k_r+1\leq k\leq T$}. \label{lemma:4:yub-a}
\EEQ
When $a > b$,
\BEQ
\sum_{k_r+1}^{T}y_{t-1}^2 & = & O_p(T^2). \label{lemma:4:y2-b}\\
\sum_{k_r+1}^{T}y_{t-1}u_t & = & O_p(T). \label{lemma:4:yu-b} \\
\sum_{k_r+1}^{k}y_{t-1}u_t & = & O_p(T) \;\mbox{uniformly over $k_r+1\leq k\leq T$.} \label{lemma:4:yua-b}
\EEQ
\end{lemma}
{\bf Proof of Lemma \ref{lemma:4}}: 
We first note that for $k_r+1\leq k\leq T$,
\BE
y_k=y_{k_r}+c_1\frac{k-k_r}{T^{\eta_1}}+\sum_{j=k_r+1}^k\varepsilon_j.
\label{lemma:4:y-1}
\EE
In particular, we have $y_T=y_{k_r}+c_1(T-k_r)/T^{\eta_1}+\sum_{j=k_r+1}^T\varepsilon_j$; thus, we obtain \eqref{lemma:4:y} using \eqref{lemma:3:y}.

When $a < b$, the first term on the right-hand side of \eqref{lemma:4:y-1} dominates the second and third terms uniformly over $k_r+1\leq k \leq T$ and thus using \eqref{lemma:3:y},
\[
\sum_{k_r+1}^Ty_{t-1}^2\sim_a (T-k_r)y_{k_r}^2=O_p(T^2\phi_a^{2(k_c-k_e)}\phi_b^{2(k_r-k_c)})
\]
\[
\sum_{k_r+1}^{k_r+M_T}y_{t-1}^2\sim_a M_T y_{k_r}^2=O_p(TM_T\phi_a^{2(k_c-k_e)}\phi_b^{2(k_r-k_c)})
\]
and
\[
\sum_{k_r+1}^Ty_{t-1}u_t\sim_a y_{k_r}\sum_{k_r+1}^Tu_t=O_p(T\phi_a^{(k_c-k_e)}\phi_b^{(k_r-k_c)}).
\]

\eqref{lemma:4:yua-a} and \eqref{lemma:4:yub-a} are obtained similarly by the FCLT.

When $a > b$,
\BEQA
\sum_{k_r+1}^Ty_{t-1}^2
& = &
(T-k_r)y_{k_r}^2+c_1^2\sum_{k_r+1}^T\left(\frac{t-k_r-1}{T^{\eta_1}}\right)^2+\sum_{k_r+1}^T\left(\sum_{j=k_r+1}^{t-1}\varepsilon_j\right)^2 \\
& &
+2c_1y_{k_r}\sum_{k_r+1}^T\frac{t-k_r-1}{T^{\eta_1}}+2y_{k_r}\sum_{k_r+1}^T\left(\sum_{j=k_r+1}^{t-1}\varepsilon_j\right)+2c_1\sum_{k_r+1}^T\frac{t-k_r-1}{T^{\eta_1}}\left(\sum_{j=k_r+1}^{t-1}\varepsilon_j\right) \\
& = &
O_p(T^{b+1})+O_p(T^{3-2\eta_1})+O_p(T^2)+O_p(T^{b/2+2-\eta_1})+O_p(T^{(b+3)/2})+O_p(T^{3/2-\eta_1}) \\ 
& = & 
O_p(T^2),
\EEQA
 and 
\BEQA
\sum_{k_r+1}^Ty_{t-1}u_t
& = &
y_{k_r}\sum_{k_r+1}^{T}u_t+c_1\sum_{k_r+1}^T\frac{t-k_r-1}{T^{\eta_1}}u_t+\sum_{k_r+1}^T\left(\sum_{j=k_r+1}^{t-1}u_j\right)u_t \\
& = &
O_p(T^{(b+1)/2})+O_p(T^{3/2-\eta_1})+O_p(T)=
O_p(T).
\EEQA

\eqref{lemma:4:yua-b} is obtained similarly by the FCLT.$\blacksquare$

\begin{lemma}
\label{lemma:uniform:b}
Under Assumption \ref{assumption:iid}, we have for $a > b$,
\BEQ
\max_{k\in E_{b1T}}\frac{\sum_{k+1}^{k_r}{y_{t-1}u_t}}{\sum_{k+1}^{k_r}{y_{t-1}^2}}
& = &
o_p\left(\frac{1}{T^b}\right), \label{lemma:uniform:a2} \\
\max_{k\in E_{b2T}}\frac{\sum_{k_r+1}^{k}{y_{t-1}u_t}}{\sum_{k_r+1}^{k}{y_{t-1}^2}}
& = &
o_p\left(\frac{1}{T^b}\right). \label{lemma:uniform:a5}
\EEQ
\end{lemma}
{\bf Proof of Lemma \ref{lemma:uniform:b}}: 
We first note that, from \eqref{lemma:3:y-1}, we can observe that
\BE
y_k\sim_a\phi_a^{(k_c-k_e)}\phi_b^{(k-k_c)}y_{k_e}+\sum_{j=k_c+1}^k\phi_b^{(k-j)}u_j=\sum_{j=k_c+1}^k\phi_b^{(k-j)}u_j+o_p(1),
\label{lemma:uniform:yk}
\EE
uniformly over $k\in E_{b1T}$ because
\[
\max_{k\in E_{b1T}}\left|\phi_a^{(k_c-k_e)}\phi_b^{(k-k_c)}y_{k_e}\right|\leq \phi_a^{(k_c-k_e)}\phi_b^{(\underline{k}_r-k_c)}|O_p(\sqrt{T})|=o_p(1)
\]
for $a > b$. Thus, to prove \eqref{lemma:uniform:a2}, it is sufficient to show that
\BE
\max_{k\in E_{b1T}}T^b \frac{\sum_{k+1}^{k_r}{y_{t-1}u_t}}{\sum_{k+1}^{k_r}{y_{t-1}^2}}
\sim_a
\max_{k\in E_{b1T}}\frac{\displaystyle \frac{1}{(k_r-k)}\sum_{k+1}^{k_r}\sum_{j=k_c+1}^{t-1}\phi_b^{t-j-1}u_ju_t}{\displaystyle \frac{1}{T_b(k_r-k)}\sum_{k+1}^{k_r}\left(\sum_{j=k_c+1}^{t-1}\phi_b^{t-j-1}u_j\right)^2}
=o_p(1).
\label{lemma:uniform:a2-1}
\EE
We first show that the denominator converges to $\sigma^2/(2c_b)$ uniformly over $k\in E_{b1T}$. Note that
\BE
\sum_{k+1}^{k_r}\left(\sum_{j=k_c+1}^{t-1}\phi_b^{t-j-1}u_j\right)^2
= 
\sum_{k+1}^{k_r}\sum_{j=k_c+1}^{t-1}\phi_b^{2(t-j-1)}u_j^2+2\sum_{k+1}^{k_r}\sum_{j=k_c+1}^{t-2}\sum_{\ell=j+1}^{t-1}\phi_b^{t-j-1}\phi_b^{t-\ell-1}u_ju_{\ell}.
\label{lemma:uniform:a2-1:d1}
\EE
The first term on the right-hand side of \eqref{lemma:uniform:a2-1:d1} is further decomposed into
\BE
\sum_{j=k_c+1}^{k}\sum_{t=k+1}^{k_r}\phi_b^{2(t-j-1)}u_j^2+\sum_{j=k+1}^{k_r-1}\sum_{t=j+1}^{k_r}\phi_b^{2(t-j-1)}u_j^2\eqqcolon D_1+D_2
\label{lemma:uniform:a2-1:d2}
\EE
by changing the order of the summations. Note that
\BEQA
D_1
& = &
\sum_{j=k_c+1}^k\frac{\phi_b^{2(k-j)}-\phi_b^{2(k_r-j)}}{1-\phi_b^2}u_j^2 \\
& \leq &
\sum_{j=k_c+1}^k\frac{\phi_b^{2(k-j)}}{1-\phi_b^2}u_j^2 \\
& \sim_a &
\frac{T^b}{2c_b}\left[\sigma^2\sum_{j=k_c+1}^{k}\phi_b^{2(k-j)}+\sum_{j=k_c+1}^{k}\phi_b^{2(k-j)}(u_j^2-\sigma^2)\right] \\
& \sim_a &
\frac{T^b}{2c_b}\left[O(T^b)+\sum_{j=k_c+1}^{k}\phi_b^{2(k-j)}(u_j^2-\sigma^2)\right]
\EEQA
uniformly over $k\in E_{b1T}$. Because
\[
\sum_{j=k_c+1}^{k}\phi_b^{2(k-j)}(u_j^2-\sigma^2)=\phi_b^{2(k-k_c)}\sum_{j'=1}^{k-k_c}\phi_b^{-2j'}(u_{k_c+j'}^2-\sigma^2),
\]
and $\phi_b^{2(k-k_c)}/(k_r-k)=\phi_b^{2(k-k_c)}/[(k_r-k_c)-(k-k_c)]$ can be shown to be a decreasing function of $k-k_c$ for all sufficiently large $T$, we have, using the HRI,
\BEQA
\lefteqn{
P\left(\max_{k\in E_{b1T}}\frac{\phi_b^{2(k-k_c)}}{k_r-k}\left|\sum_{j'=1}^{k-k_c}\phi_b^{-2j'}(u_{k_c+j'}^2-\sigma^2)\right| \geq N\right)
} \\
& \leq &
\frac{C}{N^2}\left(\frac{\phi_b^{4(\underline{k}_r-k_c)}}{(k_r-\underline{k}_r)^2}\sum_{j'=1}^{\underline{k}_r-k_c}\phi_b^{-4j'}+\sum_{j'=\underline{k}_r-k_c+1}^{k_r-k_c-M_{bT}}\frac{1}{(k_r-k_c-j')^2}\phi_b^{4j'}\phi_b^{-4j'}\right) \\
& \leq &
\frac{C}{N^2}\left[\frac{\phi_b^{4(\underline{k}_r-k_c)}}{(k_r-\underline{k}_r)^2}\frac{1-\phi_b^{-4(\underline{k}_r-k_c)}}{\phi_b^4-1}+\sum_{l=M_{bT}}^{k_r-\underline{k}_r}\frac{1}{l^2}\right]
\leq
\frac{C}{N^2}\left[O\left(\frac{1}{T^{2-b}}\right)+O\left(\frac{1}{M_{bT}}\right)\right],
\EEQA
because $\underline{k}_r-k_c\leq k-k_c\leq k_r-k_c-M_{bT}$ for $k\in E_{b1T}$, where we used the relation of $\sum_{l=M_{bT}}^{M'}1/l^2=O(1/M_{bT})$ for $M' > M_{bT}$. Thus, the second term in the last expression of $D_1$ divided by $k_r-k$ is $o_p(1)$ uniformly over $k\in E_{b1T}$. Because $k_r-k\geq M_{bT}$ for $k\in E_{b1T}$, we can observe that
\BE
\max_{k\in E_{b1T}}\frac{1}{T_b(k_r-k)}D_1\leq \left[O\left(\frac{T^b}{M_{bT}}\right)+o_p(1)\right]=o_p(1).
\label{lemma:uniform:D1}
\EE
On the other hand, $D_2$ is decomposed into
\BEQA
D_2
& = &
\sum_{j=k+1}^{k_r-1}\frac{1-\phi_b^{2(k_r-j)}}{1-\phi_b^2}u_j^2 \\
& \sim_a &
\frac{T^b}{2c_b}\left[\sigma^2(k_r-k-1)+\sum_{j=k+1}^{k_r-1}(u_j^2-\sigma^2)-\sigma^2\sum_{j=k+1}^{k_r-1}\phi_b^{2(k_r-j)}-\sum_{j=k+1}^{k_r-1}\phi_b^{2(k_r-j)}(u_j^2-\sigma^2)\right].
\EEQA
Clearly, the first term in the brackets divided by $k_r-k$ goes to $\sigma^2$, while the third term is $O(T^b)$ uniformly over $k\in E_{b1T}$. For the second term, by applying the HRI in the reverse order, we have
\BEQA
P\left(\max_{k\in E_{b1T}}\frac{1}{k_r-k} \sum_{j=k+1}^{k_r-1}(u_j^2-\sigma^2) \geq N\right) 
& \leq &
\frac{C}{N^2}\left[\frac{M_{bT}-1}{M_{bT}^2}+\sum_{j=M_{bT}}^{k_r-\underline{k}_r}\frac{1}{j^2}\right]\leq \frac{C}{N^2}\frac{1}{M_{bT}}
\EEQA
and thus the second term divided by $k_r-k$ uniformly converges in probability to zero. Similarly, the fourth term in the brackets divided by $k_r-k$ is show to be $o_p(1)$ uniformly by applying the HRI in the reverse order. As a result,
\BE
\frac{1}{T^b(k_r-k)}D_2\CP\frac{\sigma^2}{2c_b}
\label{lemma:uniform:D2}
\EE
uniformly over $k\in E_{b1T}$. By \eqref{lemma:uniform:D1} and \eqref{lemma:uniform:D2}, for the first term on the right hand side of \eqref{lemma:uniform:a2-1:d1}, we have
\[
\frac{1}{T^b(k_r-k)}\sum_{k+1}^{k_r}\sum_{j=k_c+1}^{t-1}\phi_b^{2(t-j-1)}u_j^2\CP \frac{\sigma^2}{2c_b}
\]
uniformly over $k\in E_{b1T}$.

Next, we will show that the second term on the right-hand side of \eqref{lemma:uniform:a2-1:d1} divided by $T^b(k_r-k)$ converges in probability to 0 uniformly. By changing the order of the summations, we can observe that
\BEQ
\lefteqn{
\sum_{t=k+1}^{k_r}\sum_{j=k_c+1}^{t-2}\sum_{\ell=j+1}^{t-1}\phi_b^{t-j-1}\phi_b^{t-\ell-1}u_ju_{\ell}
} \nonumber \\
&= &
\sum_{j=k_c+1}^{k-1}\sum_{t=k+1}^{k_r}\sum_{\ell=j+1}^{t-1}\phi_b^{2t-j-\ell-2}u_ju_{\ell}+\sum_{j=k}^{k_r-2}\sum_{t=j+2}^{k_r}\sum_{\ell=j+1}^{t-1}\phi_b^{2t-j-\ell-2}u_ju_{\ell} \nonumber \\
& = &
\sum_{j=k_c+1}^{k-1}\sum_{\ell=j+1}^{k}\sum_{t=k+1}^{k_r}\phi_b^{2t-j-\ell-2}u_ju_{\ell}
+\sum_{j=k_c+1}^{k-1}\sum_{\ell=k+1}^{k_r-1}\sum_{t=\ell+1}^{k_r}\phi_b^{2t-j-\ell-2}u_ju_{\ell} \nonumber \\
& &
+\sum_{j=k}^{k_r-2}\sum_{\ell=j+1}^{k_r-1}\sum_{t=\ell+1}^{k_r}\phi_b^{2t-j-\ell-2}u_ju_{\ell}.
\label{lemma:uniform:a2-1:d1-1}
\EEQ
The first term on the right-hand side of \eqref{lemma:uniform:a2-1:d1-1} becomes
\BEQ
\lefteqn{
\sum_{j=k_c+1}^{k-1}\sum_{\ell=j+1}^{k}\sum_{t=k+1}^{k_r}\phi_b^{2t-j-\ell-2}u_ju_{\ell}
} \nonumber \\
& = &
\sum_{j=k_c+1}^{k-1}\sum_{\ell=j+1}^{k}\frac{\phi_b^{2k-j-\ell}-\phi_b^{2k_r-j-\ell}}{1-\phi_b^2}u_ju_{\ell} \nonumber \\
& \sim_a &
\frac{T^b}{2c_b}\left[\sum_{\ell=k_c+2}^{k}\phi_b^{2k-\ell}u_{\ell}\sum_{j=k_c+1}^{\ell-1}\phi_b^{-j}u_j-\sum_{\ell=k_c+2}^{k}\phi_b^{k_r-\ell}u_{\ell}\sum_{j=k_c+1}^{\ell-1}\phi_b^{k_r-j}u_j\right],\label{lemma:uniform:a2-1:d1-2}
\EEQ
where the last relation is obtained by changing the order of the summations. By letting $\xi_{1\ell}\coloneqq u_{\ell}\sum_{j=k_c+1}^{\ell-1}\phi_b^{-j}u_j$, the first term in the brackets can be expressed as
\BEQA
\sum_{\ell=k_c+2}^{k}\phi_b^{2k-\ell}\xi_{1\ell}
& = &
\phi_b^{2(k-k_c)}\sum_{\ell'=1}^{k-k_c-1}\phi_b^{k_c-\ell'-1}\xi_{1k_c+1+\ell'}.
\EEQA
Noting that $\phi_b^{k_c-\ell'-1}\xi_{1k_c+\ell'}$ forms a martingale difference sequence and that $\underline{k}_r-k_c\leq k-k_c\leq k_r-k_c-M_{bT}$ for $k\in E_{b1T}$, we have
\BEQA
\lefteqn{
P\left(\max_{k\in E_{b1T}}\left|\frac{\phi_b^{2(k-k_c)}}{k_r-k}\sum_{\ell'=1}^{k-k_c-1}\phi_b^{k_c-\ell'-1}\xi_{1k_c+1+\ell'}\right| \geq N\right)
} \\
& \leq &
\frac{1}{N^2}\left[
\frac{\phi_b^{4(\underline{k}_r-k_c)}}{(k_r-\underline{k}_r)^2}\sum_{\ell'=1}^{\underline{k}_r-k_c-1}\phi_b^{2(k_c-\ell'-1)}E[\xi_{k_c+1+\ell'}^2]
+
\sum_{\ell'=\underline{k}_r-k_c}^{k_r-k_c-M_{bT}}\frac{\phi_b^{4\ell'}}{(k_r-k_c-\ell')^2}\phi_b^{2(k_c-\ell'-1)}E[\xi_{1k_c+1+\ell'}^2]
\right] \\
& \leq &
\frac{C}{N^2}\left[
\frac{\phi_b^{4(\underline{k}_r-k_c)}}{(k_r-\underline{k}_r)^2}\sum_{\ell'=1}^{\underline{k}_r-k_c-1}\phi_b^{2(k_c-\ell'-1)}T^b\phi_b^{-2(k_c+\ell'+1)}
+
\sum_{\ell'=\underline{k}_r-k_c}^{k_r-k_c-M_{bT}}\frac{\phi_b^{4\ell'}}{(k_r-k_c-\ell')^2}\phi_b^{2(k_c-\ell'-1)}T^b\phi_b^{-2(k_c+\ell'+1)}
\right] \\
& \leq &
\frac{C}{N^2}\left[\frac{T^{2b}}{(k_r-\underline{k}_r)^2}+\frac{T^b}{M_{bT}}\right]=o(1),
\EEQA
where the first inequality follows by the HRI by noting that $\phi_b^{2(k-k_c)}/(k_r-k)=\phi_b^{2(k-k_c)}/[(k_r-k_c)-(k-k_c)]$ is a decreasing function of $k-k_c$ for all sufficiently large $T$ and the second inequality holds because $E[\xi_{k_c+1+\ell'}^2]\leq CT^b\phi_b^{-2(k_c+\ell'+1)}$. This implies that the first term in the brackets in \eqref{lemma:uniform:a2-1:d1-2} divided by $k_r-k$ converges in probability to zero uniformly over $k\in E_{b1T}$. By a similar argument, the second term divided by $k_r-k$ is shown to be $o_p(1)$. As a result, for the first term on the right hand side of \eqref{lemma:uniform:a2-1:d1-1}, we have
\[
\frac{1}{T^b(k_r-k)}\sum_{j=k_c+1}^{k-1}\sum_{\ell=j+1}^{k}\sum_{t=k+1}^{k_r}\phi_b^{2t-j-\ell-2}u_ju_{\ell}\CP 0
\]
uniformly over $k\in E_{1T}$.

We next investigate the second term of \eqref{lemma:uniform:a2-1:d1-1}. Note that
\BE
\sum_{j=k_c+1}^{k-1}\sum_{\ell=k+1}^{k_r-1}\sum_{t=\ell+1}^{k_r}\phi_b^{2t-j-\ell-2}u_ju_{\ell}
\sim_a
\frac{T_b}{2c_b}\left[
\sum_{j=k_c+1}^{k-1}\sum_{\ell=k+1}^{k_r-1}\phi_b^{\ell-j}u_ju_{\ell}-\sum_{j=k_c+1}^{k-1}\sum_{\ell=k+1}^{k_r-1}\phi_b^{2k_r-j-\ell}u_ju_{\ell}
\right].
\label{lemma:uniform:a2-1:d1-3}
\EE
We first show that the first term on the right hand side of the brackets divided by $k_r-k$ is uniformly $o_p(1)$ outside of the $T^{2b}$ neighborhood of $k_r$, and then refinement will be performed. Let $M_{b1T}\to \infty$ with $T^{2b}/M_{b1T}\to 0$. Then, we can observe that
\BEQ
\lefteqn{
P\left(\max_{\underline{k}_r+1\leq k \leq k_r-M_{b1T}}\frac{1}{k_r-k}\left|\sum_{j=k_c+1}^{k-1}\sum_{\ell=k+1}^{k_r-1}\phi_{b}^{\ell-j}u_ju_{\ell}\right|\geq N\right)
} \nonumber \\
& \leq &
\sum_{k=\underline{k}_r+1}^{k_r-M_{b1T}}P\left(\frac{1}{k_r-k}\left|\sum_{j=k_c+1}^{k-1}\sum_{\ell=k+1}^{k_r-1}\phi_{b}^{\ell-j}u_ju_{\ell}\right|\geq N\right) \nonumber  \\
& \leq &
\frac{C}{N^2}\sum_{k=\underline{k}_r+1}^{k_r-M_{b1T}}\frac{1}{(k_r-k)^2}T^{2b} \nonumber \\
& \leq &
\frac{C}{N^2}\frac{T^{2b}}{M_{b1T}}=o(1).
\label{lemma:uniform:a2-1:d1-1a}
\EEQ
Next,  let $V_{k}\coloneqq \sum_{j=k_c+1}^{k}\phi_b^{-j}u_j$ and $W_{k}=\sum_{\ell=k}^{k_r-1}\phi_b^{\ell}u_{\ell}$. Because we obtained \eqref{lemma:uniform:a2-1:d1-1a}, we focus on the rage of $k$ given by $k_r-C_2T^{2b}\leq k \leq k_r-M_{b2T}$ for a sufficiently large $C_2$ where $M_{b2T}\to\infty$ and $T^{3b/2}/M_{b2T}\to 0$. Then,
\BEQ
\lefteqn{
\sum_{j=k_c+1}^{k-1}\sum_{\ell=k+1}^{k_r-1}\phi_{b}^{\ell-j}u_ju_{\ell}
} \nonumber \\
& = &
V_{k-1}W_{k+1} \nonumber \\
& = &
V_{k-1}(W_{k+1}-W_{k})+(V_{k-1}-V_{k-2})W_{k}+V_{k-2}W_{k} \nonumber \\
& = &
\sum_{k_r-C_2T^{2b}-1}^{k-1}V_t(W_{t+2}-W_{t+1})+\sum_{k_r-C_2T^{2b}-1}^{k-1}(V_t-V_{t-1})W_{t+1}+V_{k_r-C_2T^{2b}-2}W_{k_r-C_2T^{2b}}.
\nonumber 
\EEQ
The first term becomes, by noting that $W_{t+2}-W_{t+1}=-\phi_b^{(t+1)} u_{t+1}$,
\[
\sum_{k_r-C_2T^{2b}-1}^{k-1}V_t(W_{t+2}-W_{t+1})
=
-\sum_{k_r-C_2T^{2b}-1}^{k-1}u_{t+1}(\sum_{j=k_c+1}^{t-1}\phi_b^{t+1-j}u_j)
\]
 and thus by the HRI,
\[
P\left(\max_{k_r-C_2T^{2b}\leq k \leq k_r-M_{b2T}}\left|\sum_{k_r-C_2T^{2b}-1}^{k-1}V_t(W_{t+2}-W_{t+1})\right| \geq N\right)
\leq 
\frac{C}{N^2}\sum_{k_r-C_2T^{2b}-1}^{k_r-M_{b2T}}T^b\leq \frac{C}{N^2}T^{3b}.
\]
This implies that 
\[
\max_{k_r-CT^{2b}\leq k \leq k_r-M_{b2T}}\frac{1}{k_r-k}\left|\sum_{k_r-CT^{2b}-1}^{k-1}V_t(W_{t+2}-W_{t+1})\right|
=
O_p\left(\frac{T^{3b/2}}{M_{b2T}}\right)=o_p(1).
\]
Similarly, because
\BEQA
\lefteqn{
\max_{k_r-C_2T^{2b}\le k \leq k_r-M_{b2T}}\left|\sum_{k_r-C_2T^{2b}-1}^{k-1}(V_t-V_{t-1})W_{t+1}\right|
} \\
& \leq &
\left|\sum_{k_r-C_2T^{2b}-1}^{k_r-M_{b2T}-1}(V_t-V_{t-1})W_{t+1}\right|+
\max_{k_r-C_2T^{2b}\le k \leq k_r-M_{b2T}}\left|\sum_{k}^{k_r-M_{2bT}-1}(V_t-V_{t-1})W_{t+1}\right|
\EEQA
and $V_t-V_{t-1}=\phi_b^{-t}u_t$, we can show that, by using the HRI in the reverse order,
\[
\max_{k_r-C_2T^{2b}\leq k \leq k_r-M_{b2T}}\frac{1}{k_r-k}\left|\sum_{k_r-C_2T^{2b}-1}^{k-1}(V_t-V_{t-1})W_{t+1}\right|=o_p(1).
\]
On the other hand, from the definitions of $V_t$ and $W_t$, we can easily show that
\[
E\left[\left(V_{k_r-C_2T^{2b}-2}W_{k_r-C_2T^{2b}}\right)^2\right]=O(T^{2b})
\]
and thus $W_{k_r-C_2T^{2b}-2}W_{k_r-C_2T^{2b}}/M_{bT}=o_p(1)$. Therefore, by combining these results with \eqref{lemma:uniform:a2-1:d1-1a}, we have
\BE
\max_{\underline{k}_r+1\leq k\leq k_r-M_{b2T}}\frac{1}{k-k_r}\left|\sum_{j=k_c+1}^{k-1}\sum_{\ell=k+1}^{k_r-1}\phi_{b}^{\ell-j}u_ju_{\ell}\right|=o_p(1).
\label{lemma:uniform:a2-1:d1-1b}
\EE

Next, we focus on the rage of $k$ given by $k_r-C_3T^{3b/2}\leq k \leq k_r-M_{b3T}$ for a sufficiently large $C_3$, where $M_{b3T}\to\infty$ and $T^{5b/4}/M_{b3T}\to 0$. Then, by using the same argument, we can observe that \eqref{lemma:uniform:a2-1:d1-1b} holds with $M_{b2T}$ replaced by $M_{b3T}$. By repeating the same argument, we can observe that 
\BE
\max_{k\in E_{b2T}}\frac{1}{k-k_r}\left|\sum_{j=k_c+1}^{k-1}\sum_{\ell=k+1}^{k_r-1}\phi_{b}^{\ell-j}u_ju_{\ell}\right|=o_p(1).
\label{lemma:uniform:a2-1:d1-1c}
\EE
where $M_{bT}\to \infty$ and $T^{b+\epsilon}/M_{bT}\to 0$ for a given $\epsilon >0$.

On the other hand, the order of the second term of \eqref{lemma:uniform:a2-1:d1-3} is shown to be $o_p(1)$ because
\BEQA
\lefteqn{
P\left(\max_{\underline{k}_r+1\leq k \leq k_r-M_{bT}}\frac{1}{k_r-k}\left|\sum_{j=k_c+1}^{k-1}\sum_{\ell=k+1}^{k_r-1}\phi_b^{2k_r-j-\ell}u_ju_{\ell}\right|\geq N\right)
} \\
& \leq &
\sum_{k=\underline{k}_r+1}^{k_r-M_{bT}}P\left(\frac{1}{k_r-k}\left|\sum_{j=k_c+1}^{k-1}\sum_{\ell=k+1}^{k_r-1}\phi_b^{2k_r-j-\ell}u_ju_{\ell}\right|\geq N\right) \\
& \leq &
\frac{C}{N^2}\sum_{k=\underline{k}_r+1}^{k_r-M_{bT}}\frac{1}{(k_r-k)^2}\frac{\phi_b^{2(k_r-k+1)}-\phi_b^{2(k_r-k_c)}}{1-\phi_b^2}\frac{\phi_b^2-\phi_b^{2(k_r-k)}}{1-\phi_b^2} \\
& \leq &
\frac{C}{N^2}\frac{T^{2b}}{M_{bT}}\phi_b^{2M_{bT}}=o(1).
\EEQA
Therefore, for the second term on the right hand side of \eqref{lemma:uniform:a2-1:d1-1},
\[
\frac{1}{T^b(k_r-k)} \sum_{j=k_c+1}^{k-1}\sum_{\ell=k+1}^{k_r-1}\sum_{t=\ell+1}^{k_r}\phi_b^{2t-j-\ell-2}u_ju_{\ell} \CP 0
\]
 uniformly over $k\in E_{1T}$.

 The third term on the right hand side of \eqref{lemma:uniform:a2-1:d1-1} becomes
 \BEQ
 \lefteqn{
 \sum_{j=k}^{k_r-2}\sum_{\ell=j+1}^{k_r-1}\sum_{t=\ell+1}^{k_r}\phi_b^{2t-j-\ell-2}u_ju_{\ell}
 } \nonumber \\
 & = &
  \sum_{j=k}^{k_r-2}\sum_{\ell=j+1}^{k_r-1}\frac{\phi_b^{\ell-j}-\phi_b^{2k_r-j-\ell}}{1-\phi_b^2}u_ju_{\ell}
  \nonumber \\
& \sim_a &
\frac{T^b}{2c_b}\left[  \sum_{j=k}^{k_r-2}\sum_{\ell=j+1}^{k_r-1}\phi_b^{\ell-j}u_ju_{\ell}
-
\sum_{j=k}^{k_r-2}\sum_{\ell=j+1}^{k_r-1}\phi_b^{2k_r-j-\ell}u_ju_{\ell}
\right].
\label{lemma:uniform:as2-1_d1-1-1}
\EEQ
We express the first term in the brackets as
\BEQA
\sum_{j=k}^{k_r-2}\sum_{\ell=j+1}^{k_r-1}\phi_b^{\ell-j}u_ju_{\ell}
& = &
\sum_{t=1}^{k_r-k-1} u_{k_r-t-1}(\phi_b u_{k_r-t}+\phi_b^2u_{k_r-t+1}+\cdots+\phi_b^tu_{k_r-1}) \\
& \eqqcolon &
\sum_{t=1}^{k_r-k-1} \tilde{\xi}_{1t}.
\EEQA
Because $\tilde{\xi}_{1t}$ is a martingale difference sequence with the filtration in the reserve order and $E[\tilde{\xi}_{1t}^2]\leq CT^b$, by the HRI, we have
\BEQA
P\left(\max_{M_{bT}\leq k_r-k\leq k_r-\underline{k}_r}\frac{1}{k_r-k}\left|
\sum_{t=1}^{k_r-k-1} \tilde{\xi}_{1t}\right| \geq N\right)
& \leq &
\frac{C}{N^2}\left(\frac{1}{M_{bT}^2}\sum_{t=1}^{M_{bT}-1}T^b+\sum_{t=M_{bT}}^{k_r-\underline{k}_r}\frac{T^b}{t^2}\right) \\
& \leq & 
C\frac{T^b}{M_{bT}} =o(1).
\EEQA
A similar result can be obtained for the second term in the brackets, and thus, for the third term on the right-hand side of \eqref{lemma:uniform:a2-1:d1-1},
 \[
\frac{1}{T^b(k_r-k)} \sum_{j=k}^{k_r-2}\sum_{\ell=j+1}^{k_r-1}\sum_{t=\ell+1}^{k_r}\phi_b^{2t-j-\ell-2}u_ju_{\ell}
 \CP 0
 \]
uniformly over $k\in E_{1T}$. By combining these result, we can observe that the denominator of \eqref{lemma:uniform:a2-1} converges to $\sigma^2/(2c_b)$ uniformly over $k\in E_{b1T}$. 
 
The numerator of \eqref{lemma:uniform:a2-1} is expressed as, by changing the order of the summations,
\[
\sum_{k+1}^{k_r}\sum_{j=k_c+1}^{t-1}\phi_b^{t-j-1}u_ju_t
=
\sum_{j=k_c+1}^{k}\sum_{t=k+1}^{k_r}\phi_b^{t-j-1}u_ju_t+\sum_{j=k+1}^{k_r-1}\sum_{t=j+1}^{k_r}\phi_b^{t-j-1}u_ju_t.
\]
The first term on the right-hand side has the same structure as the first term in the brackets of \eqref{lemma:uniform:a2-1:d1-3}, and we have shown \eqref{lemma:uniform:a2-1:d1-1c}. On the other hand, the second term on the right-hand side is essentially the same as the first term in the brackets in \eqref{lemma:uniform:as2-1_d1-1-1}, which, divided by $k_r-k$, is $o_p(1)$ uniformly. Therefore, the numerator of \eqref{lemma:uniform:a2-1} is $o_p(1)$ uniformly and thus we obtain \eqref{lemma:uniform:a2-1}.

To prove \eqref{lemma:uniform:a5}, note that
\BE
 \max_{k\in E_{b2T}}\frac{\sum_{k_r+1}^{k}{y_{t-1}u_t}}{\sum_{k_r+1}^{k}{y_{t-1}^2}}
=
 \max_{k\in E_{b2T}}\frac{\displaystyle\frac{1}{(k-k_r)^2}\sum_{k_r+1}^{k}{y_{t-1}u_t}}{\displaystyle\frac{1}{(k-k_r)^2}\sum_{k_r+1}^{k}{y_{t-1}^2}}.
\label{lemma:uniform:a5-1:equiv}
\EE
 From \eqref{lemma:4:y-1}, we can observe that, for $a > b$,
 \[
 y_{k}\sim_a\sum_{j=k_c+1}^k u_j
 \]
 uniformly over $k\in E_{b2T}$ because $y_{k_r}=O_p(T^{b/2})$ by \eqref{lemma:3:y} for $a > b$. Thus, the denominator of \eqref{lemma:uniform:a5-1:equiv} becomes
 \[
 \frac{1}{(k-k_r)^2}\sum_{k_r+1}^{k}y_{t-1}^2\sim_a \frac{1}{(k-k_r)^2}\sum_{k_r+1}^{k-1}\left(\sum_{j=k_r+1}^tu_j\right)^2
 \eqqcolon
 \tilde{U}_{k}\geq 0.
 \]
 The tedious but direct calculation shows that 
\BE
E\left[\tilde{U}_{k}\right]=\frac{\sigma^2}{2}+o(1)
\quad\mbox{and}\quad
E\left[\tilde{U}_{k}^2\right]\leq C < \infty,
\label{lemma:uniform:a5:moment}
\EE
 where the $o(1)$ term is uniform over $k\in E_{b2T}$. In this case, the following relation holds:
\BE
\min_{k\in E_{b2T}}\tilde{U}_{k}\geq O_p(1).
\label{lemma:uniform:a5-1}
\EE
To prove this, we use the contradiction argument. Suppose that \eqref{lemma:uniform:a5-1} does not hold. Then, there exists $\tilde{k}\in E_{b2T}$ such that $\tilde{U}_{\tilde{k}}\CP 0$. This implies that, because $\tilde{U}_{\tilde{k}}$ is uniformly integrable by \eqref{lemma:uniform:a5:moment}, 
\[
\lim_{T\to \infty} E[\tilde{U}_{\tilde{k}}]=E\left[\lim_{T\to \infty} \tilde{U}_{\tilde{k}}\right]=0,
\]
which contradicts $E[\tilde{U}_{\tilde{k}}]=\sigma^2/2+o(1) > 0$. Thus, \eqref{lemma:uniform:a5-1}  must hold. This implies that
\BE
\frac{1}{\tilde{U}_{k}}\leq O_p(1)
\label{lemma:uniform:a5-2}
\EE
uniformly over $k\in E_{b2T}$. We next investigate the numerator of \eqref{lemma:uniform:a5-1:equiv}. Note that
\[
\frac{1}{(k-k_r)^2}\sum_{k_r+1}^{k}{y_{t-1}u_t}
\sim_a
\frac{1}{2(k-k_r)^2}\left(y_{k}^2-y_{k_r}^2-\sum_{k_r+1}^ku_t^2\right).
\]
From \eqref{lemma:4:y-1}, we have
\[
\frac{1}{k-k_r}y_k=\frac{1}{k-k_r}y_{k_r}+\frac{c_1}{T^{\eta_1}}+\frac{1}{k-k_r}\sum_{j=k_r+1}^k\varepsilon_j.
\]
By noting that for $a > b$,
\[
\max_{k\in E_{b2T}}\frac{1}{k-k_r}|y_{k_r}|\leq O_p\left(\frac{T^{b/2}}{M_{bT}}\right)=o_p\left(\frac{1}{T^{b/2}}\right),
\]
$c_1/T^{\eta_1}=o(1/T^{b/2})$ because $0<b<1$ and $\eta >1/2$, and
\BEQA
P\left(\max_{k\in E_{b2T}}\frac{1}{k-k_r}\left|\sum_{k_r+1}^k \varepsilon_t\right| \geq N\right)
&  \leq &
\frac{C}{N^2}\left(\frac{1}{M_{bT}^2}\sum_{1}^{M_{bT}}\sigma^2+\sum_{M_{bT}+1}^{T-\overline{k}_r}\frac{\sigma^2}{t^2}\right) \\
& \leq &
\frac{C}{N^2}\frac{1}{M_{bT}},
\EEQA
we can observe that
\[
\max_{k\in E_{b2T}}\frac{1}{k-k_r}|y_k|=o_p\left(\frac{1}{T^{b/2}}\right).
\]
Thus, \eqref{lemma:uniform:a5-1:equiv} is bounded by
\[
 \sup_{k\in E_{2T}}\frac{\frac{1}{(k-k_r)^2}\sum_{k_r+1}^{k}{y_{t-1}u_t}}{\frac{1}{(k-k_r)^2}\sum_{k_r+1}^{k}{y_{t-1}^2}}
 \leq o_p\left(\frac{1}{T^b}\right).
 \]
Therefore, we obtain \eqref{lemma:uniform:a5}$\blacksquare$.

\newpage
\renewcommand{\theequation}{\Alph{section}.\arabic{equation}}
\setcounter{equation}{0}
\section{Proofs in the Serially Correlated Case}

In this appendix, we first show that Lemmas \ref{lemma:1}--\ref{lemma:4} hold for model \eqref{model:0:correl} under Assumption \ref{assumption:correl}. As a result, we obtain the same result as Theorem \ref{theorem:k}.

It is well known in the econometric literature that Lemma 1 holds for a linear process with \eqref{lemma:1:y} replaced by
\[
\frac{1}{\sqrt{T}}y_{k_e}\Rightarrow \psi W(\tau_e)
\]
and we omit the details.

To prove \eqref{lemma:2:y} in Lemma \ref{lemma:2}, from \eqref{pf:lemma2:y} it is sufficient to prove that
\BEQ
\sum_{j'=1}^{k-k_e}\phi_a^{-j'}u_{k_e+j'}
& = &
\sum_{j'=1}^{k-k_e}\phi_a^{-j'}\varepsilon_{k_e+j'}
\nonumber \\
& = &
\psi\sum_{j'=1}^{k-k_e}\phi_a^{-j'}v_{k_e+j'}
+
\sum_{j'=1}^{k-k_e}\phi_a^{-j'}(\tilde{v}_{k_e+j'-1}-\tilde{v}_{k_e+j'})
\label{lemma:2:y:correl:0} \\
& = &
o_p(T^{1/2}).
\label{lemma:2:y:correl}
\EEQ
The first term of \eqref{lemma:2:y:correl:0} is $O_p(T^{a/2})$ uniformly as proved in \eqref{lemma:2:y}, while the second term can be expanded as
\[
\sum_{j'=1}^{k-k_e}\phi_a^{-j'}(\tilde{v}_{k_e+j'-1}-\tilde{v}_{k_e+j'})
=
\phi_a^{-1}\tilde{v}_{k_e}+\sum_{j'=1}^{k-k_e-1}(\phi_a^{-j'-1}-\phi_a^{-j'})\tilde{v}_{k_e+j'}-\phi_a^{-(k-k_e)}\tilde{v}_k.
\]
The first term is $O_p(1)$ while the third terms are $o_p(T^{a/2}) $ because $\phi_a^{-(k-k_e)} <1$ and $\max_k|\tilde{v}_k|=o_p(T^{a/2})$ as proved in Proposition A3 in \citet{PM2007b}. Similarly, the second term is
\BEQA
\left|\sum_{j'=1}^{k-k_e-1}(\phi_a^{-j'-1}-\phi_a^{-j'})\tilde{v}_{k_e+j'}\right|
& \leq &
\frac{c_a}{T^a}\sum_{j'=1}^{k-k_e-1}\phi_a^{-j'-1}\max_k|\tilde{v}_k| \\
& = &
 o_p(T^{a/2}).
\EEQA
Thus, we obtain \eqref{lemma:2:y:correl}.

\eqref{lemma:2:y2} and \eqref{lemma:2:y2c} follow because \eqref{lemma:2:y} holds.

For \eqref{lemma:2:yua}, we note that
\BEQ
\sum_{k_e+1}^{k}y_{t-1}u_t
& \sim_a &
y_{k_e}\sum_{k_e+1}^{k}\phi_a^{(t-k_e-1)}u_t 
\nonumber \\
& = &
y_{k_e}\left\{\psi\sum_{k_e+1}^{k}\phi_a^{(t-k_e-1)}v_t+\sum_{k_e+1}^{k}\phi_a^{(t-k_e-1)}(\tilde{v}_{t-1}-\tilde{v}_t)\right\}.
\label{lemma:2yua:correl:1}
\EEQ
The first term in the parentheses is $O_p(T^{a/2}\phi_a^{(k_c-k_e)})$ uniformly as proved in Appendix B, whereas the second term of \eqref{lemma:2yua:correl:1} becomes
\[
\sum_{k_e+1}^{k}\phi_a^{(t-k_e-1)}(\tilde{v}_{t-1}-\tilde{v}_t)
=
\tilde{v}_{k_e}+\frac{c_a}{T^a}\sum_{k_e+1}^{k-1}\phi_a^{(t-k_e-1)}\tilde{v}_t-\phi_a^{(k-k_e-1)}\tilde{v}_k.
\]
Because $\max_t|\tilde{v}_t|=o_p(T^{a/2})$, we can see that the second term of \eqref{lemma:2yua:correl:1} is $o_p(T^{a/2}\phi_a^{(k_c-k_e)})$ uniformly and thus \eqref{lemma:2:yua} holds.

\eqref{lemma:2:yu}, \eqref{lemma:2:yub}, and \eqref{lemma:2:yuc} can be proved similarly and we omit the details.

In exactly the same manner, Lemmas \ref{lemma:3} and \ref{lemma:4} hold for the model with serially correlated shocks using the BN decomposition given by \eqref{BNdecomposition}.

Theorem \ref{theorem:ke} in the serially correlate case is obtained in \citet{pang2018structural}.

For the proof of $\hat{k}_r$, let $k=k_r+vT^b$ for a given $v$ and we proceed with the proof in the same way as Theorem \ref{theorem:kr} but we need to deal with the serially correlated shocks carefully. First, consider the case where $v < 0$ with \eqref{Chong:B2}. Using \eqref{BNdecomposition}, it can be shown that $\Lambda_T(k/T)$ is $o_p(1)$ and that \eqref{limit:kr:1a} holds, whereas we need to modify \eqref{limit:kr:2a} and \eqref{limit:kr:3a}. Note that
\BEQ
y_k
& = &
\phi_b^{(k-k_e)}y_{k_c}+\sum_{j=k_c+1}^{k}\phi_b^{(k-j)}\left(\psi v_j+\tilde{v}_{j-1}-\tilde{v}_j\right) \nonumber \\
& \sim_a &
\psi\sum_{j=k_c+1}^k\phi_b^{(k-j)}v_j+\sum_{j=k_c+1}^{k}\phi_b^{(k-j)}(\tilde{v}_{j-1}-\tilde{v}_j) \nonumber \\
& = &
\psi\sum_{j=k_c+1}^k\phi_b^{(k-j)}v_j
+\phi_b^{(k-k_c-1)}\tilde{v}_{k_c}+(1-\phi_b)\sum_{j=k_c+1}^{k-1}\phi_b^{(k-j-1)}\tilde{v}_j-\tilde{v}_k.
\label{correl:y}
\EEQ
Then, we have
\BEQA
\frac{1}{T^b}\sum_{k+1}^{k_r}y_{t-1}u_t
& \sim_a &
\frac{\psi}{T^b}\sum_{k_r-|v|T^b+1}^{k_r}\left(\sum_{j=k_c+1}^{t-1}\phi_b^{(t-1-j)}v_j\right)\left(\psi v_t+\tilde{v}_{t-1}-\tilde{v}_{t}\right) \\
& & 
+\tilde{v}_{k_c}\frac{1}{T^b}\sum_{k_r-|v|T^b+1}^{k_r}\phi_b^{(t-2-k_c)}\left(\psi v_t+\tilde{v}_{t-1}-\tilde{v}_{t}\right) \\
& &
+\frac{1}{T^b}\sum_{k_r-|v|T^b+1}^{k_r}\left(\frac{c_b}{T^b}\sum_{j=k_c+1}^{t-2}\phi_b^{(t-2-j)}\tilde{v}_{j}\right)\left(\psi v_t+\tilde{v}_{t-1}-\tilde{v}_{t}\right) \\
& &
-\frac{1}{T^b}\sum_{k_r-|v|T^b+1}^{k_r}\tilde{v}_{t-1}\left(\psi v_t+\tilde{v}_{t-1}-\tilde{v}_{t}\right) \\
& \eqqcolon &
I+II+III+IV.
\EEQA
For $I$, we have, as derived in the proof of Theorem \ref{theorem:kr},
\[
\frac{\psi}{T^b}\sum_{k_r-|v|T^b+1}^{k_r}\left(\sum_{j=k_c+1}^{t-1}\phi_b^{(t-1-j)}v_j\right)\psi v_t
\Rightarrow
\sigma^2\psi^2\int_0^{|v|}\tilde{B}_{c_b}(s)dB_1(s),
\]
while, by letting $G_t\coloneqq \sum_{j=k_c+1}^{t}\phi_b^{t-j}v_j$,
\BEQA
\lefteqn{
\frac{\psi}{T^b}\sum_{k_r-|v|T^b+1}^{k_r}\left(\sum_{j=k_c+1}^{t-1}\phi_b^{(t-1-j)}v_j\right)(\tilde{v}_{t-1}-\tilde{v}_{t}) 
} \\
& = &
\frac{\psi}{T^b}\sum_{k_r-|v|T^b+1}^{k_r}G_{t-1}(\tilde{v}_{t-1}-\tilde{v}_{t})  \\
& = &
\frac{\psi}{T^b}\left\{G_{k_r-|v|T^b}\tilde{v}_{k_r-|v|T^b}+\sum_{k_r-|v|T^b+1}^{k_r-1}(G_t-G_{t-1})\tilde{v}_t+G_{k_r-1}\tilde{v}_{k_r}\right\}.
\EEQA
Because $G_t-G_{t-1}=v_t-(c_b/T^b)\sum_{j=k_c+1}^{t-1}\phi_b^{t-1-j}v_j=v_t+O_p(1/T^{b/2})$ while it can be shown that the first and third terms in the curly braces are $o_p(T^b)$, we have, by the WLLN,
\BEQA
\frac{\psi}{T^b}\sum_{k_r-|v|T^b+1}^{k_r}G_{t-1}(\tilde{v}_{t-1}-\tilde{v}_{t}) 
& = &
\frac{\psi}{T^b}\sum_{k_r-|v|T^b+1}^{k_r}v_t\tilde{v}_t +o_p(1) \\
& \CP &
\sigma^2\psi\tilde{\psi}_0|v|.
\EEQA
These results yield
\[
I
\Rightarrow 
\sigma^2\psi^2\int_0^{|v|}\tilde{B}_{c_b}(s)dB_1(s)+\sigma^2\psi\tilde{\psi}_0|v|.
\]
Similarly, it can be shown that $II$ and $III$ are $o_p(1)$. For $IV$, we have, by the WLLN,
\[
IV
\CP
-|v|(E[\tilde{v}_{t-1}^2]-E[\tilde{v}_{t-1}\tilde{v}_t])=-\sigma^2|v|\left(\sum_{j=0}^{\infty}\tilde{\psi}_{j}^2-\sum_{j=0}^{\infty}\tilde{\psi}_{j}\tilde{\psi}_{j+1}\right).
\]
These results imply that
\BEQA
-2(1-\phi_b)\frac{\sum_{k_r+1}^{T}{y_{t-1}^2}\sum_{k+1}^{k_r}{y_{t-1}u_t}}{\sum_{k+1}^{T}{y_{t-1}^2}}
& = &
-\frac{2c_b}{T^b}\sum_{k_r-|v|T^b+1}^{k_r}y_{t-1}u_t(1+o_p(1)) \nonumber \\
& \Rightarrow &
-\sigma^2\psi^2c_b\left(2\int_0^{|v|}\tilde{B}_{c_b}(s)dB_1(s)+\frac{\check{\psi}|v|}{2}\right),
\EEQA
where
\BE
\check{\psi}\coloneqq \frac{4}{\psi^2}\left(\psi\tilde{\psi}_0-\sum_{j=0}^{\infty}\tilde{\psi}_{j}^2+\sum_{j=0}^{\infty}\tilde{\psi}_{j}\tilde{\psi}_{j+1}\right).
\label{psicheck}
\EE
Similarly, it is shown using \eqref{correl:y} that
\BEQA
(1-\phi_b)^2\frac{\sum_{k_r+1}^{T}{y_{t-1}^2}\sum_{k+1}^{k_r}{y_{t-1}^2}}{\sum_{k+1}^{T}{y_{t-1}^2}}
& = &
\frac{c_b^2}{T^{2b}}\sum_{k_r-|v|T^{b}+1}^{k_r}y_{t-1}^2(1+o_p(1)) \nonumber \\
& \Rightarrow &
\sigma^2\psi^2c_b\left\{c_b\int_0^{|v|} \left(\tilde{B}_{c_b}^2(s)-\frac{1}{2c_b}\right)ds+\frac{|v|}{2}\right\}.
\EEQA
These results yield
\BEQA
\lefteqn{
SSR_2(k/T)-SSR_2(\tau_r)
} \\
& \Rightarrow &
-\sigma^2\psi^2c_b\left(2\int_0^{|v|}\tilde{B}_{c_b}(s)dB_1(s)-c_b\int_0^{|v|}\left(\tilde{B}_{c_b}^2(s)-\frac{1}{2c_b}\right)ds-\frac{(1-\check{\psi})|v|}{2}\right)
\EEQA
for $v < 0$. We obtain the result in the theorem by defining $\psi^*\coloneqq 1-\check{\psi}$.

For $k=k_r+vT^b$ with $v> 0$, it can be shown in expression \eqref{Chong:B4} that $\Lambda_T(k/T)$ is $o_p(1)$ and that \eqref{limit:kr:2b} holds, whereas we need to modify \eqref{limit:kr:1b} and \eqref{limit:kr:3b}. Note that $u_k=c_1/T^{\eta_1}+\psi v_k+\tilde{v}_{k-1}-\tilde{v}_k$ and that
\[
y_{k-1}
=
y_{k_r}+c_1\frac{k-k_r-1}{T^{\eta_1}}+\psi\sum_{j=k_r+1}^{k-1}v_j+\tilde{v}_{k_r}-\tilde{v}_{k-1}.
\]
Then, it can be shown that
\BEQA
\lefteqn{
2(1-\phi_b)\frac{\sum_{k_r+1}^{k}{y_{t-1}u_t}\sum_{k_c+1}^{k_r}{y_{t-1}^2}}{\sum_{k_c+1}^{k}{y_{t-1}^2}}
} \nonumber \\
& = &
\frac{2c_b}{T^b}\sum_{k_r+1}^{k_r+vT^b}y_{t-1}u_t(1+o_p(1)) \nonumber \\
& = &
\frac{c_b}{T^b}\left(y_{k_r+vT^b}^2-y_{k_r}^2-\sum_{k_r+1}^{k_r+vT^b}u_t^2\right)(1+o_p(1)) \nonumber \\
& \Rightarrow &
c_b\left\{\sigma^2\psi^2\left(\tilde{B}_{c_b}(0)+B_2(v)\right)^2-\sigma^2\psi^2\tilde{B}_{c_b}^2(0)-v\sigma^2\sum_{j=0}^{\infty}\psi_j^2\right\} \\
& = &
2\sigma^2\psi^2c_b^2\left\{\frac{1}{c_b}\int_0^v\left(\tilde{B}_{c_b}(0)+B_2(s)\right)dB_2(s)+\frac{v}{2c_b}\left(1-\frac{\sum_{j=0}^{\infty}\psi_j^2}{\psi^2}\right)\right\}.
\EEQA
On the other hand, we can observe that
\BEQA
(1-\phi_b)^2\frac{\sum_{k_r+1}^{k}{y_{t-1}^2}\sum_{k_c+1}^{k_r}{y_{t-1}^2}}{\sum_{k_c+1}^{k}{y_{t-1}^2}}
& = &
\frac{c_b^2}{T^{2b}}\sum_{k_r+1}^{k_r+vT^b}y_{t-1}^2(1+o_p(1)) \\
& \Rightarrow &
\sigma^2\psi^2c_b^2\int_0^v\left(\tilde{B}_{c_b}(0)+B_2(s)\right)^2ds.
\EEQA
Combining these results, we have
\[
SSR_2(k/T)-SSR_2(\tau_r)
\Rightarrow
-2\sigma^2\psi^2c_b^2\tilde{B}_{c_b}^2(0)\left\{C_{c_b}^*(v)-\frac{v}{2}\left(1+\frac{1-\sum_{j=0}^{\infty}\psi_j^2/\psi}{c_b\tilde{B}_{c_b}^2(0)}\right)\right\}
\]
for $v\geq 0$. By defining $\psi^*\coloneqq 1+(1-\sum_{j=0}^{\infty}\psi_j^2/\psi^2)/(c_b\tilde{B}_{c_b}^2(0))$, we obtain the result.$\blacksquare$

\newpage
\renewcommand{\thefigure}{\Alph{section}.\arabic{figure}}
\setcounter{figure}{0}

\section{Additional Figures}
\begin{figure}[h!]%
\begin{center}%
\subfigure[$T=400$, $\phi_a=1.01$, $\phi_b=0.96$]{\includegraphics[width=0.45\linewidth]{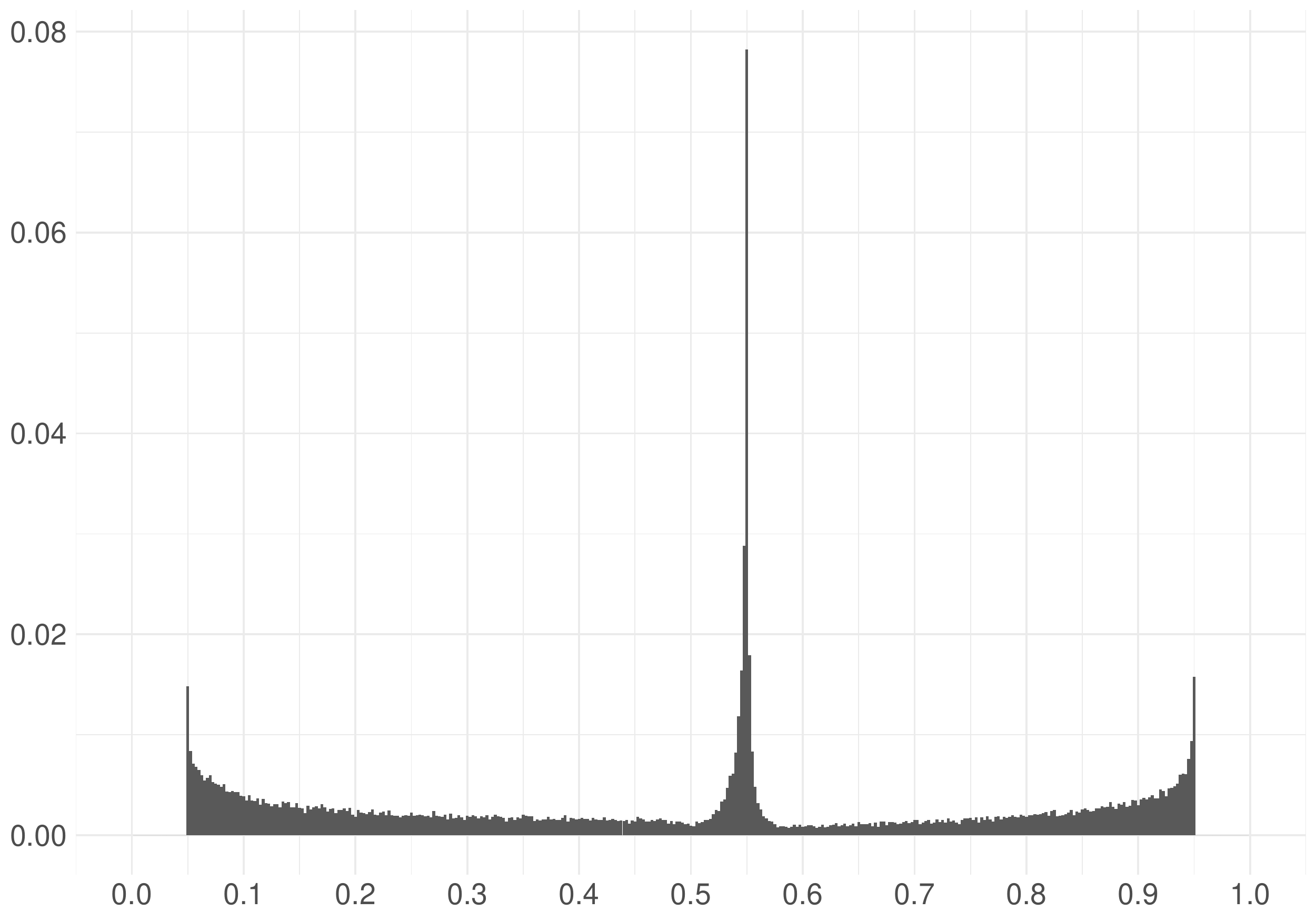}\label{fig:10:1}}
\subfigure[$T=800$, $\phi_a=1.01$, $\phi_b=0.96$]{\includegraphics[width=0.45\linewidth]{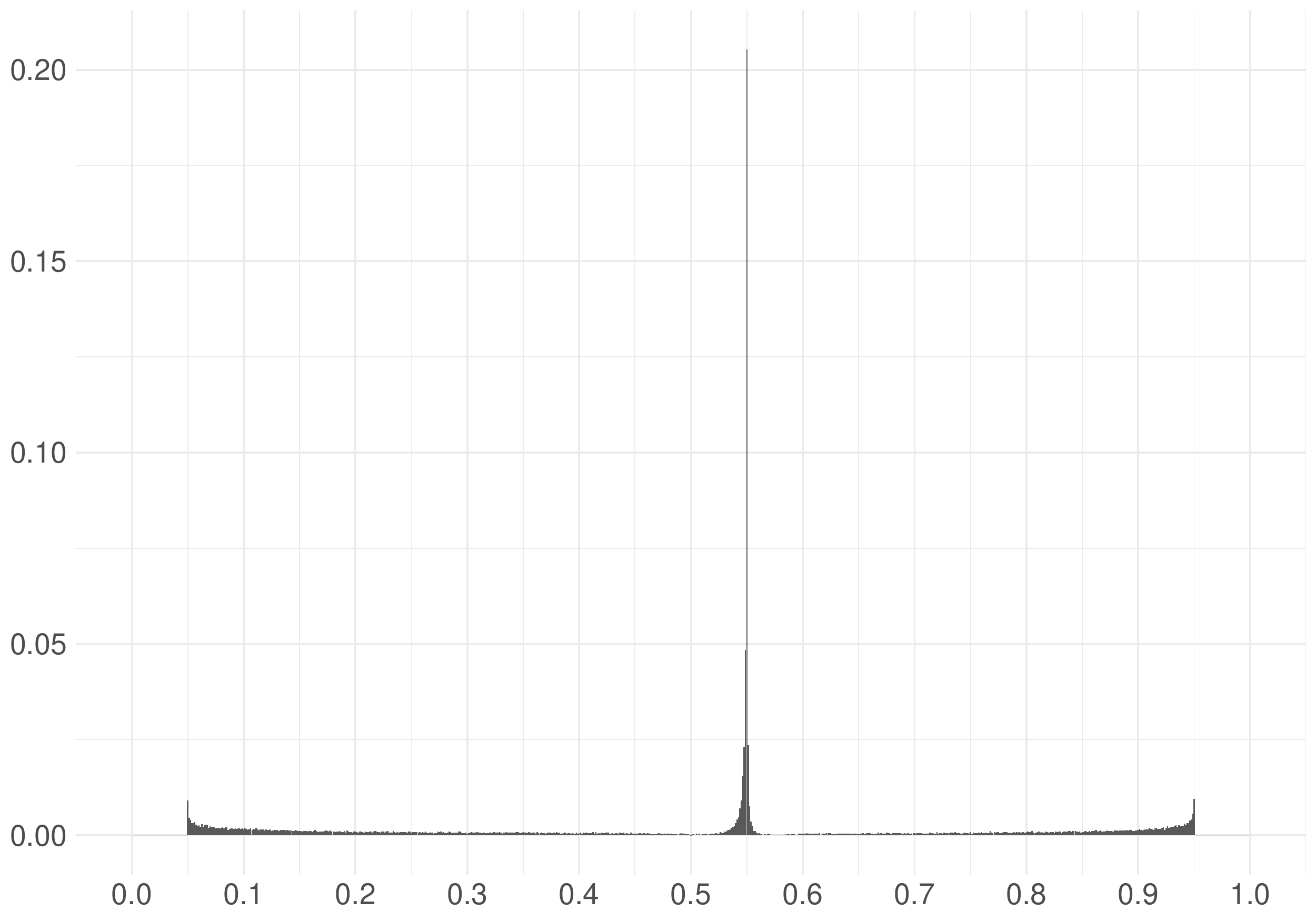}\label{fig:10:2}}\\
\subfigure[$T=400$, $\phi_a=1.05$, $\phi_b=0.96$]{\includegraphics[width=0.45\linewidth]{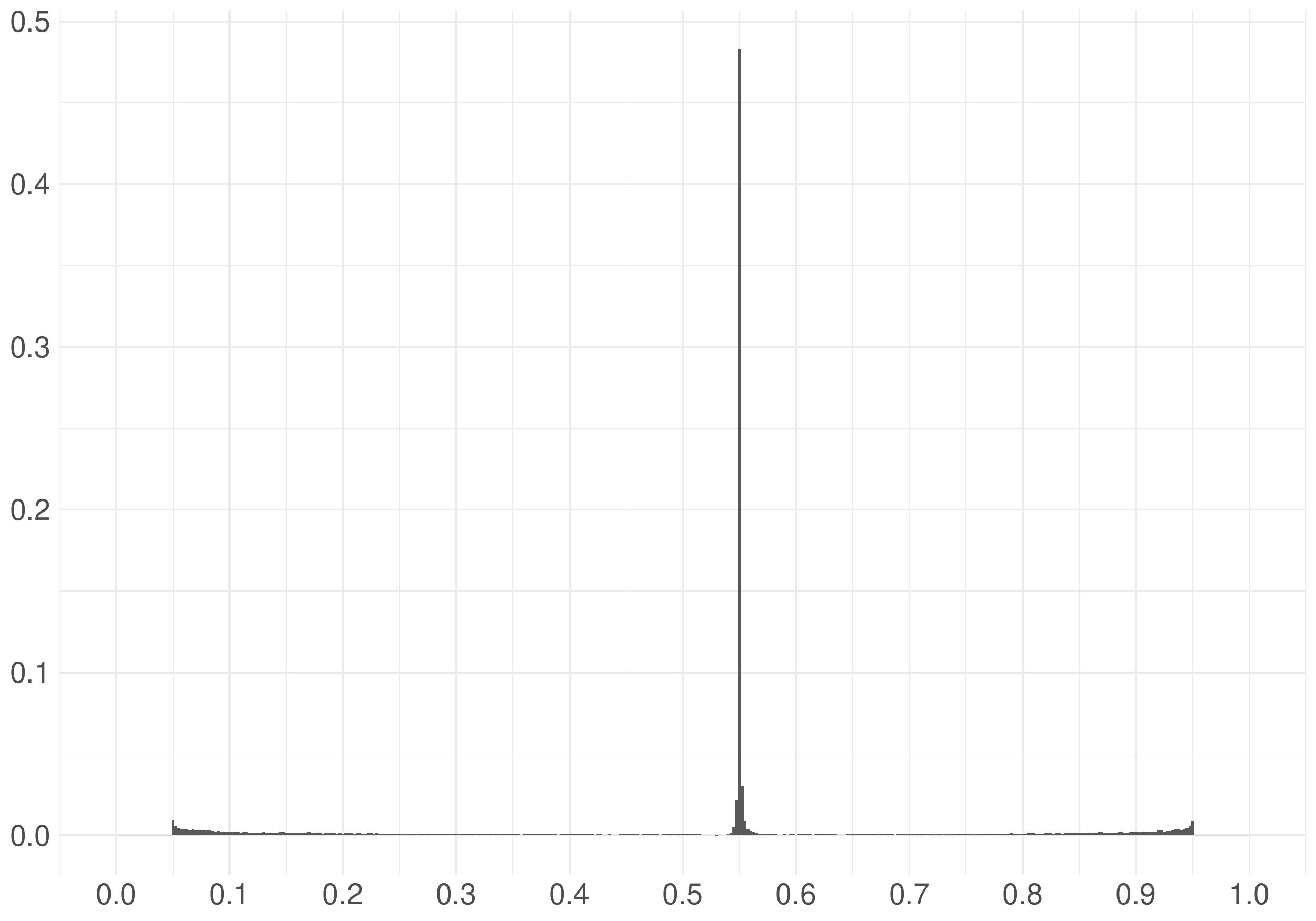}\label{fig:10:3}}
\subfigure[$T=800$, $\phi_a=1.05$, $\phi_b=0.96$]{\includegraphics[width=0.45\linewidth]{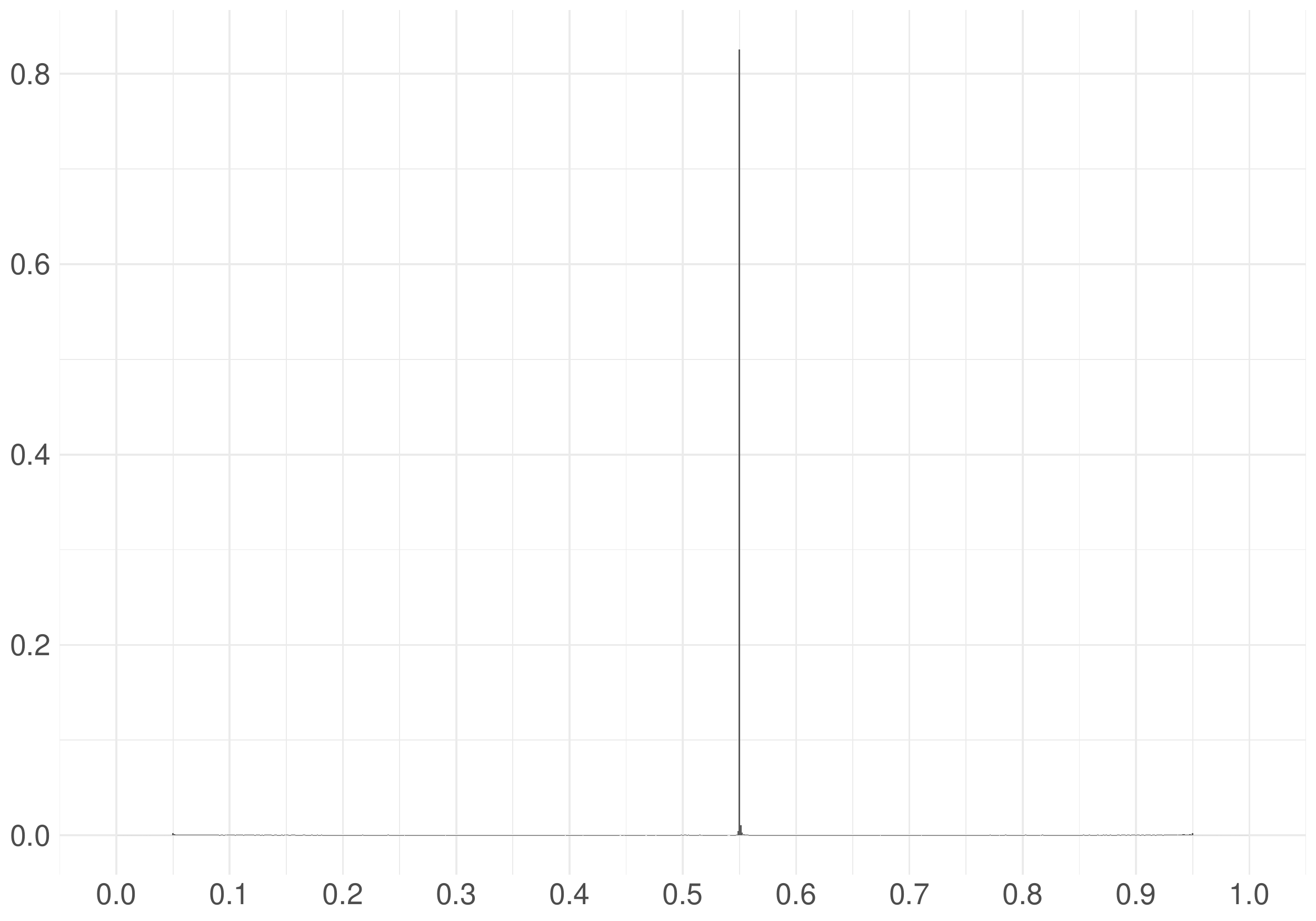}\label{fig:10:4}}\\
\subfigure[$T=400$, $\phi_a=1.09$, $\phi_b=0.96$]{\includegraphics[width=0.45\linewidth]{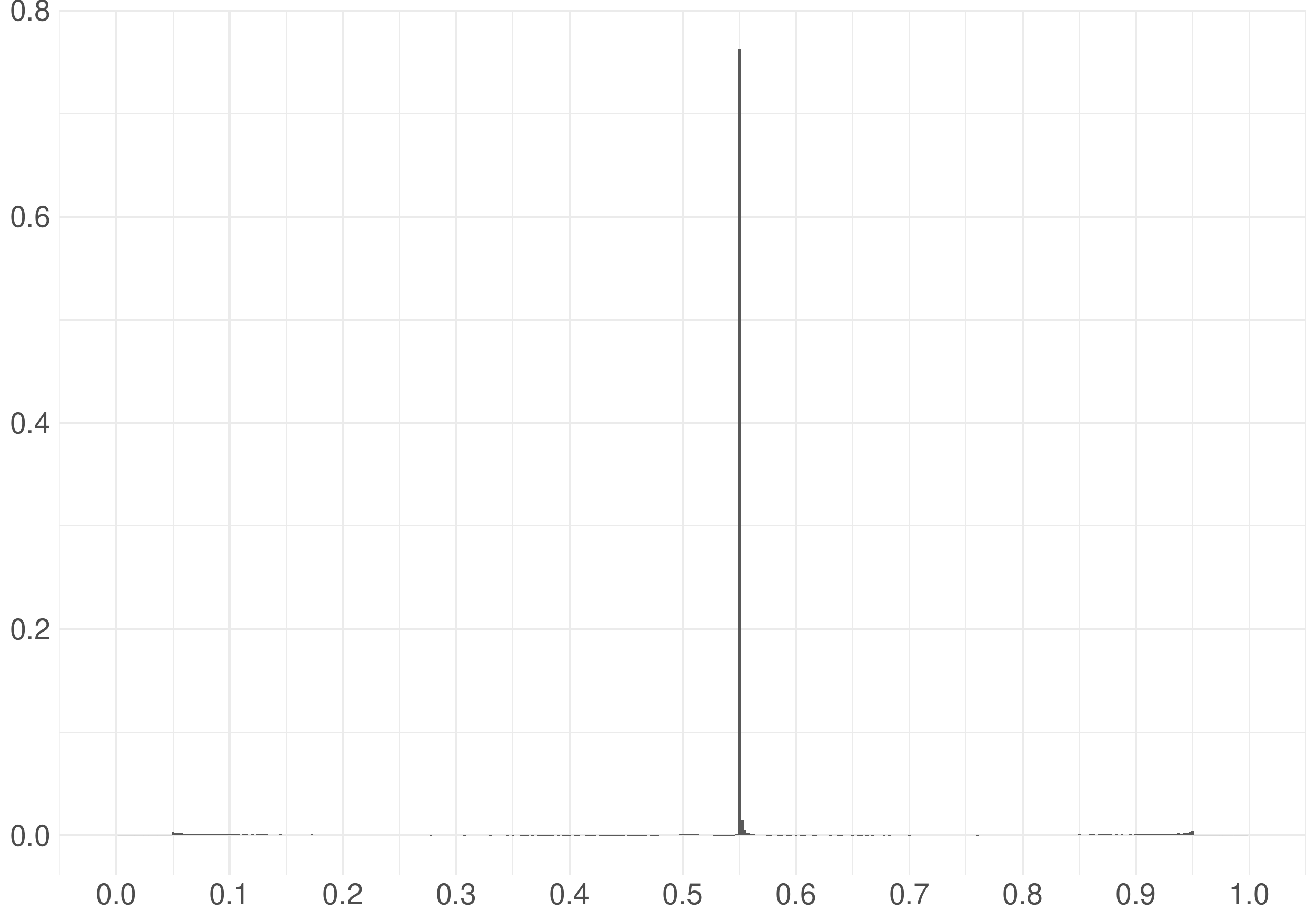}\label{fig:10:5}}
\subfigure[$T=800$, $\phi_a=1.09$, $\phi_b=0.96$]{\includegraphics[width=0.45\linewidth]{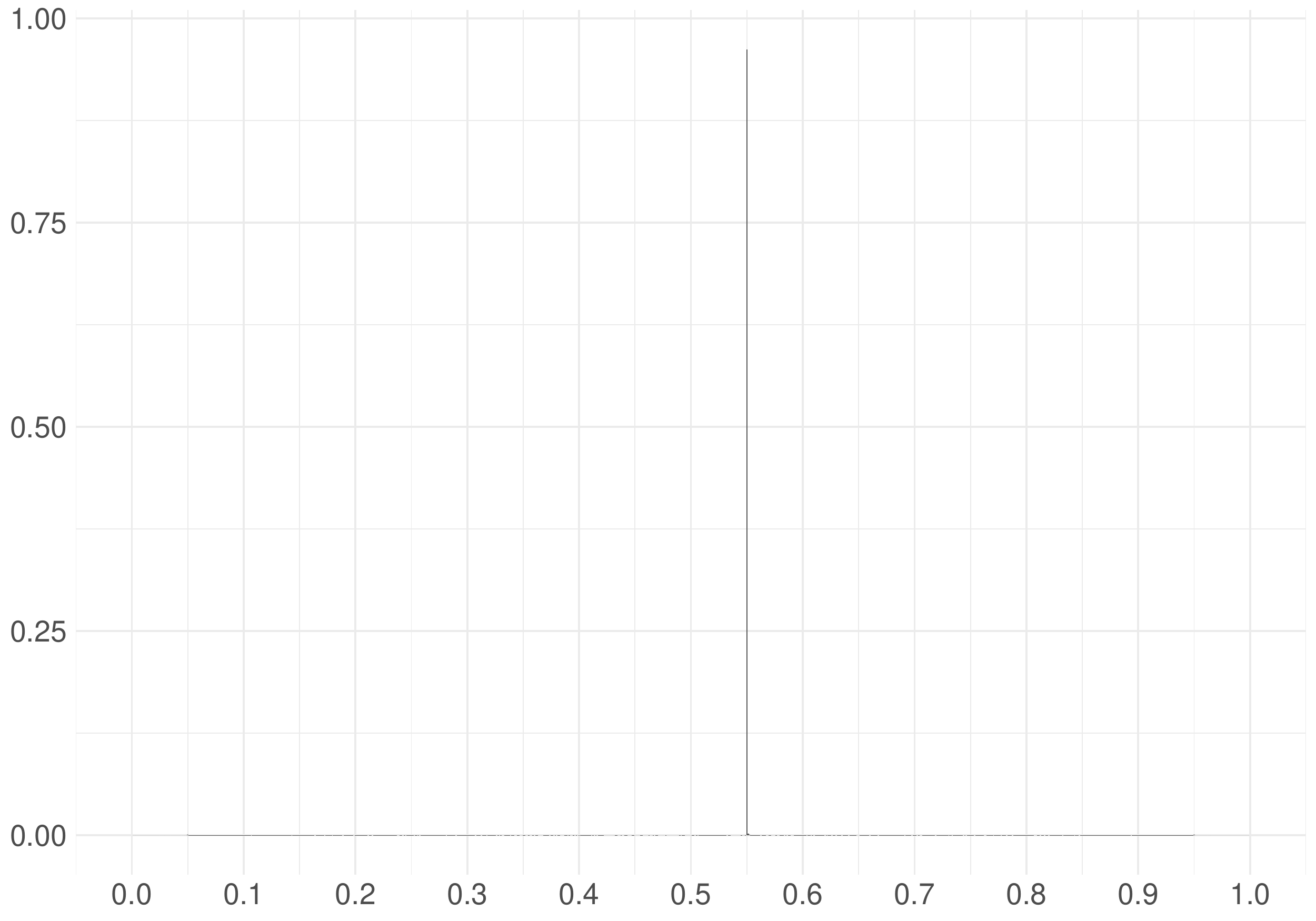}\label{fig:10:6}}
\end{center}%
\caption{Histograms of $\hat{k}_c$ % (left) and $\hat{k}_r$ (right) 
for $(\tau_e,\tau_c,\tau_r)=(0.5,0.55,0.6)$}
\label{fig10}
%\centering
%\footnotesize{OLS}
\end{figure}

\newpage

\begin{figure}[h!]%
\begin{center}%
\subfigure[$T=400$, $\phi_a=1.05$, $\phi_b=0.98$]{\includegraphics[width=0.45\linewidth]{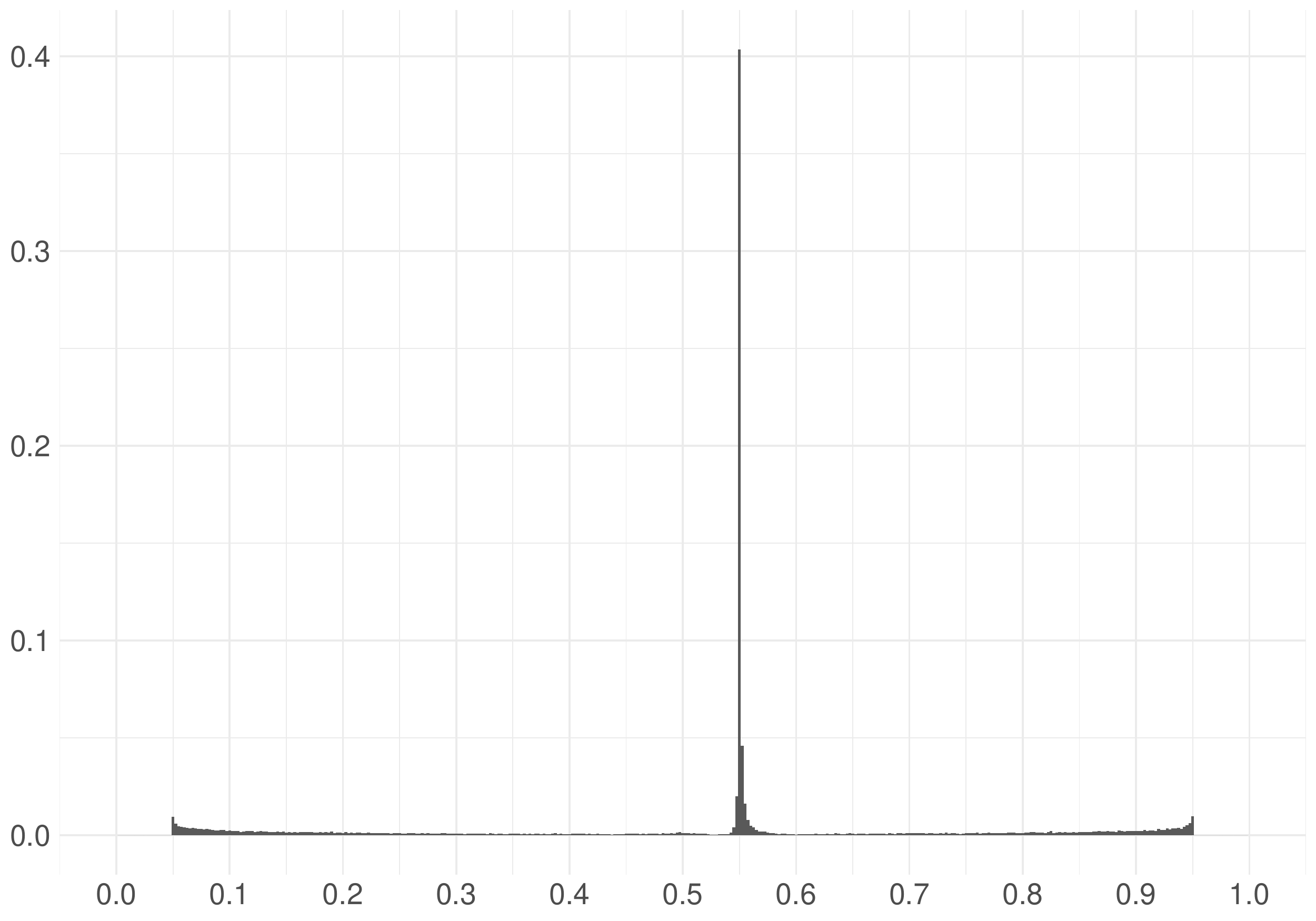}\label{fig:11:1}}
\subfigure[$T=800$, $\phi_a=1.05$, $\phi_b=0.98$]{\includegraphics[width=0.45\linewidth]{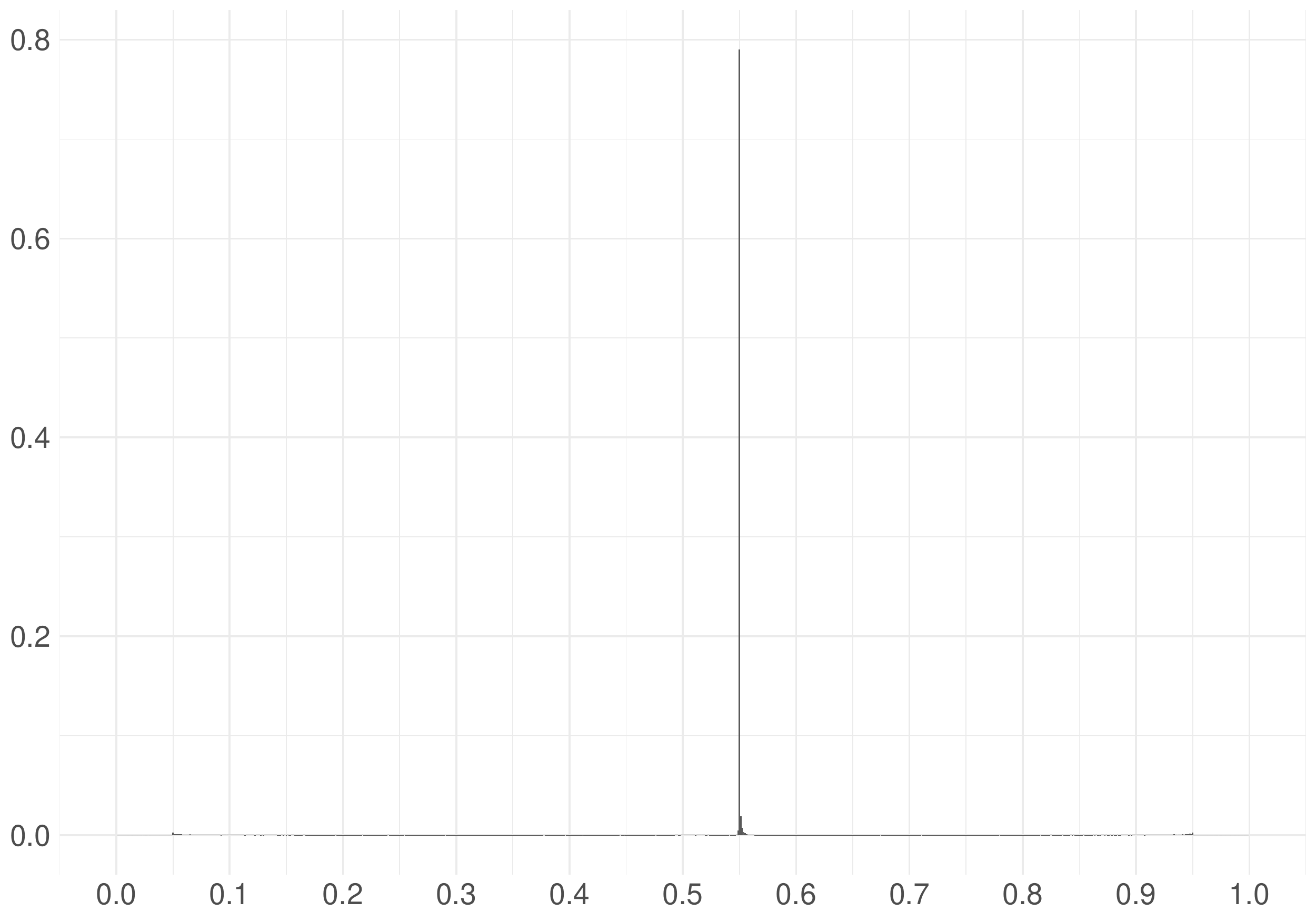}\label{fig:11:2}}\\
\subfigure[$T=400$, $\phi_a=1.05$, $\phi_b=0.96$]{\includegraphics[width=0.45\linewidth]{graph/NVt_k_c_T=400_1.05_0.96_Model1s0.s11.pdf}\label{fig:11:3}}
\subfigure[$T=800$, $\phi_a=1.05$, $\phi_b=0.96$]{\includegraphics[width=0.45\linewidth]{graph/NVt_k_c_T=800_1.05_0.96_Model1s0.s11.pdf}\label{fig:11:4}}\\
\subfigure[$T=400$, $\phi_a=1.05$, $\phi_b=0.94$]{\includegraphics[width=0.45\linewidth]{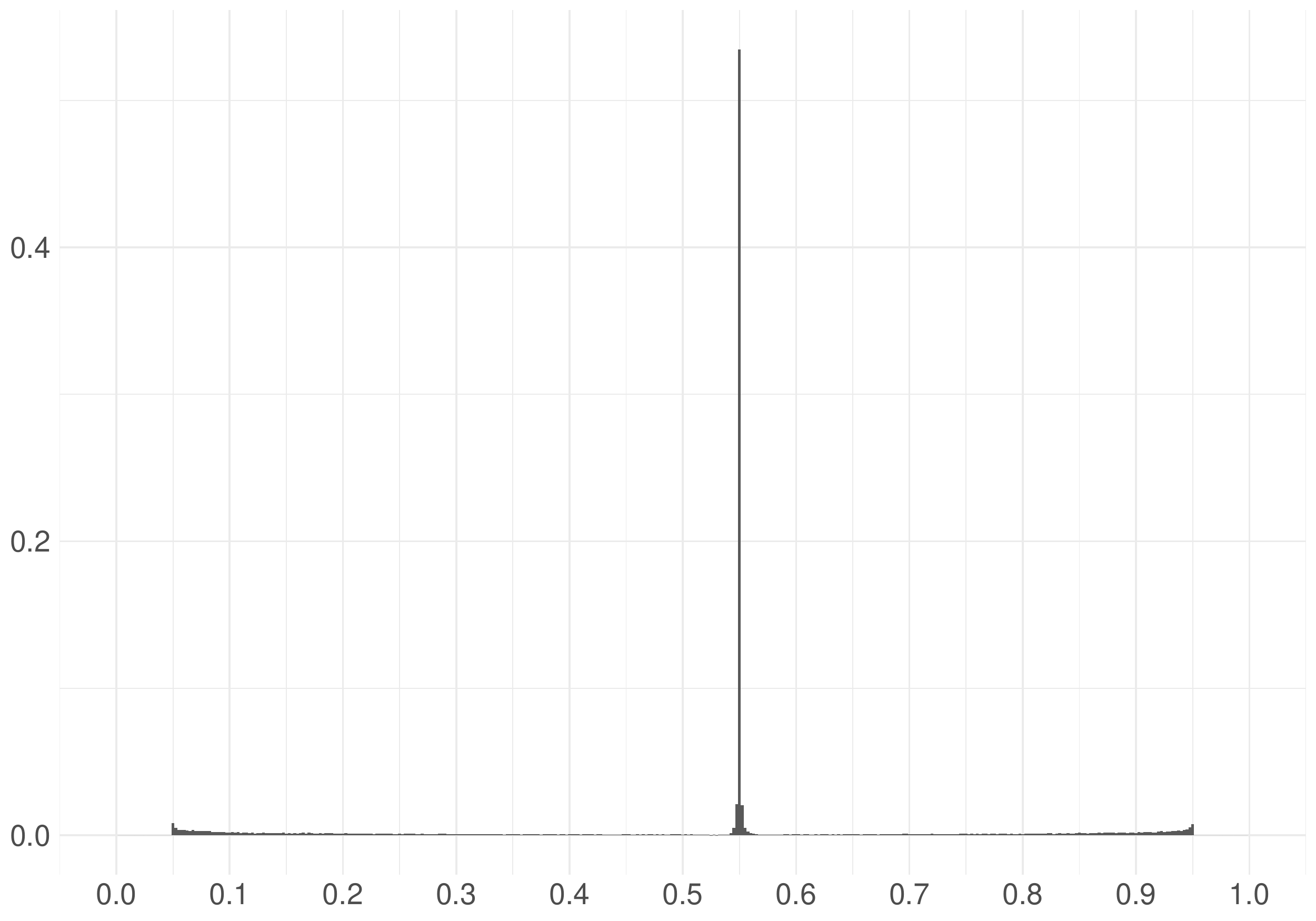}\label{fig:11:5}}
\subfigure[$T=800$, $\phi_a=1.05$, $\phi_b=0.94$]{\includegraphics[width=0.45\linewidth]{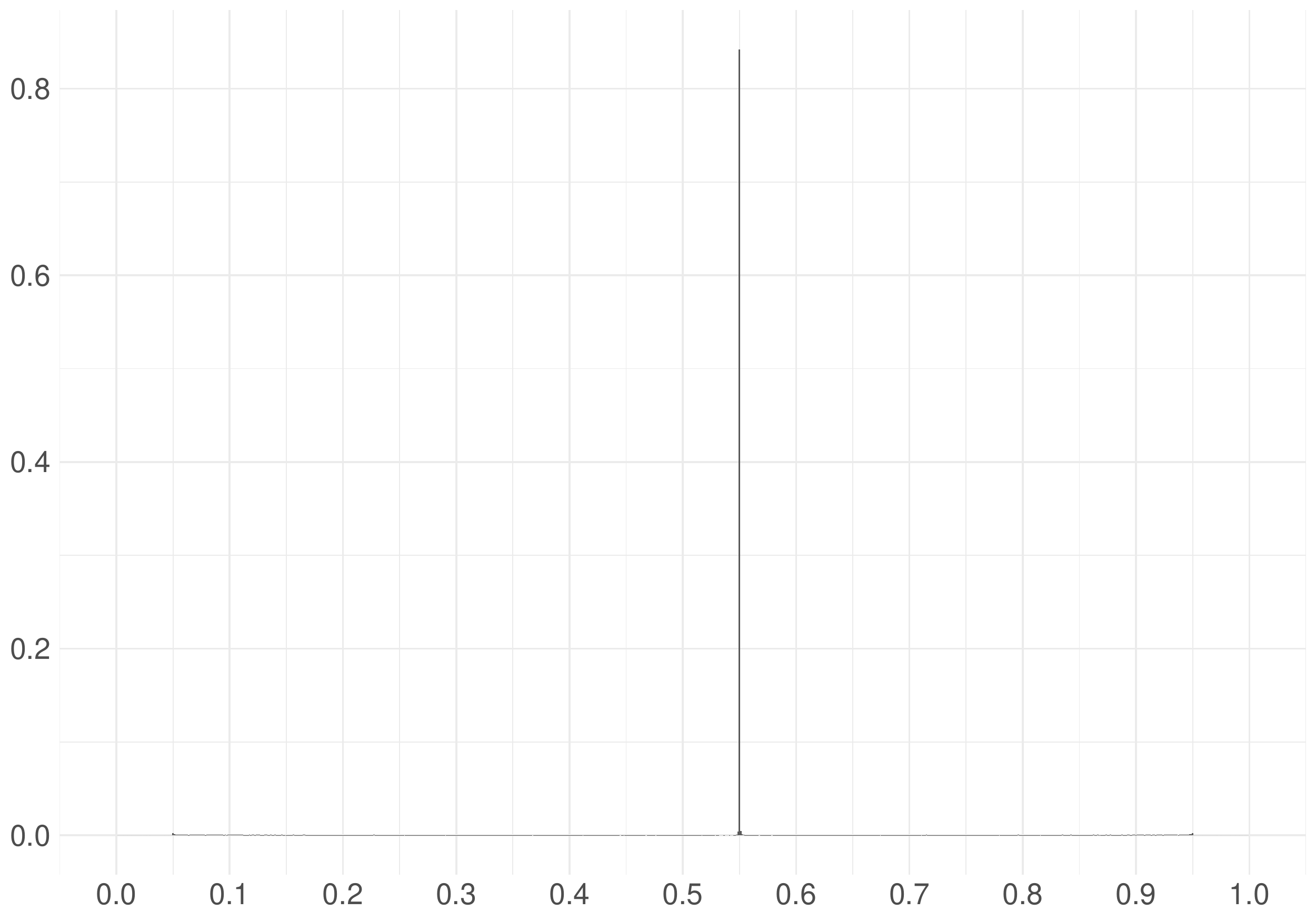}\label{fig:11:6}}
\end{center}%
\caption{Histograms of $\hat{k}_c$ % (left) and $\hat{k}_r$ (right) 
for $(\tau_e,\tau_c,\tau_r)=(0.5,0.55,0.6)$}
\label{fig11}
%\centering
%\footnotesize{OLS}
\end{figure}

\newpage

%%k_r

\begin{figure}[h!]%
\begin{center}%
\subfigure[$T=400$, $\phi_a=1.01$, $\phi_b=0.96$]{\includegraphics[width=0.45\linewidth]{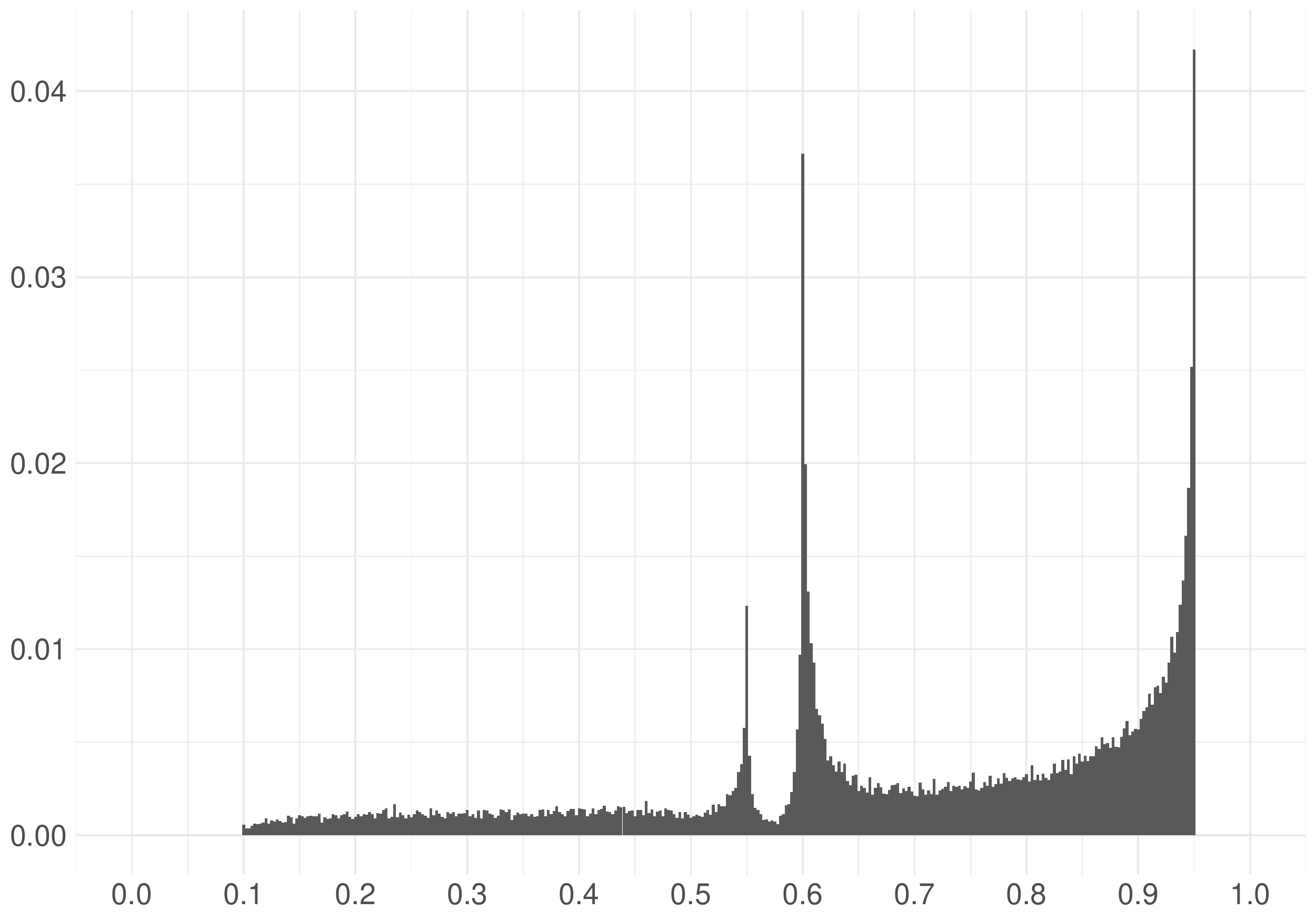}\label{fig:12:1}}
\subfigure[$T=800$, $\phi_a=1.01$, $\phi_b=0.96$]{\includegraphics[width=0.45\linewidth]{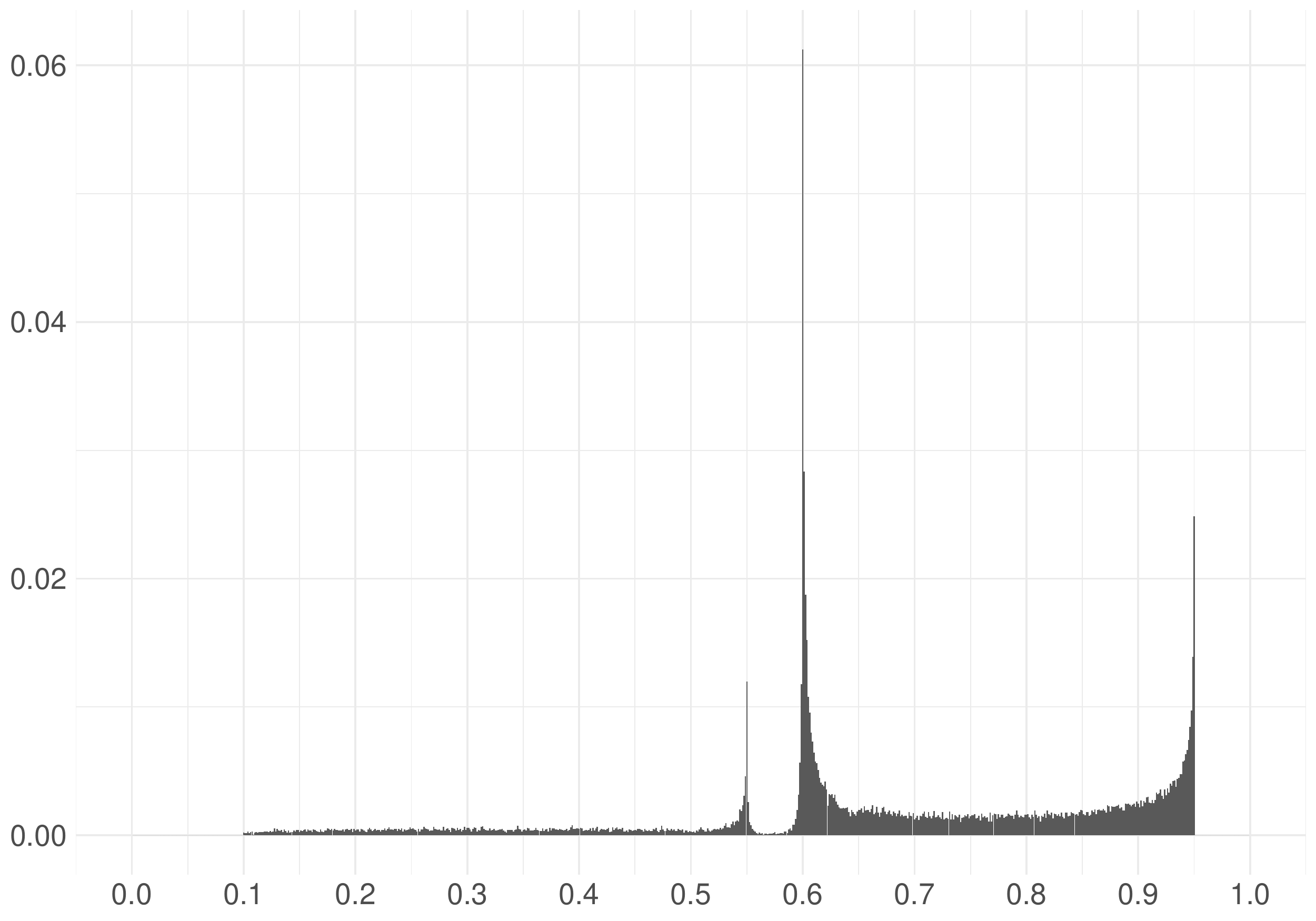}\label{fig:12:2}}\\
\subfigure[$T=400$, $\phi_a=1.05$, $\phi_b=0.96$]{\includegraphics[width=0.45\linewidth]{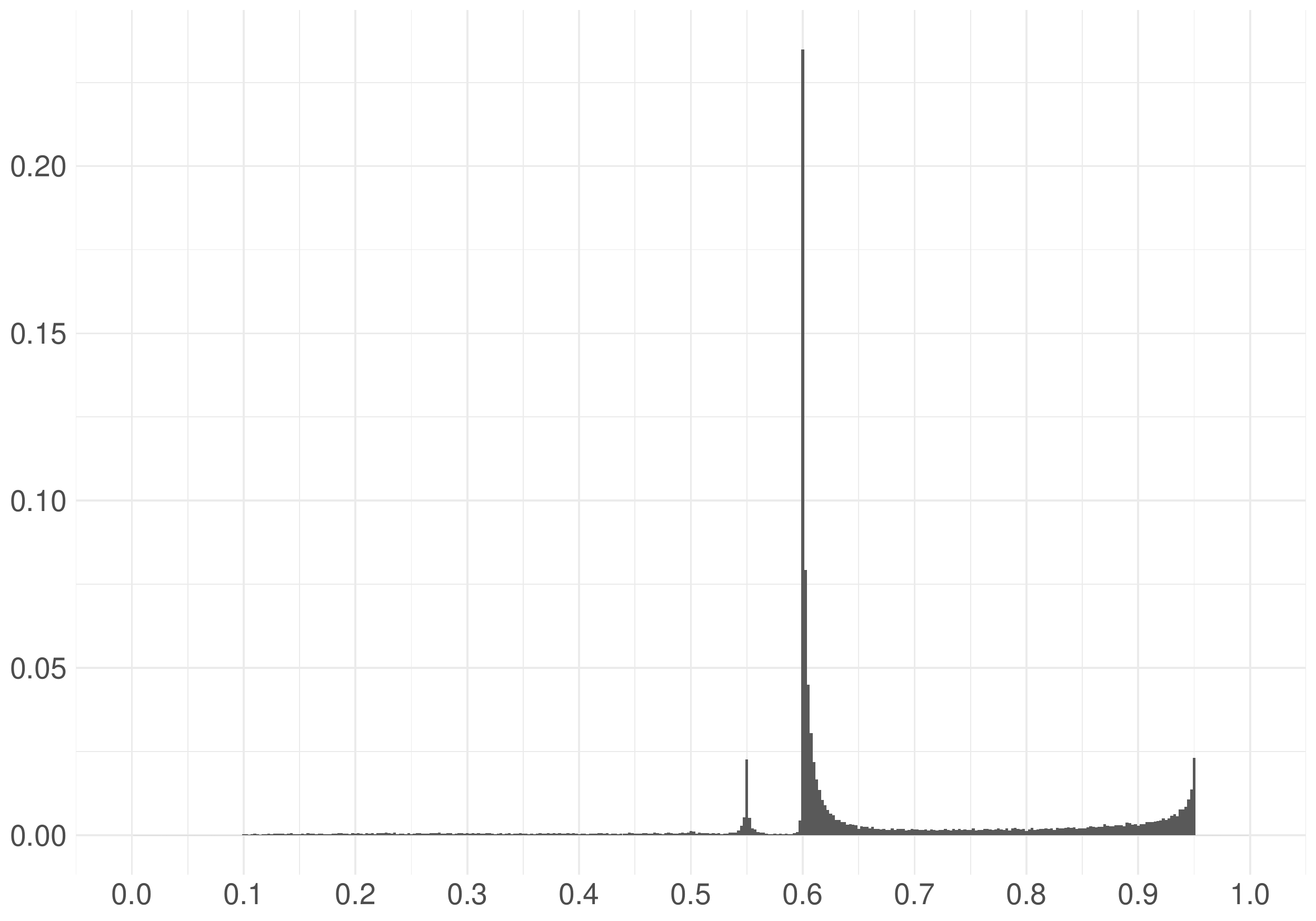}\label{fig:12:3}}
\subfigure[$T=800$, $\phi_a=1.05$, $\phi_b=0.96$]{\includegraphics[width=0.45\linewidth]{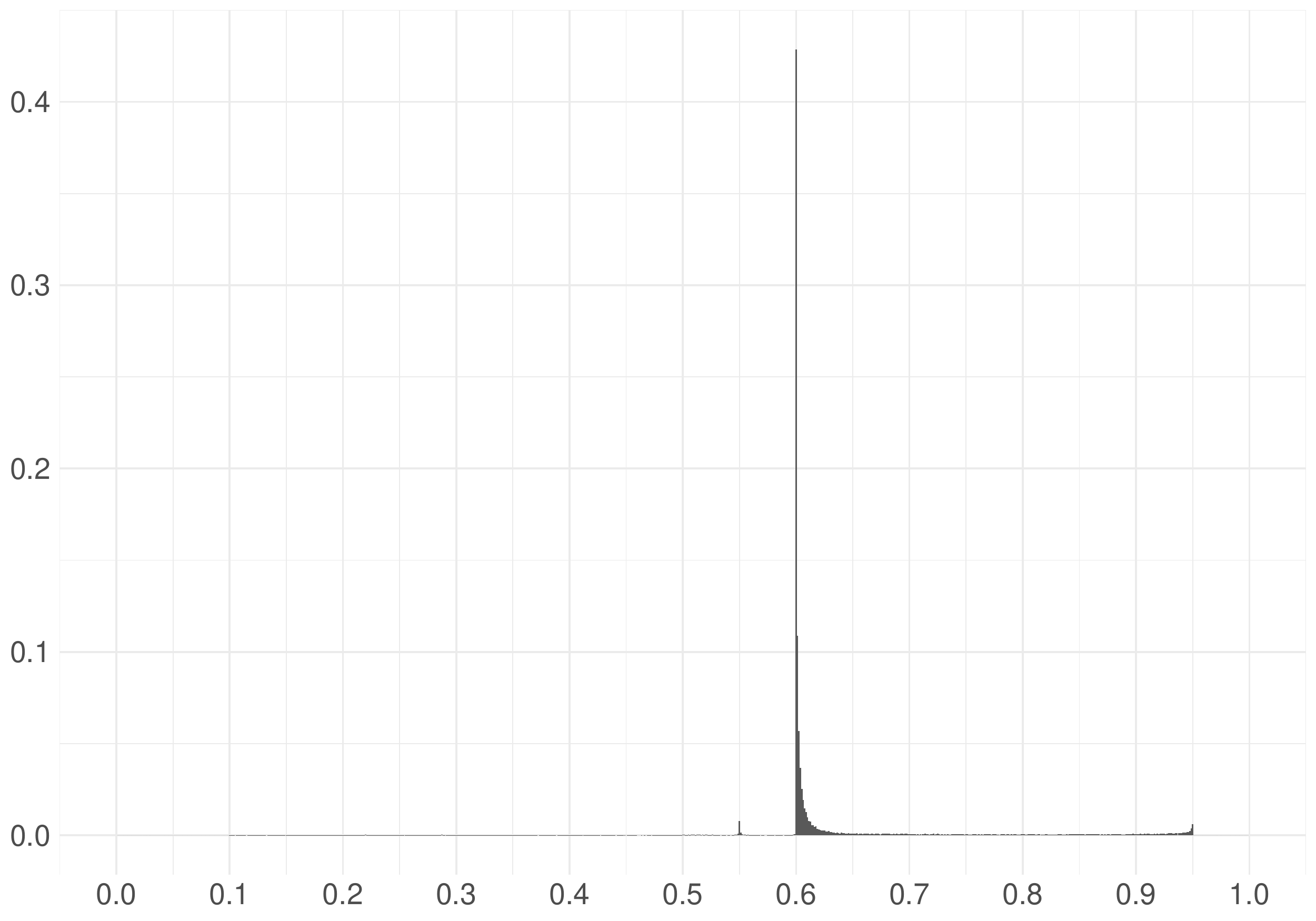}\label{fig:12:4}}\\
\subfigure[$T=400$, $\phi_a=1.09$, $\phi_b=0.96$]{\includegraphics[width=0.45\linewidth]{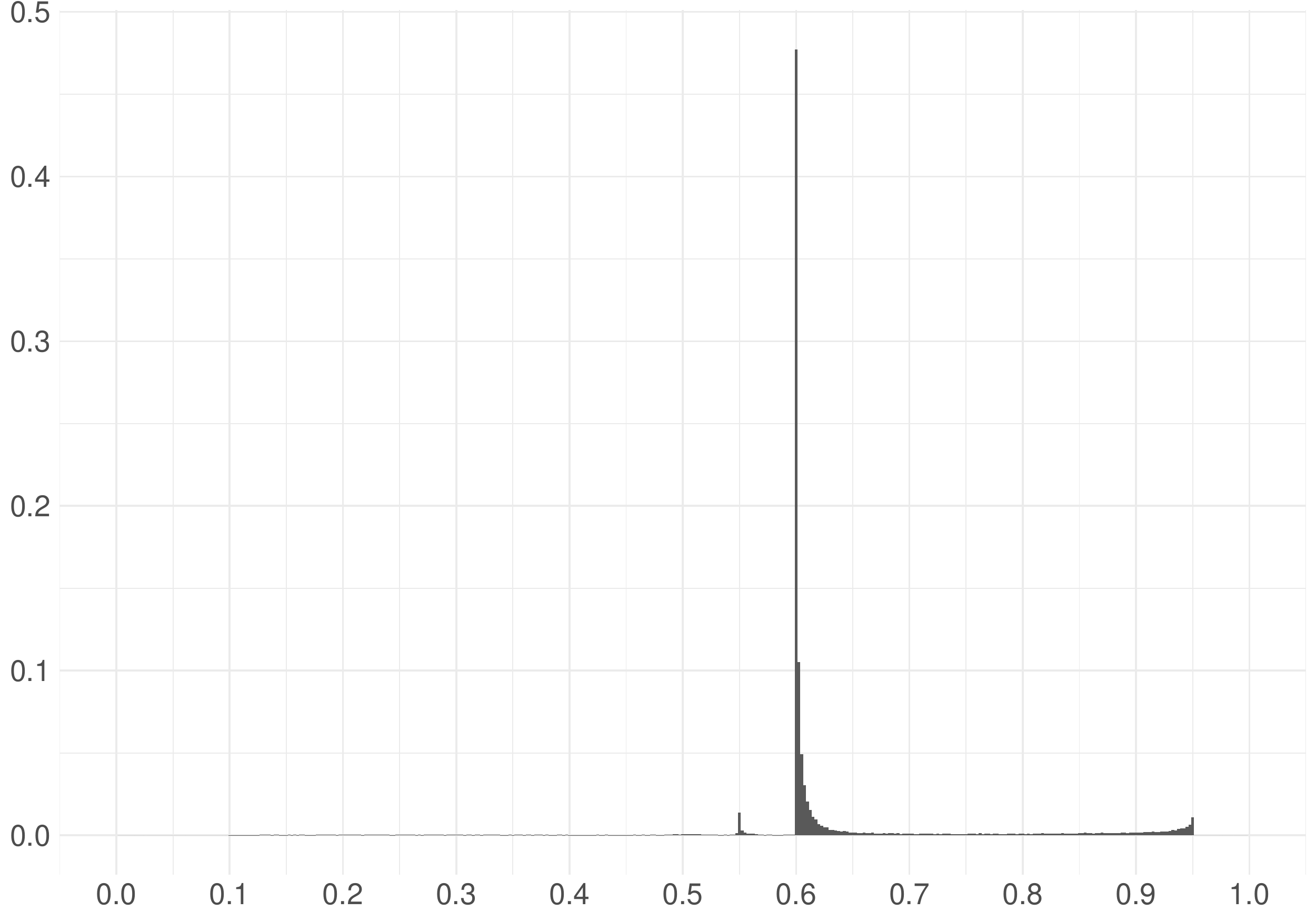}\label{fig:12:5}}
\subfigure[$T=800$, $\phi_a=1.09$, $\phi_b=0.96$]{\includegraphics[width=0.45\linewidth]{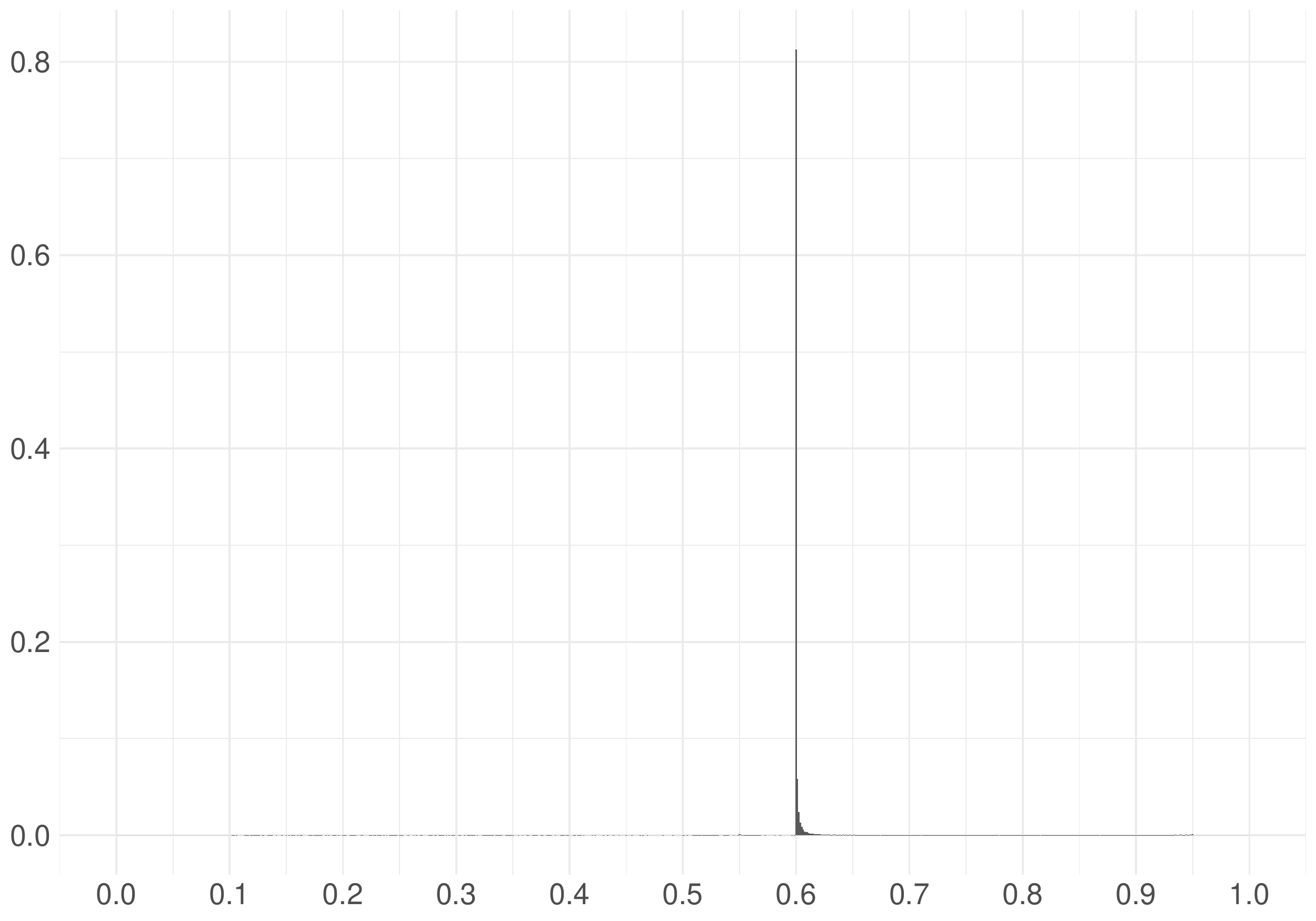}\label{fig:12:6}}
\end{center}%
\caption{Histograms of $\hat{k}_r$ % (left) and $\hat{k}_r$ (right) 
for $(\tau_e,\tau_c,\tau_r)=(0.5,0.55,0.6)$}
\label{fig12}
%\centering
%\footnotesize{OLS}
\end{figure}

\newpage

\begin{figure}[h!]%
\begin{center}%
\subfigure[$T=400$, $\phi_a=1.05$, $\phi_b=0.98$]{\includegraphics[width=0.45\linewidth]{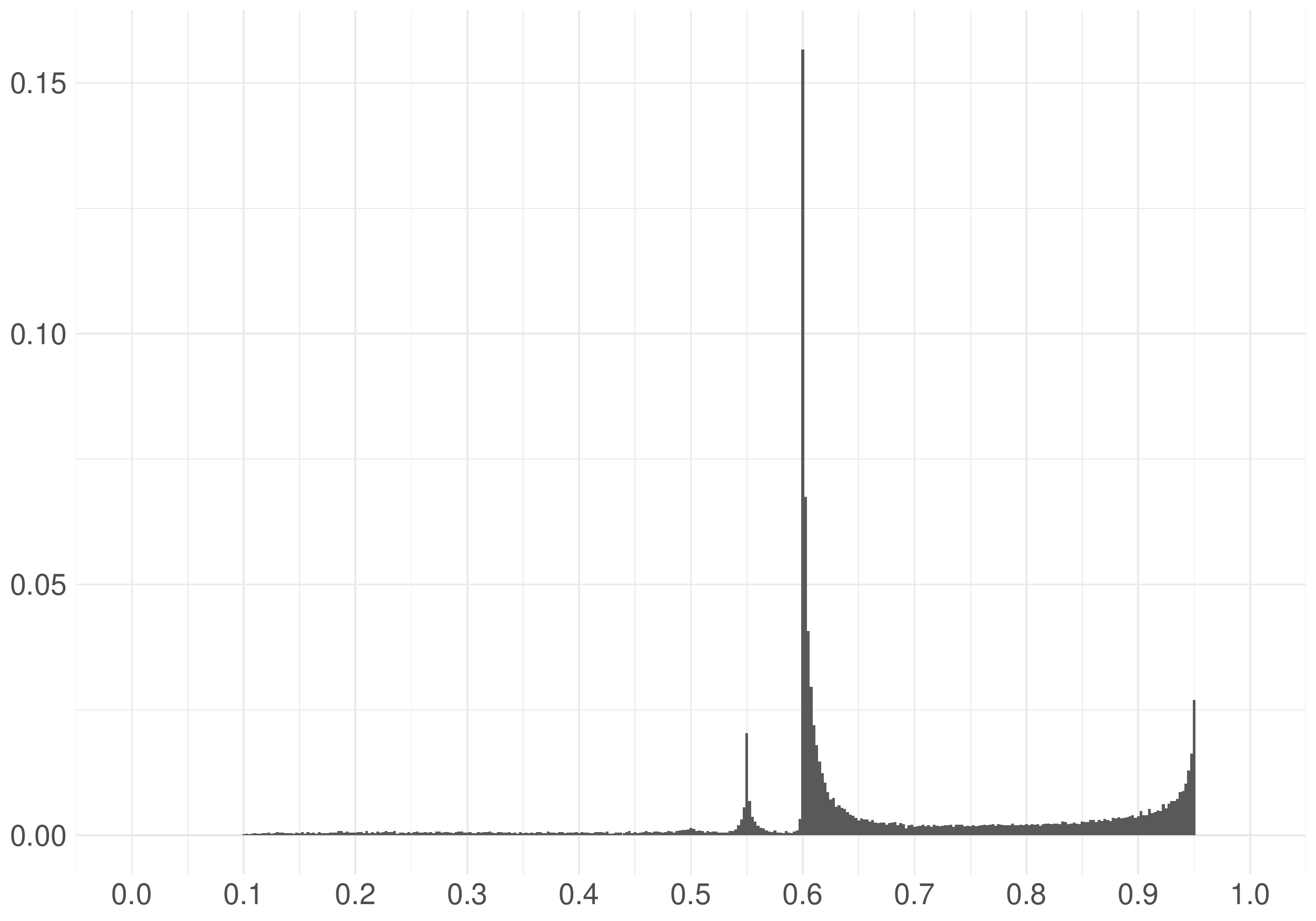}\label{fig:13:1}}
\subfigure[$T=800$, $\phi_a=1.05$, $\phi_b=0.98$]{\includegraphics[width=0.45\linewidth]{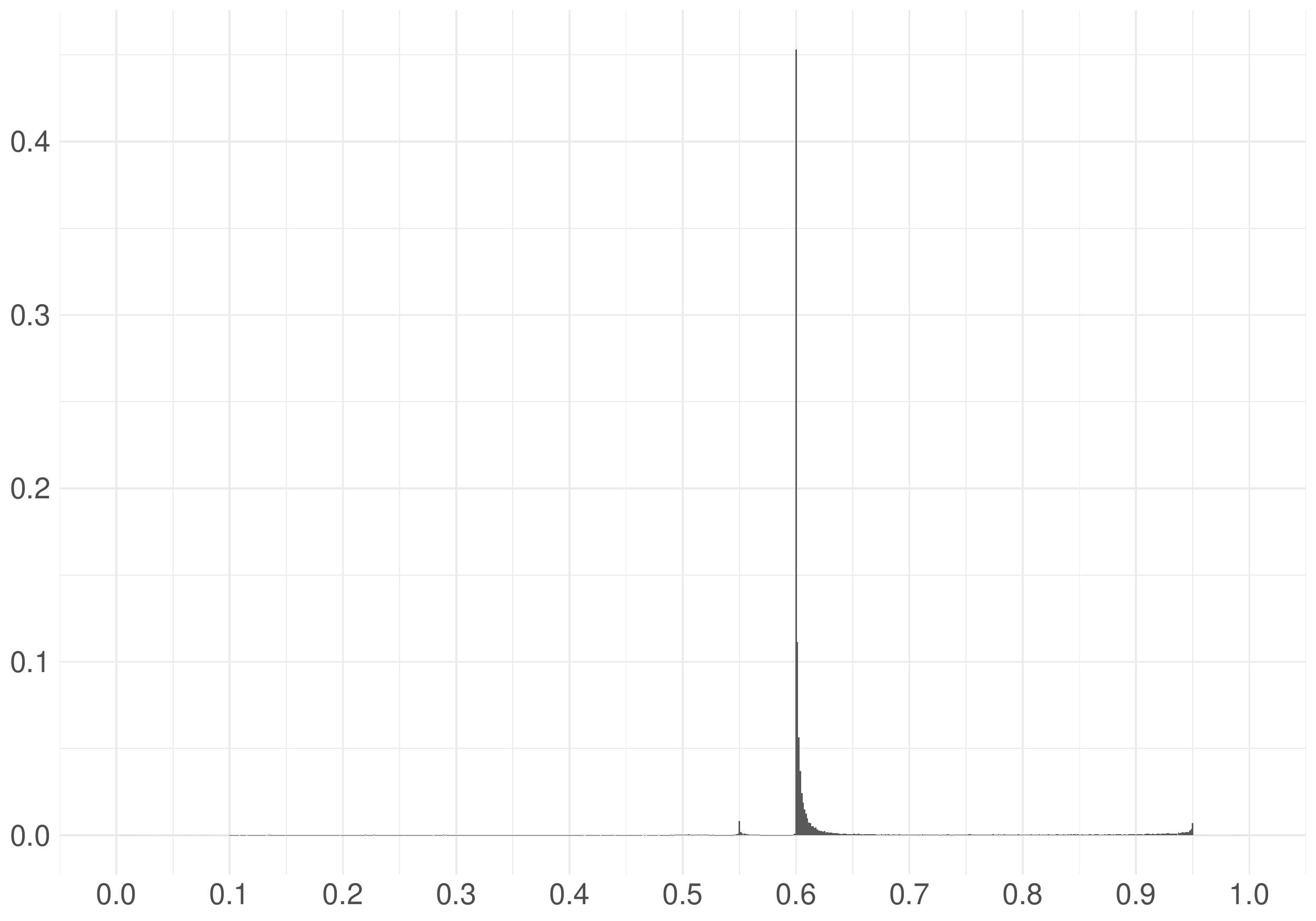}\label{fig:13:2}}\\
\subfigure[$T=400$, $\phi_a=1.05$, $\phi_b=0.96$]{\includegraphics[width=0.45\linewidth]{graph/NVt_k_r_T=400_1.05_0.96_Model1s0.s11.pdf}\label{fig:13:3}}
\subfigure[$T=800$, $\phi_a=1.05$, $\phi_b=0.96$]{\includegraphics[width=0.45\linewidth]{graph/NVt_k_r_T=800_1.05_0.96_Model1s0.s11.pdf}\label{fig:13:4}}\\
\subfigure[$T=400$, $\phi_a=1.05$, $\phi_b=0.94$]{\includegraphics[width=0.45\linewidth]{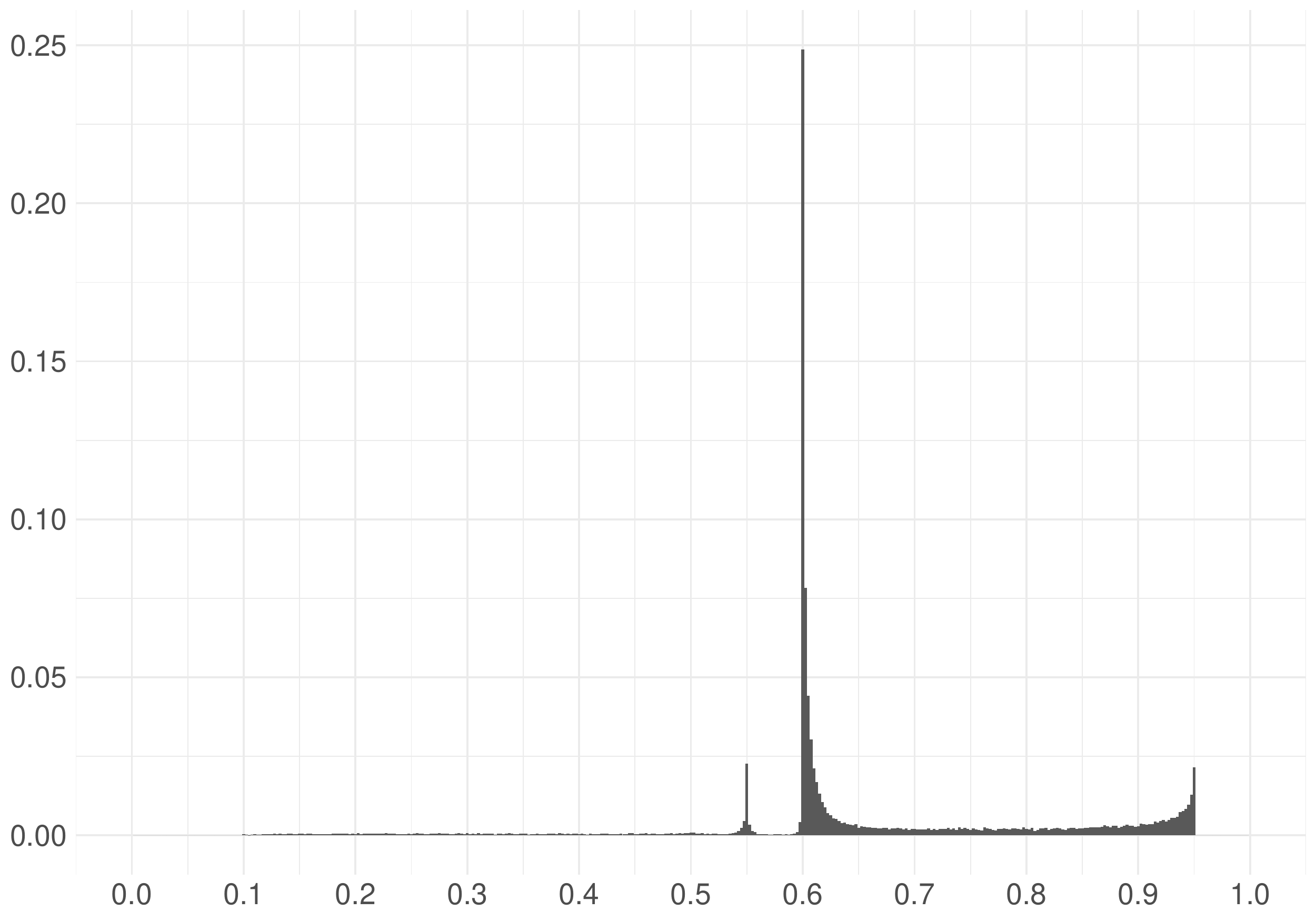}\label{fig:13:5}}
\subfigure[$T=800$, $\phi_a=1.05$, $\phi_b=0.94$]{\includegraphics[width=0.45\linewidth]{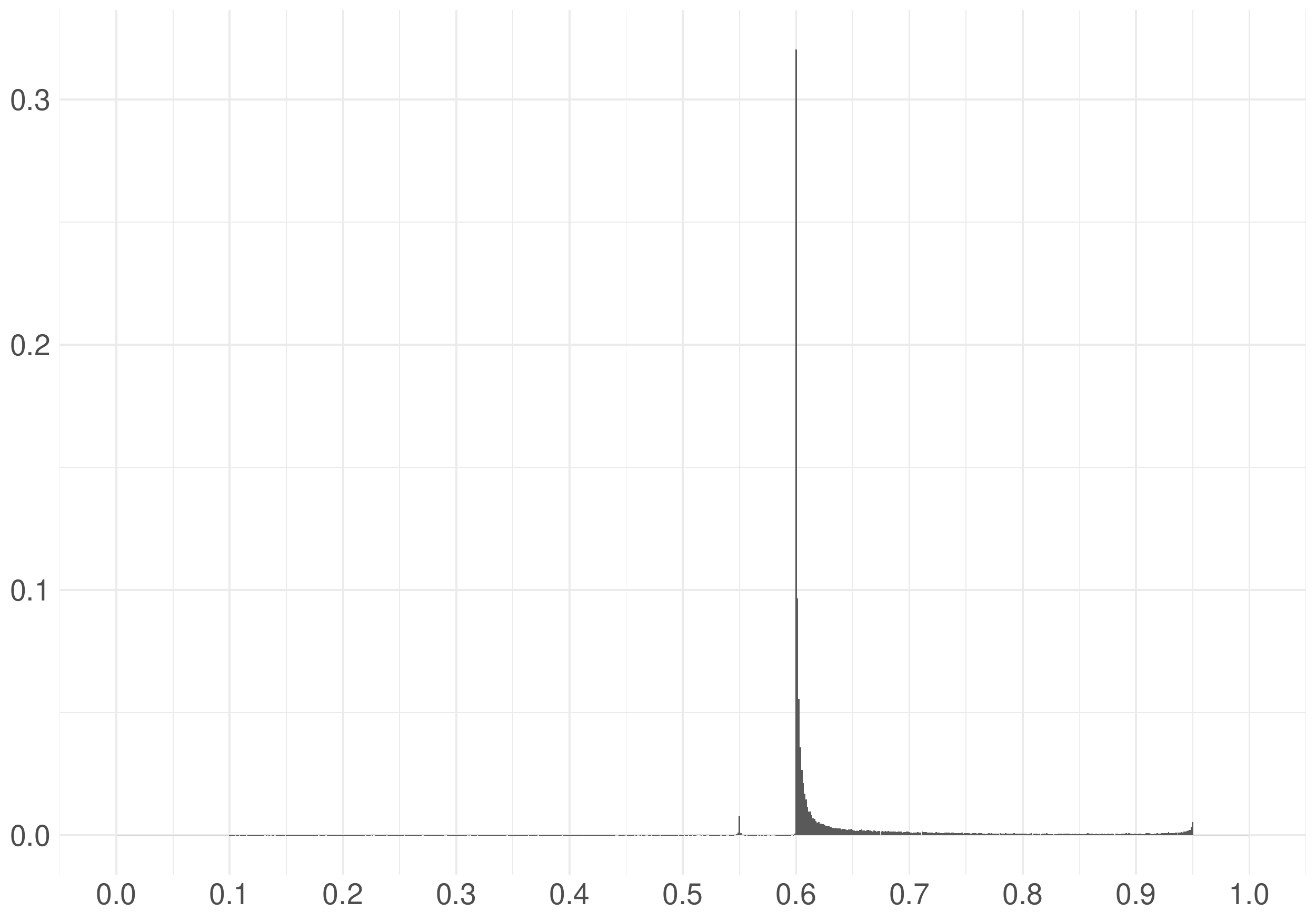}\label{fig:13:6}}
\end{center}%
\caption{Histograms of $\hat{k}_r$ % (left) and $\hat{k}_r$ (right) 
for $(\tau_e,\tau_c,\tau_r)=(0.5,0.55,0.6)$}
\label{fig13}
%\centering
%\footnotesize{OLS}
\end{figure}

\begin{figure}[h!]%
\begin{center}%
\subfigure[$T=400$, $\phi_a=1.01$, $\phi_b=0.96$]{\includegraphics[width=0.45\linewidth]{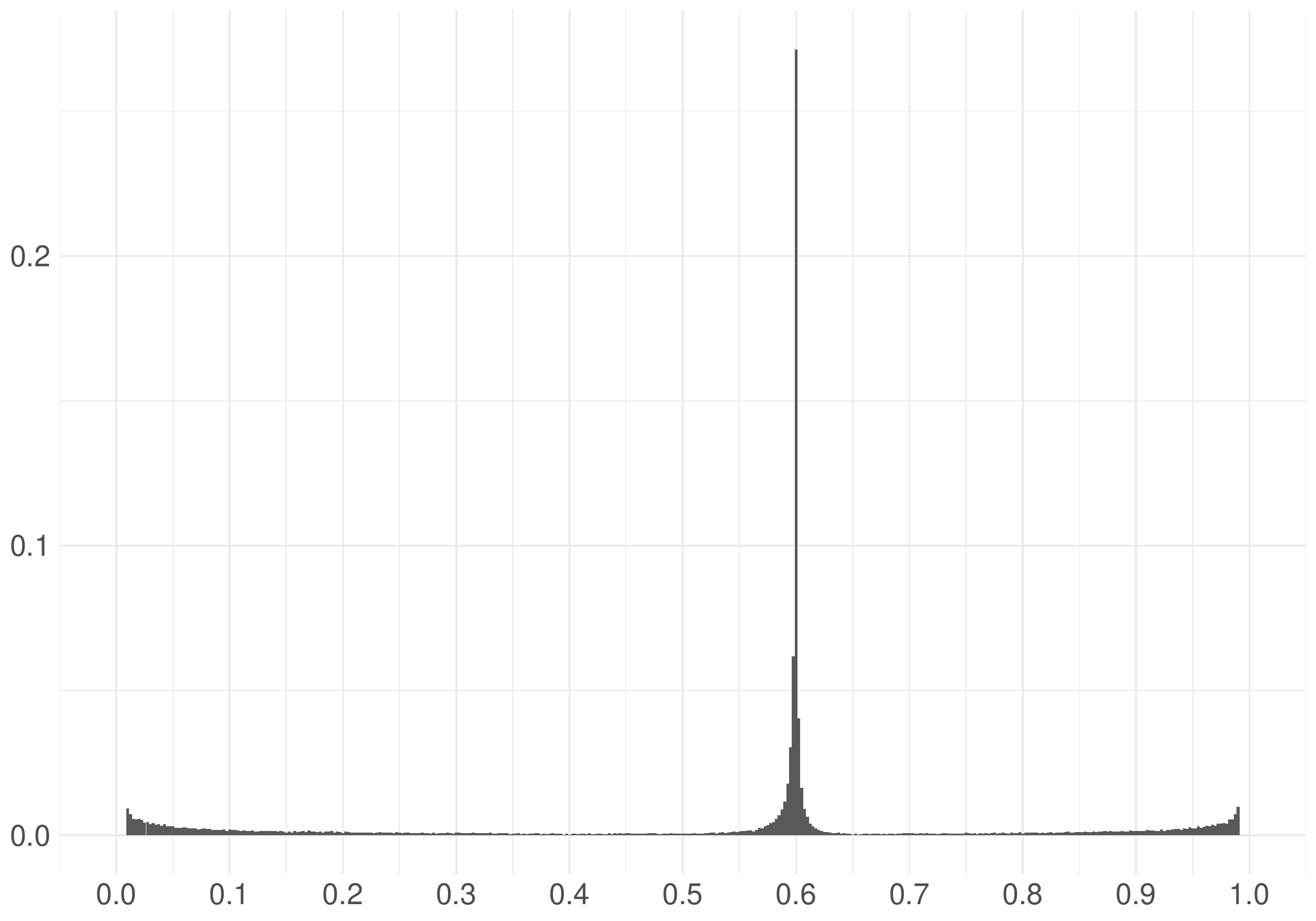}\label{fig:10:1}}
\subfigure[$T=800$, $\phi_a=1.01$, $\phi_b=0.96$]{\includegraphics[width=0.45\linewidth]{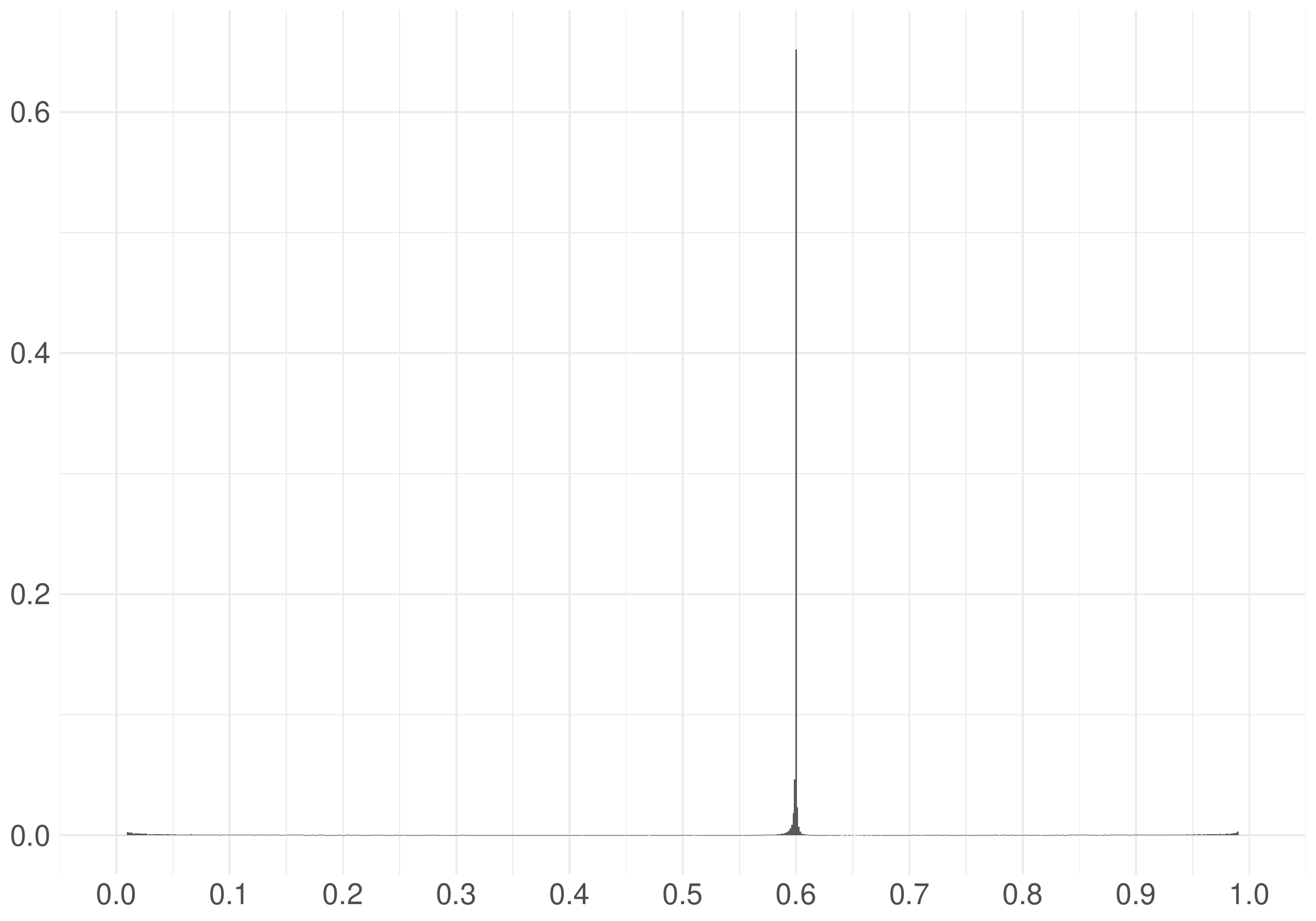}\label{fig:10:2}}\\
\subfigure[$T=400$, $\phi_a=1.05$, $\phi_b=0.96$]{\includegraphics[width=0.45\linewidth]{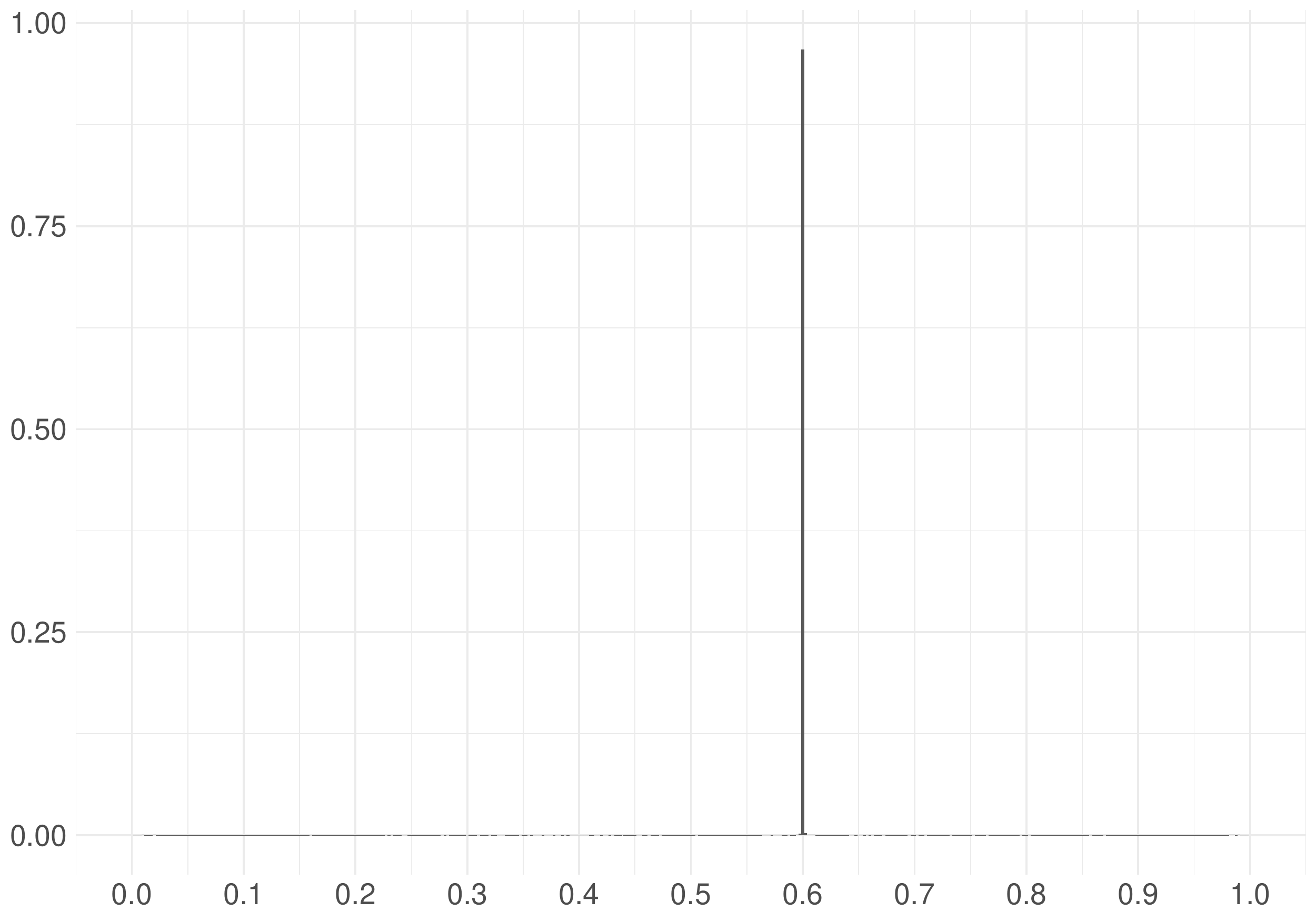}\label{fig:10:3}}
\subfigure[$T=800$, $\phi_a=1.05$, $\phi_b=0.96$]{\includegraphics[width=0.45\linewidth]{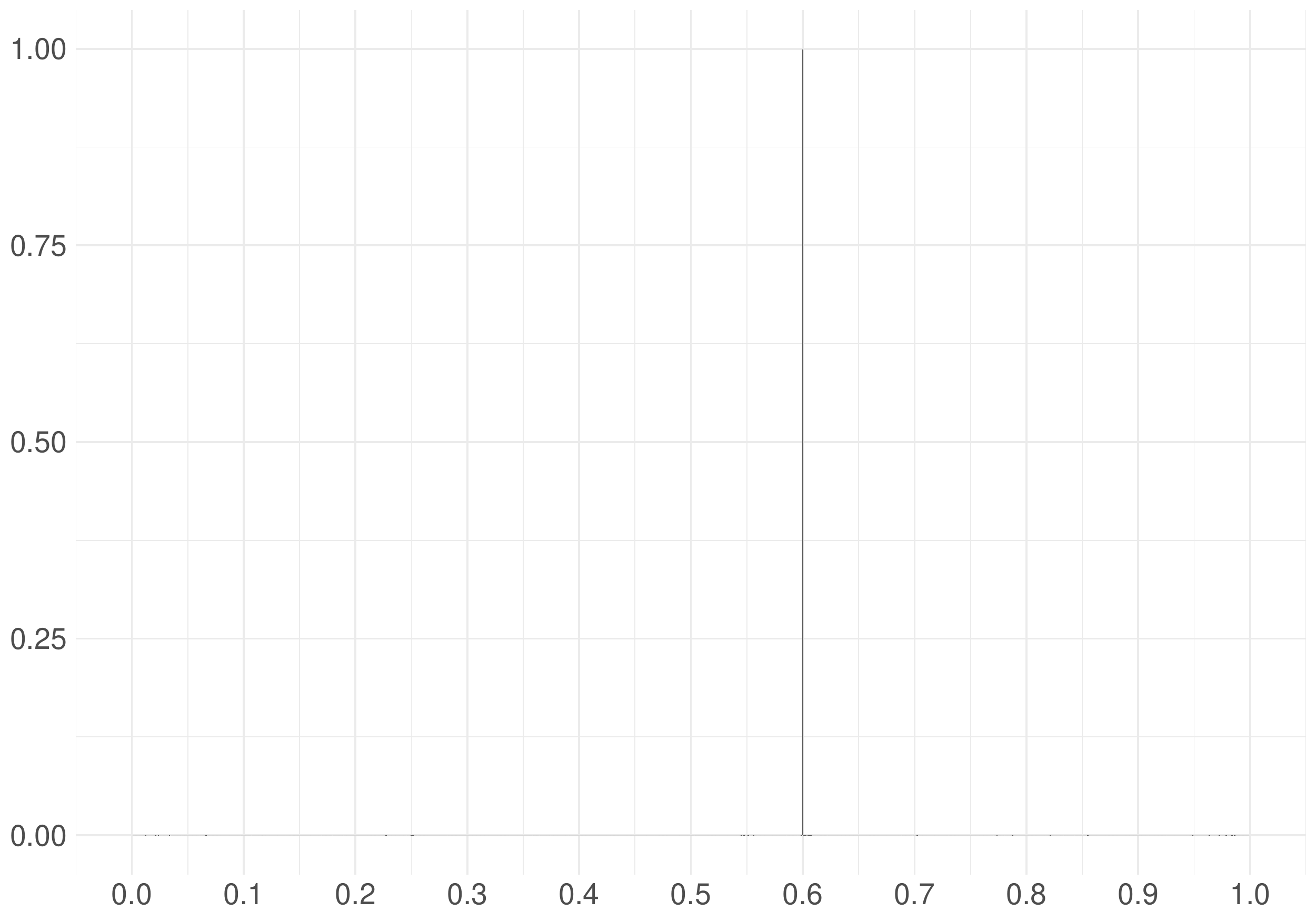}\label{fig:10:4}}\\
\subfigure[$T=400$, $\phi_a=1.09$, $\phi_b=0.96$]{\includegraphics[width=0.45\linewidth]{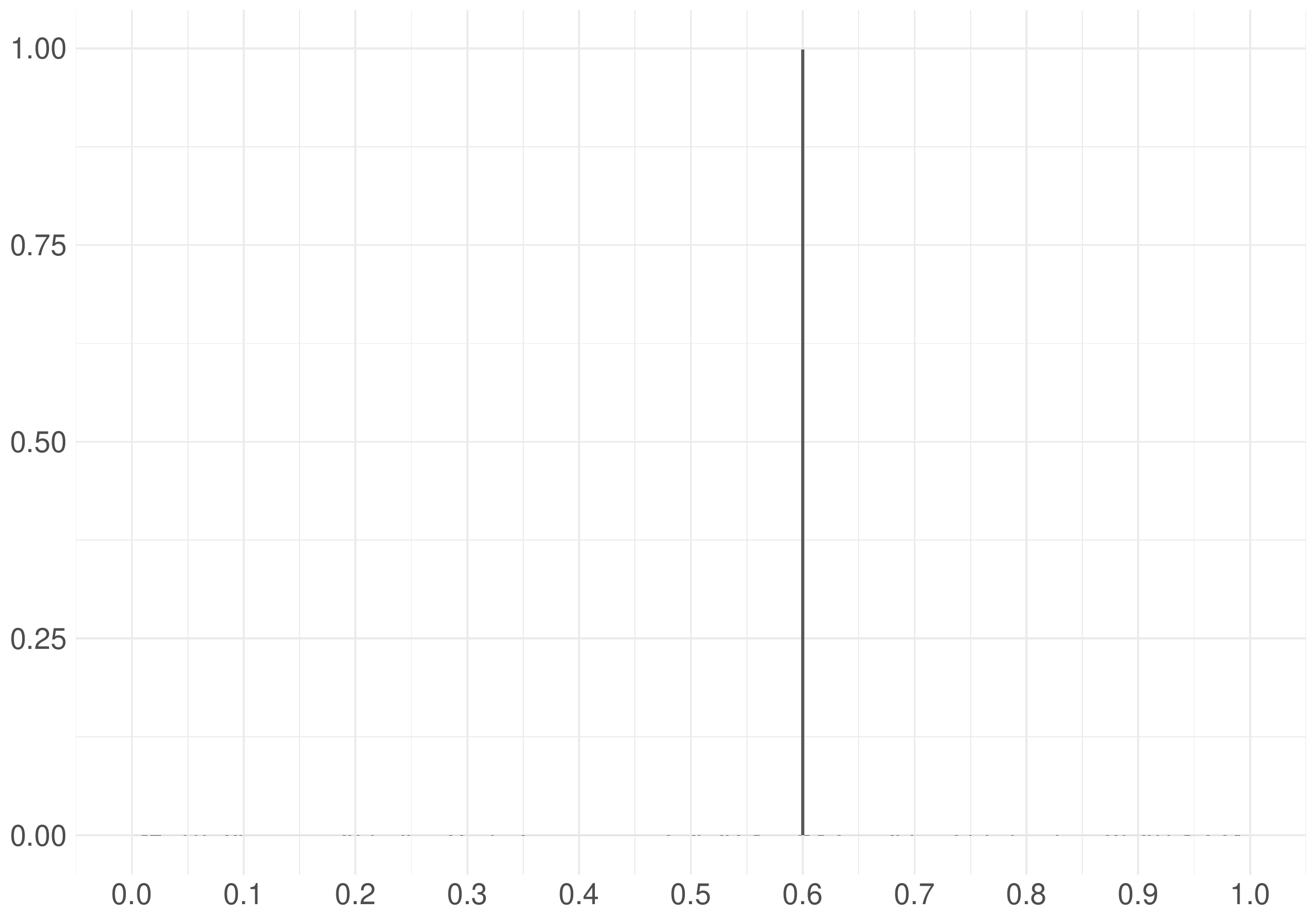}\label{fig:10:5}}
\subfigure[$T=800$, $\phi_a=1.09$, $\phi_b=0.96$]{\includegraphics[width=0.45\linewidth]{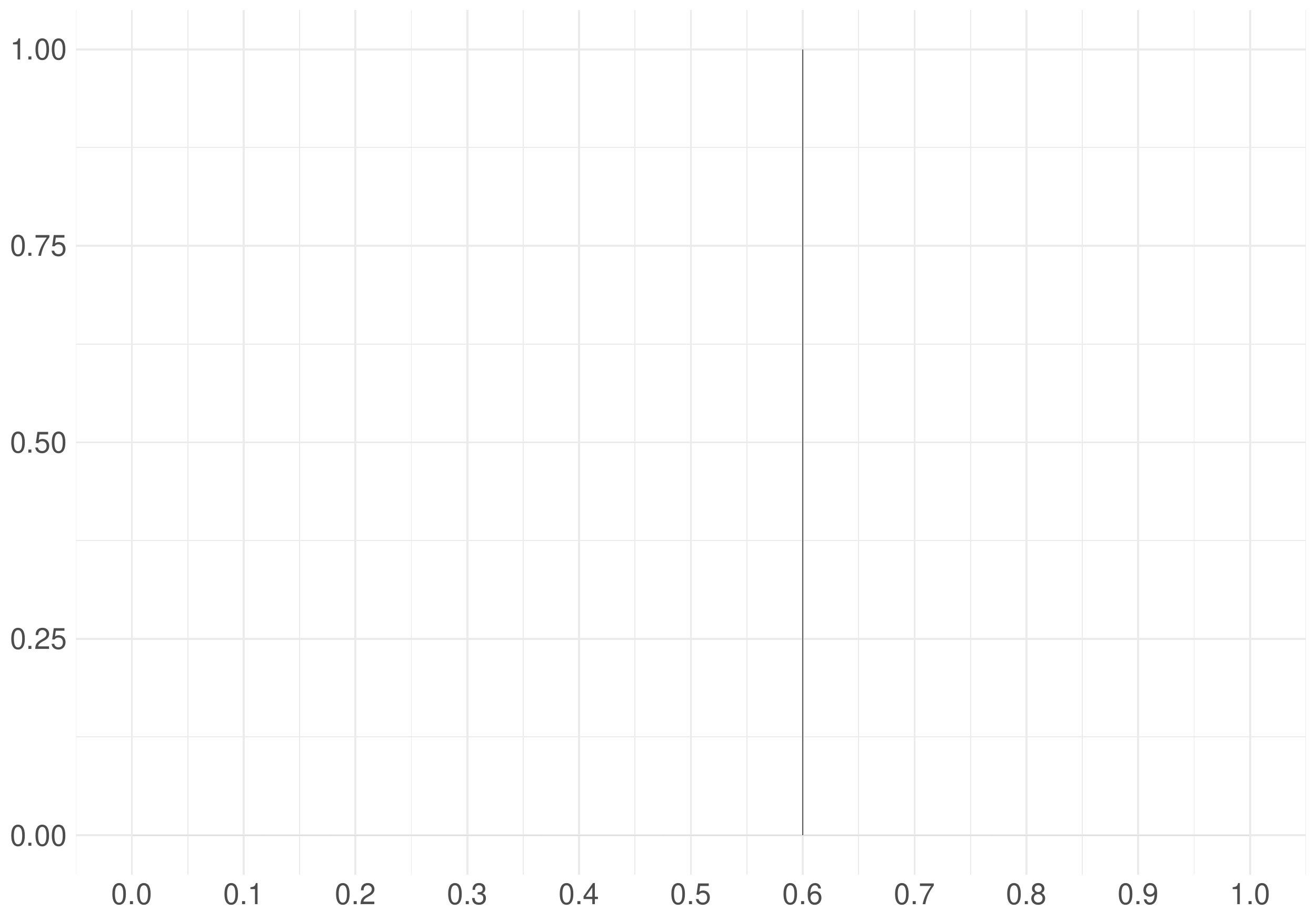}\label{fig:10:6}}
\end{center}%
\caption{Histograms of $\hat{k}_c$ % (left) and $\hat{k}_r$ (right) 
for $(\tau_e,\tau_c,\tau_r)=(0.4,0.6,0.7)$ with 1\% trimming}
\label{fig10}
%\centering
%\footnotesize{OLS}
\end{figure}

\newpage

\begin{figure}[h!]%
\begin{center}%
\subfigure[$T=400$, $\phi_a=1.05$, $\phi_b=0.98$]{\includegraphics[width=0.45\linewidth]{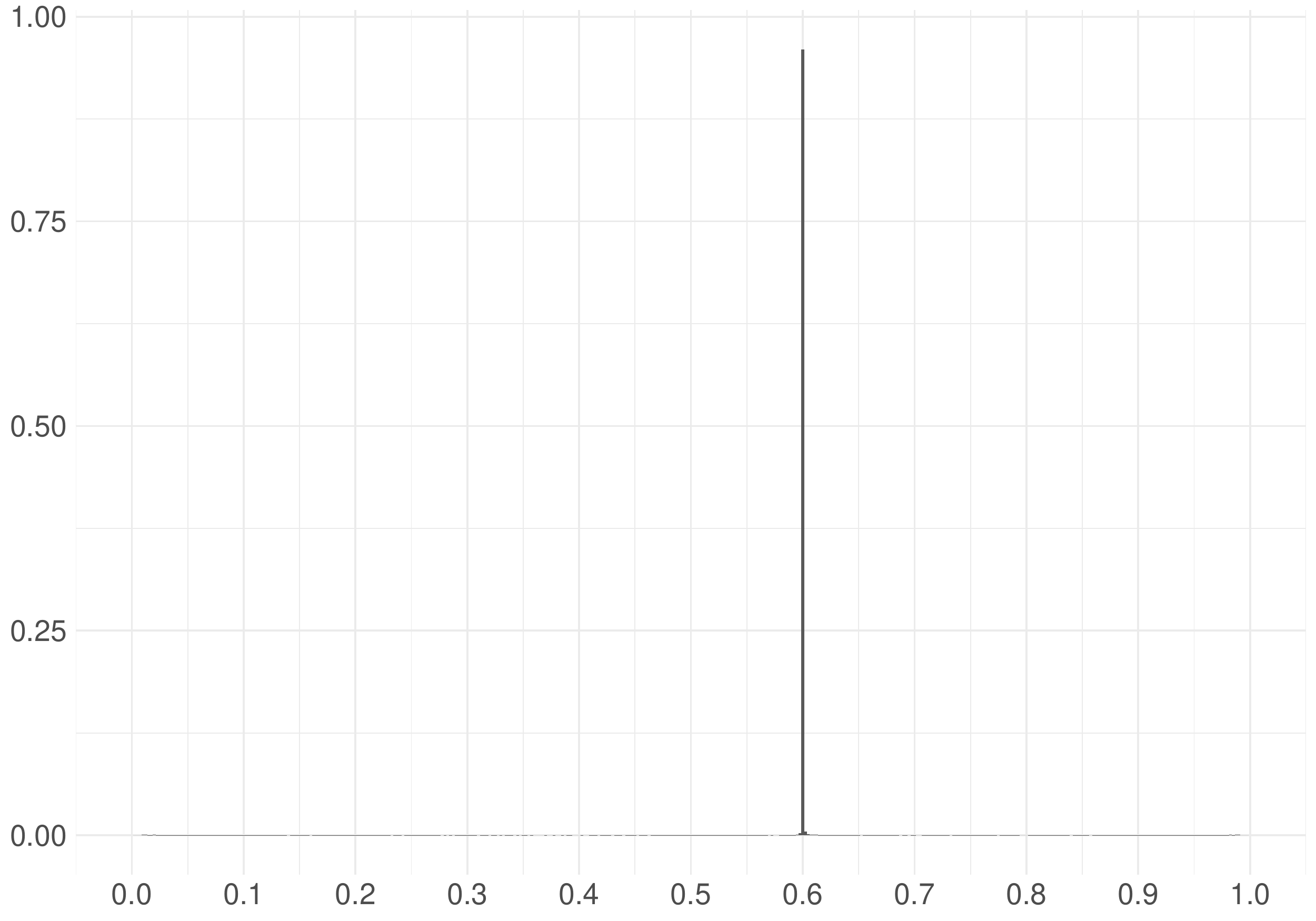}\label{fig:11:1}}
\subfigure[$T=800$, $\phi_a=1.05$, $\phi_b=0.98$]{\includegraphics[width=0.45\linewidth]{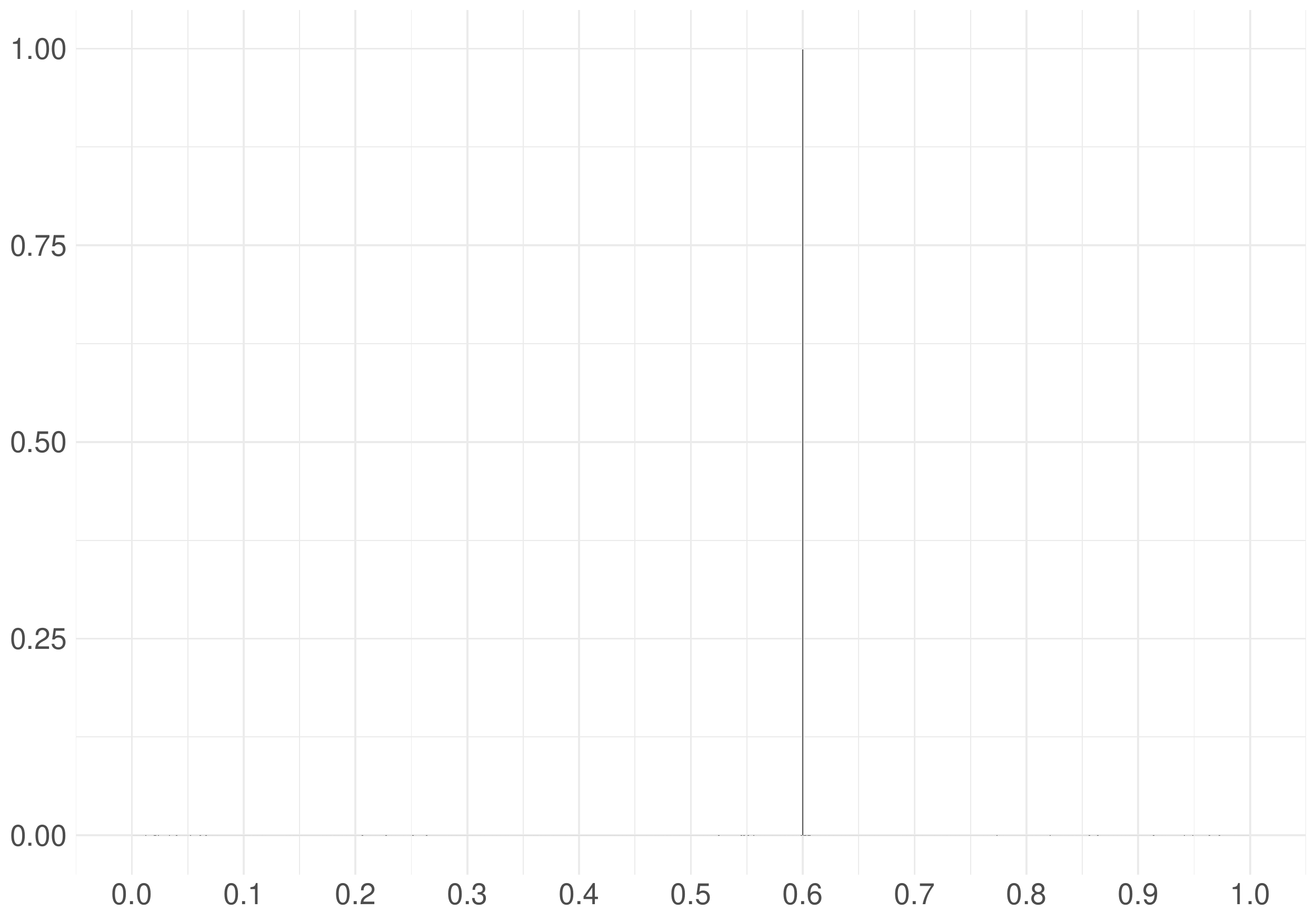}\label{fig:11:2}}\\
\subfigure[$T=400$, $\phi_a=1.05$, $\phi_b=0.96$]{\includegraphics[width=0.45\linewidth]{graph/NV_0.01_k_c_T=400_1.05_0.96_Model1s0.s11.pdf}\label{fig:11:3}}
\subfigure[$T=800$, $\phi_a=1.05$, $\phi_b=0.96$]{\includegraphics[width=0.45\linewidth]{graph/NV_0.01_k_c_T=800_1.05_0.96_Model1s0.s11.pdf}\label{fig:11:4}}\\
\subfigure[$T=400$, $\phi_a=1.05$, $\phi_b=0.94$]{\includegraphics[width=0.45\linewidth]{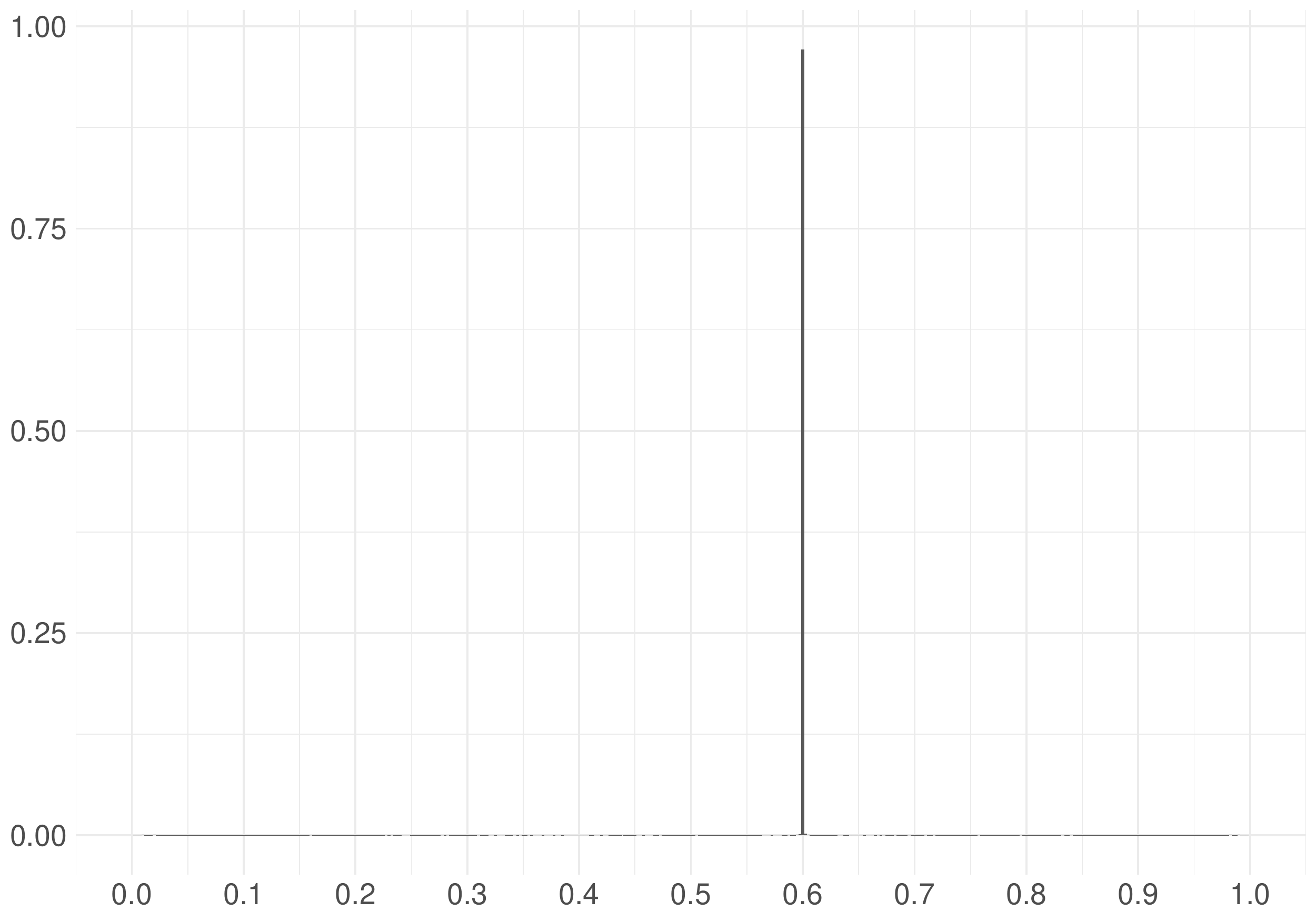}\label{fig:11:5}}
\subfigure[$T=800$, $\phi_a=1.05$, $\phi_b=0.94$]{\includegraphics[width=0.45\linewidth]{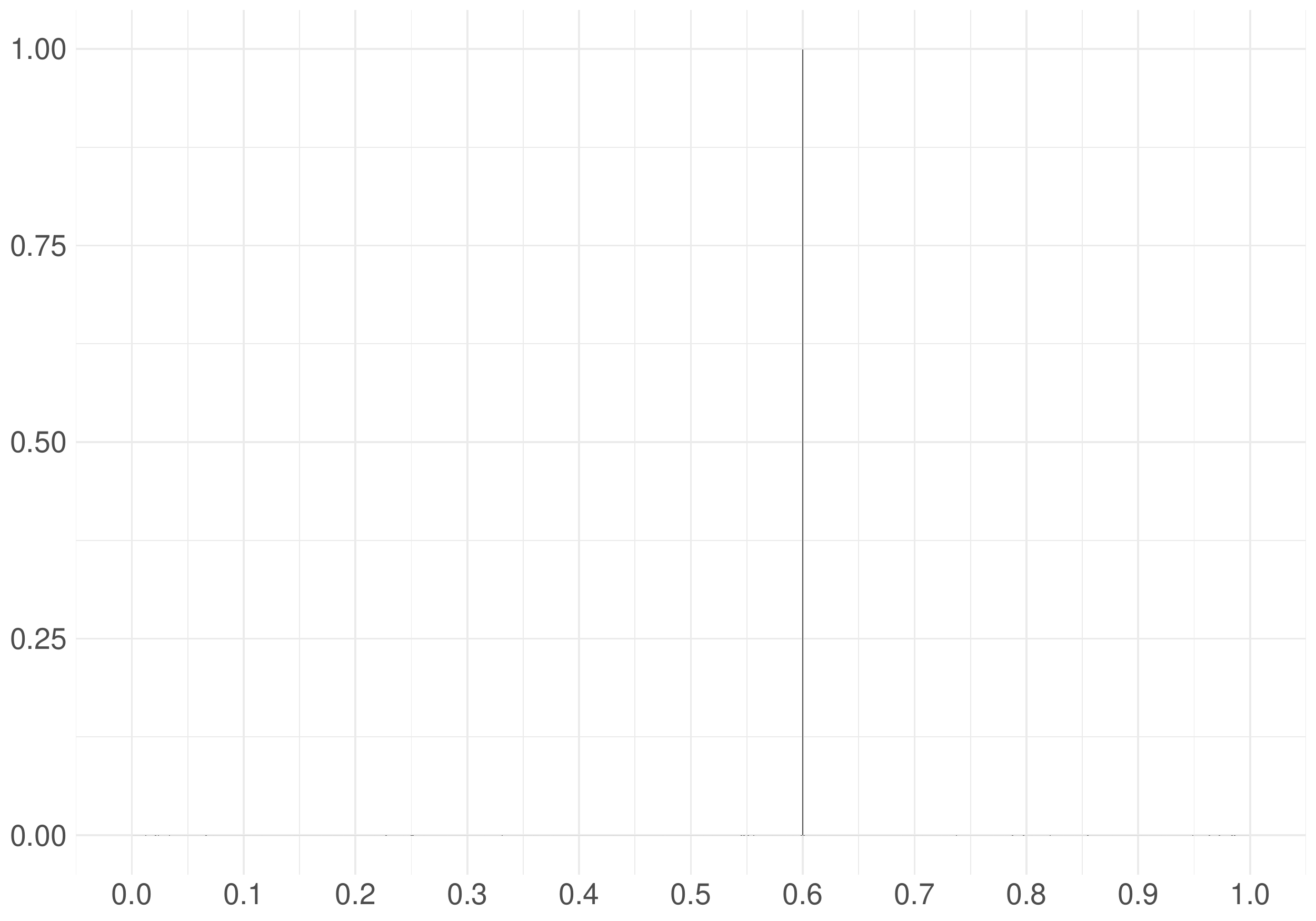}\label{fig:11:6}}
\end{center}%
\caption{Histograms of $\hat{k}_c$ % (left) and $\hat{k}_r$ (right) 
for $(\tau_e,\tau_c,\tau_r)=(0.4,0.6,0.7)$ with 1\% trimming}
\label{fig11}
%\centering
%\footnotesize{OLS}
\end{figure}

\newpage
%%k_r

\begin{figure}[h!]%
\begin{center}%
\subfigure[$T=400$, $\phi_a=1.01$, $\phi_b=0.96$]{\includegraphics[width=0.45\linewidth]{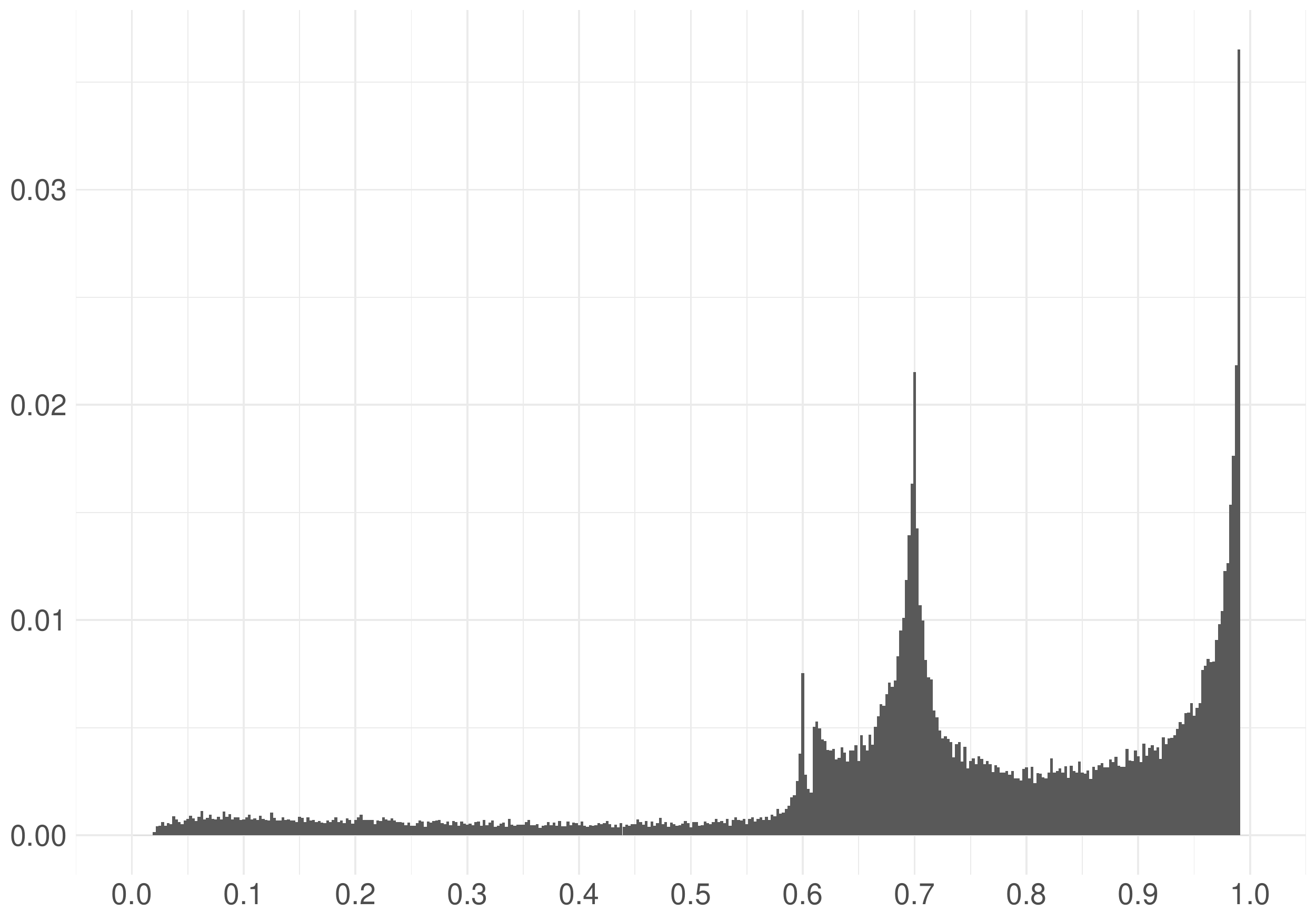}\label{fig:12:1}}
\subfigure[$T=800$, $\phi_a=1.01$, $\phi_b=0.96$]{\includegraphics[width=0.45\linewidth]{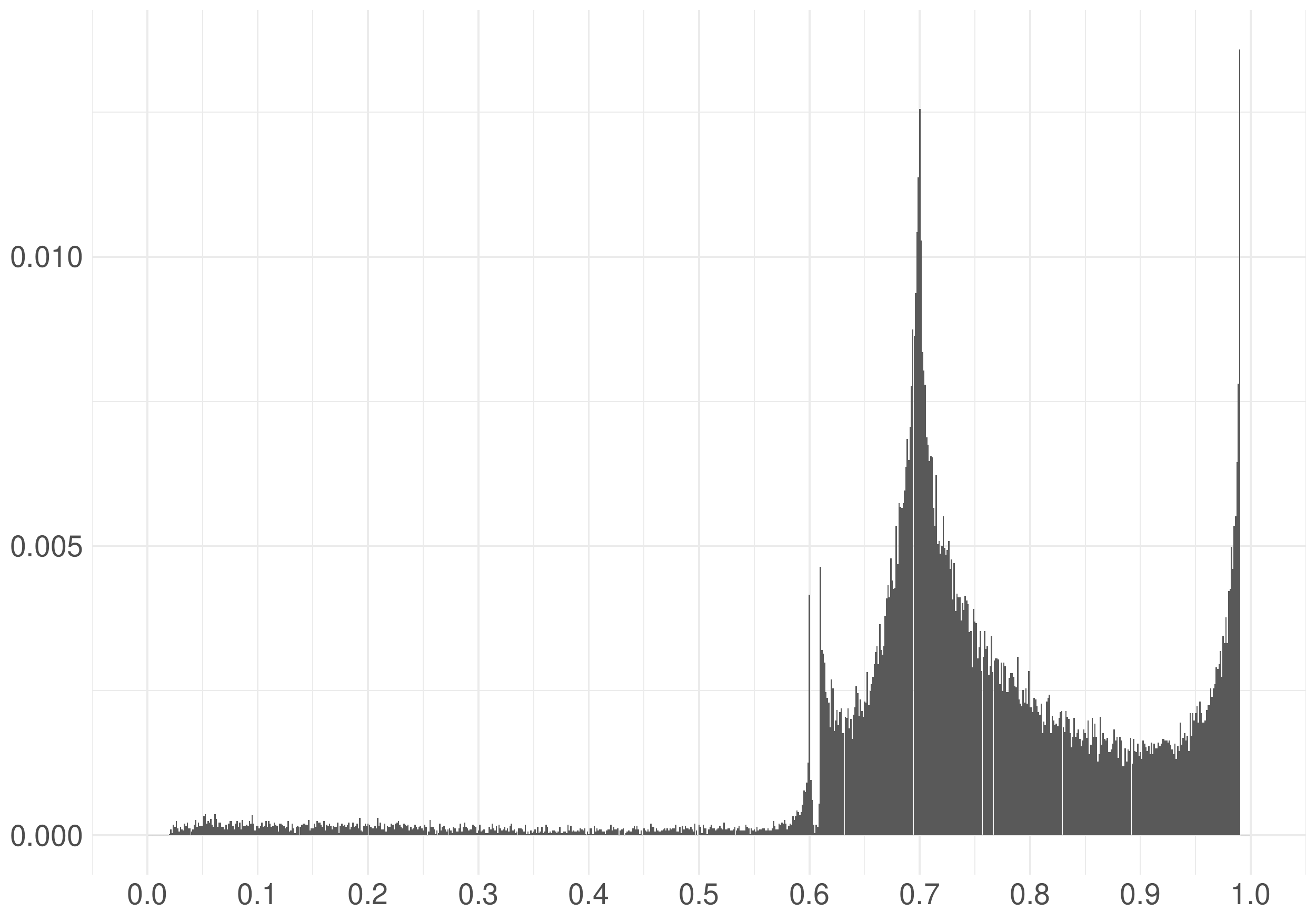}\label{fig:12:2}}\\
\subfigure[$T=400$, $\phi_a=1.05$, $\phi_b=0.96$]{\includegraphics[width=0.45\linewidth]{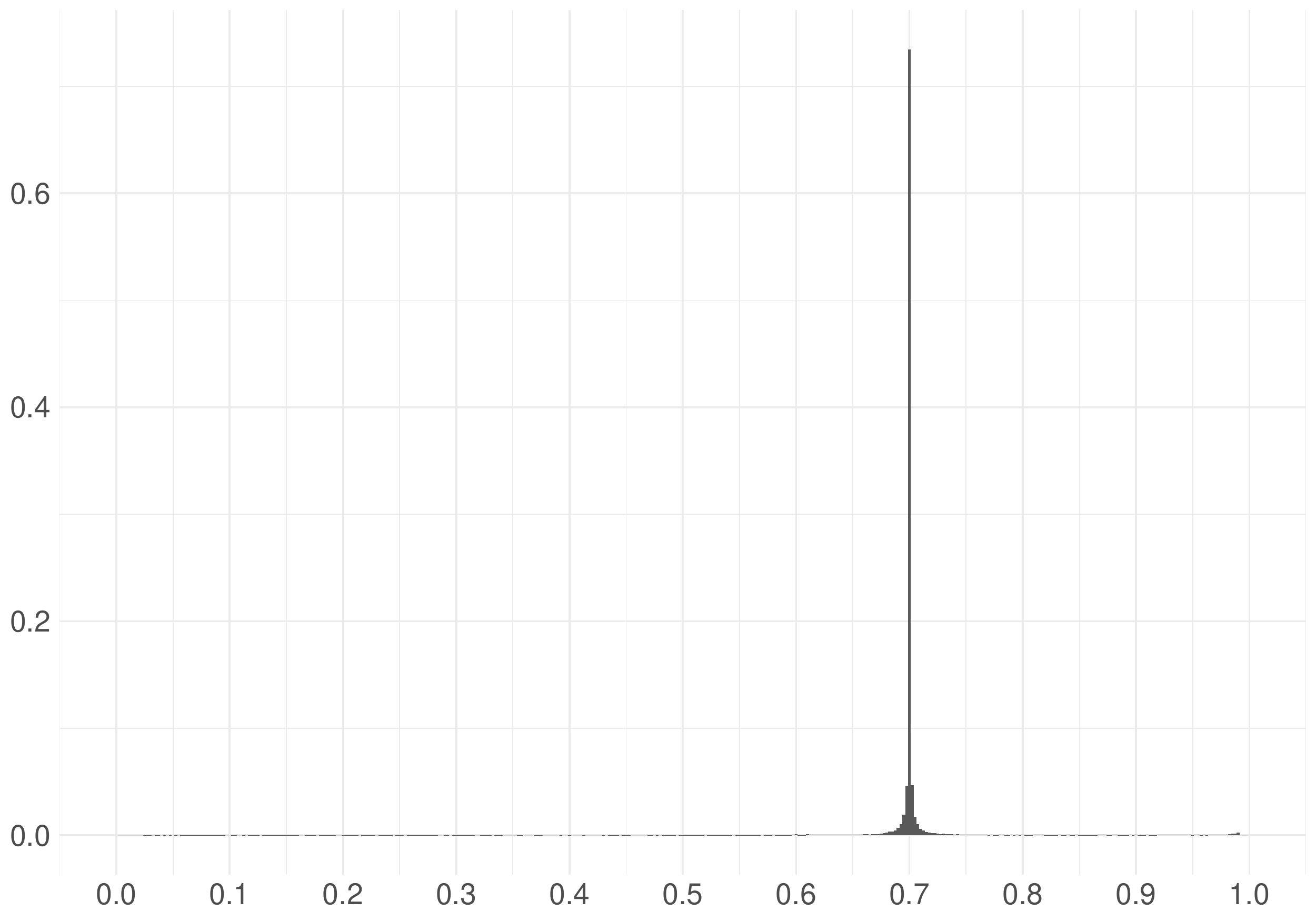}\label{fig:12:3}}
\subfigure[$T=800$, $\phi_a=1.05$, $\phi_b=0.96$]{\includegraphics[width=0.45\linewidth]{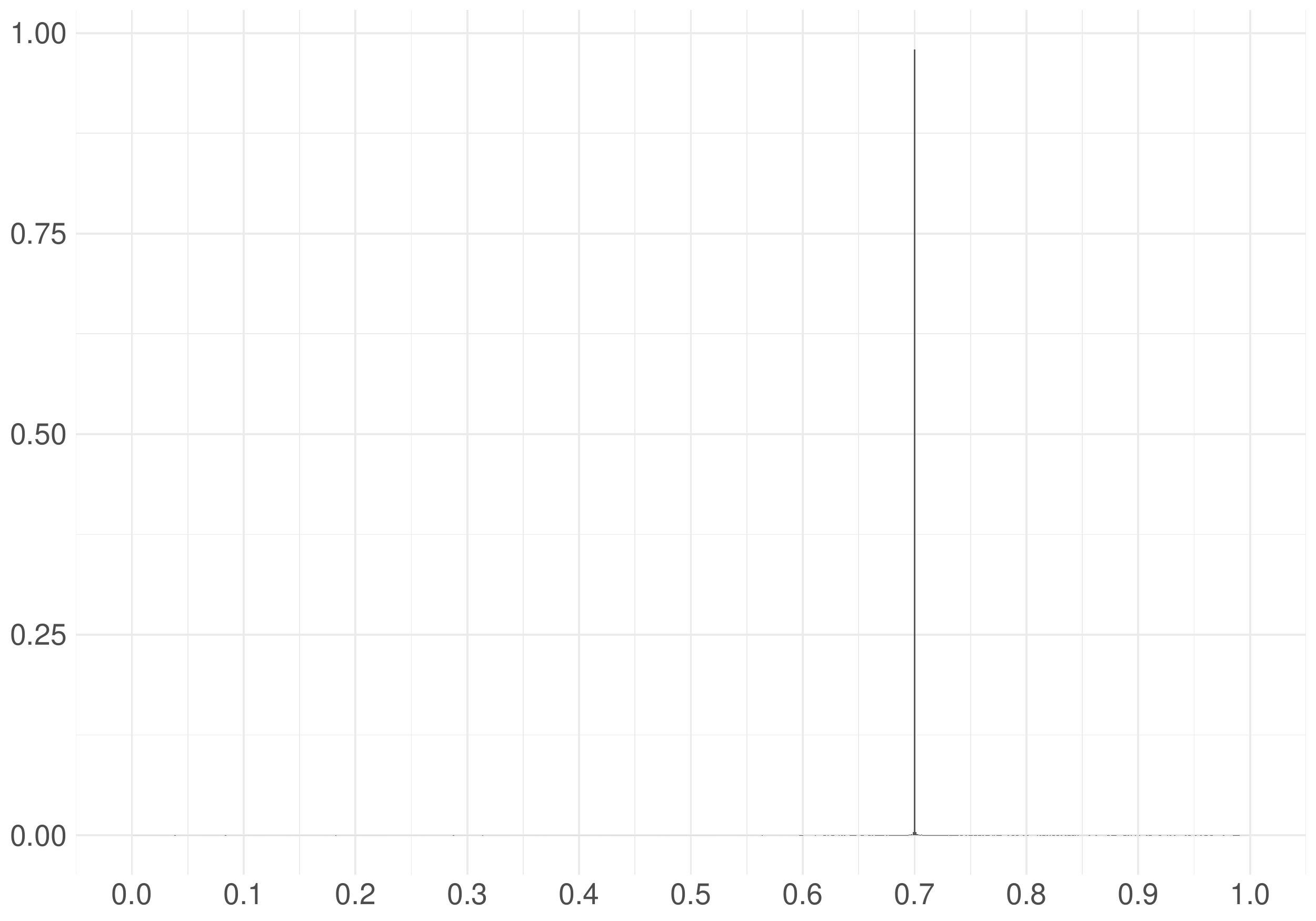}\label{fig:12:4}}\\
\subfigure[$T=400$, $\phi_a=1.09$, $\phi_b=0.96$]{\includegraphics[width=0.45\linewidth]{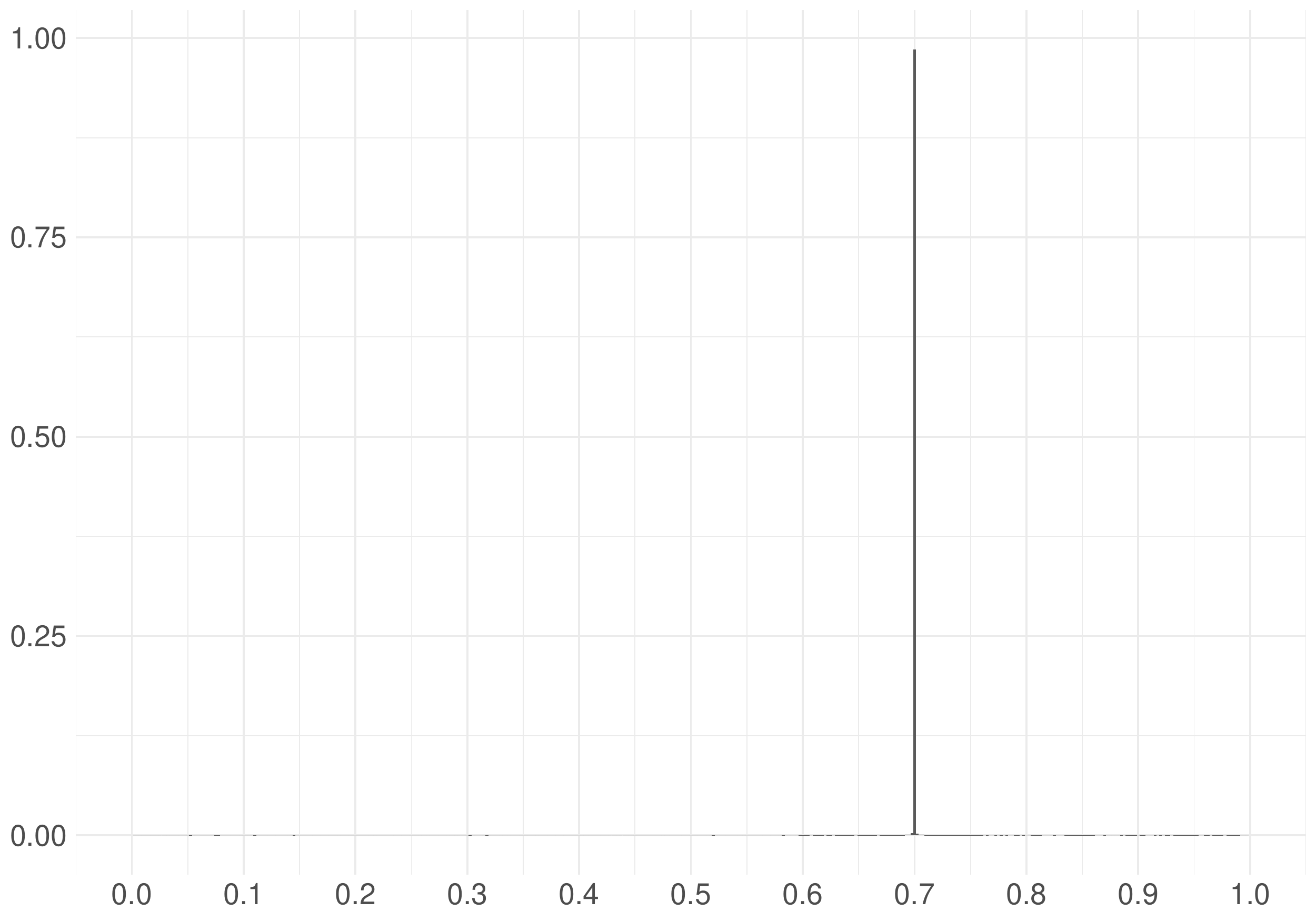}\label{fig:12:5}}
\subfigure[$T=800$, $\phi_a=1.09$, $\phi_b=0.96$]{\includegraphics[width=0.45\linewidth]{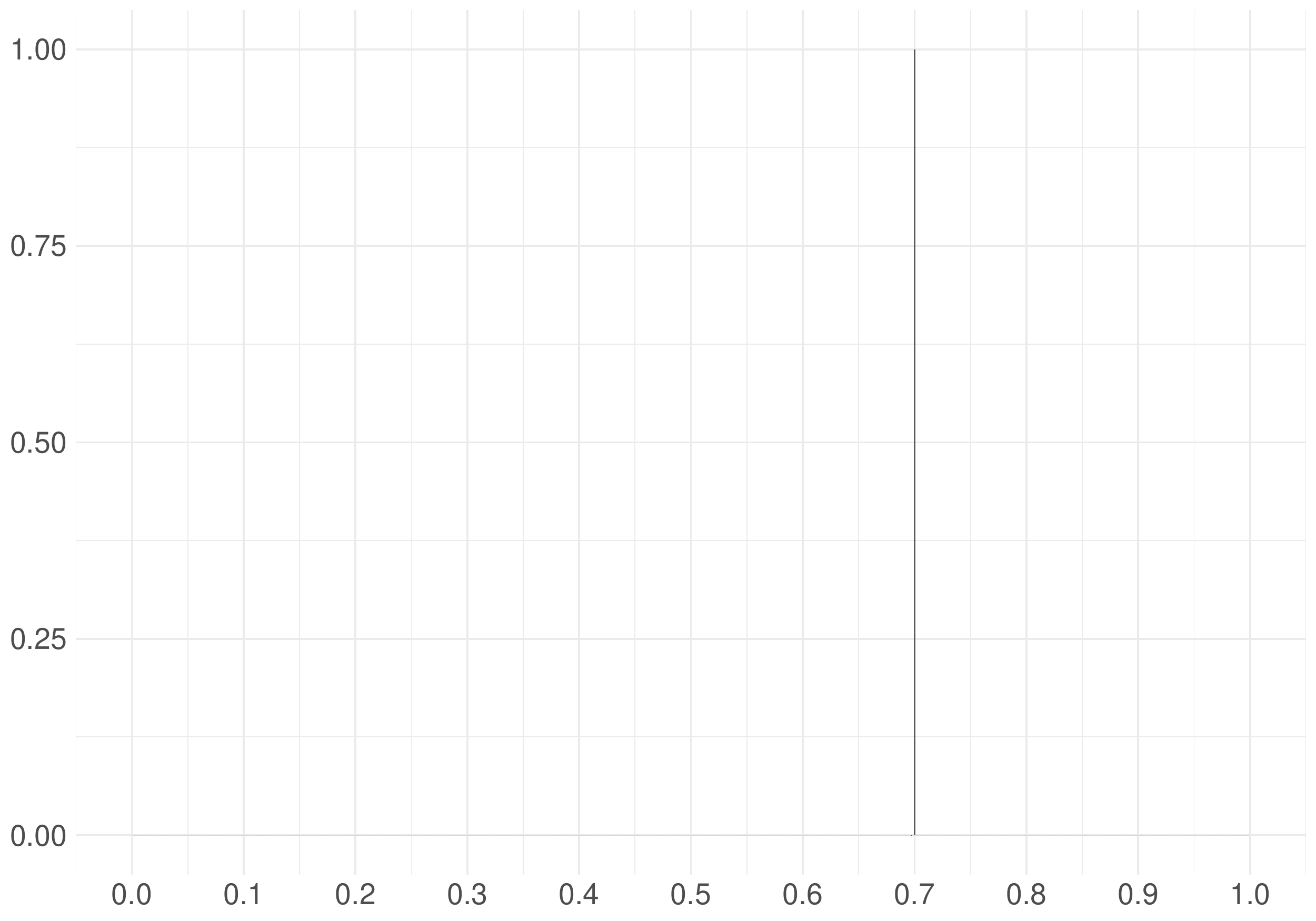}\label{fig:12:6}}
\end{center}%
\caption{Histograms of $\hat{k}_r$ % (left) and $\hat{k}_r$ (right) 
for $(\tau_e,\tau_c,\tau_r)=(0.4,0.6,0.7)$ with 1\% trimming}
\label{fig12}
%\centering
%\footnotesize{OLS}
\end{figure}

\newpage

\begin{figure}[h!]%
\begin{center}%
\subfigure[$T=400$, $\phi_a=1.05$, $\phi_b=0.98$]{\includegraphics[width=0.45\linewidth]{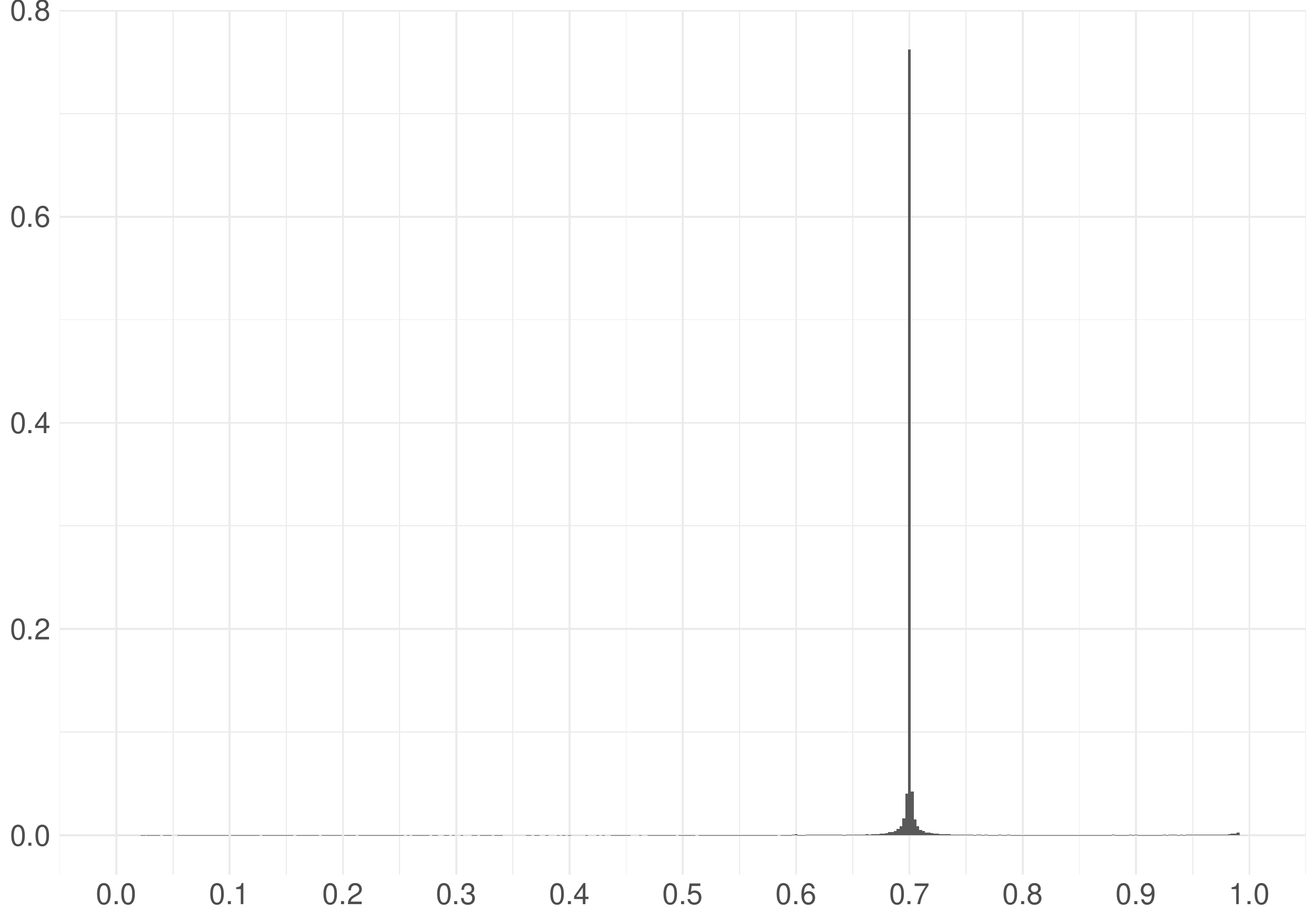}\label{fig:13:1}}
\subfigure[$T=800$, $\phi_a=1.05$, $\phi_b=0.98$]{\includegraphics[width=0.45\linewidth]{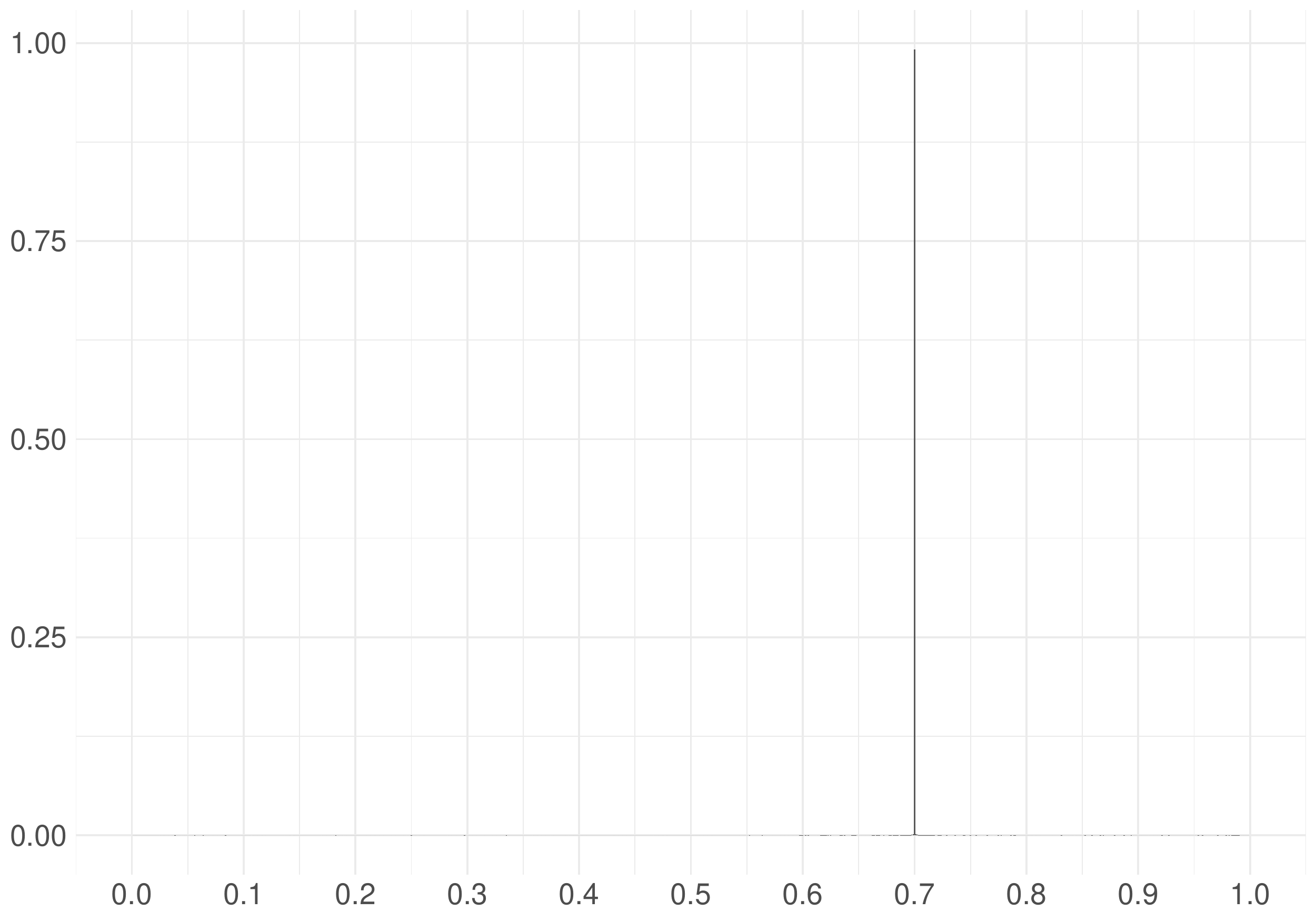}\label{fig:13:2}}\\
\subfigure[$T=400$, $\phi_a=1.05$, $\phi_b=0.96$]{\includegraphics[width=0.45\linewidth]{graph/NV_0.01_k_r_T=400_1.05_0.96_Model1s0.s11.pdf}\label{fig:13:3}}
\subfigure[$T=800$, $\phi_a=1.05$, $\phi_b=0.96$]{\includegraphics[width=0.45\linewidth]{graph/NV_0.01_k_r_T=800_1.05_0.96_Model1s0.s11.pdf}\label{fig:13:4}}\\
\subfigure[$T=400$, $\phi_a=1.05$, $\phi_b=0.94$]{\includegraphics[width=0.45\linewidth]{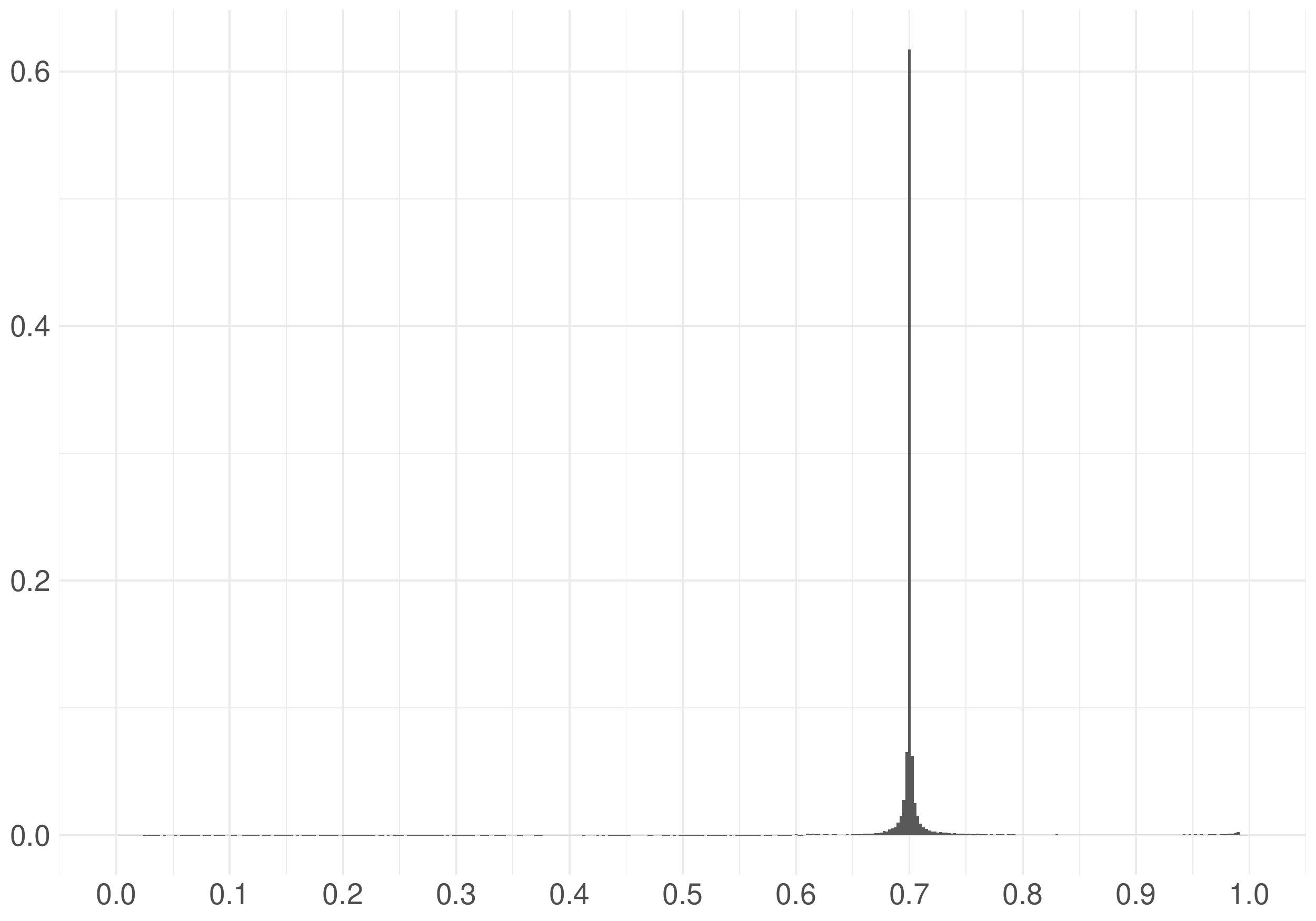}\label{fig:13:5}}
\subfigure[$T=800$, $\phi_a=1.05$, $\phi_b=0.94$]{\includegraphics[width=0.45\linewidth]{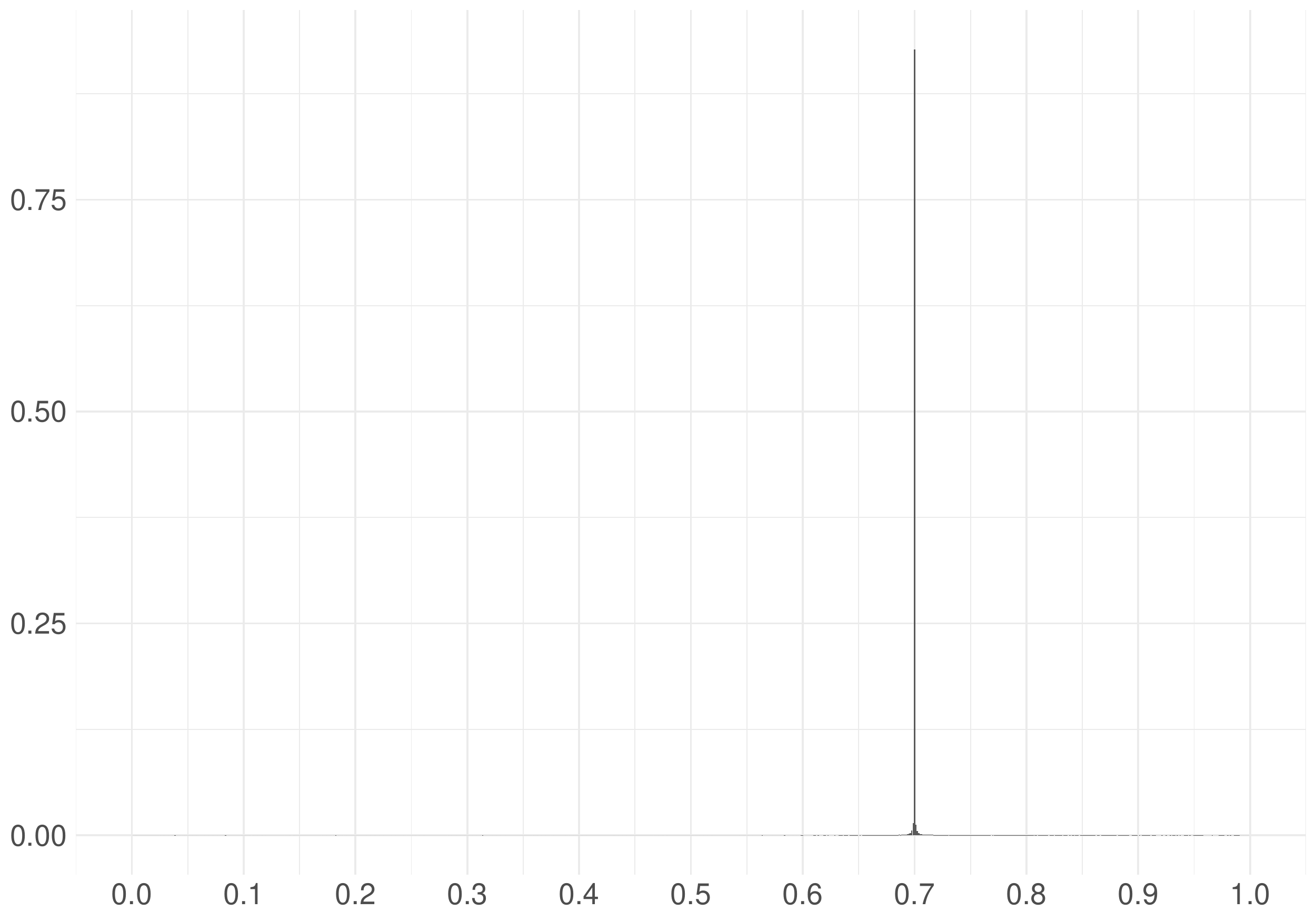}\label{fig:13:6}}
\end{center}%
\caption{Histograms of $\hat{k}_r$ % (left) and $\hat{k}_r$ (right) 
for $(\tau_e,\tau_c,\tau_r)=(0.4,0.6,0.7)$ with 1\% trimming}
\label{fig13}
%\centering
%\footnotesize{OLS}
\end{figure}

\newpage
%%%
%s_0/s_1=1/3
%%%

\begin{figure}[h]%
\begin{center}%
\subfigure[$T=400$, $\phi_a=1.01$, $\phi_b=0.96$]{\includegraphics[width=0.45\linewidth]{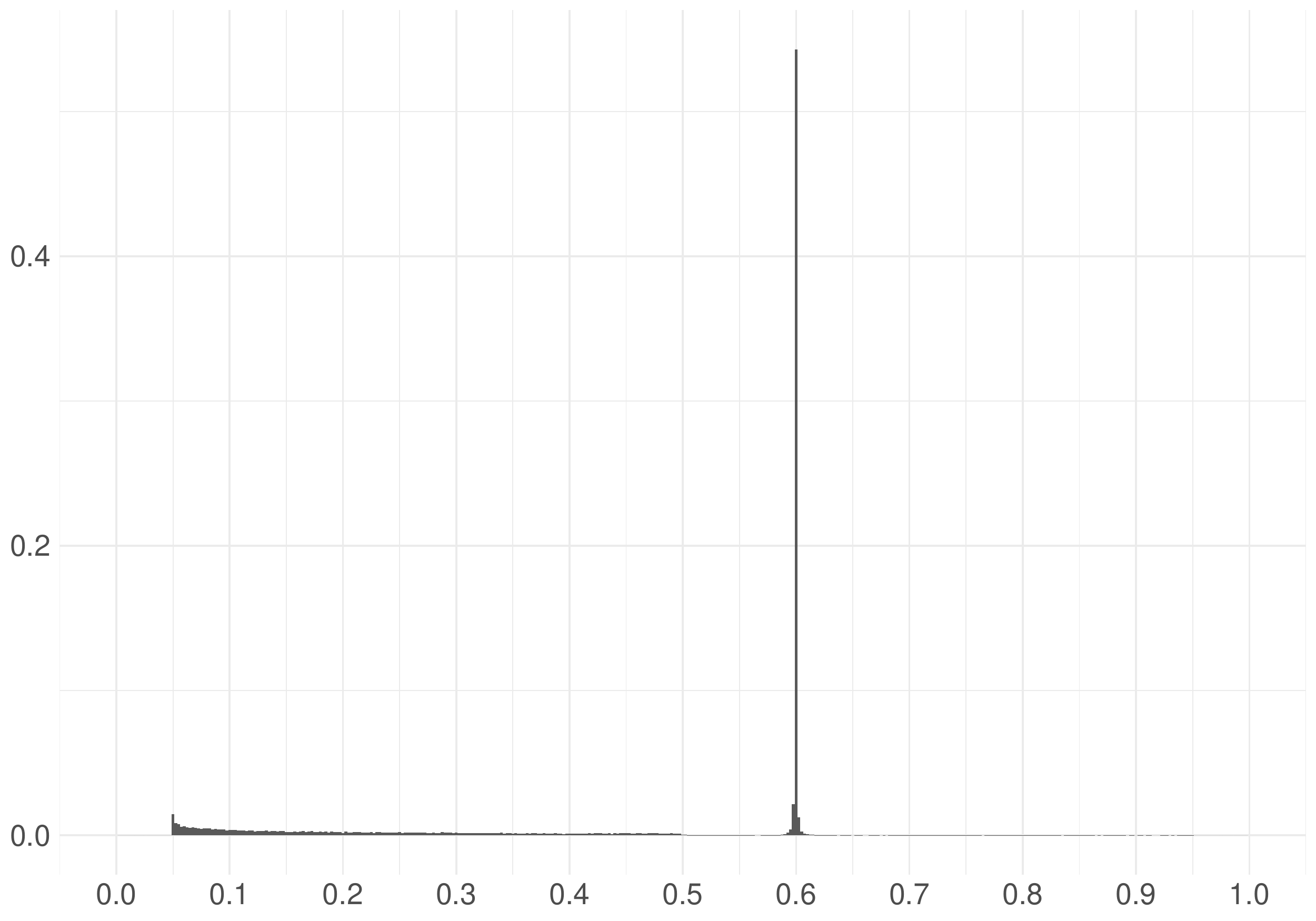}\label{fig:14:1}}
\subfigure[$T=800$, $\phi_a=1.01$, $\phi_b=0.96$]{\includegraphics[width=0.45\linewidth]{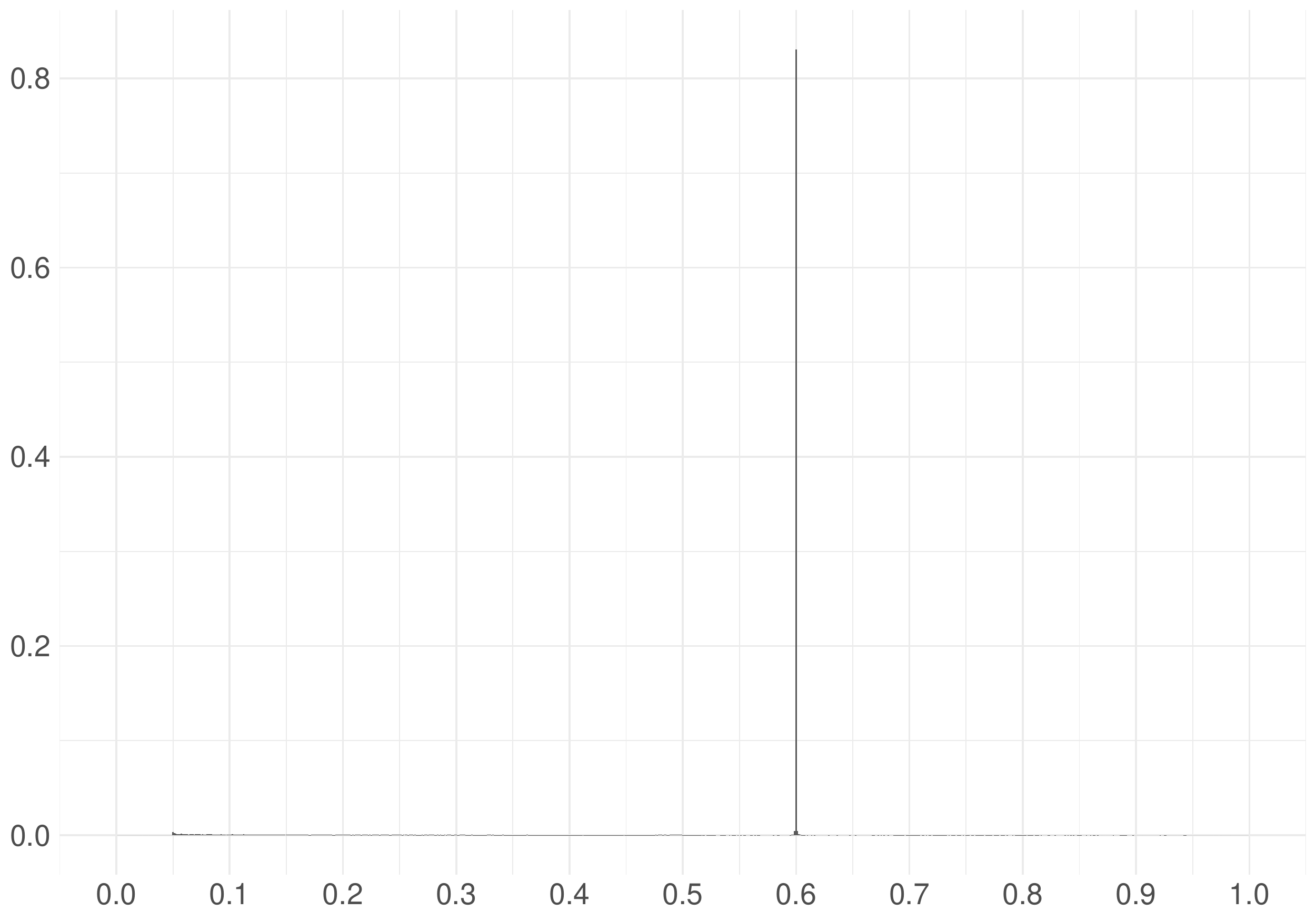}\label{fig:14:2}}\\
\subfigure[$T=400$, $\phi_a=1.05$, $\phi_b=0.96$]{\includegraphics[width=0.45\linewidth]{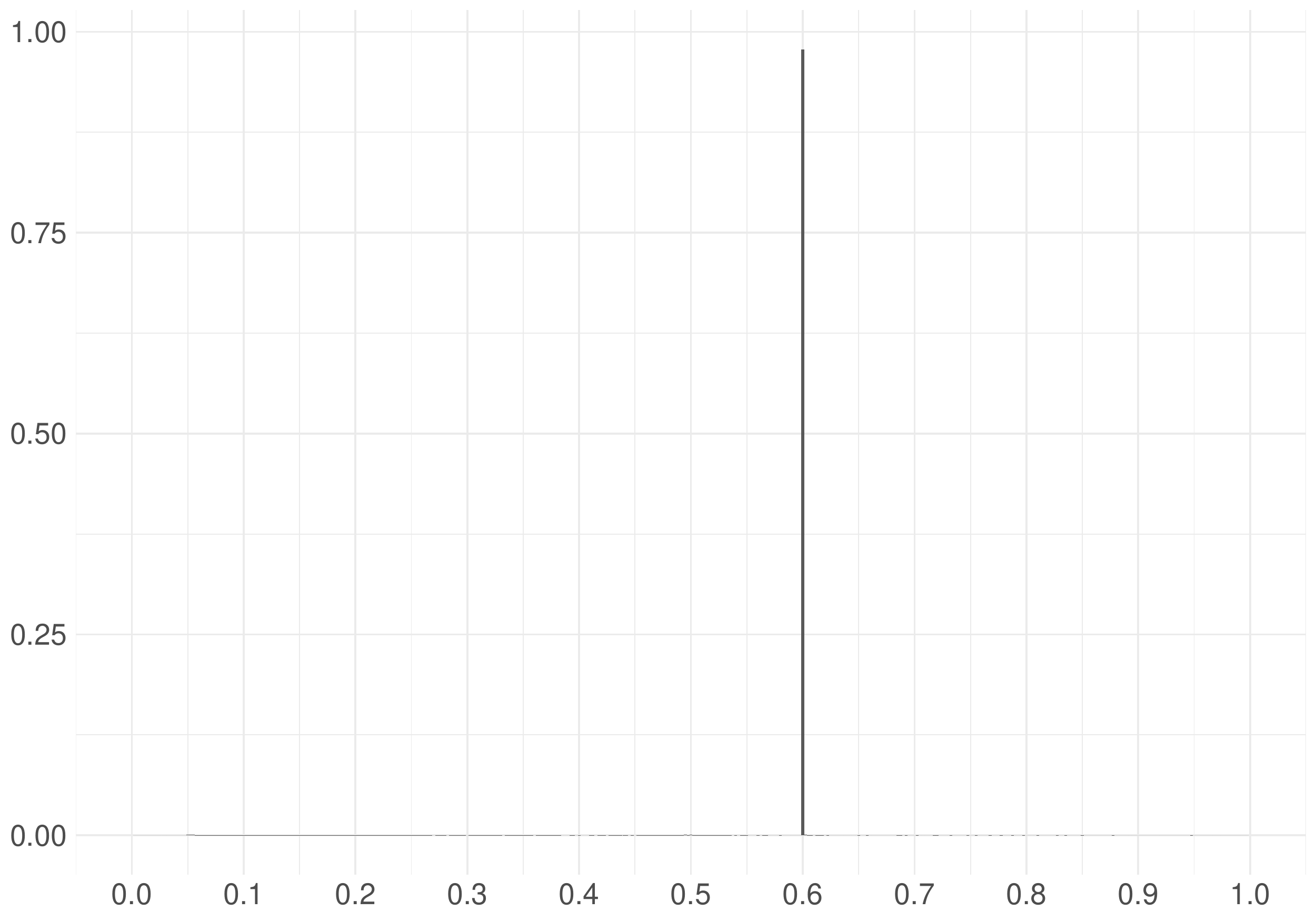}\label{fig:14:3}}
\subfigure[$T=800$, $\phi_a=1.05$, $\phi_b=0.96$]{\includegraphics[width=0.45\linewidth]{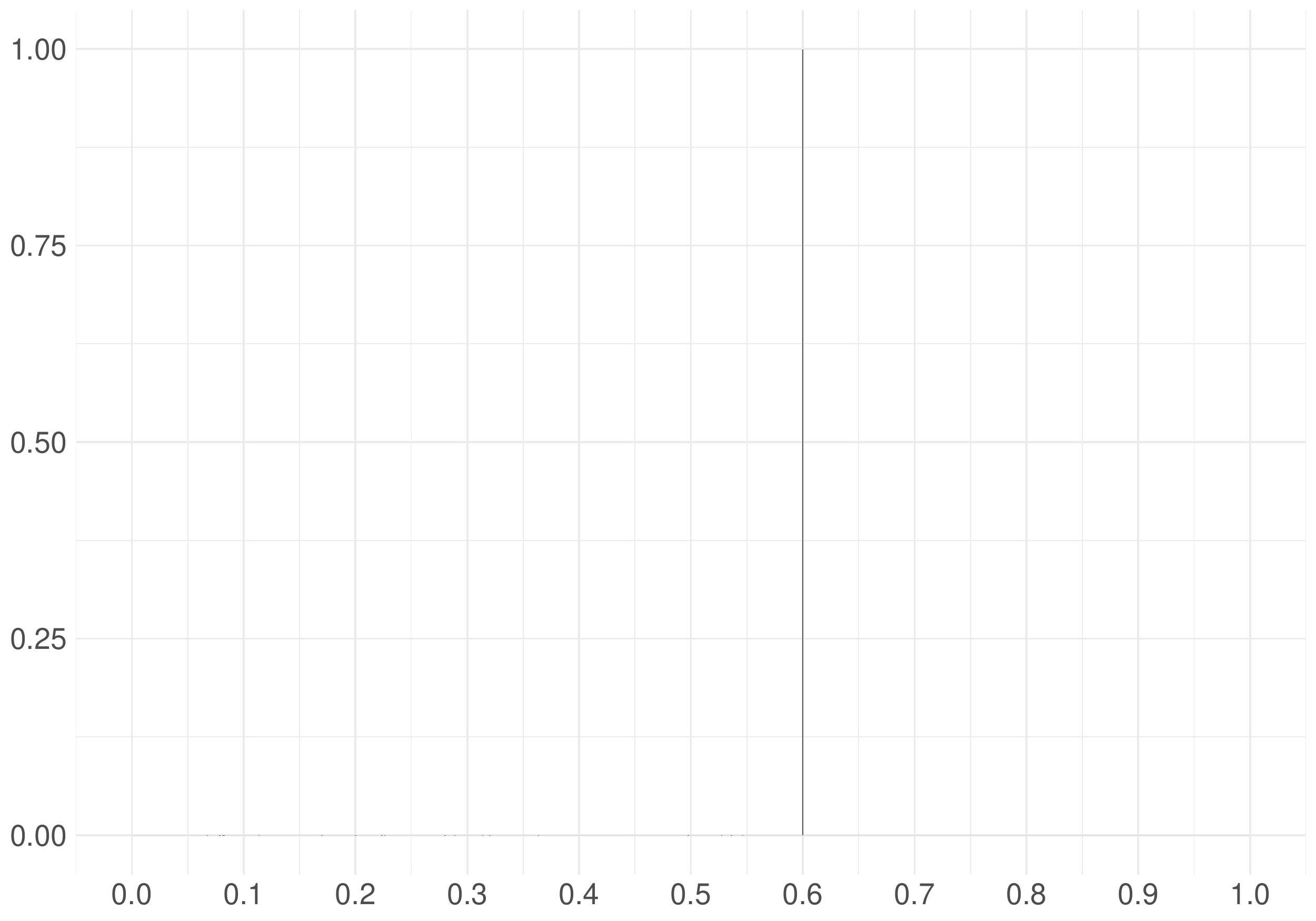}\label{fig:14:4}}\\
\subfigure[$T=400$, $\phi_a=1.09$, $\phi_b=0.96$]{\includegraphics[width=0.45\linewidth]{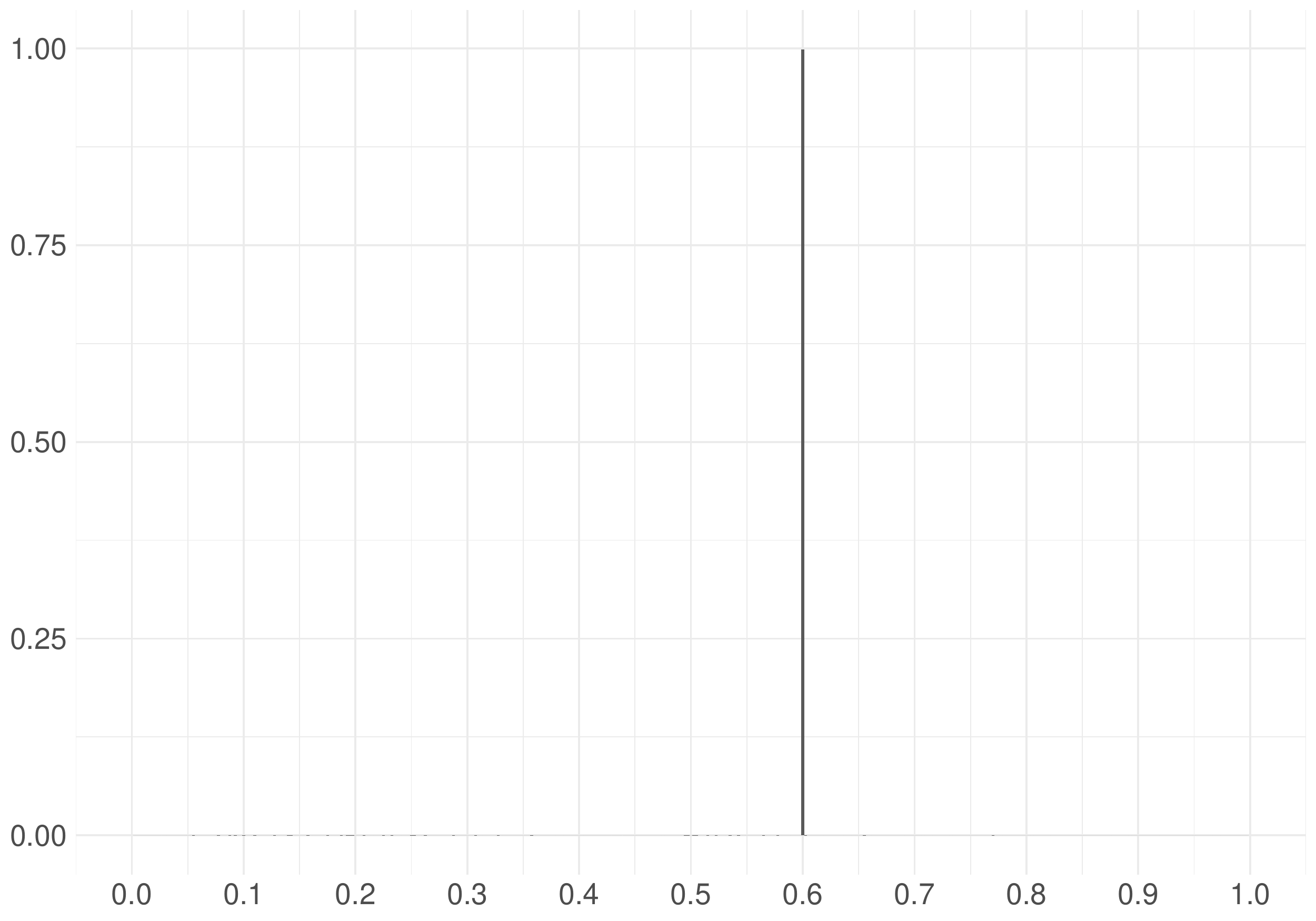}\label{fig:14:5}}
\subfigure[$T=800$, $\phi_a=1.09$, $\phi_b=0.96$]{\includegraphics[width=0.45\linewidth]{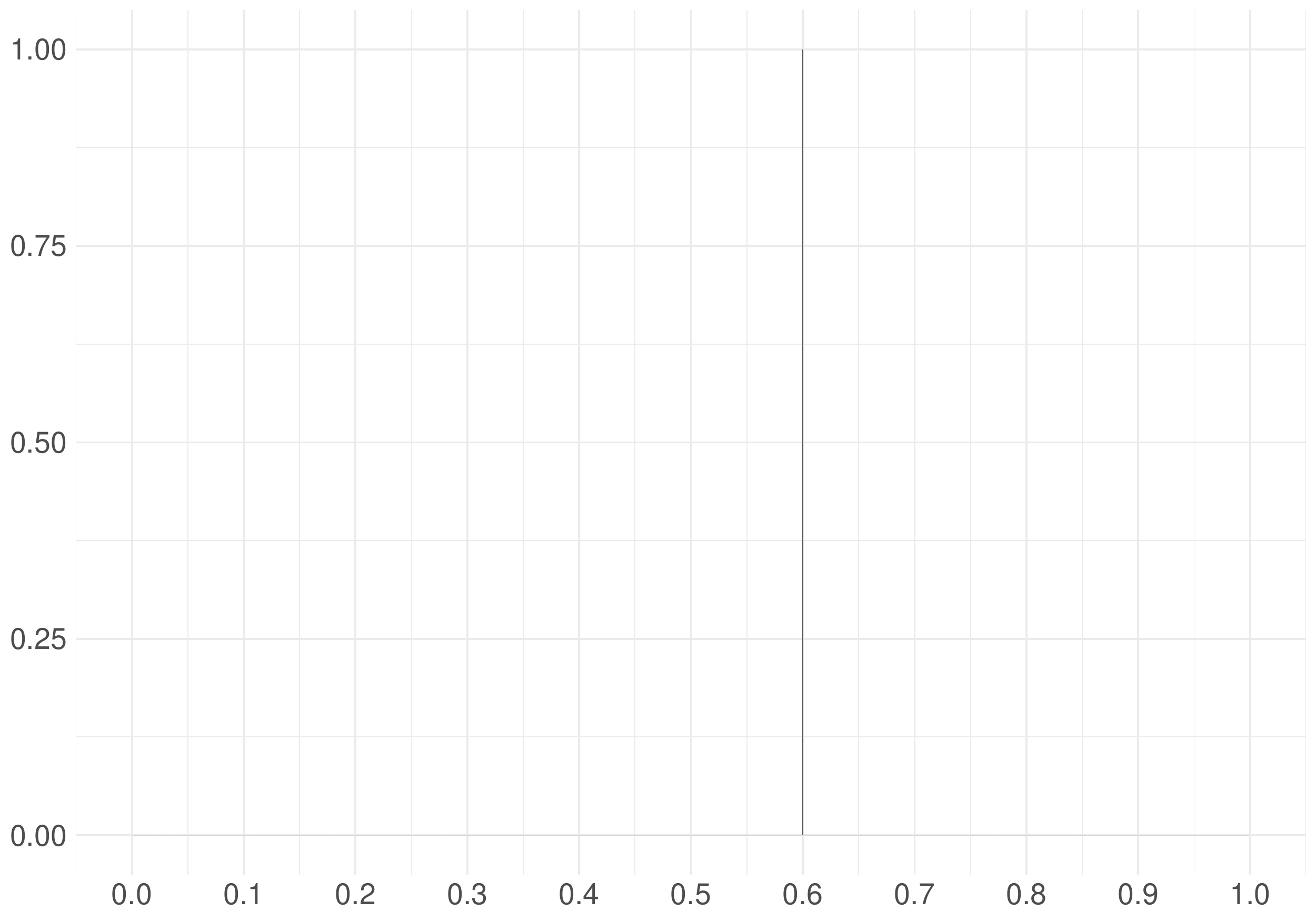}\label{fig:14:6}}
\end{center}%
\caption{Histograms of $\hat{k}_c$ % (left) and $\hat{k}_r$ (right) 
for $(\tau_e,\tau_c,\tau_r)=(0.4,0.6,0.7)$ with $\sigma_1/\sigma_0=1/3$}
\label{fig14}
%\centering
%\footnotesize{OLS}
\end{figure}

\begin{figure}[h]%
\begin{center}%
\subfigure[$T=400$, $\phi_a=1.05$, $\phi_b=0.98$]{\includegraphics[width=0.45\linewidth]{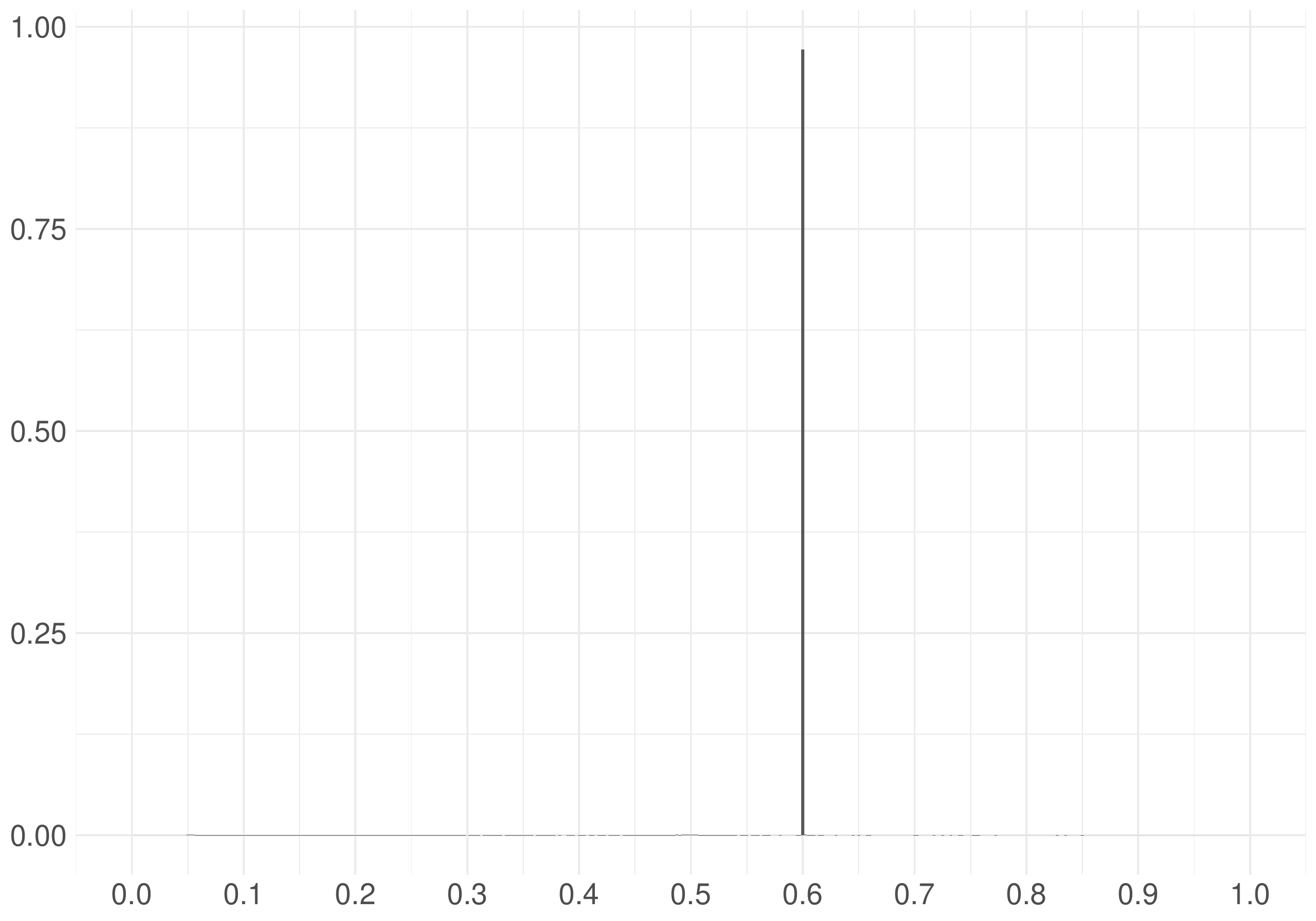}\label{fig:15:1}}
\subfigure[$T=800$, $\phi_a=1.05$, $\phi_b=0.98$]{\includegraphics[width=0.45\linewidth]{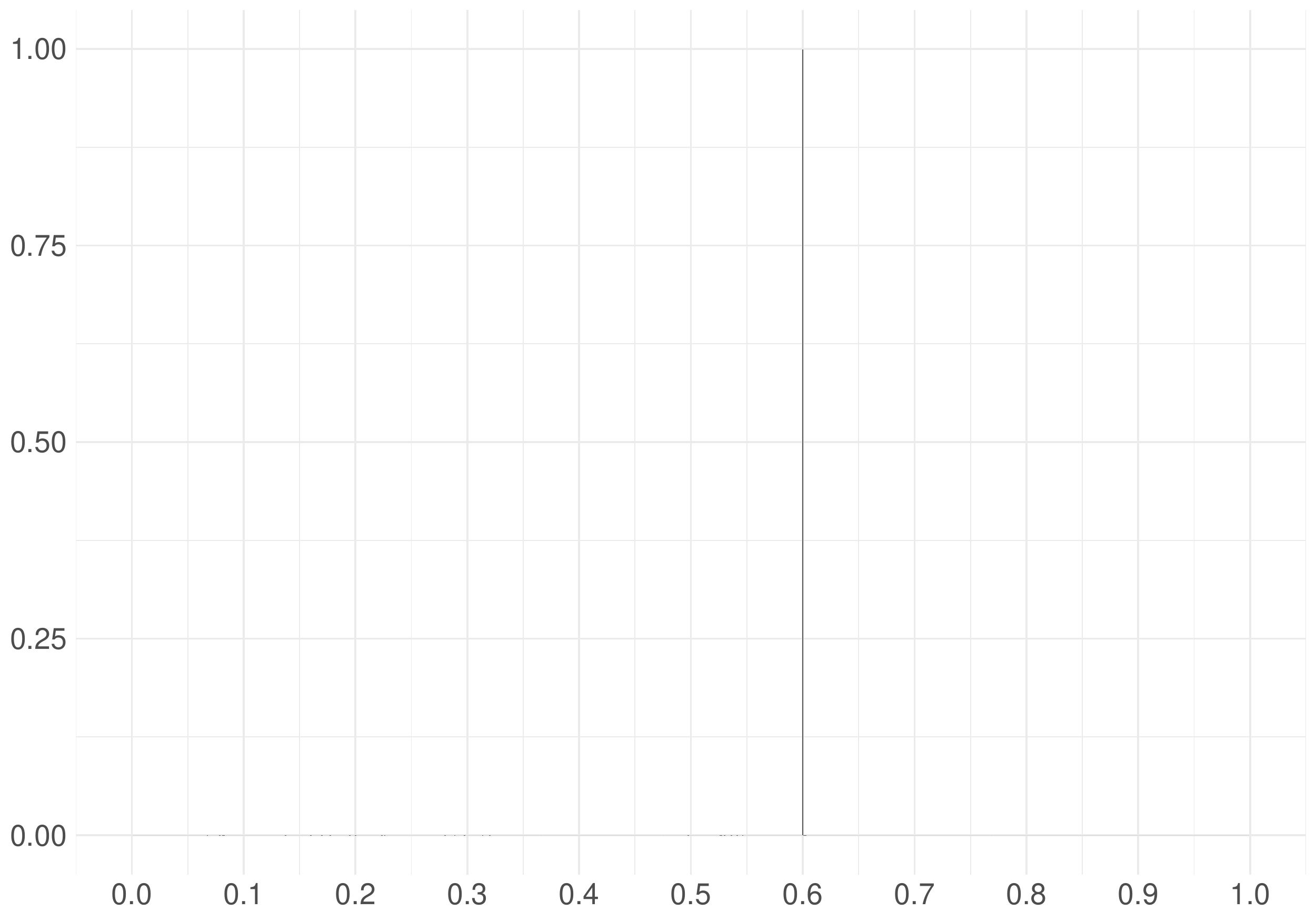}\label{fig:15:2}}\\
\subfigure[$T=400$, $\phi_a=1.05$, $\phi_b=0.96$]{\includegraphics[width=0.45\linewidth]{graph/NV_k_c_T=400_1.05_0.96_Model1s0.s13.pdf}\label{fig:15:3}}
\subfigure[$T=800$, $\phi_a=1.05$, $\phi_b=0.96$]{\includegraphics[width=0.45\linewidth]{graph/NV_k_c_T=800_1.05_0.96_Model1s0.s13.pdf}\label{fig:15:4}}\\
\subfigure[$T=400$, $\phi_a=1.05$, $\phi_b=0.94$]{\includegraphics[width=0.45\linewidth]{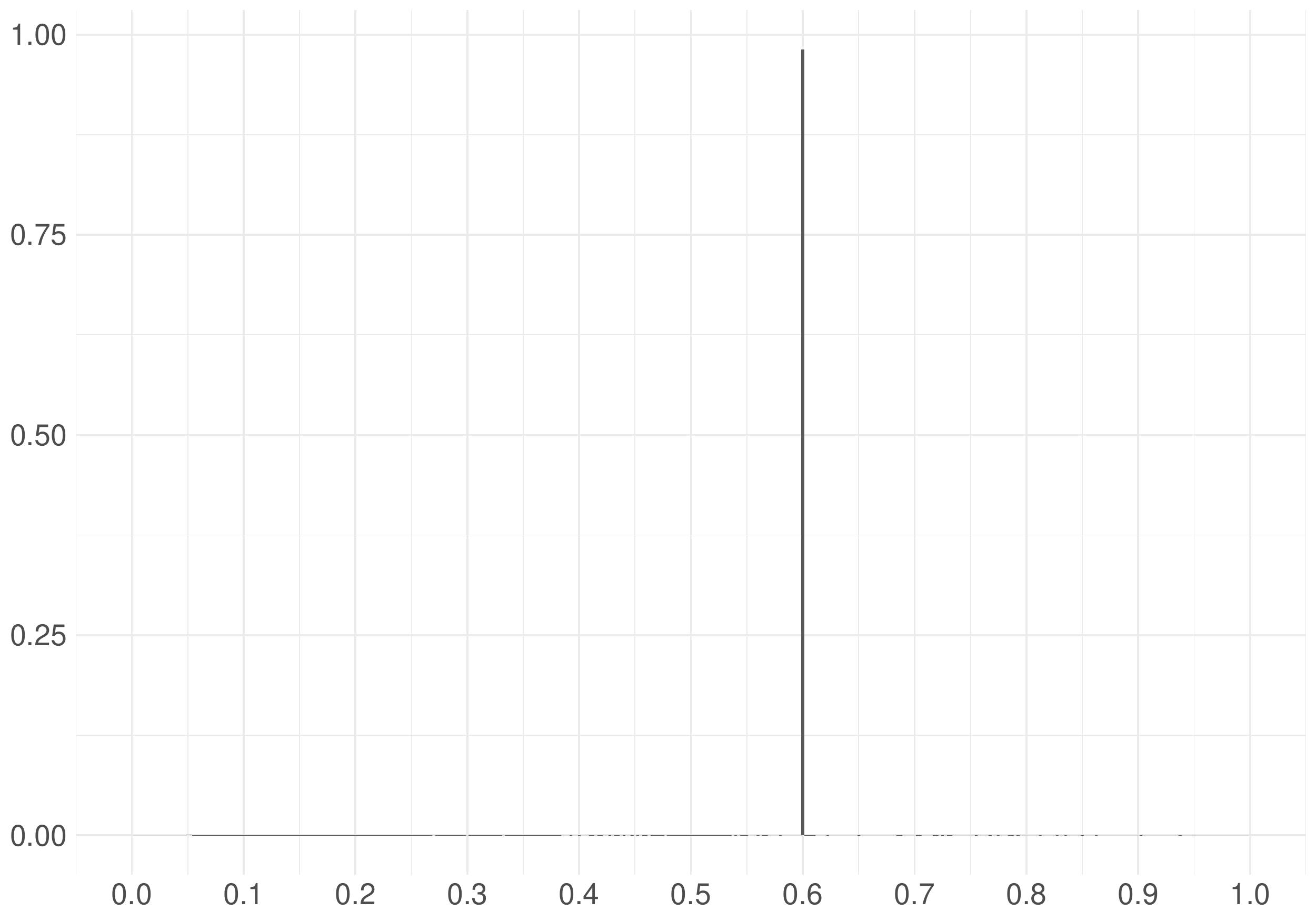}\label{fig:15:5}}
\subfigure[$T=800$, $\phi_a=1.05$, $\phi_b=0.94$]{\includegraphics[width=0.45\linewidth]{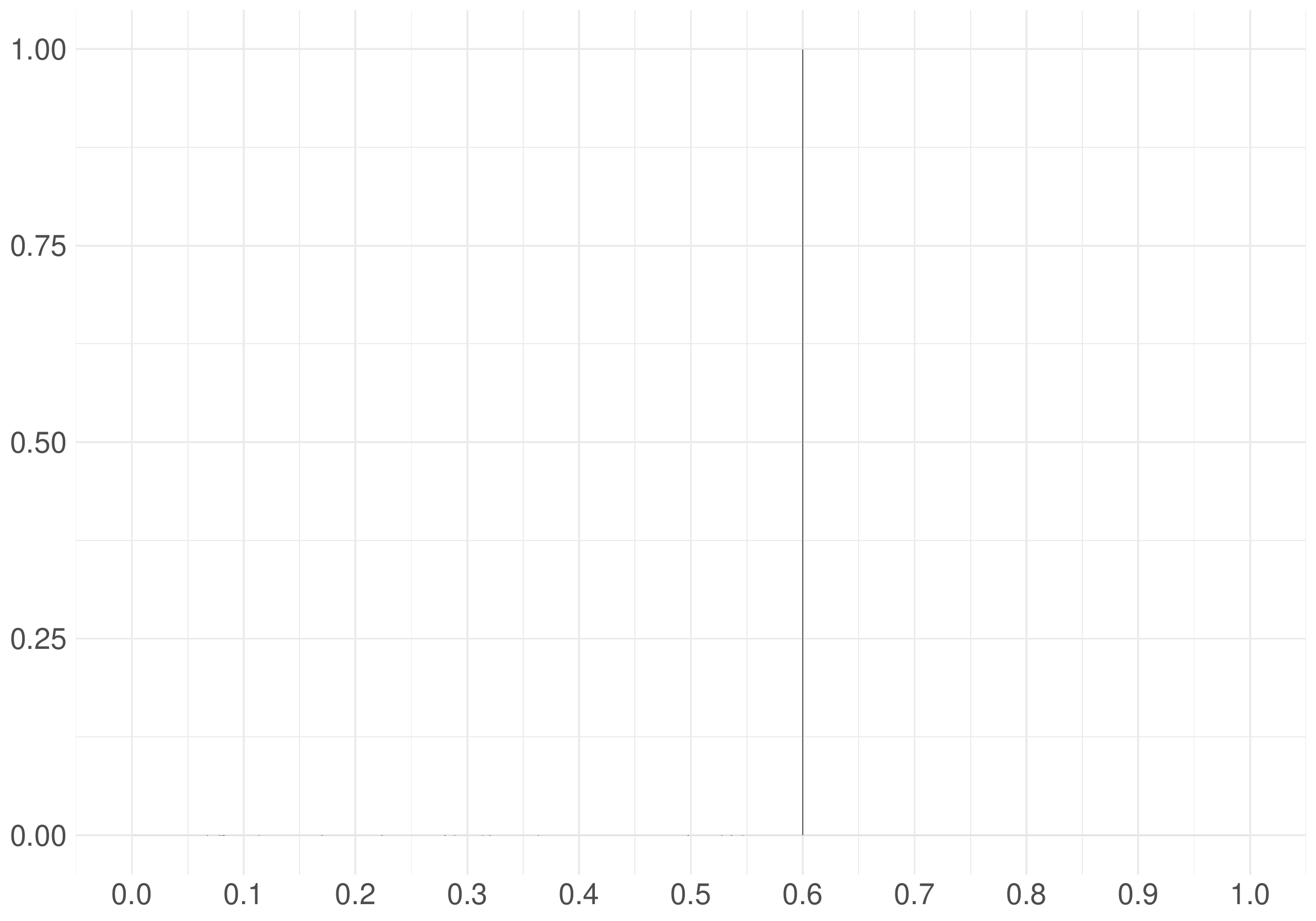}\label{fig:15:6}}
\end{center}%
\caption{Histograms of $\hat{k}_c$ % (left) and $\hat{k}_r$ (right) 
for $(\tau_e,\tau_c,\tau_r)=(0.4,0.6,0.7)$ with $\sigma_1/\sigma_0=1/3$}
\label{fig15}
%\centering
%\footnotesize{OLS}
\end{figure}

%%k_r

\begin{figure}[h]%
\begin{center}%
\subfigure[$T=400$, $\phi_a=1.01$, $\phi_b=0.96$]{\includegraphics[width=0.45\linewidth]{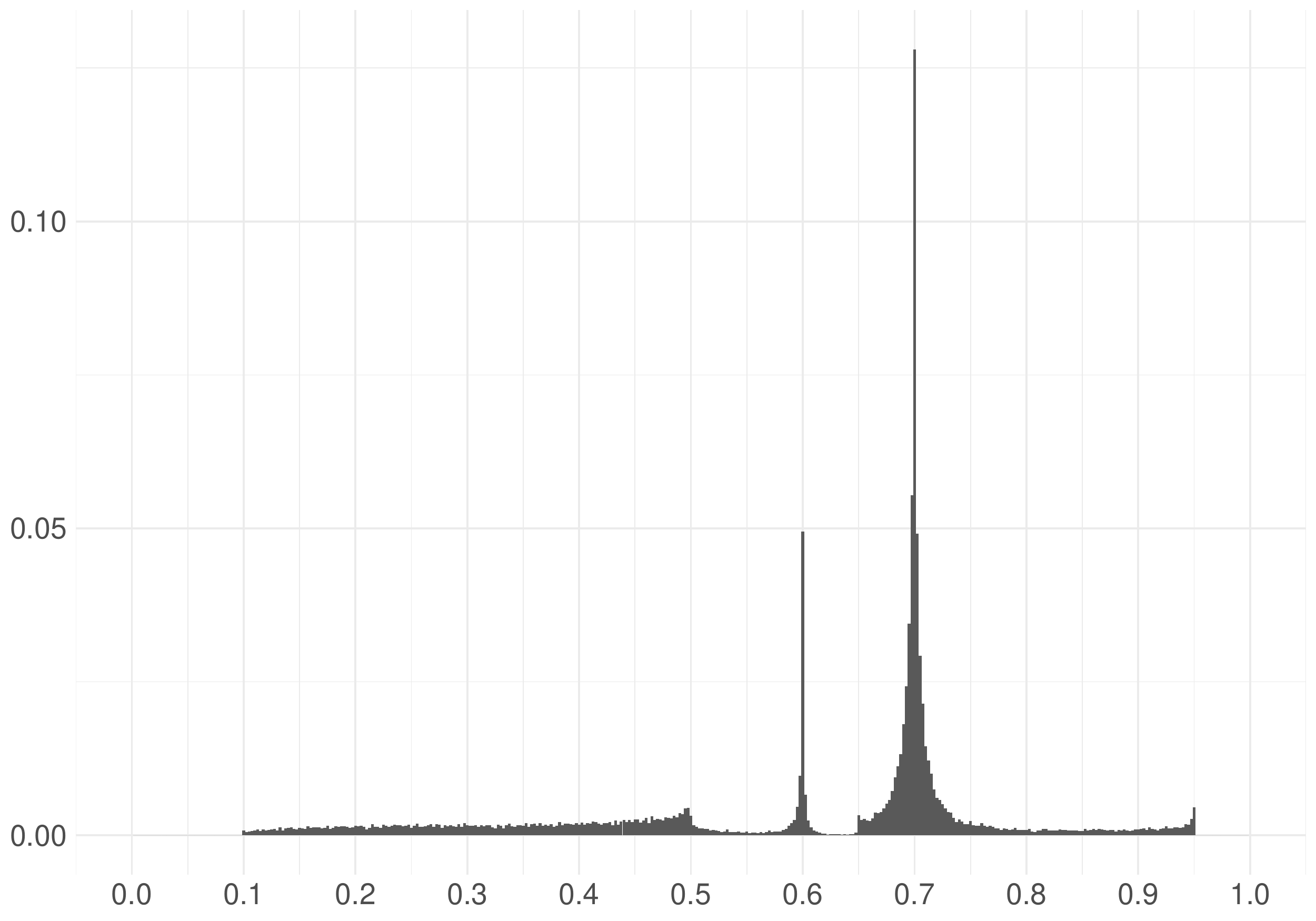}\label{fig:16:1}}
\subfigure[$T=800$, $\phi_a=1.01$, $\phi_b=0.96$]{\includegraphics[width=0.45\linewidth]{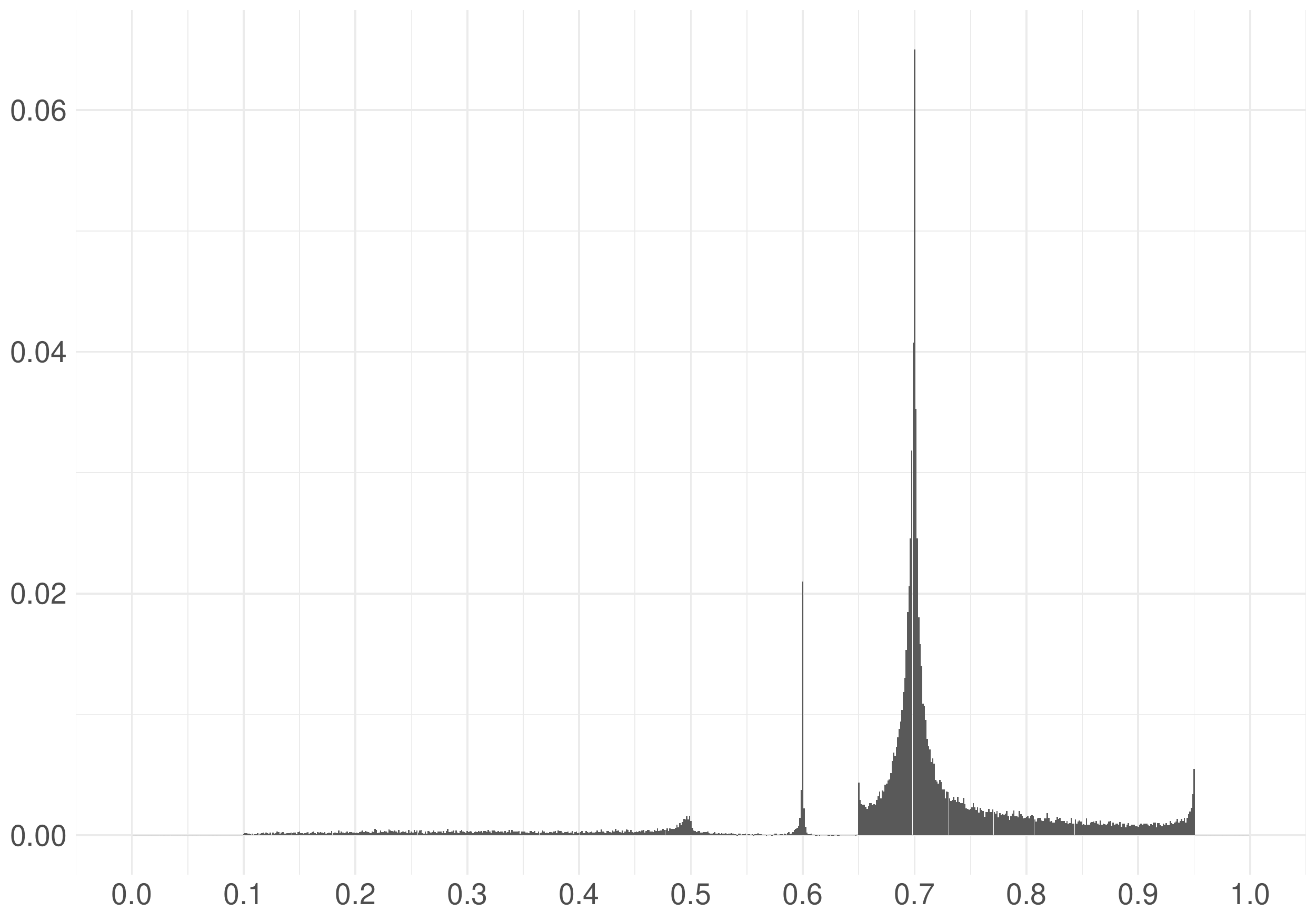}\label{fig:16:2}}\\
\subfigure[$T=400$, $\phi_a=1.05$, $\phi_b=0.96$]{\includegraphics[width=0.45\linewidth]{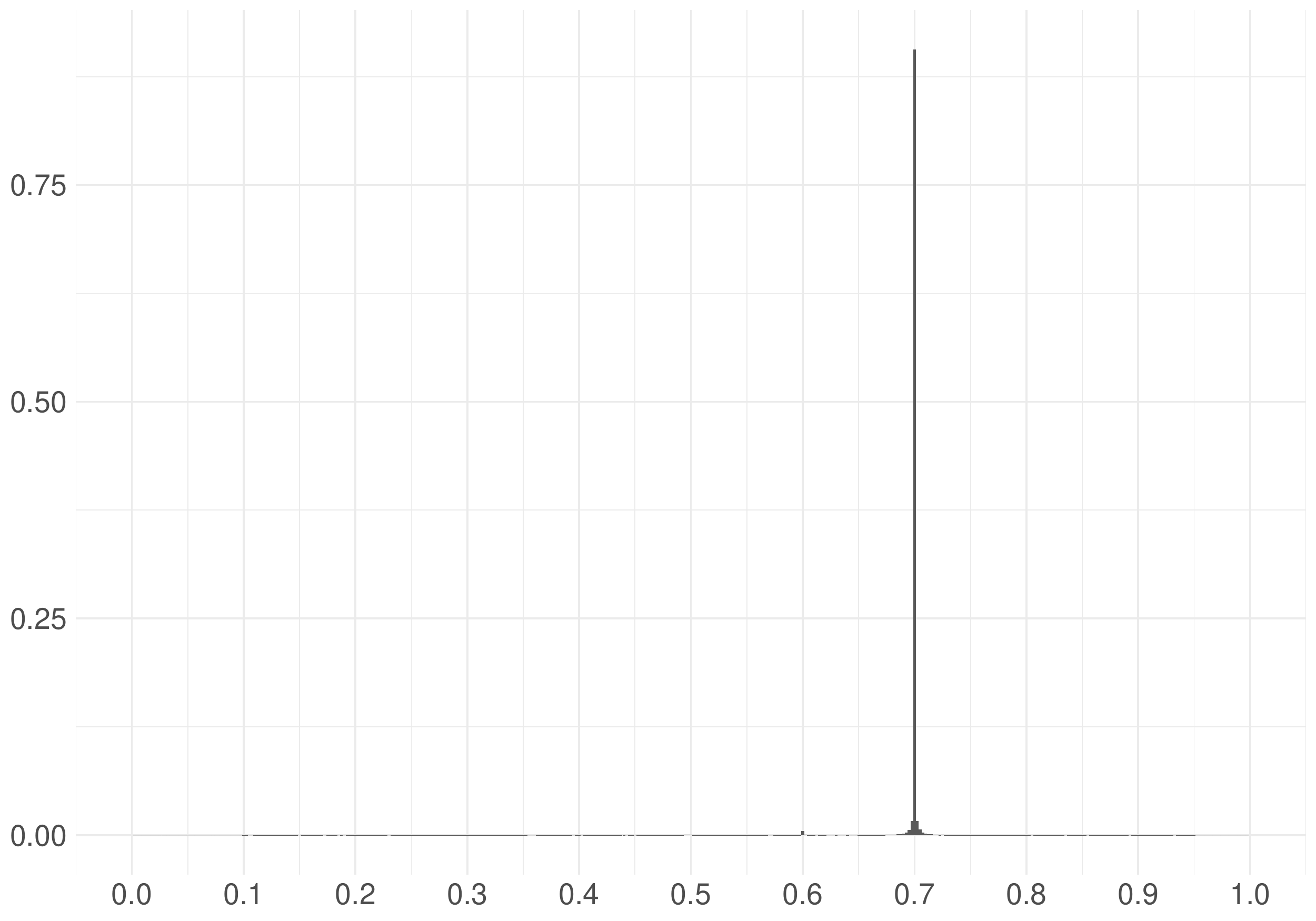}\label{fig:16:3}}
\subfigure[$T=800$, $\phi_a=1.05$, $\phi_b=0.96$]{\includegraphics[width=0.45\linewidth]{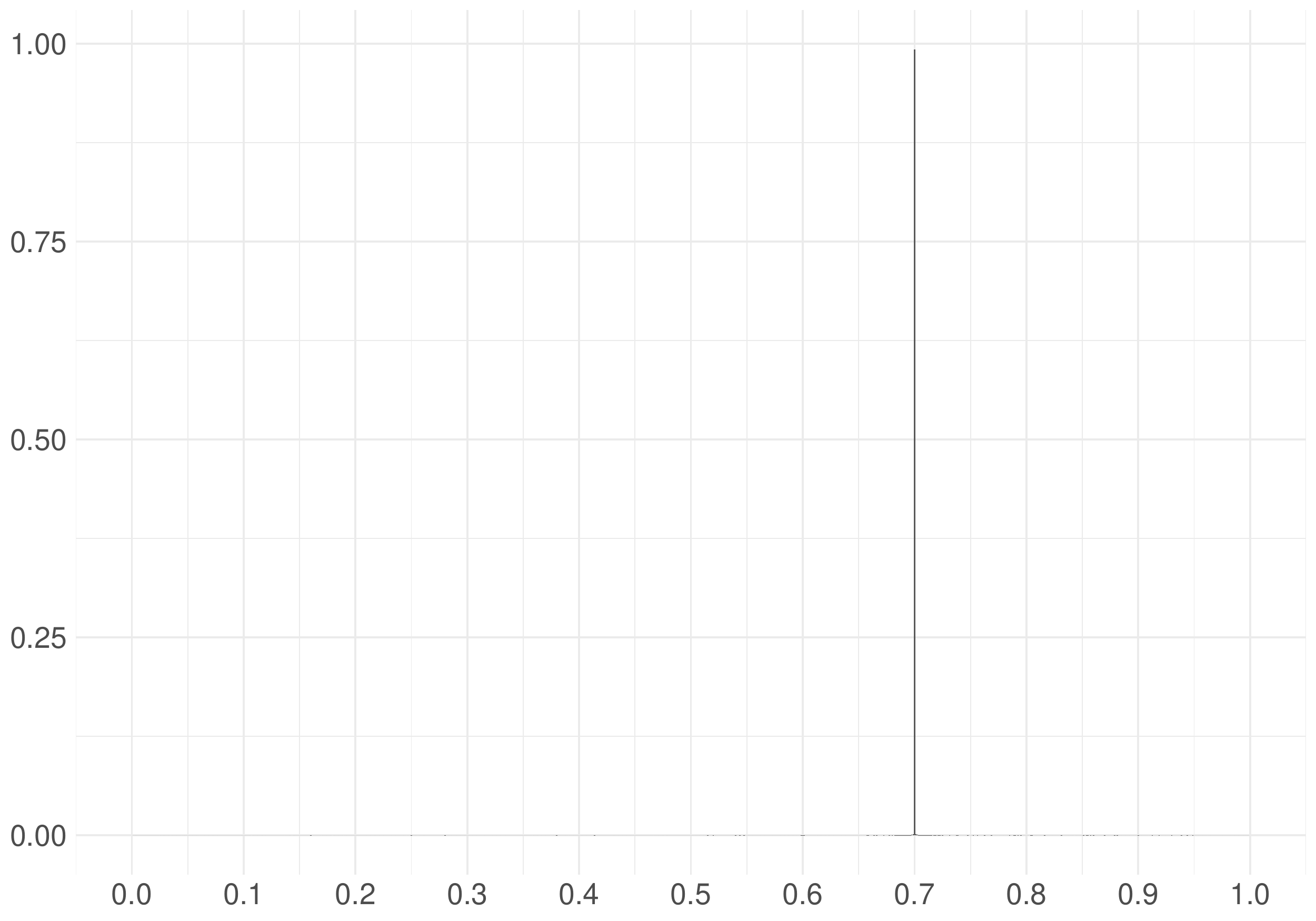}\label{fig:16:4}}\\
\subfigure[$T=400$, $\phi_a=1.09$, $\phi_b=0.96$]{\includegraphics[width=0.45\linewidth]{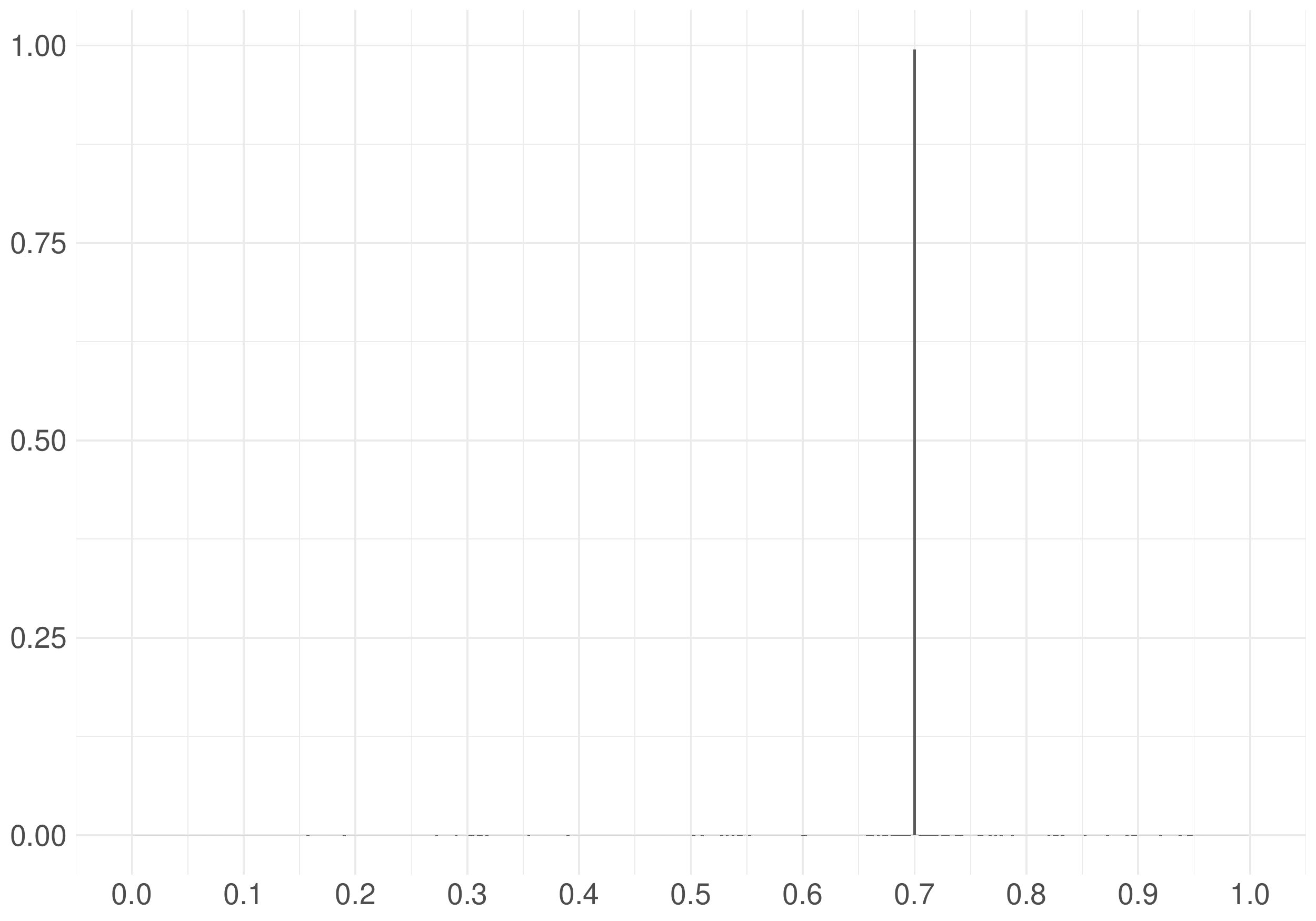}\label{fig:16:5}}
\subfigure[$T=800$, $\phi_a=1.09$, $\phi_b=0.96$]{\includegraphics[width=0.45\linewidth]{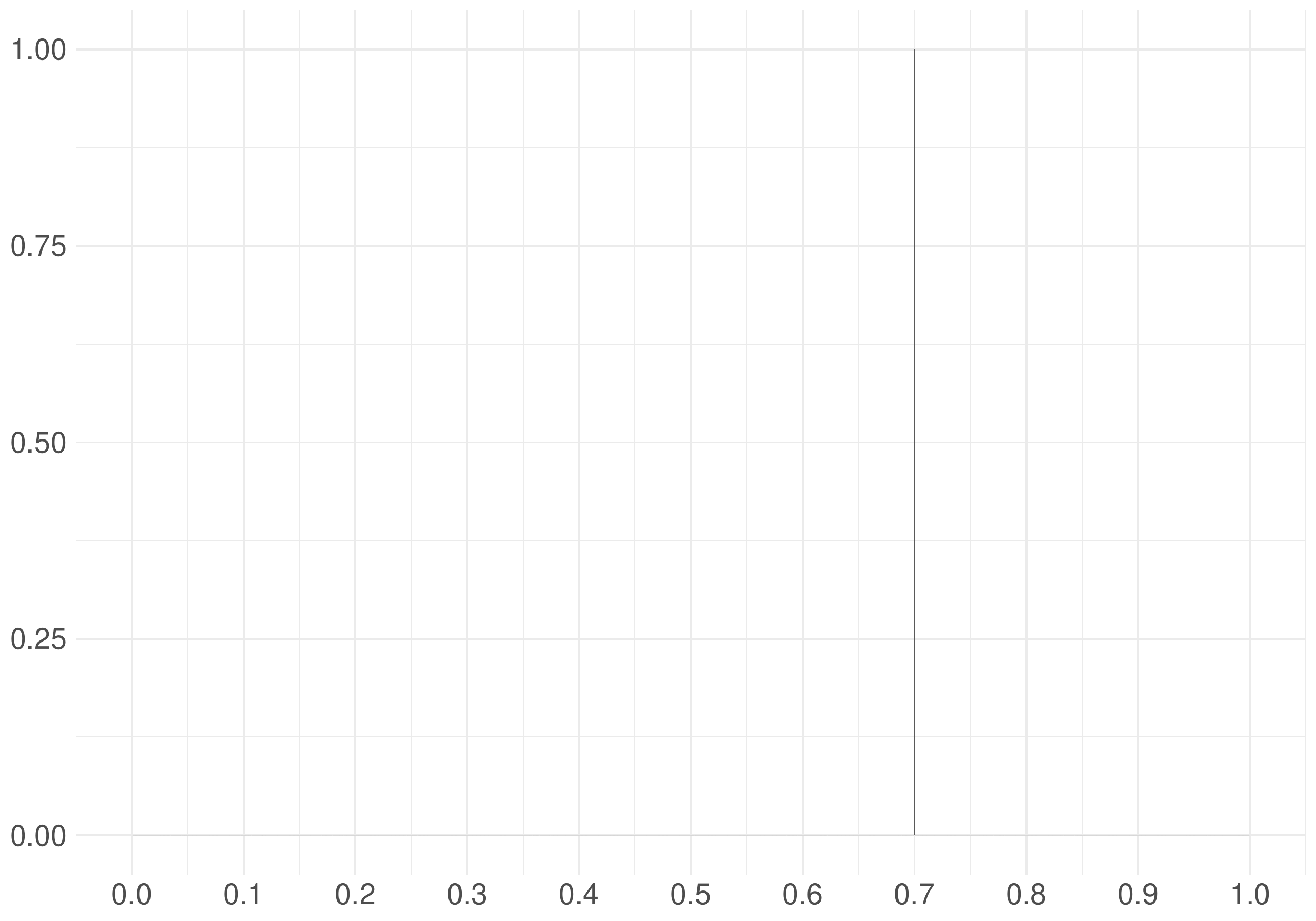}\label{fig:16:6}}
\end{center}%
\caption{Histograms of $\hat{k}_r$ % (left) and $\hat{k}_r$ (right) 
for $(\tau_e,\tau_c,\tau_r)=(0.4,0.6,0.7)$ with $\sigma_1/\sigma_0=1/3$}
\label{fig16}
%\centering
%\footnotesize{OLS}
\end{figure}

\begin{figure}[h]%
\begin{center}%
\subfigure[$T=400$, $\phi_a=1.05$, $\phi_b=0.98$]{\includegraphics[width=0.45\linewidth]{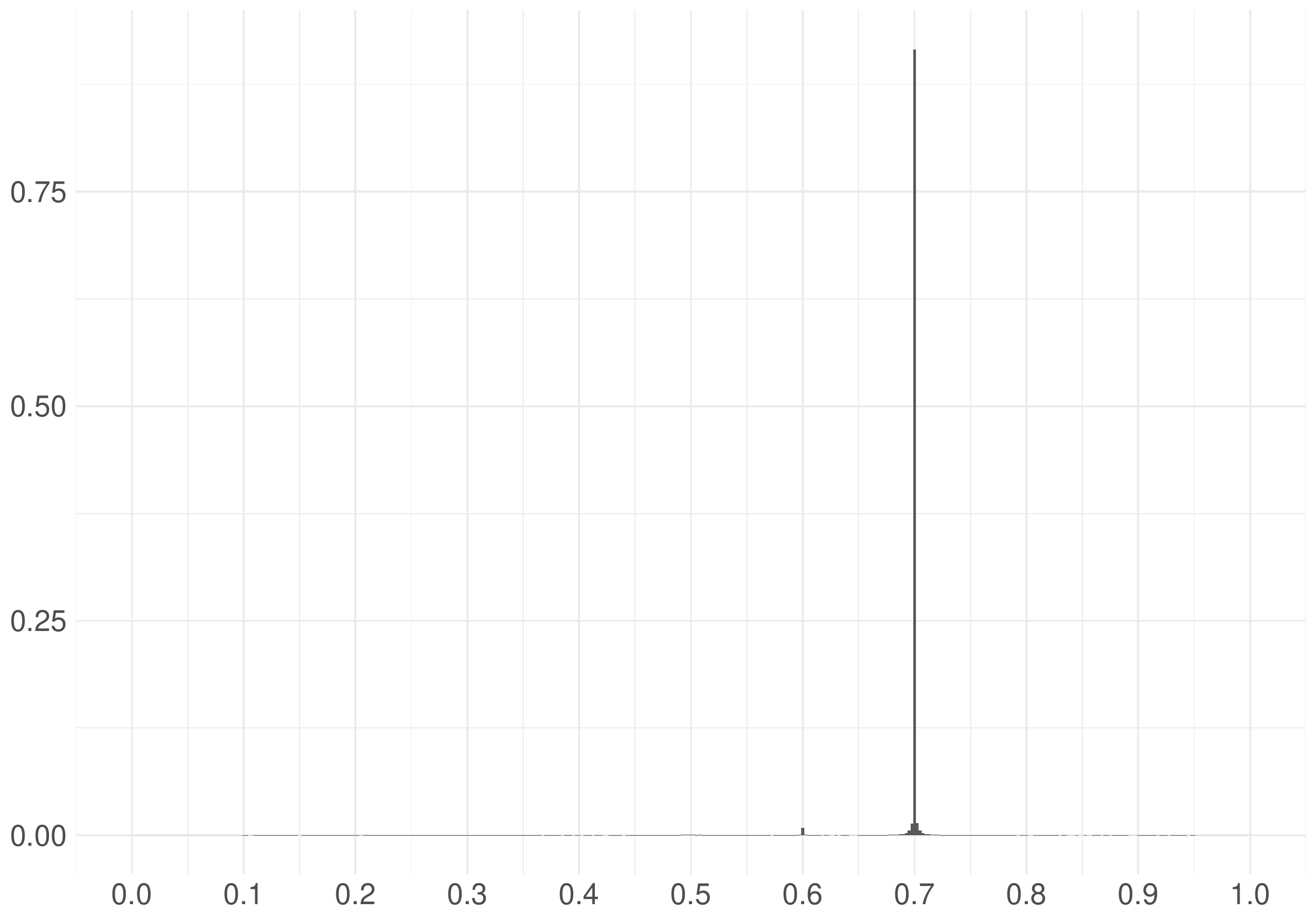}\label{fig:17:1}}
\subfigure[$T=800$, $\phi_a=1.05$, $\phi_b=0.98$]{\includegraphics[width=0.45\linewidth]{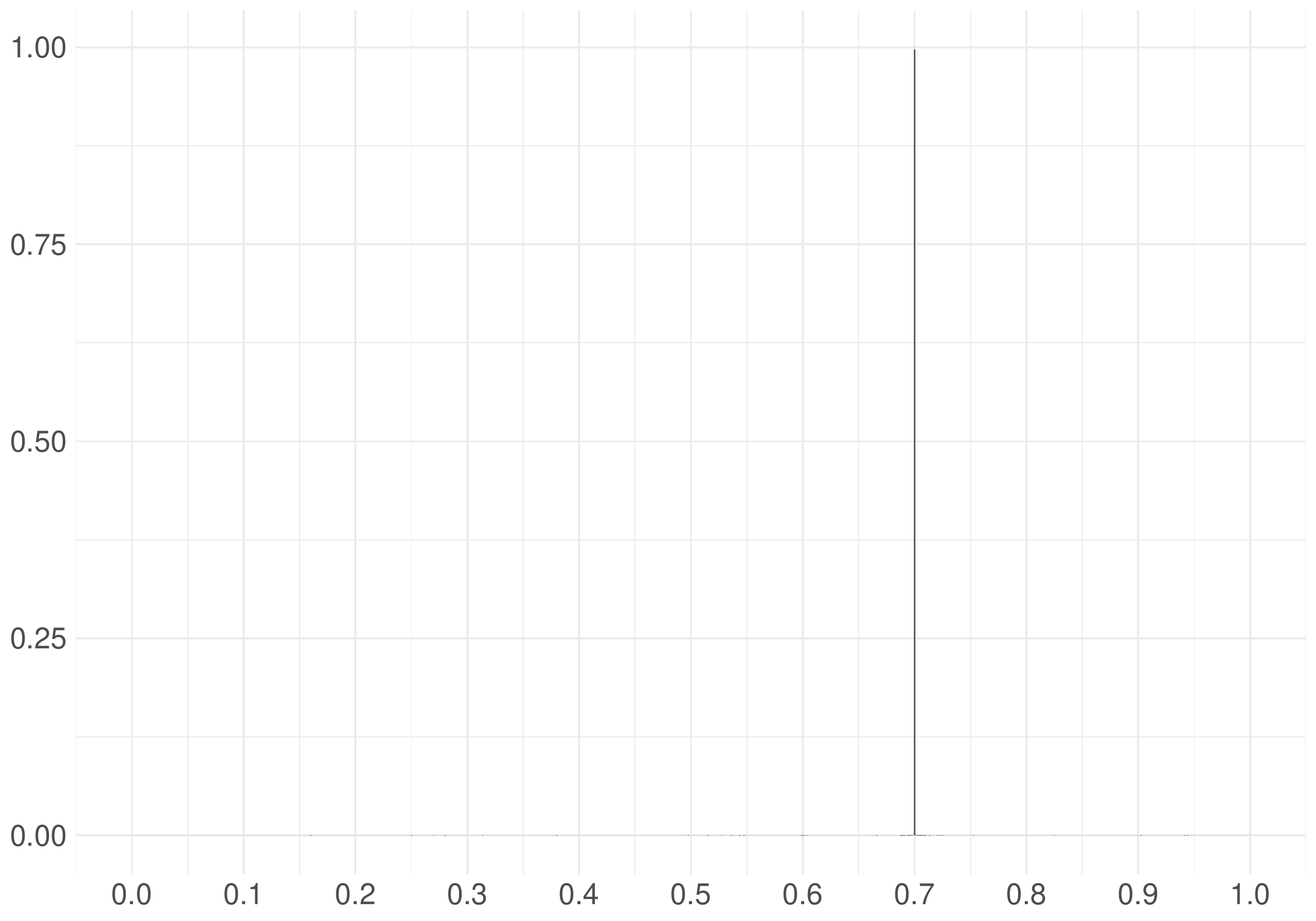}\label{fig:17:2}}\\
\subfigure[$T=400$, $\phi_a=1.05$, $\phi_b=0.96$]{\includegraphics[width=0.45\linewidth]{graph/NV_k_r_T=400_1.05_0.96_Model1s0.s13.pdf}\label{fig:17:3}}
\subfigure[$T=800$, $\phi_a=1.05$, $\phi_b=0.96$]{\includegraphics[width=0.45\linewidth]{graph/NV_k_r_T=800_1.05_0.96_Model1s0.s13.pdf}\label{fig:17:4}}\\
\subfigure[$T=400$, $\phi_a=1.05$, $\phi_b=0.94$]{\includegraphics[width=0.45\linewidth]{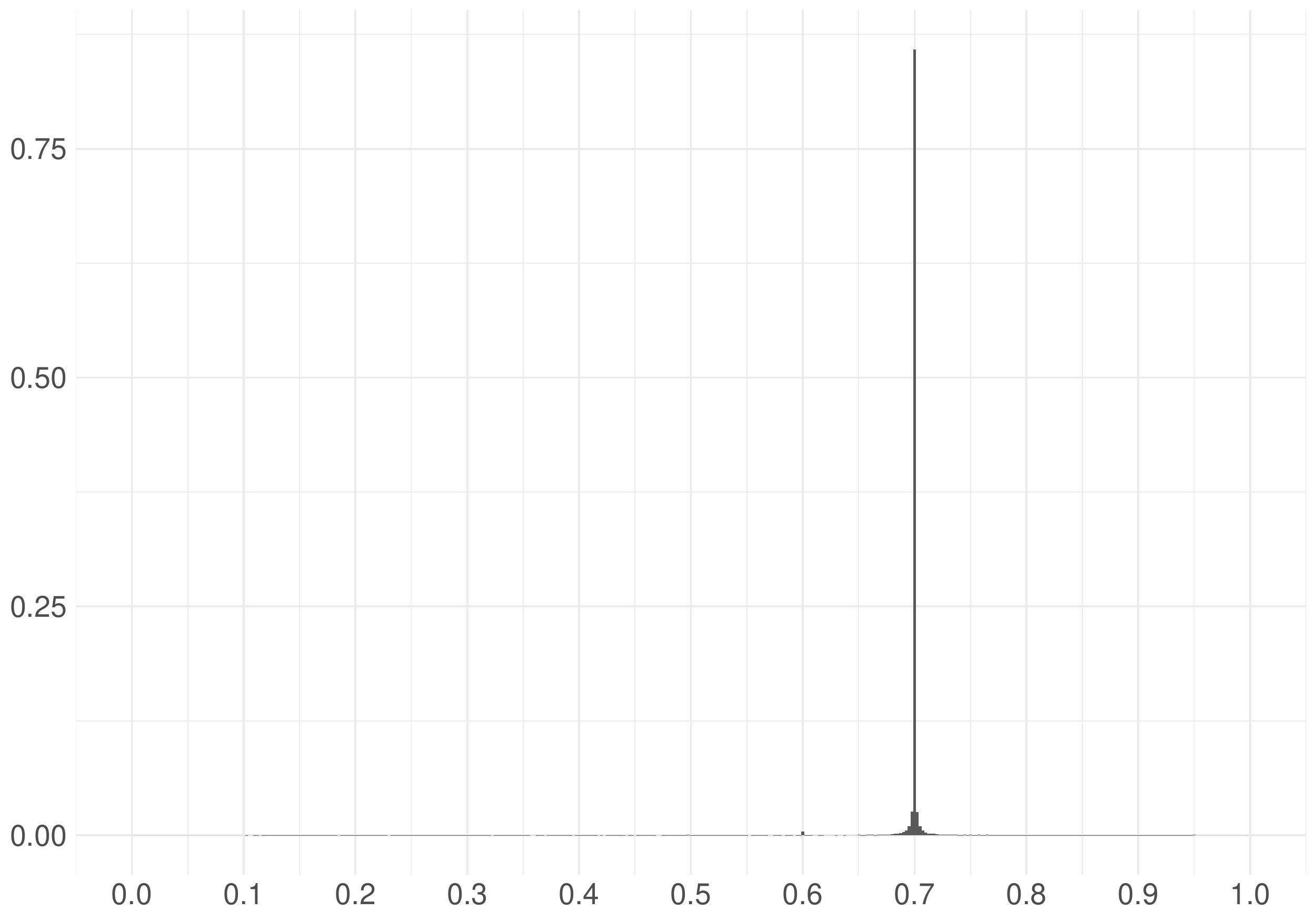}\label{fig:17:5}}
\subfigure[$T=800$, $\phi_a=1.05$, $\phi_b=0.94$]{\includegraphics[width=0.45\linewidth]{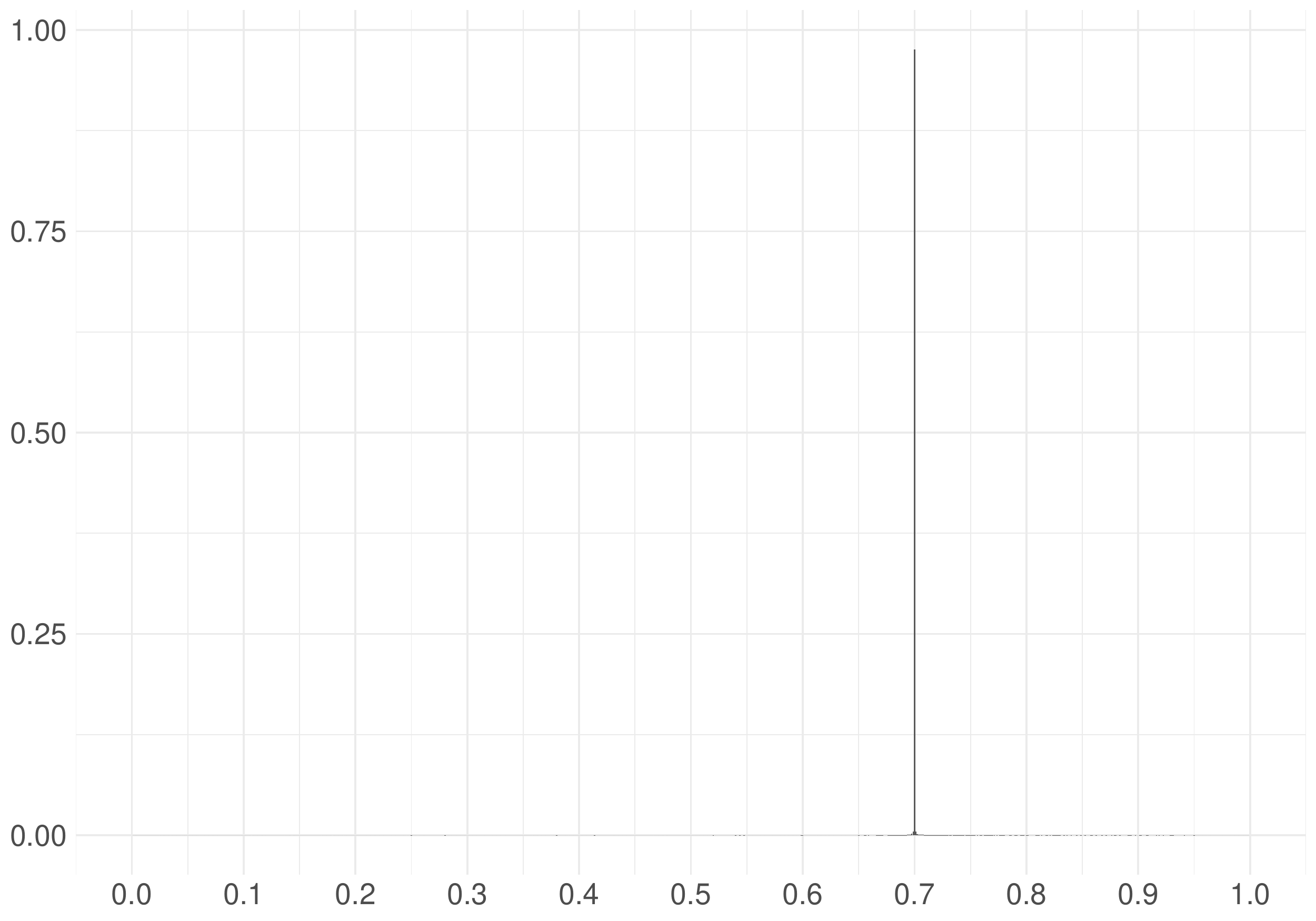}\label{fig:17:6}}
\end{center}%
\caption{Histograms of $\hat{k}_r$ % (left) and $\hat{k}_r$ (right) 
for $(\tau_e,\tau_c,\tau_r)=(0.4,0.6,0.7)$ with $\sigma_1/\sigma_0=1/3$}
\label{fig17}
%\centering
%\footnotesize{OLS}
\end{figure}

%%%
%s_0/s_1=3
%%%

\begin{figure}[h]%
\begin{center}%
\subfigure[$T=400$, $\phi_a=1.01$, $\phi_b=0.96$]{\includegraphics[width=0.45\linewidth]{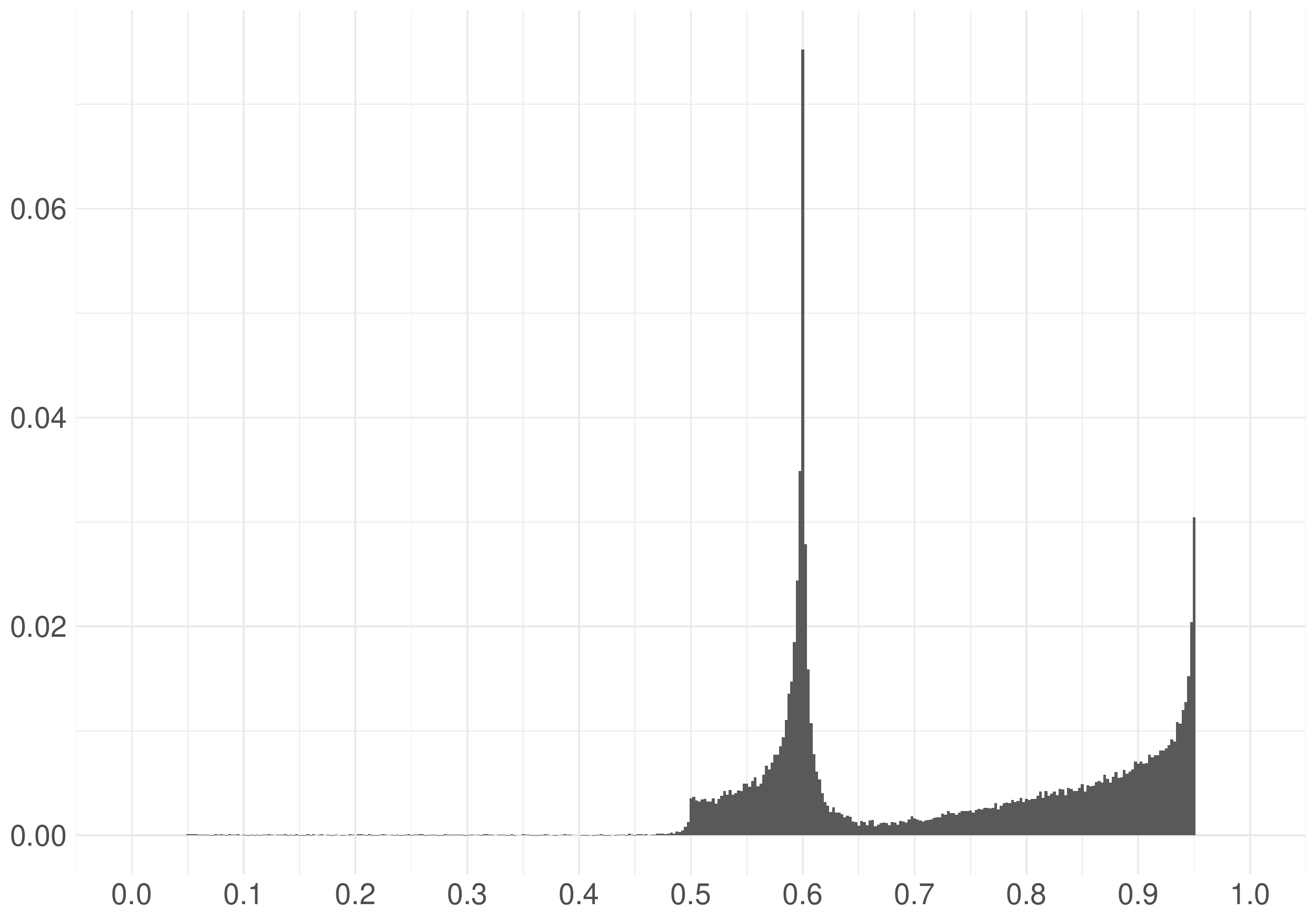}\label{fig:18:1}}
\subfigure[$T=800$, $\phi_a=1.01$, $\phi_b=0.96$]{\includegraphics[width=0.45\linewidth]{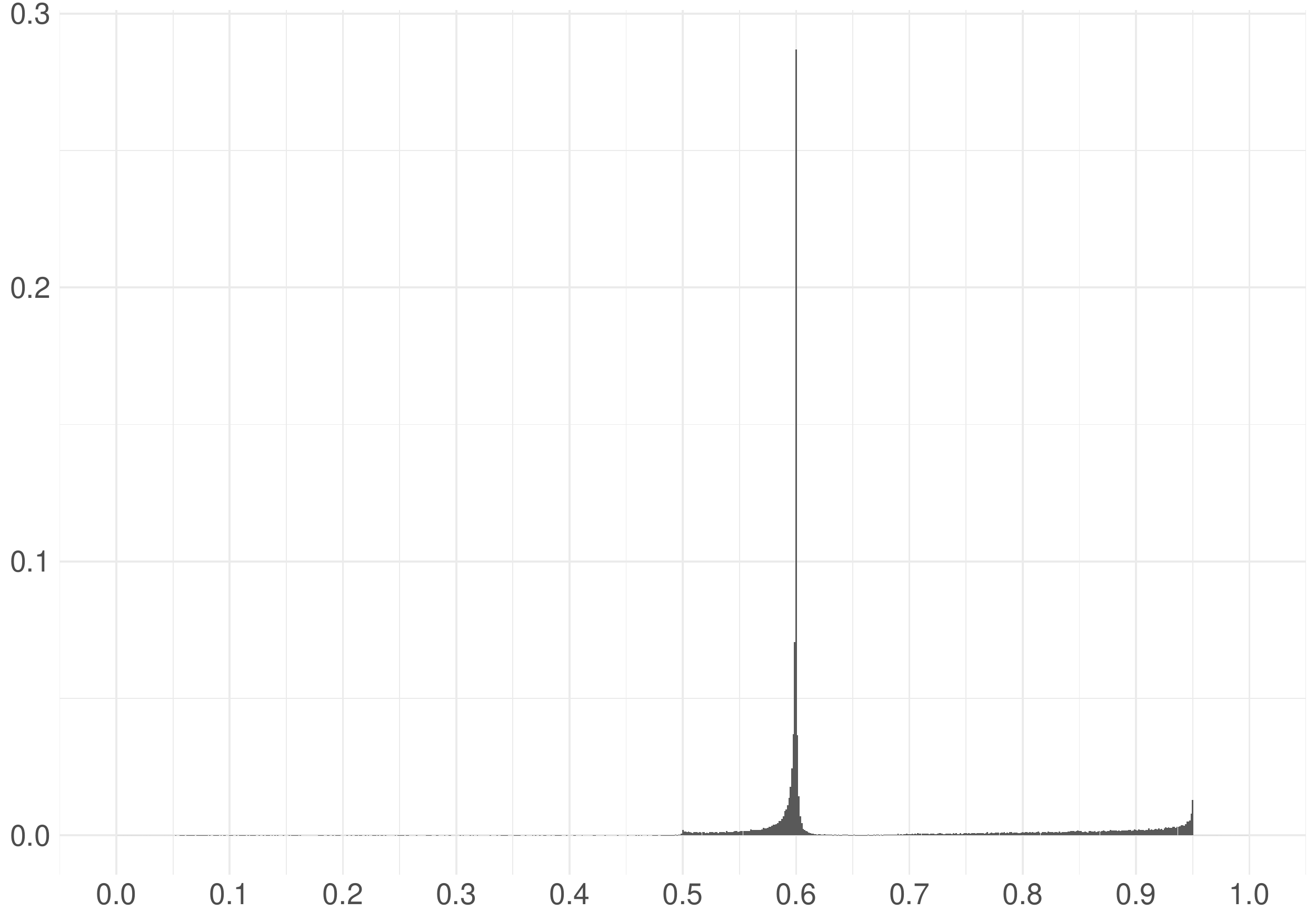}\label{fig:18:2}}\\
\subfigure[$T=400$, $\phi_a=1.05$, $\phi_b=0.96$]{\includegraphics[width=0.45\linewidth]{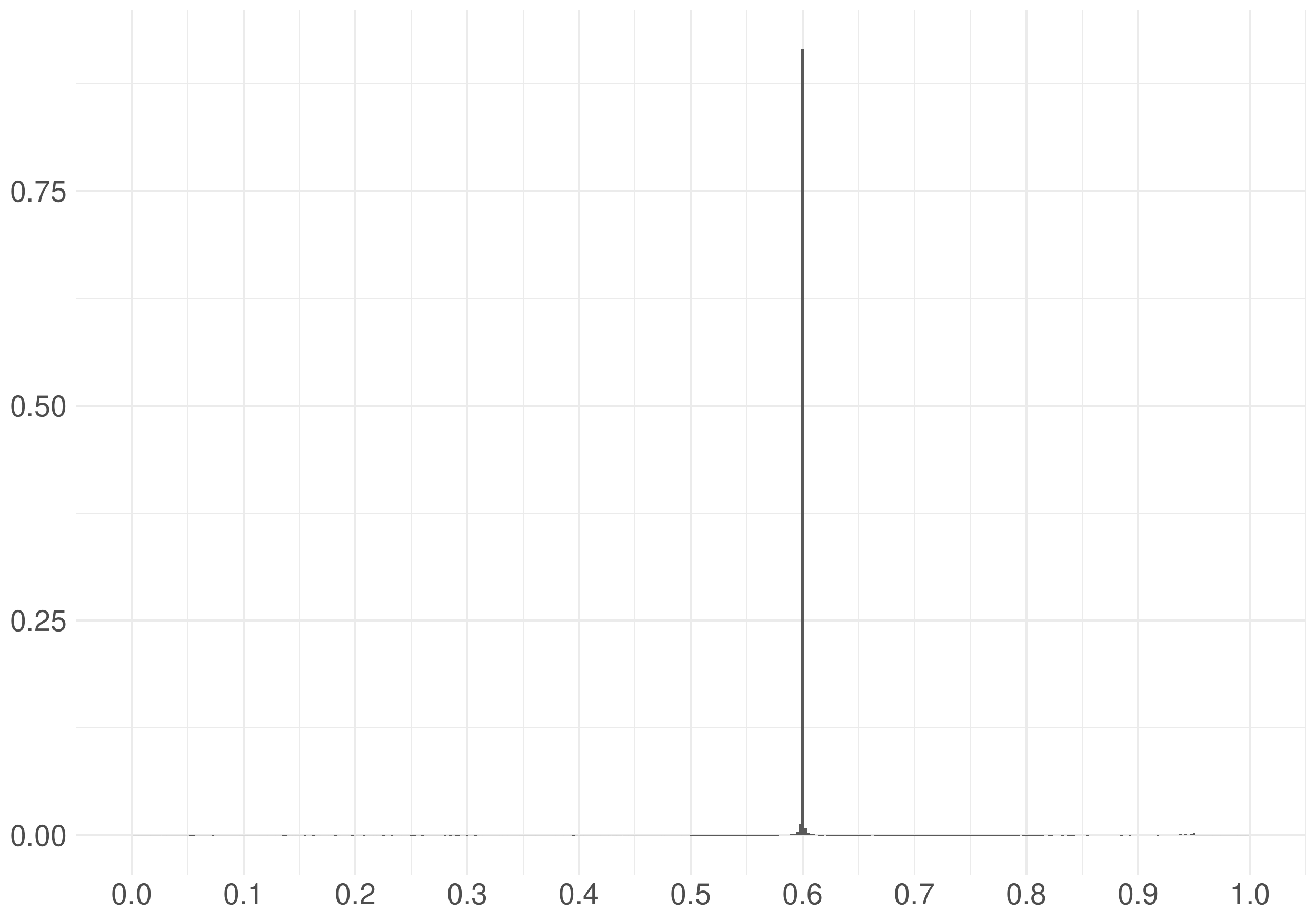}\label{fig:18:3}}
\subfigure[$T=800$, $\phi_a=1.05$, $\phi_b=0.96$]{\includegraphics[width=0.45\linewidth]{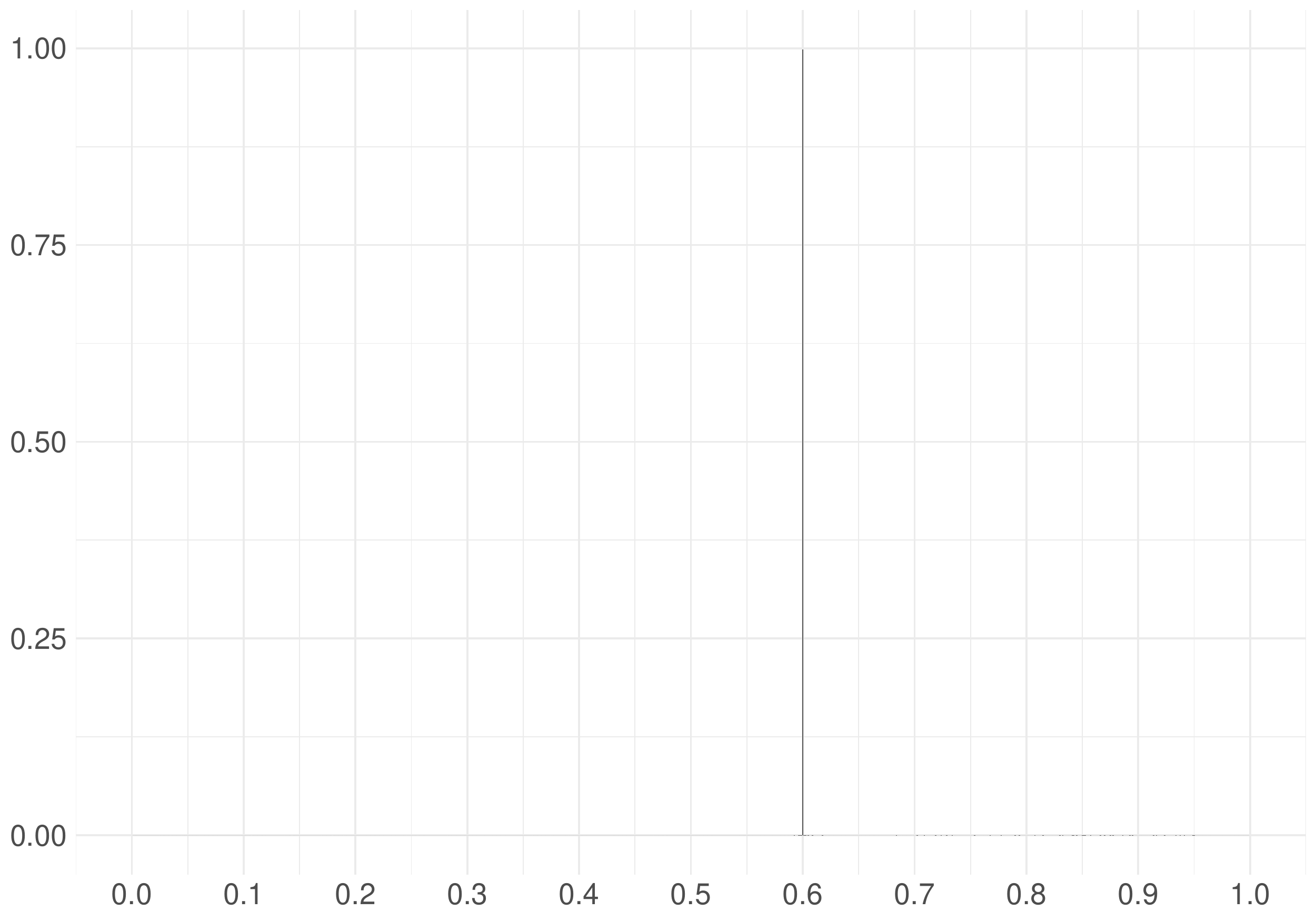}\label{fig:18:4}}\\
\subfigure[$T=400$, $\phi_a=1.09$, $\phi_b=0.96$]{\includegraphics[width=0.45\linewidth]{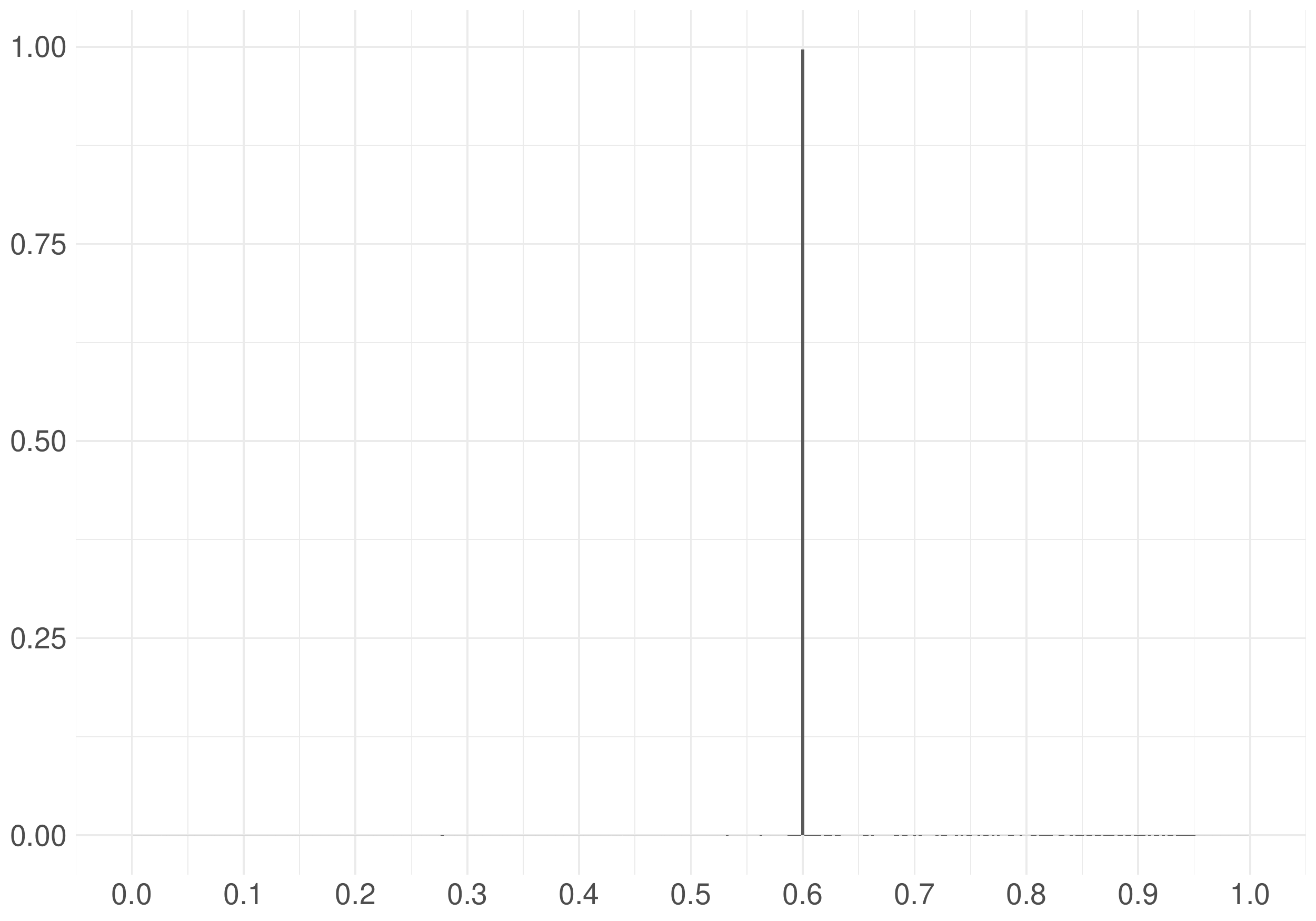}\label{fig:18:5}}
\subfigure[$T=800$, $\phi_a=1.09$, $\phi_b=0.96$]{\includegraphics[width=0.45\linewidth]{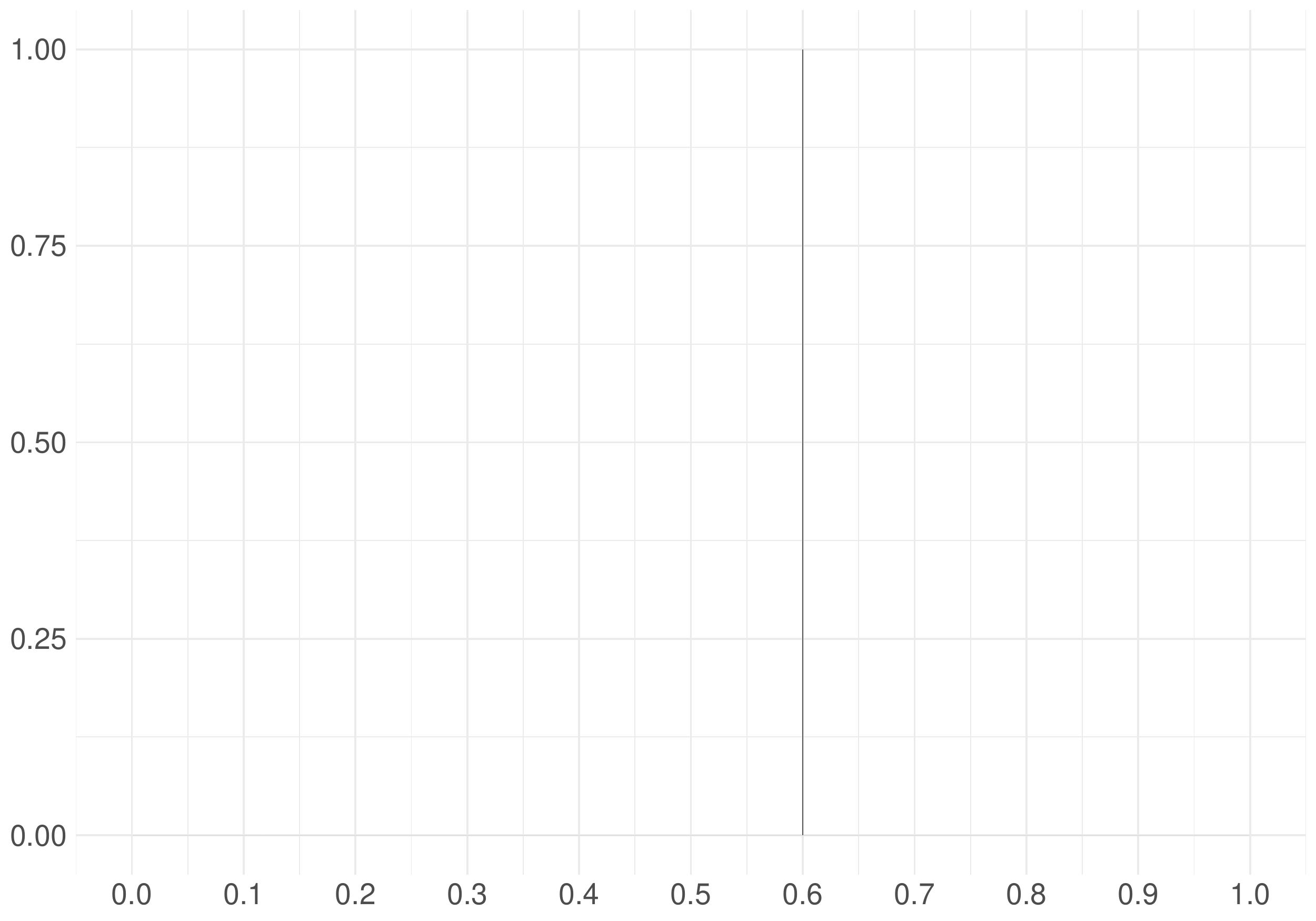}\label{fig:18:6}}
\end{center}%
\caption{Histograms of $\hat{k}_c$ % (left) and $\hat{k}_r$ (right) 
for $(\tau_e,\tau_c,\tau_r)=(0.4,0.6,0.7)$ with $\sigma_1/\sigma_0=3$}
\label{fig18}
%\centering
%\footnotesize{OLS}
\end{figure}

\begin{figure}[h]%
\begin{center}%
\subfigure[$T=400$, $\phi_a=1.05$, $\phi_b=0.98$]{\includegraphics[width=0.45\linewidth]{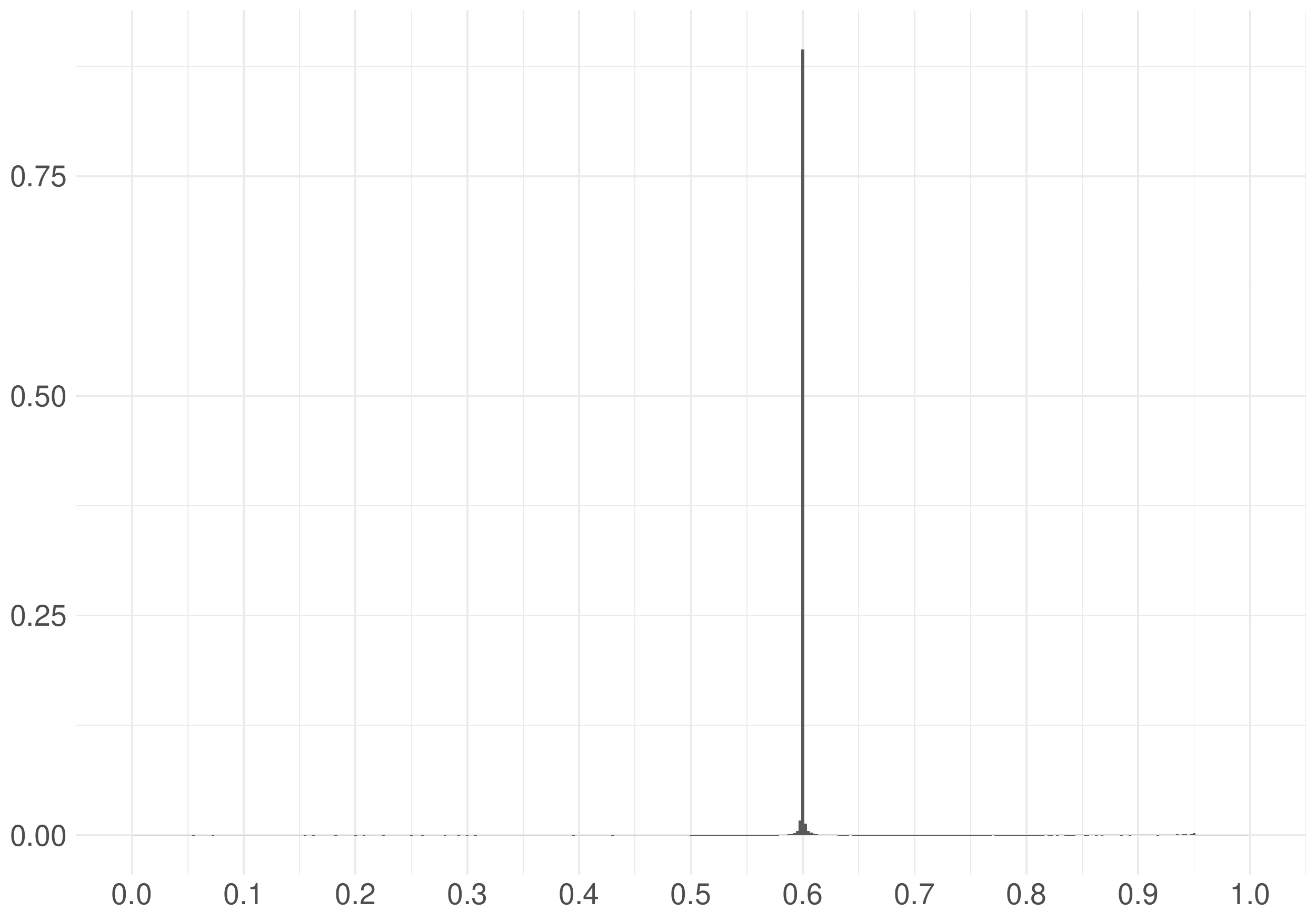}\label{fig:19:1}}
\subfigure[$T=800$, $\phi_a=1.05$, $\phi_b=0.98$]{\includegraphics[width=0.45\linewidth]{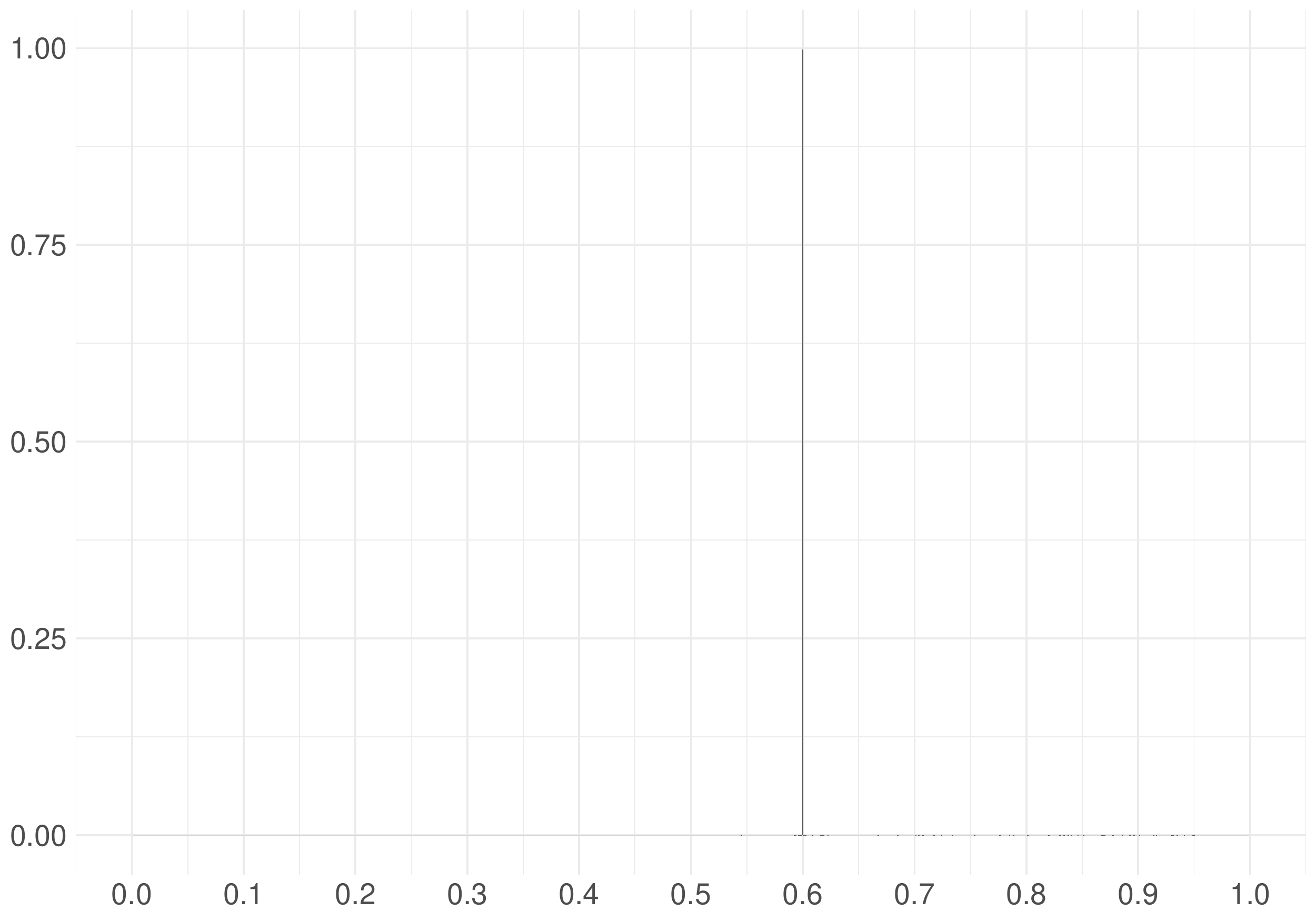}\label{fig:19:2}}\\
\subfigure[$T=400$, $\phi_a=1.05$, $\phi_b=0.96$]{\includegraphics[width=0.45\linewidth]{graph/NV_k_c_T=400_1.05_0.96_Model1s0.s10.33.pdf}\label{fig:19:3}}
\subfigure[$T=800$, $\phi_a=1.05$, $\phi_b=0.96$]{\includegraphics[width=0.45\linewidth]{graph/NV_k_c_T=800_1.05_0.96_Model1s0.s10.33.pdf}\label{fig:19:4}}\\
\subfigure[$T=400$, $\phi_a=1.05$, $\phi_b=0.94$]{\includegraphics[width=0.45\linewidth]{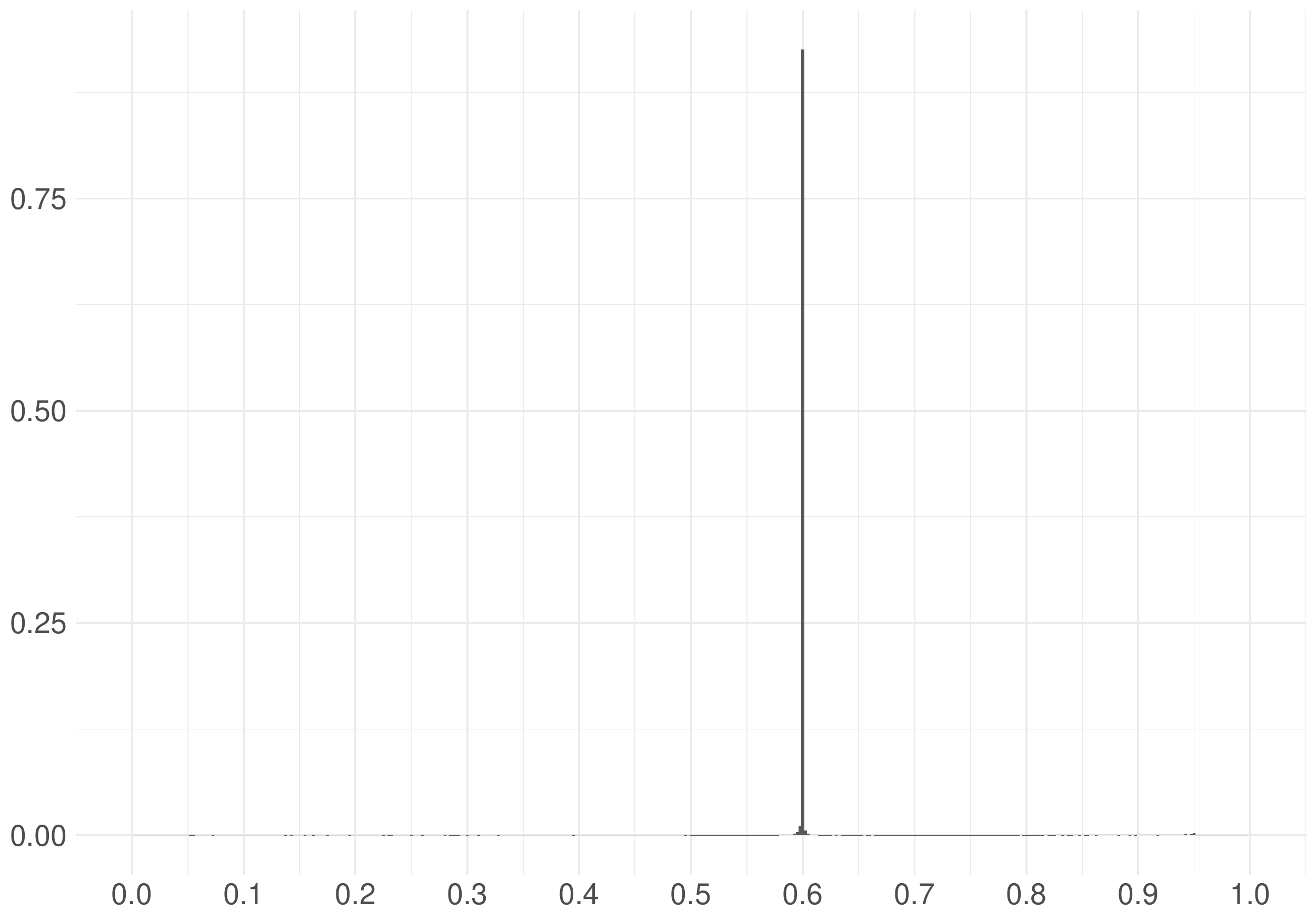}\label{fig:19:5}}
\subfigure[$T=800$, $\phi_a=1.05$, $\phi_b=0.94$]{\includegraphics[width=0.45\linewidth]{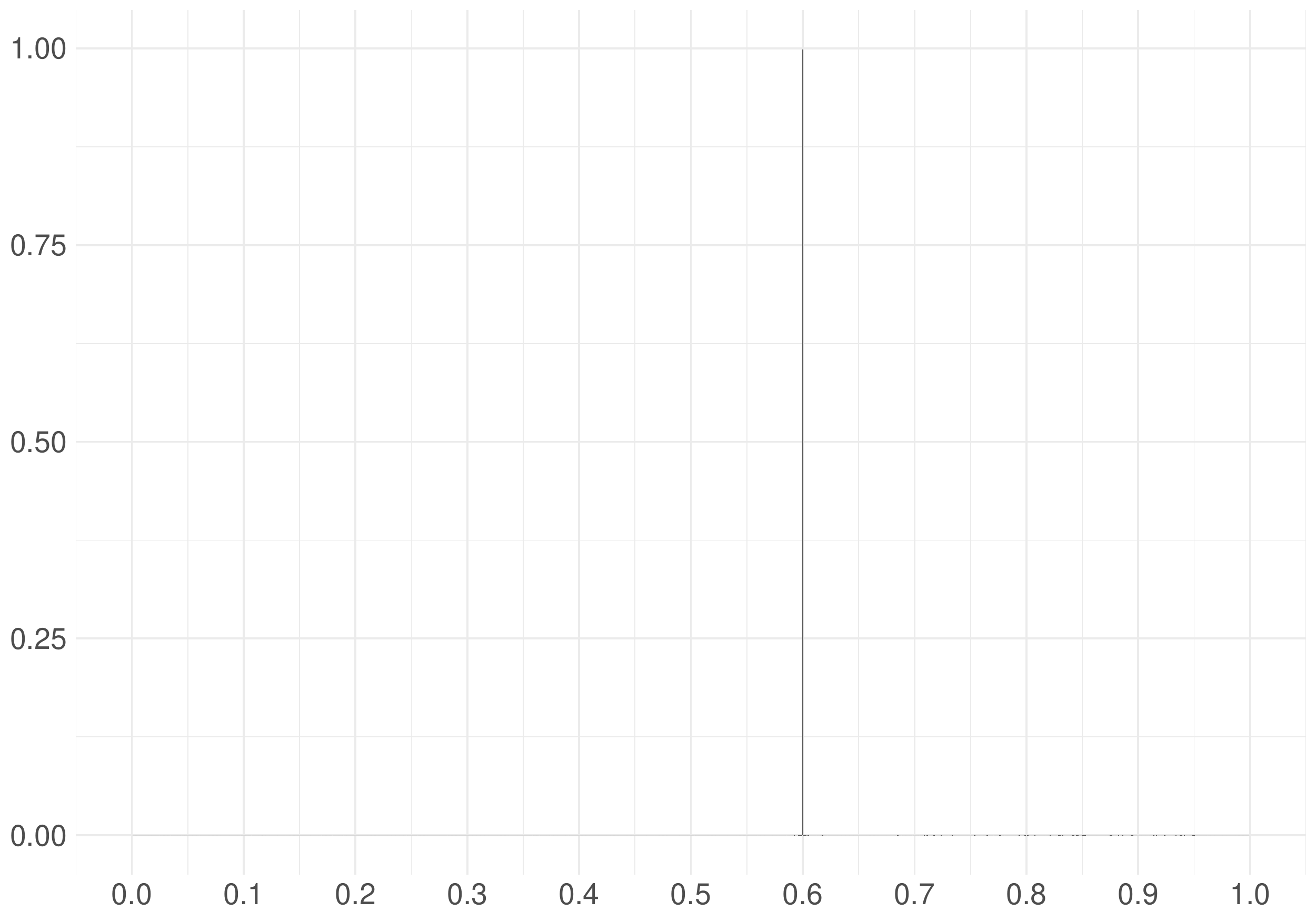}\label{fig:19:6}}
\end{center}%
\caption{Histograms of $\hat{k}_c$ % (left) and $\hat{k}_r$ (right) 
for $(\tau_e,\tau_c,\tau_r)=(0.4,0.6,0.7)$ with $\sigma_1/\sigma_0=3$}
\label{fig19}
%\centering
%\footnotesize{OLS}
\end{figure}

%%k_r

\begin{figure}[h]%
\begin{center}%
\subfigure[$T=400$, $\phi_a=1.01$, $\phi_b=0.96$]{\includegraphics[width=0.45\linewidth]{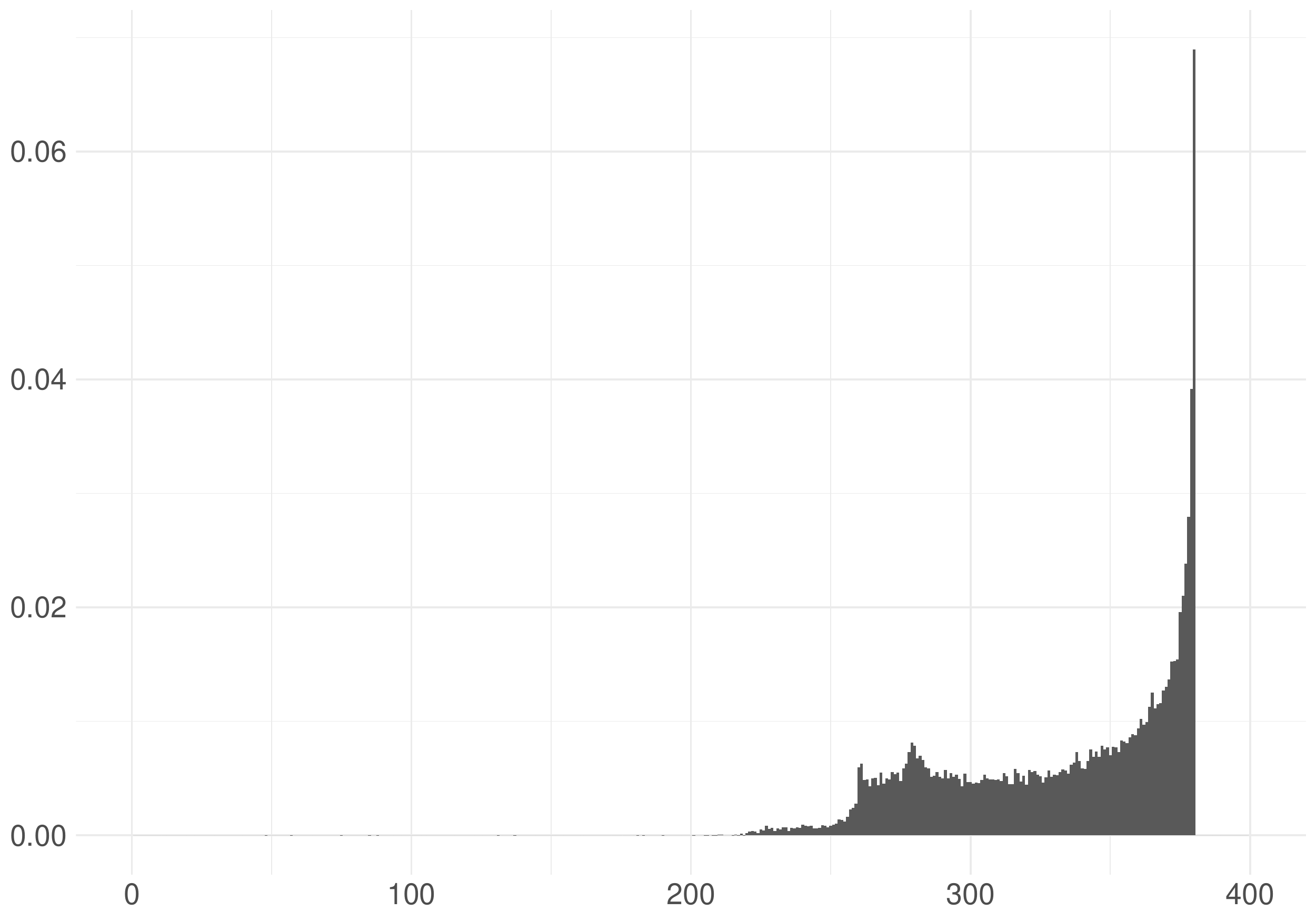}\label{fig:20:1}}
\subfigure[$T=800$, $\phi_a=1.01$, $\phi_b=0.96$]{\includegraphics[width=0.45\linewidth]{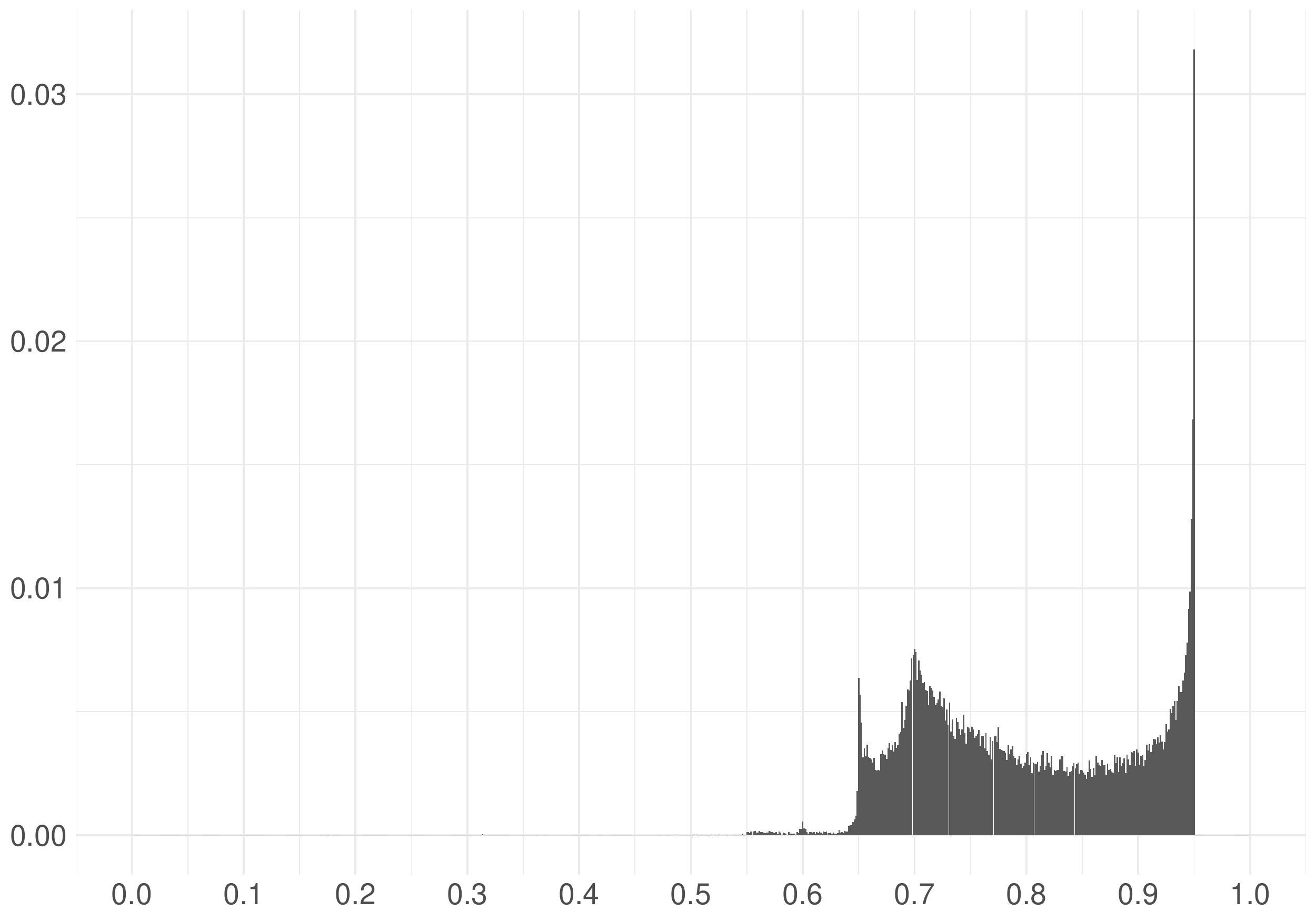}\label{fig:20:2}}\\
\subfigure[$T=400$, $\phi_a=1.05$, $\phi_b=0.96$]{\includegraphics[width=0.45\linewidth]{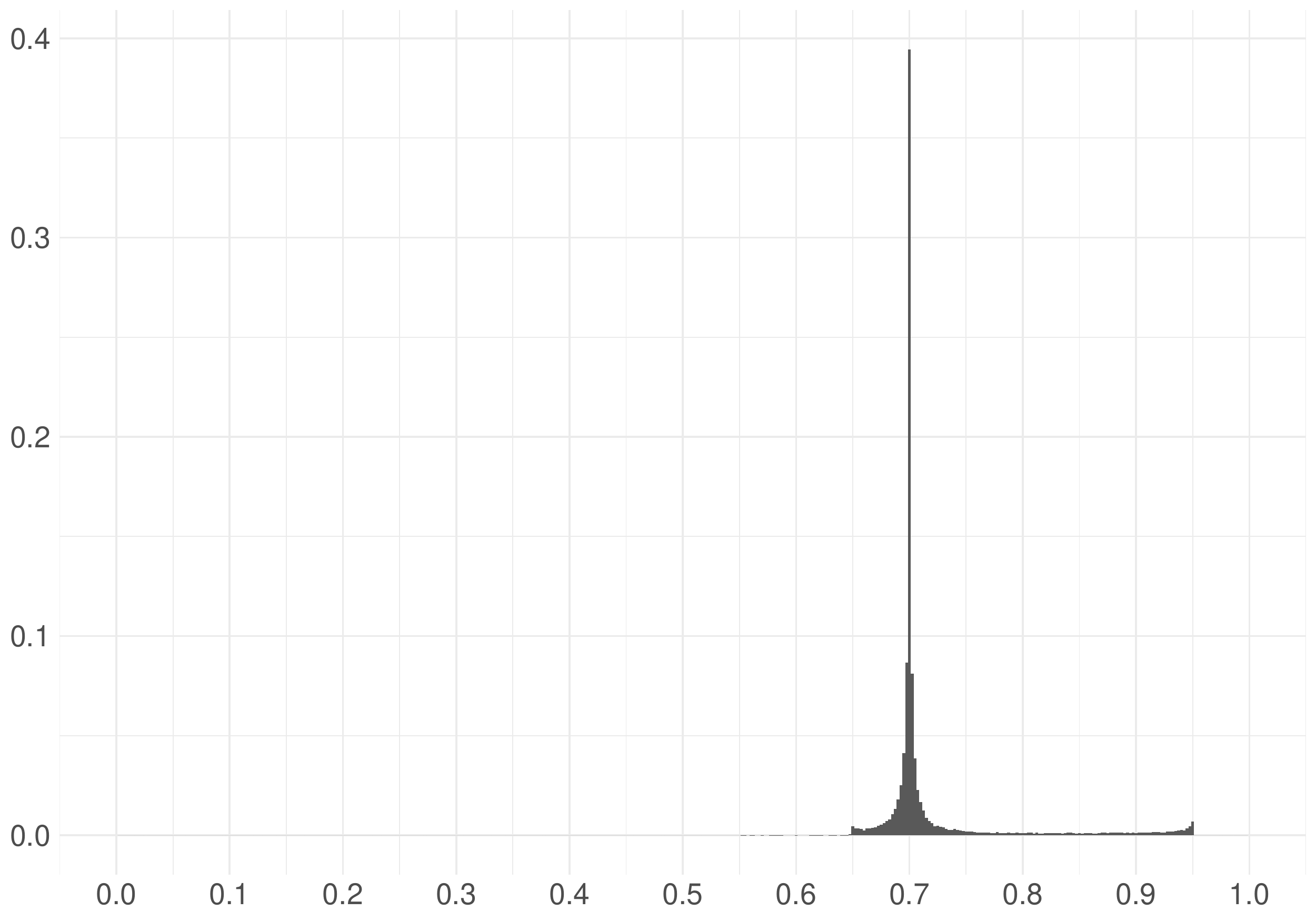}\label{fig:20:3}}
\subfigure[$T=800$, $\phi_a=1.05$, $\phi_b=0.96$]{\includegraphics[width=0.45\linewidth]{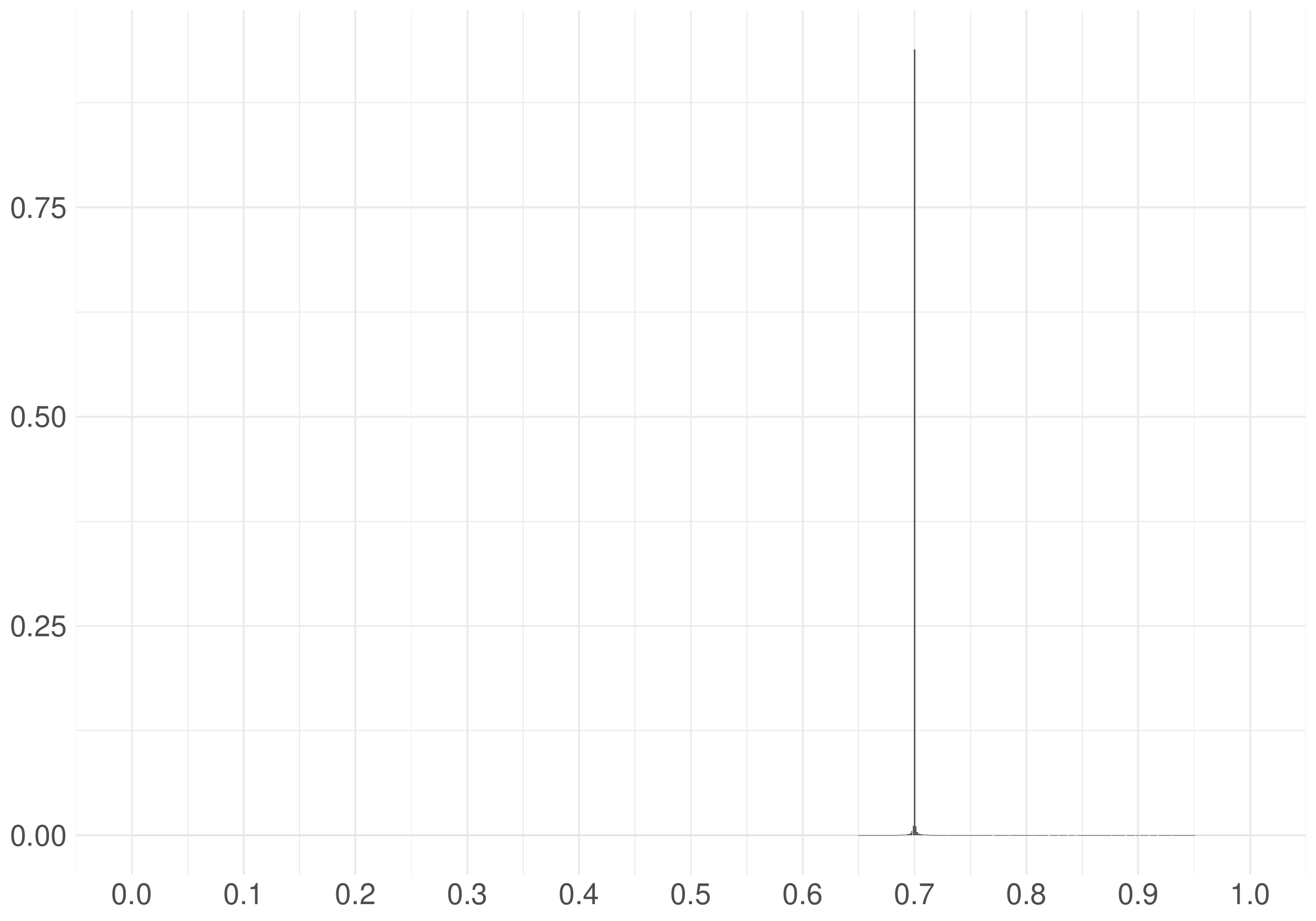}\label{fig:20:4}}\\
\subfigure[$T=400$, $\phi_a=1.09$, $\phi_b=0.96$]{\includegraphics[width=0.45\linewidth]{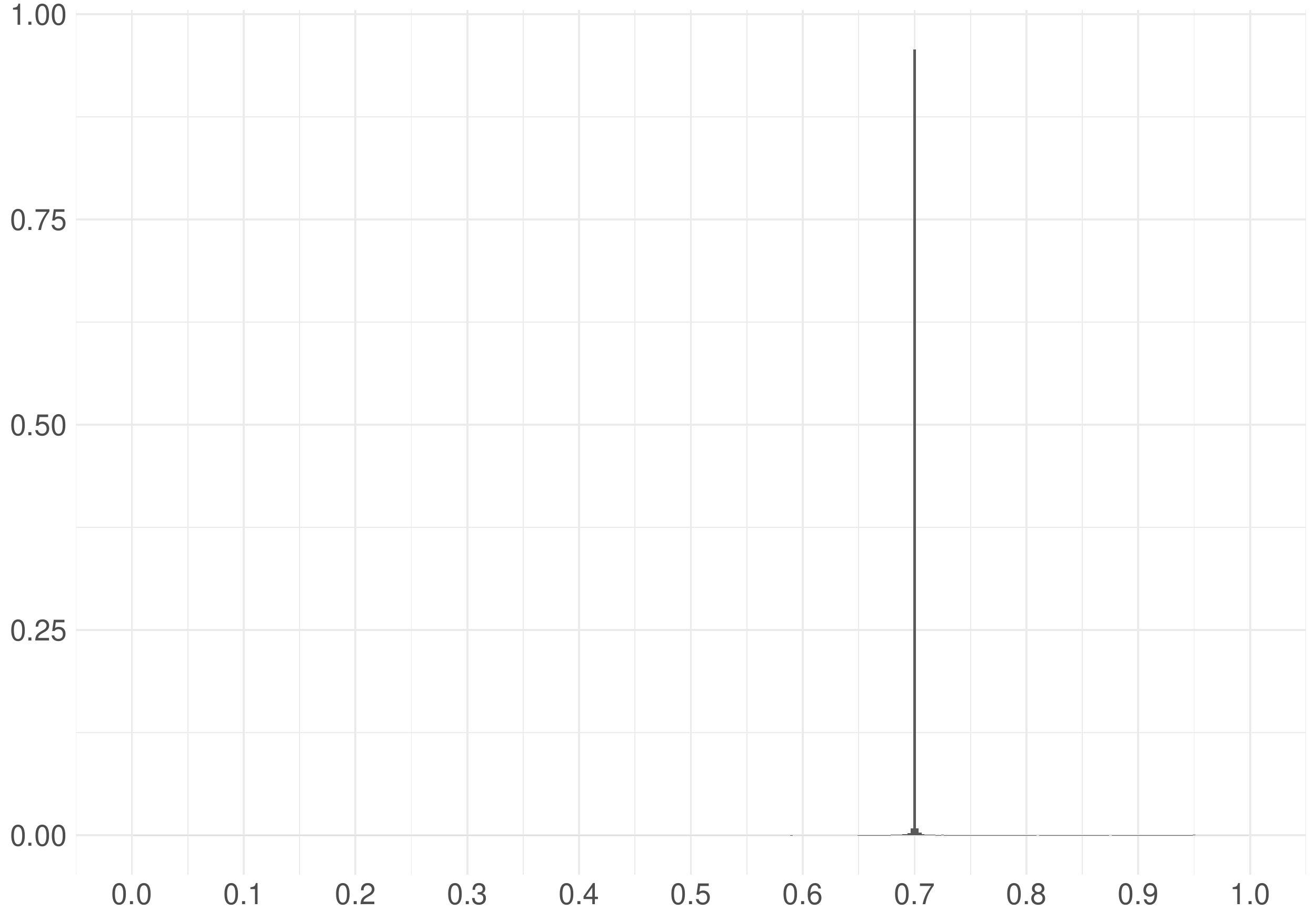}\label{fig:20:5}}
\subfigure[$T=800$, $\phi_a=1.09$, $\phi_b=0.96$]{\includegraphics[width=0.45\linewidth]{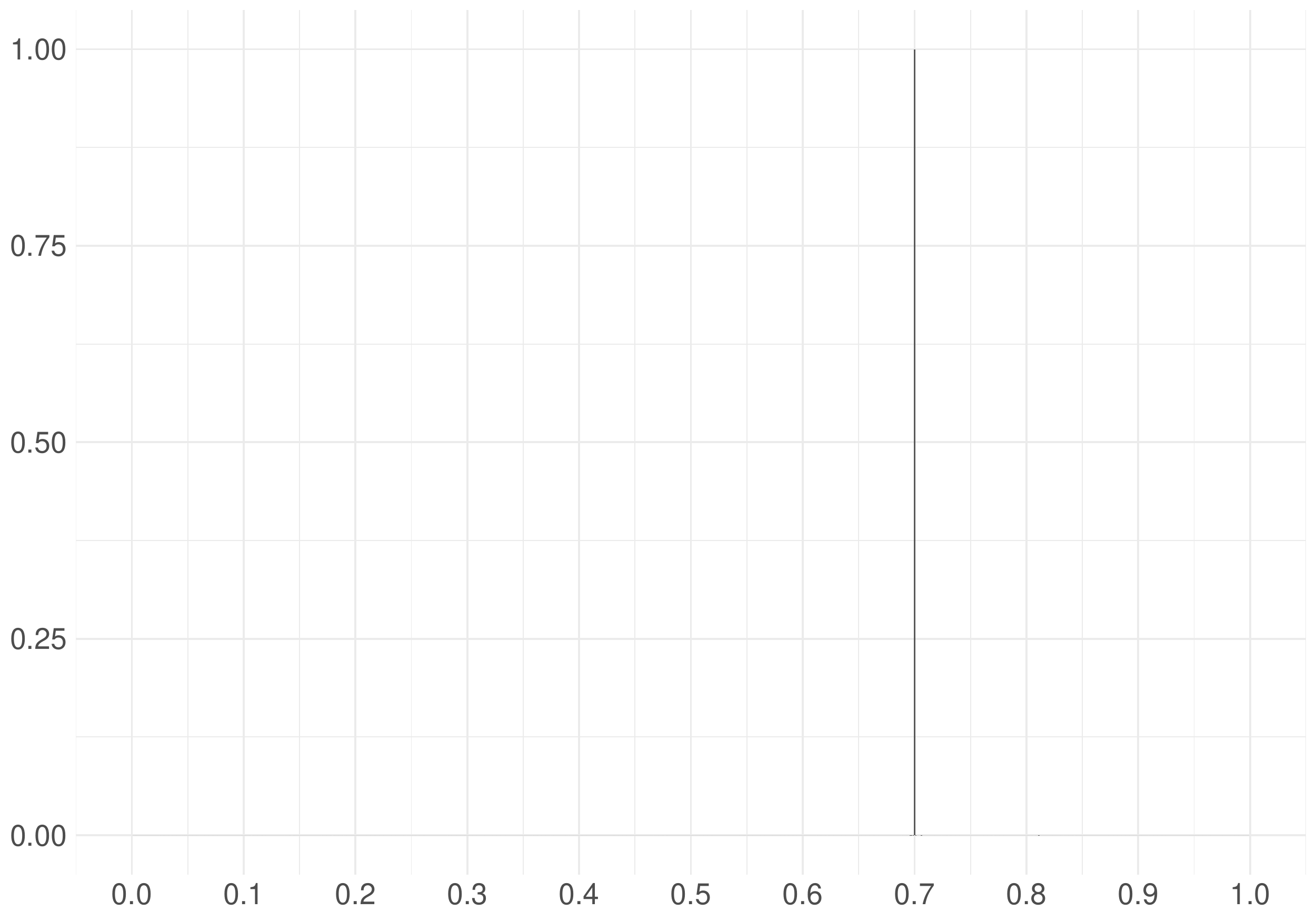}\label{fig:20:6}}
\end{center}%
\caption{Histograms of $\hat{k}_r$ % (left) and $\hat{k}_r$ (right) 
for $(\tau_e,\tau_c,\tau_r)=(0.4,0.6,0.7)$ with $\sigma_1/\sigma_0=3$}
\label{fig20}
%\centering
%\footnotesize{OLS}
\end{figure}

\begin{figure}[h]%
\begin{center}%
\subfigure[$T=400$, $\phi_a=1.05$, $\phi_b=0.98$]{\includegraphics[width=0.45\linewidth]{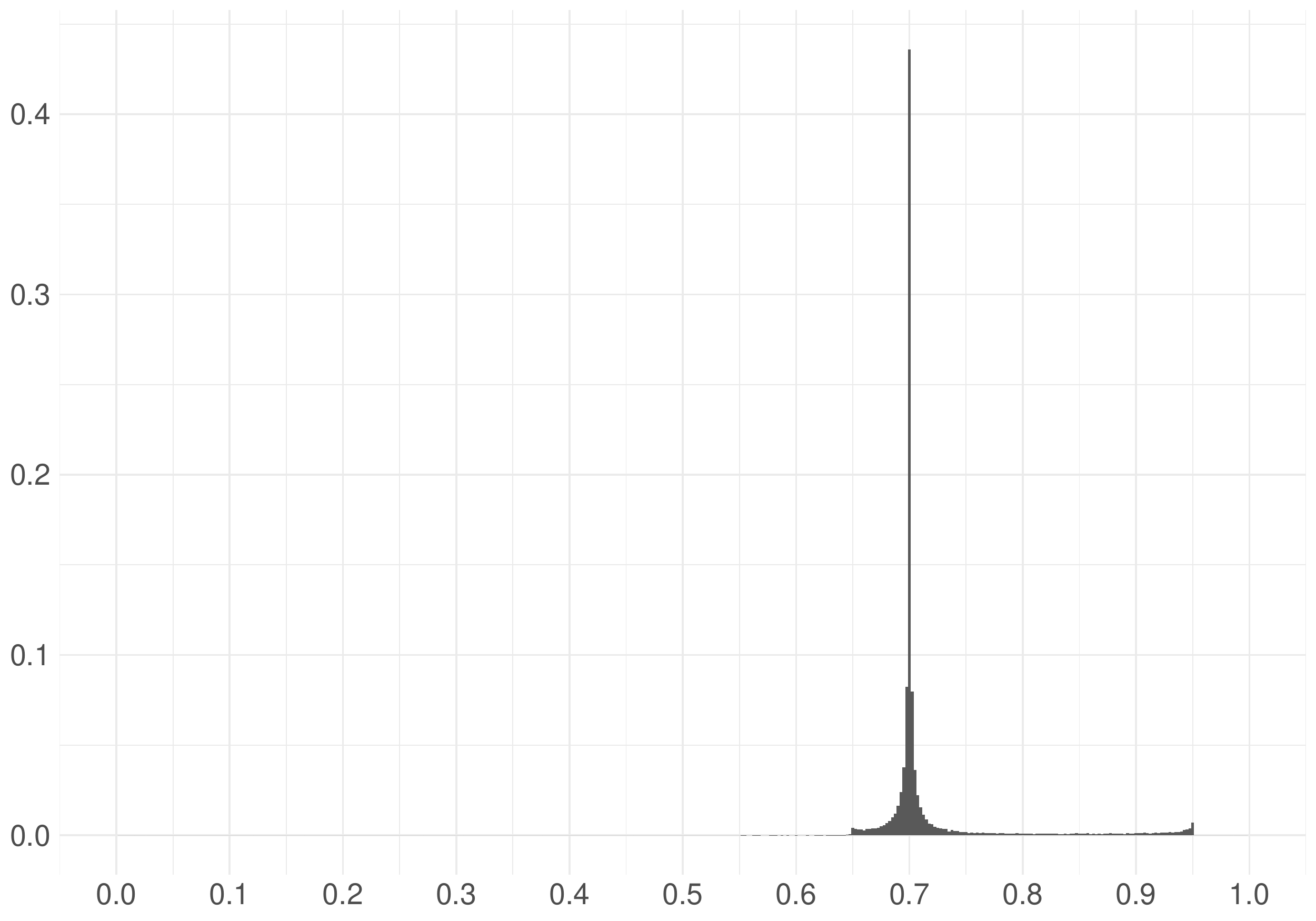}\label{fig:21:1}}
\subfigure[$T=800$, $\phi_a=1.05$, $\phi_b=0.98$]{\includegraphics[width=0.45\linewidth]{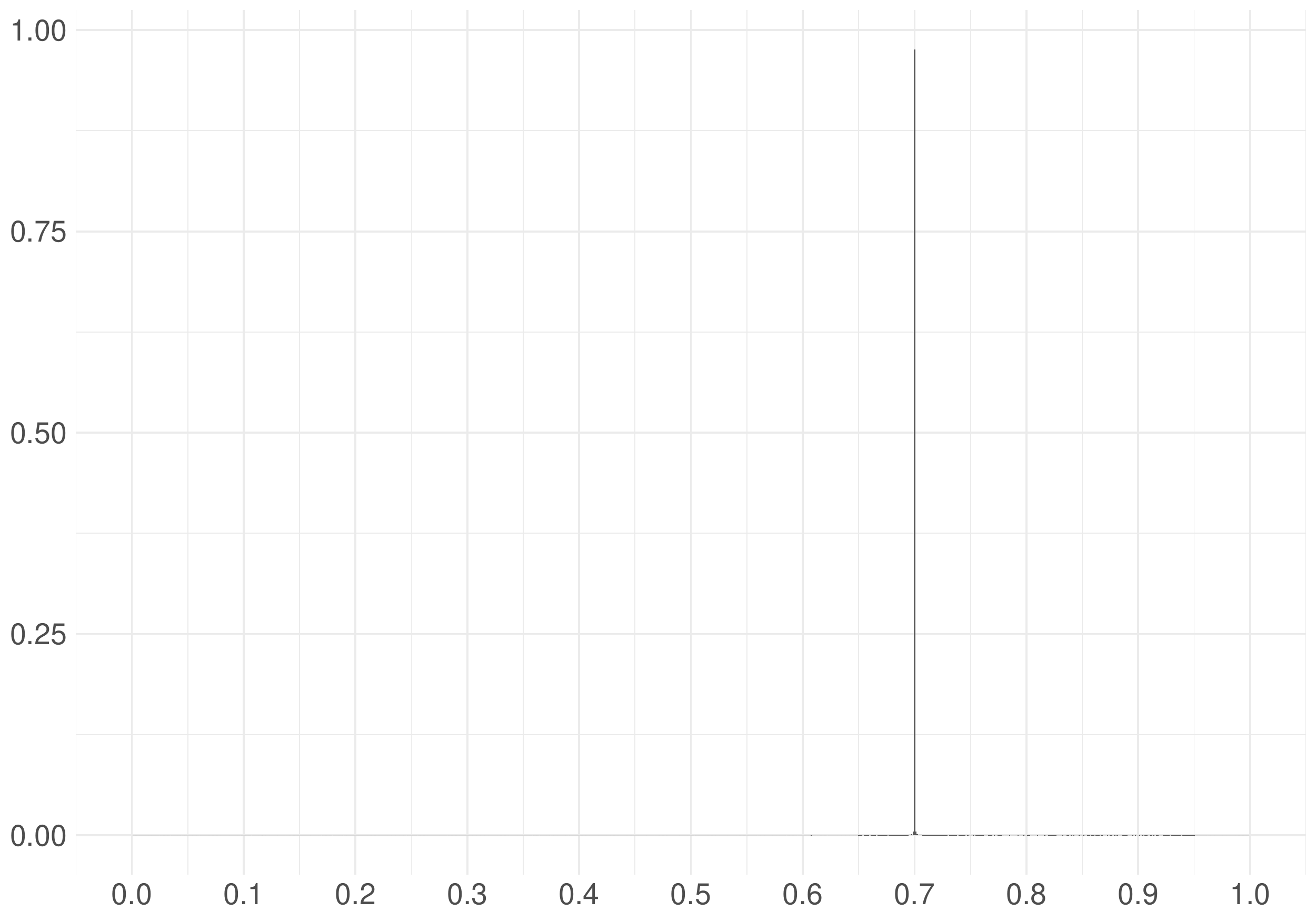}\label{fig:21:2}}\\
\subfigure[$T=400$, $\phi_a=1.05$, $\phi_b=0.96$]{\includegraphics[width=0.45\linewidth]{graph/NV_k_r_T=400_1.05_0.96_Model1s0.s10.33.pdf}\label{fig:21:3}}
\subfigure[$T=800$, $\phi_a=1.05$, $\phi_b=0.96$]{\includegraphics[width=0.45\linewidth]{graph/NV_k_r_T=800_1.05_0.96_Model1s0.s10.33.pdf}\label{fig:21:4}}\\
\subfigure[$T=400$, $\phi_a=1.05$, $\phi_b=0.94$]{\includegraphics[width=0.45\linewidth]{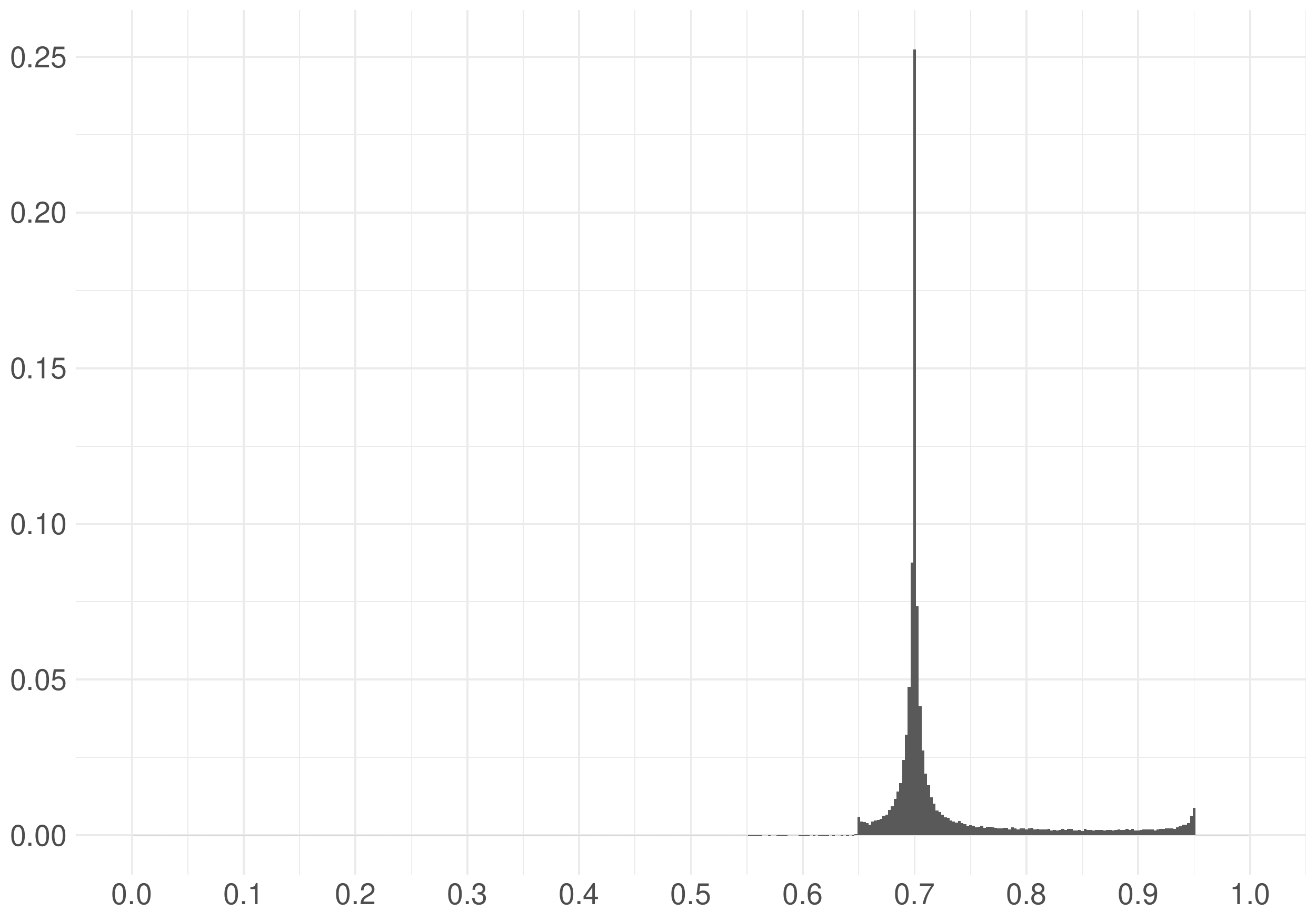}\label{fig:21:5}}
\subfigure[$T=800$, $\phi_a=1.05$, $\phi_b=0.94$]{\includegraphics[width=0.45\linewidth]{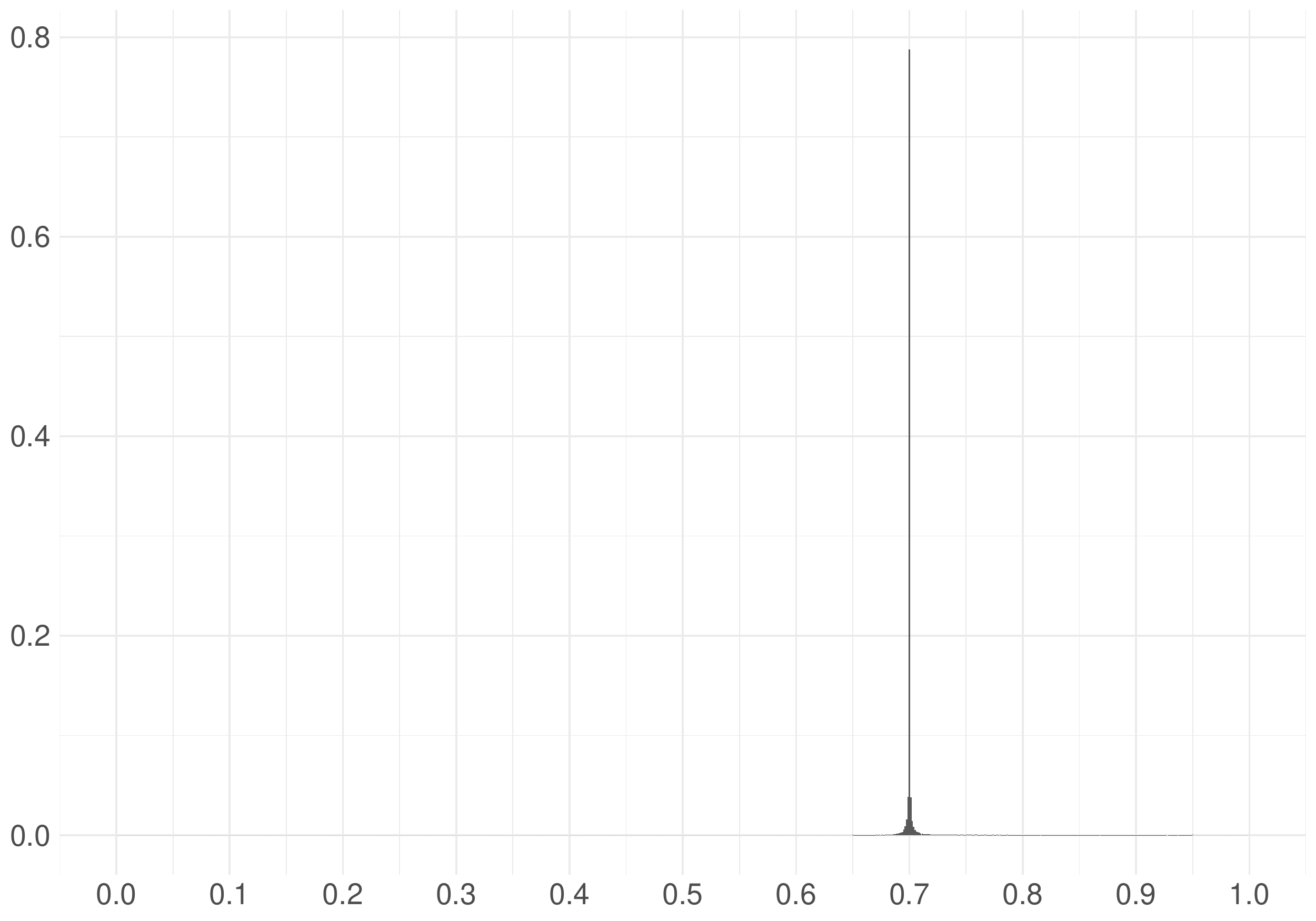}\label{fig:21:6}}
\end{center}%
\caption{Histograms of $\hat{k}_r$ % (left) and $\hat{k}_r$ (right) 
for $(\tau_e,\tau_c,\tau_r)=(0.4,0.6,0.7)$ with $\sigma_1/\sigma_0=3$}
\label{fig21}
%\centering
%\footnotesize{OLS}
\end{figure}

%----- figure_standard -----$
\newpage
\renewcommand{\thefigure}{\arabic{figure}}
\setcounter{figure}{0}
%%%
%%%
%new results
%%%
%%%
%%\pagestyle{empty}

\begin{figure}[t]%
\begin{center}%
\subfigure[$T=400$, $\phi_a=1.01$, $\phi_b=0.96$]{\includegraphics[width=0.45\linewidth]{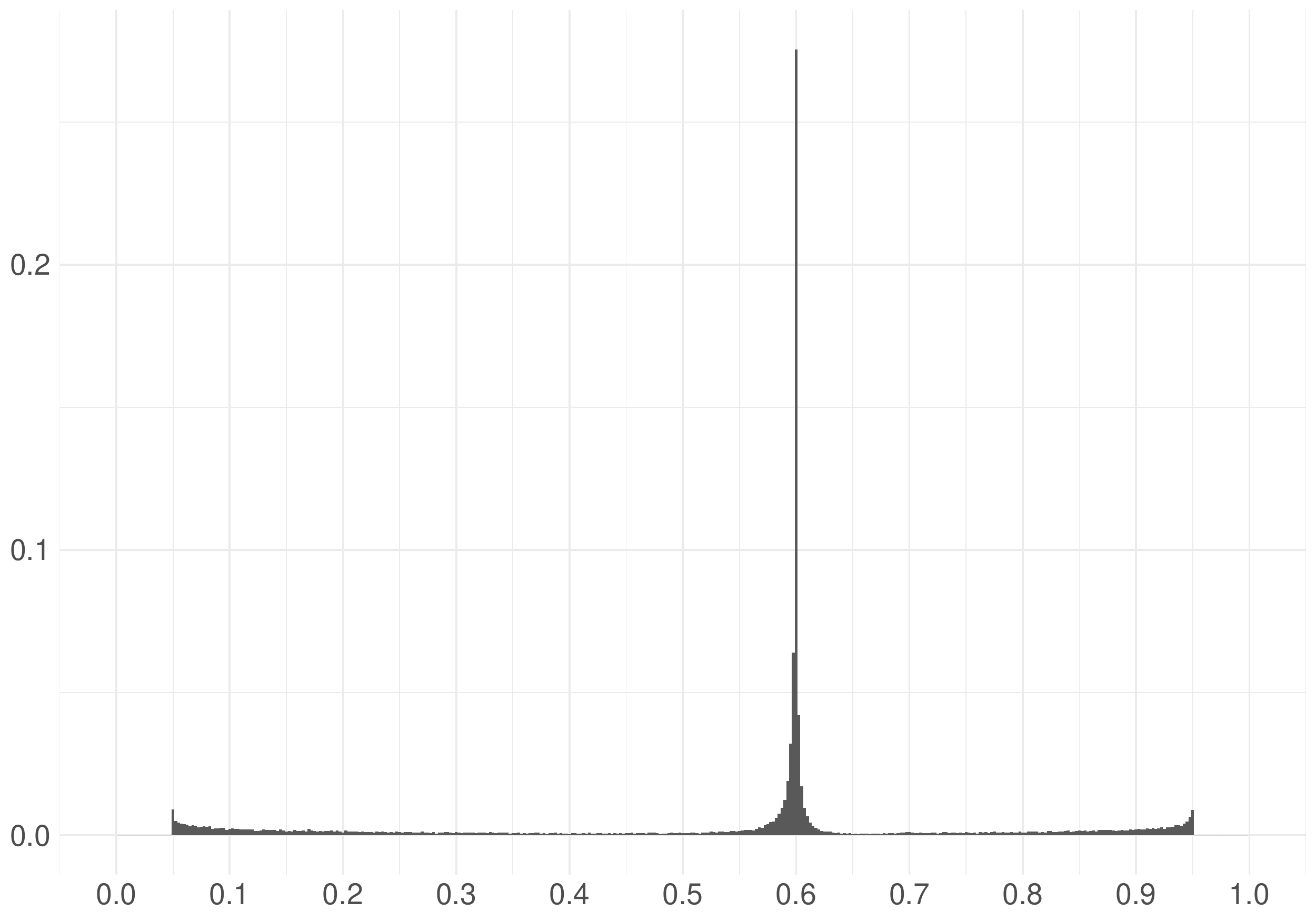}\label{fig:10:1}}
\subfigure[$T=800$, $\phi_a=1.01$, $\phi_b=0.96$]{\includegraphics[width=0.45\linewidth]{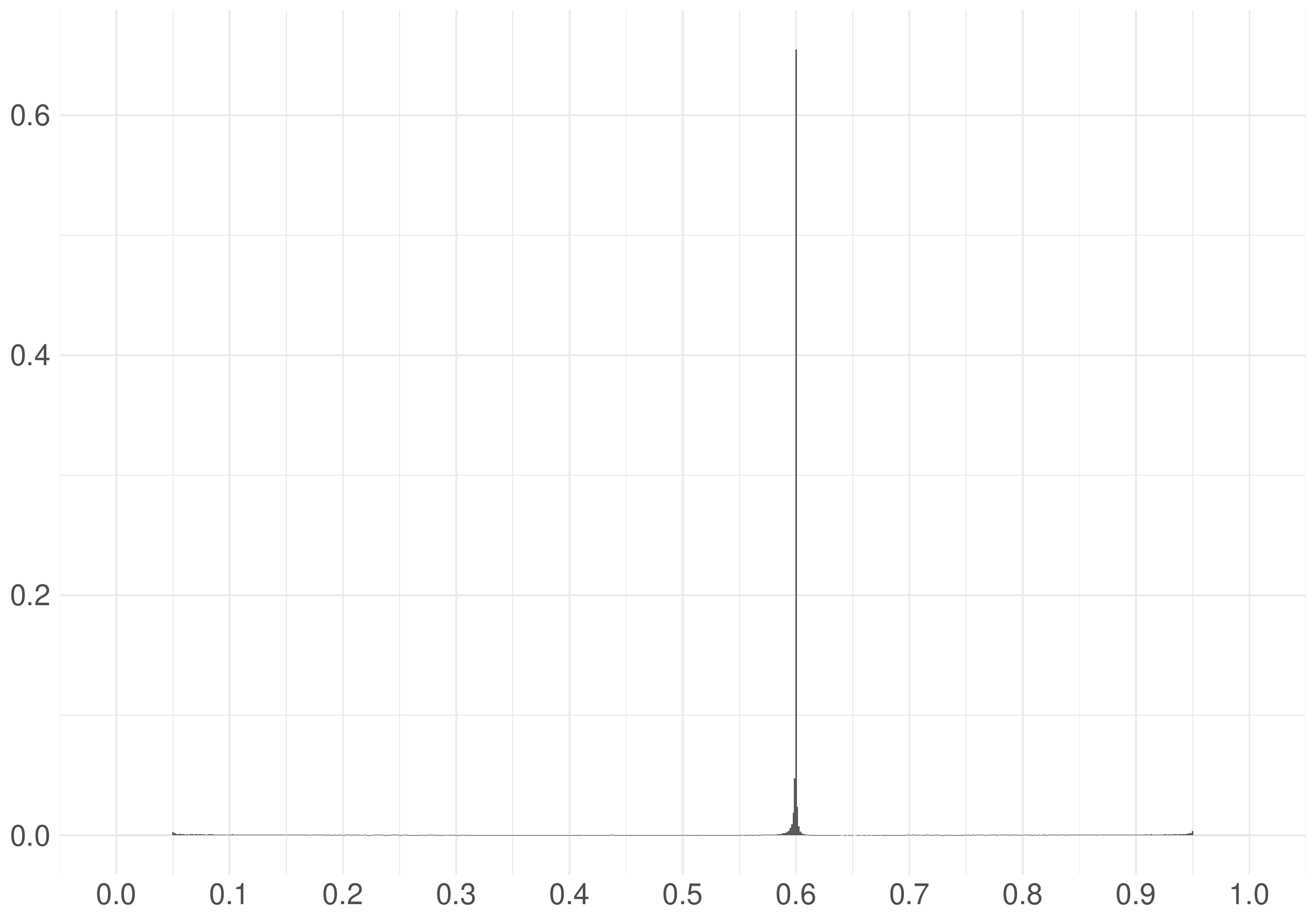}\label{fig:10:2}}\\
\subfigure[$T=400$, $\phi_a=1.05$, $\phi_b=0.96$]{\includegraphics[width=0.45\linewidth]{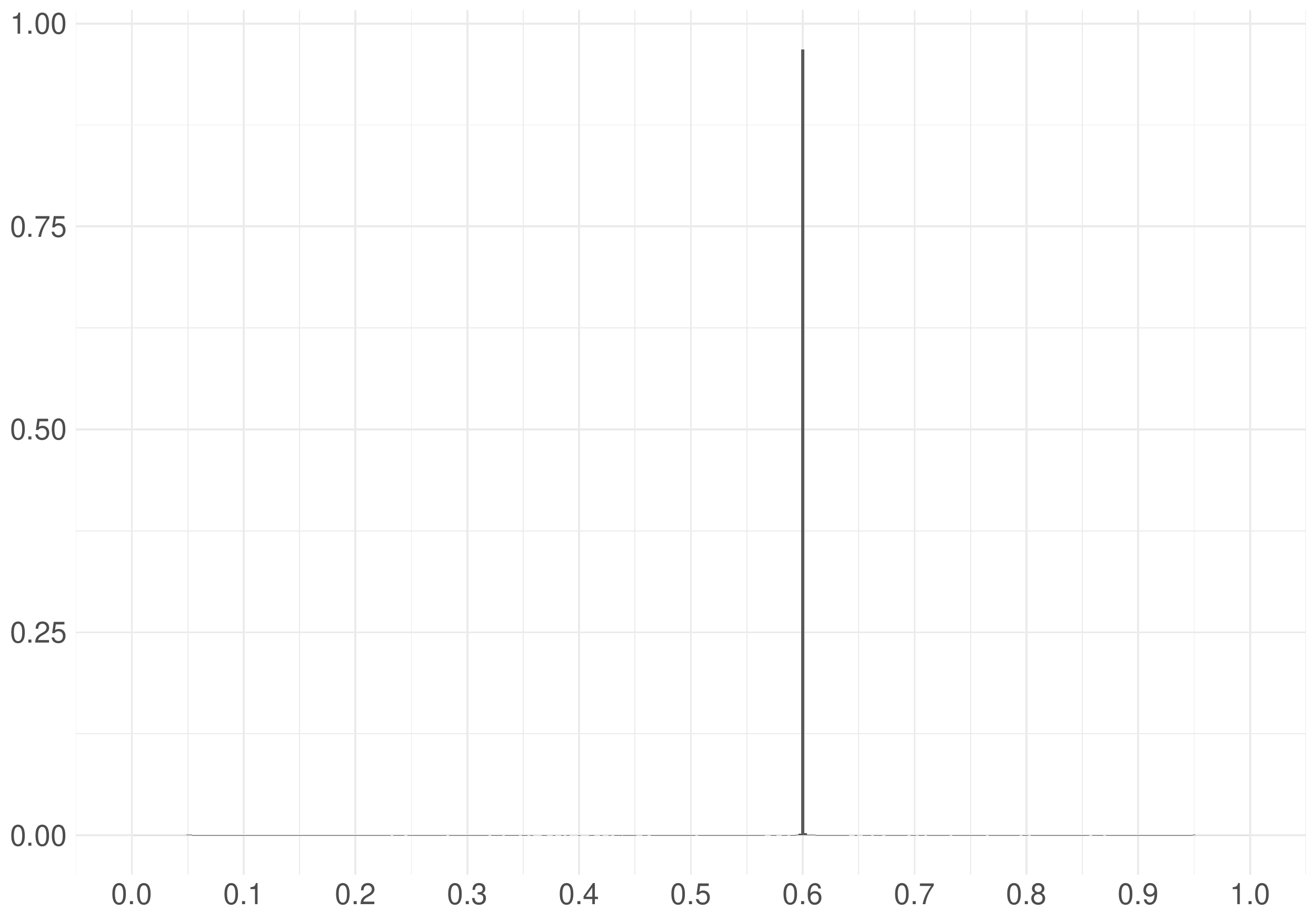}\label{fig:10:3}}
\subfigure[$T=800$, $\phi_a=1.05$, $\phi_b=0.96$]{\includegraphics[width=0.45\linewidth]{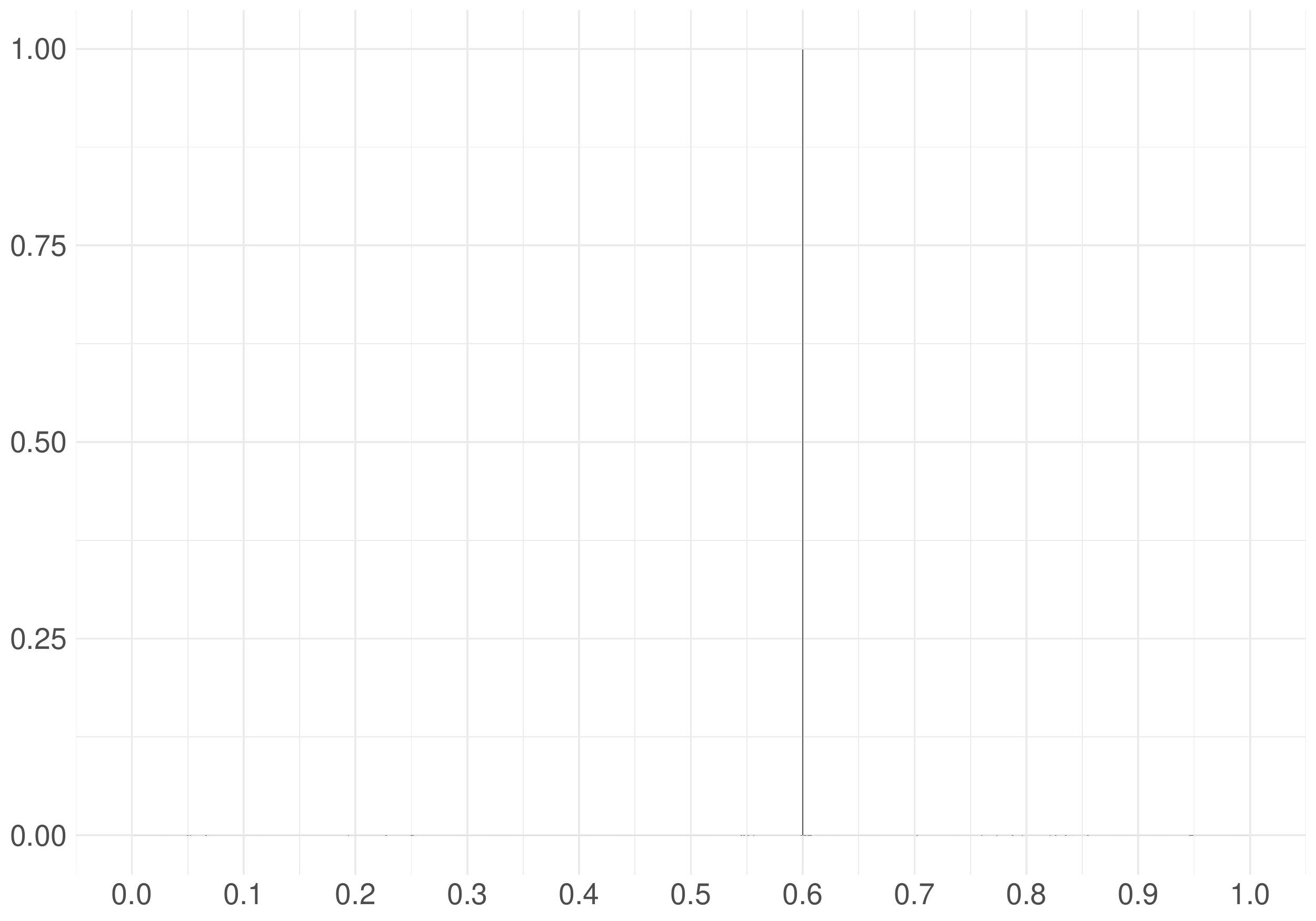}\label{fig:10:4}}\\
\subfigure[$T=400$, $\phi_a=1.09$, $\phi_b=0.96$]{\includegraphics[width=0.45\linewidth]{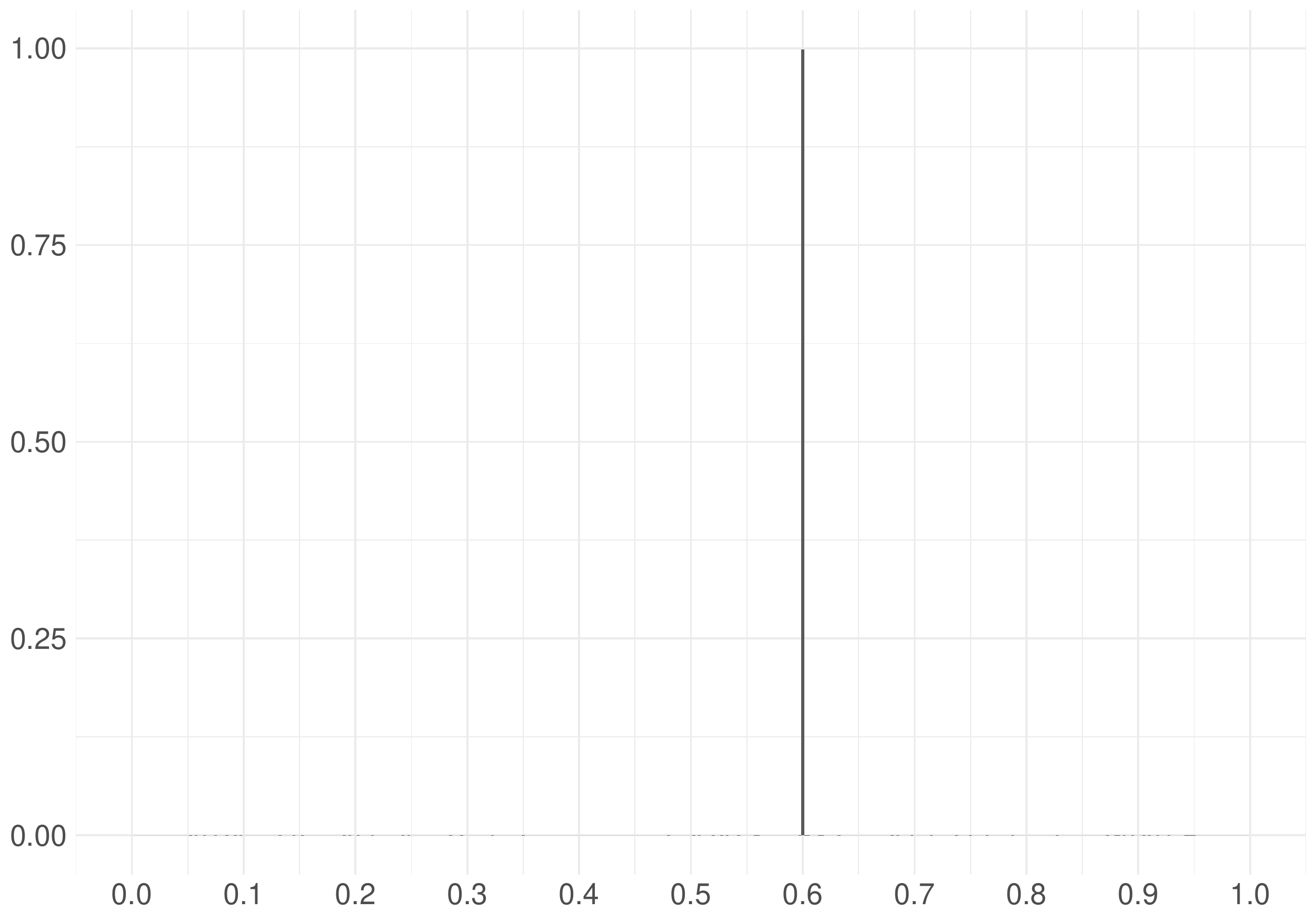}\label{fig:10:5}}
\subfigure[$T=800$, $\phi_a=1.09$, $\phi_b=0.96$]{\includegraphics[width=0.45\linewidth]{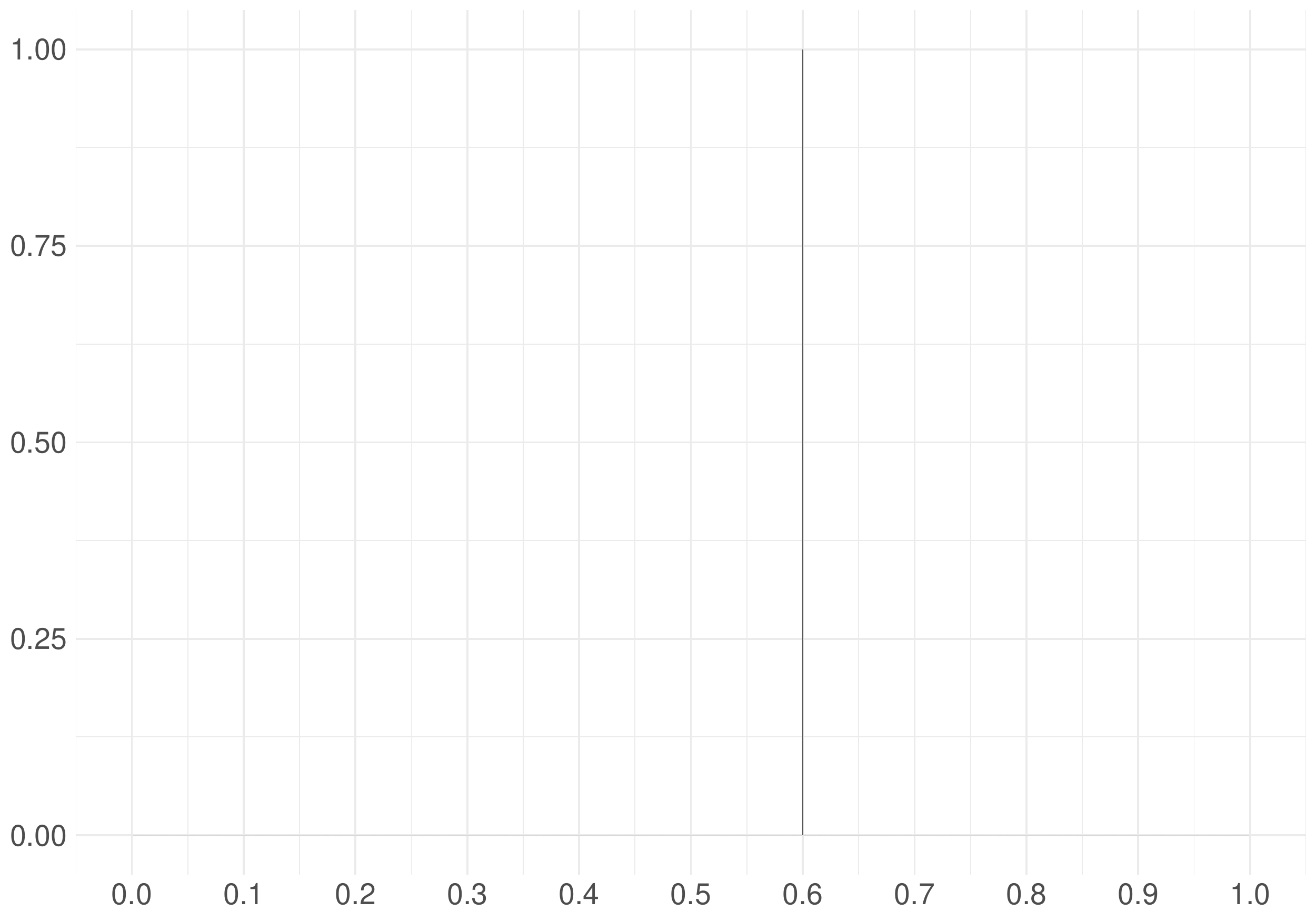}\label{fig:10:6}}
\end{center}%
\caption{Histograms of $\hat{k}_c$ % (left) and $\hat{k}_r$ (right) 
for $(\tau_e,\tau_c,\tau_r)=(0.4,0.6,0.7)$}
\label{fig10}
%\centering
%\footnotesize{OLS}
\end{figure}

\begin{figure}[h]%
\begin{center}%
\subfigure[$T=400$, $\phi_a=1.05$, $\phi_b=0.98$]{\includegraphics[width=0.45\linewidth]{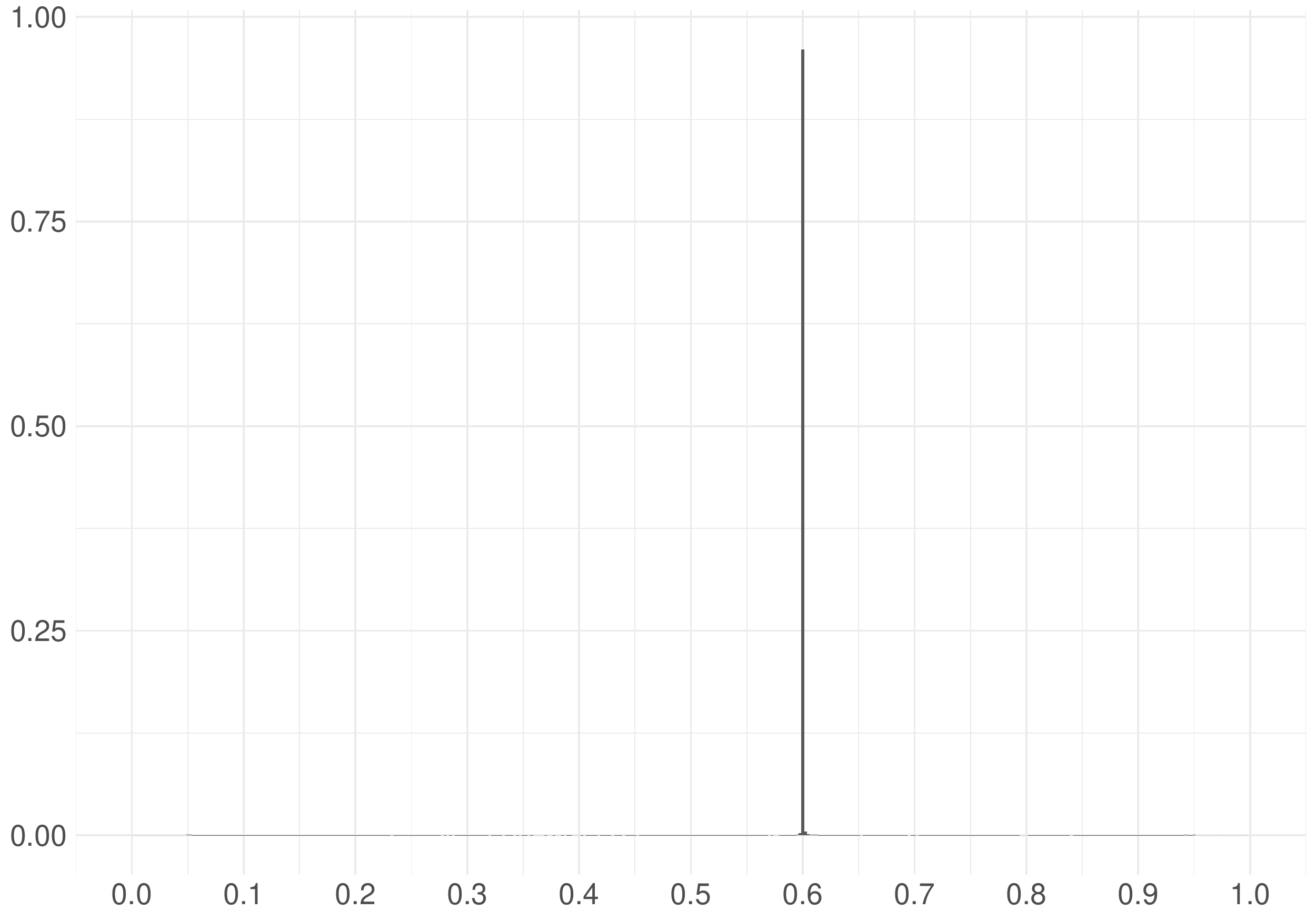}\label{fig:11:1}}
\subfigure[$T=800$, $\phi_a=1.05$, $\phi_b=0.98$]{\includegraphics[width=0.45\linewidth]{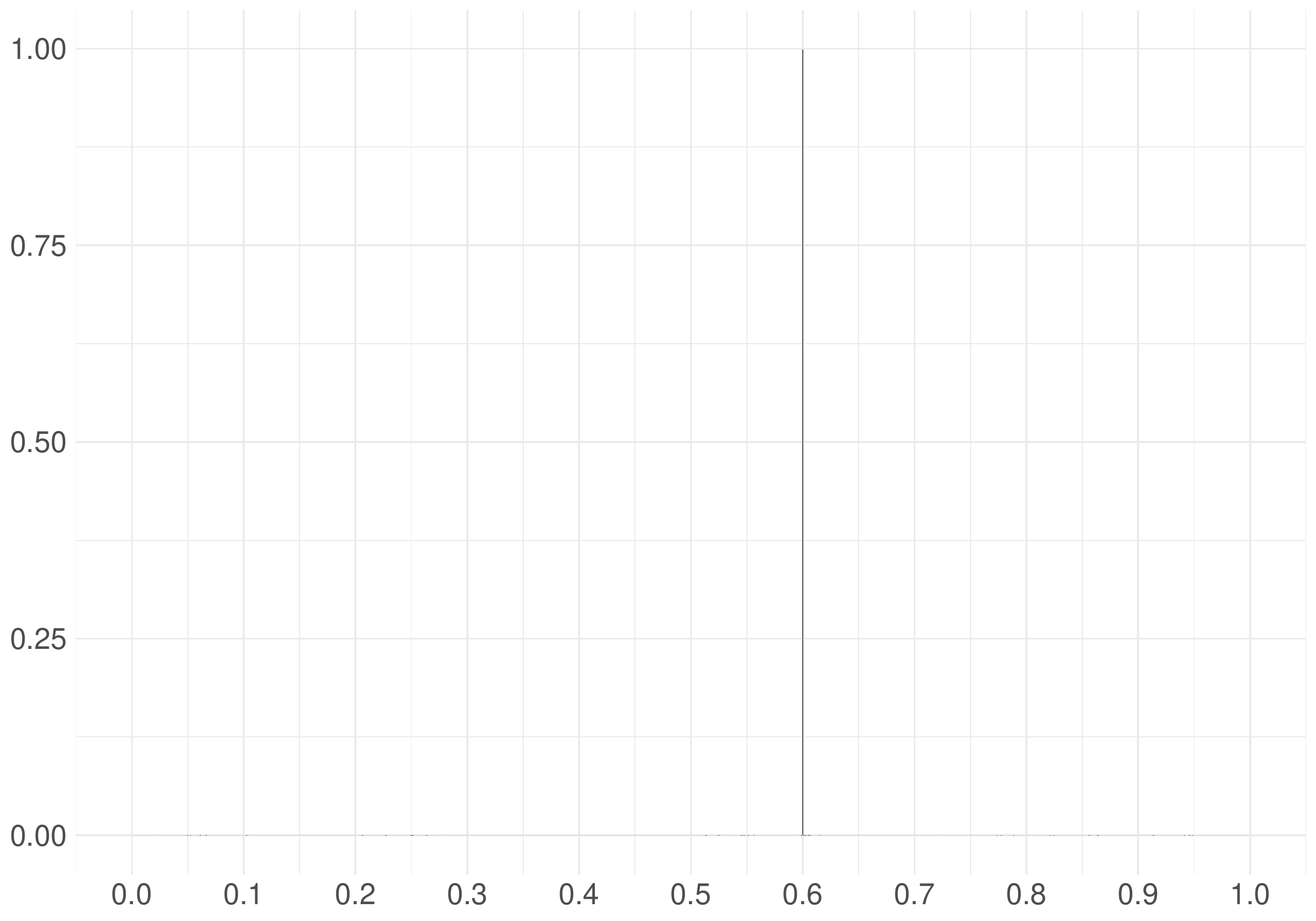}\label{fig:11:2}}\\
\subfigure[$T=400$, $\phi_a=1.05$, $\phi_b=0.96$]{\includegraphics[width=0.45\linewidth]{graph/NV_k_c_T=400_1.05_0.96_Model1s0.s11.pdf}\label{fig:11:3}}
\subfigure[$T=800$, $\phi_a=1.05$, $\phi_b=0.96$]{\includegraphics[width=0.45\linewidth]{graph/NV_k_c_T=800_1.05_0.96_Model1s0.s11.pdf}\label{fig:11:4}}\\
\subfigure[$T=400$, $\phi_a=1.05$, $\phi_b=0.94$]{\includegraphics[width=0.45\linewidth]{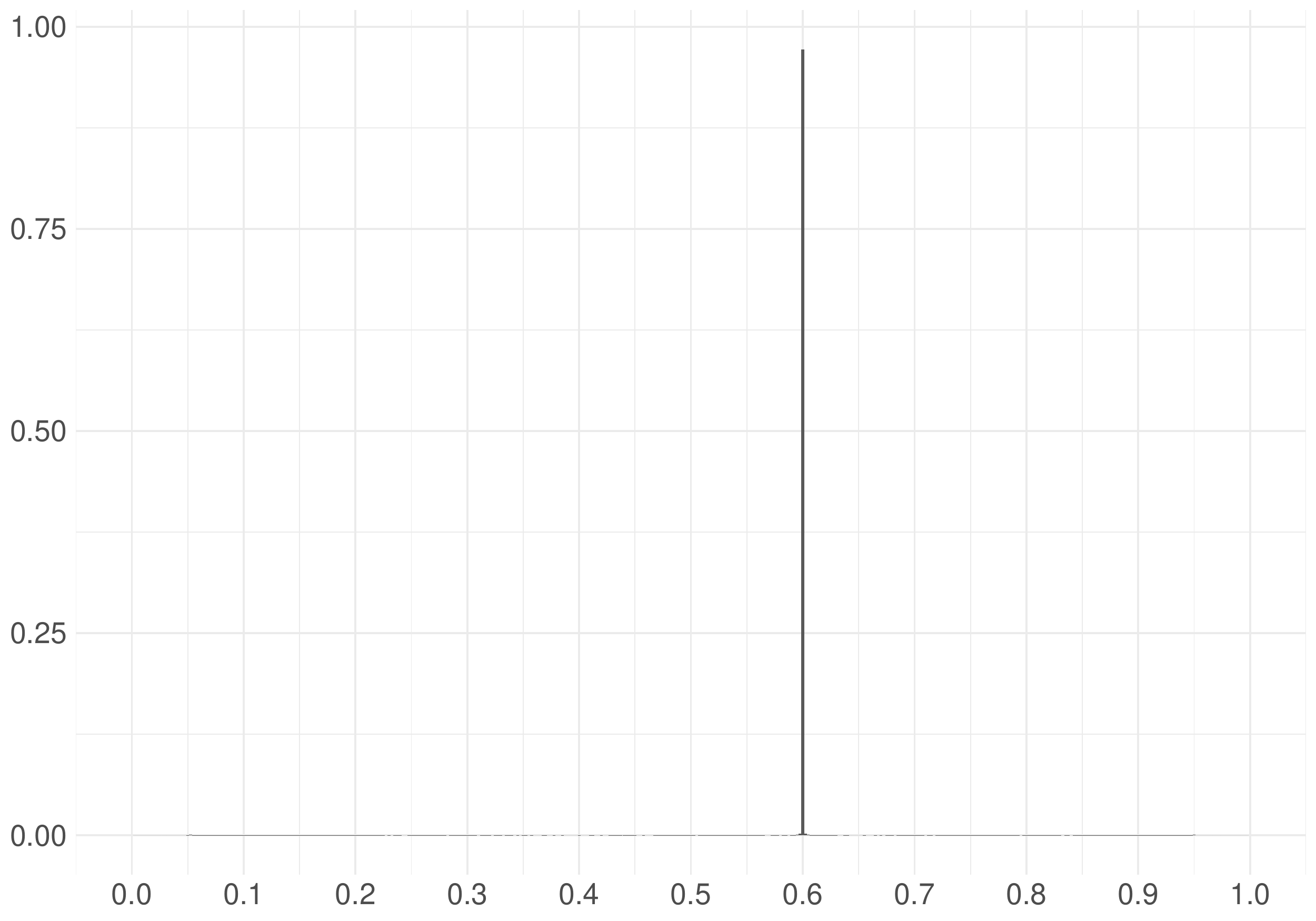}\label{fig:11:5}}
\subfigure[$T=800$, $\phi_a=1.05$, $\phi_b=0.94$]{\includegraphics[width=0.45\linewidth]{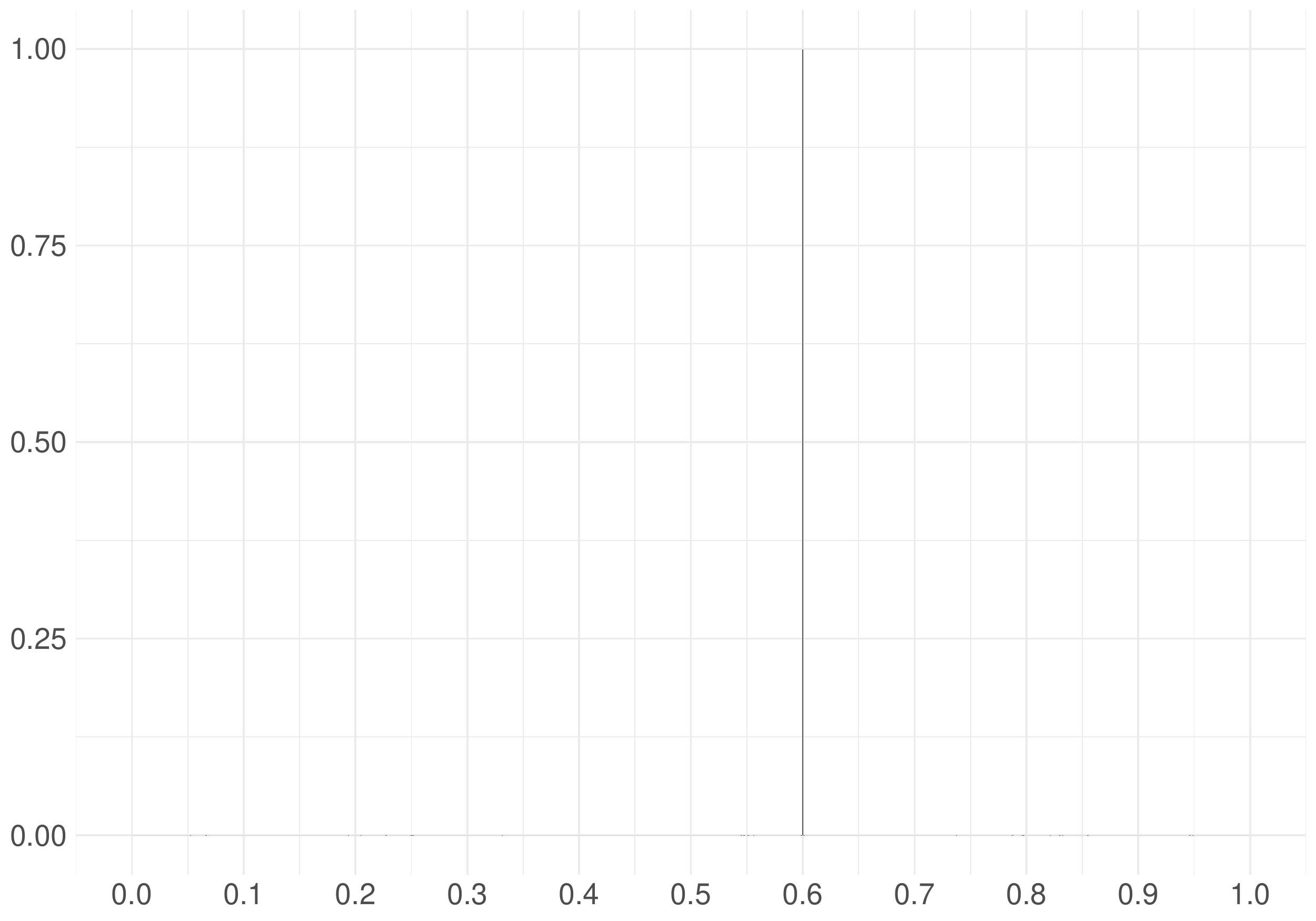}\label{fig:11:6}}
\end{center}%
\caption{Histograms of $\hat{k}_c$ % (left) and $\hat{k}_r$ (right) 
for $(\tau_e,\tau_c,\tau_r)=(0.4,0.6,0.7)$}
\label{fig11}
%\centering
%\footnotesize{OLS}
\end{figure}

%%k_r

\begin{figure}[h]%
\begin{center}%
\subfigure[$T=400$, $\phi_a=1.01$, $\phi_b=0.96$]{\includegraphics[width=0.45\linewidth]{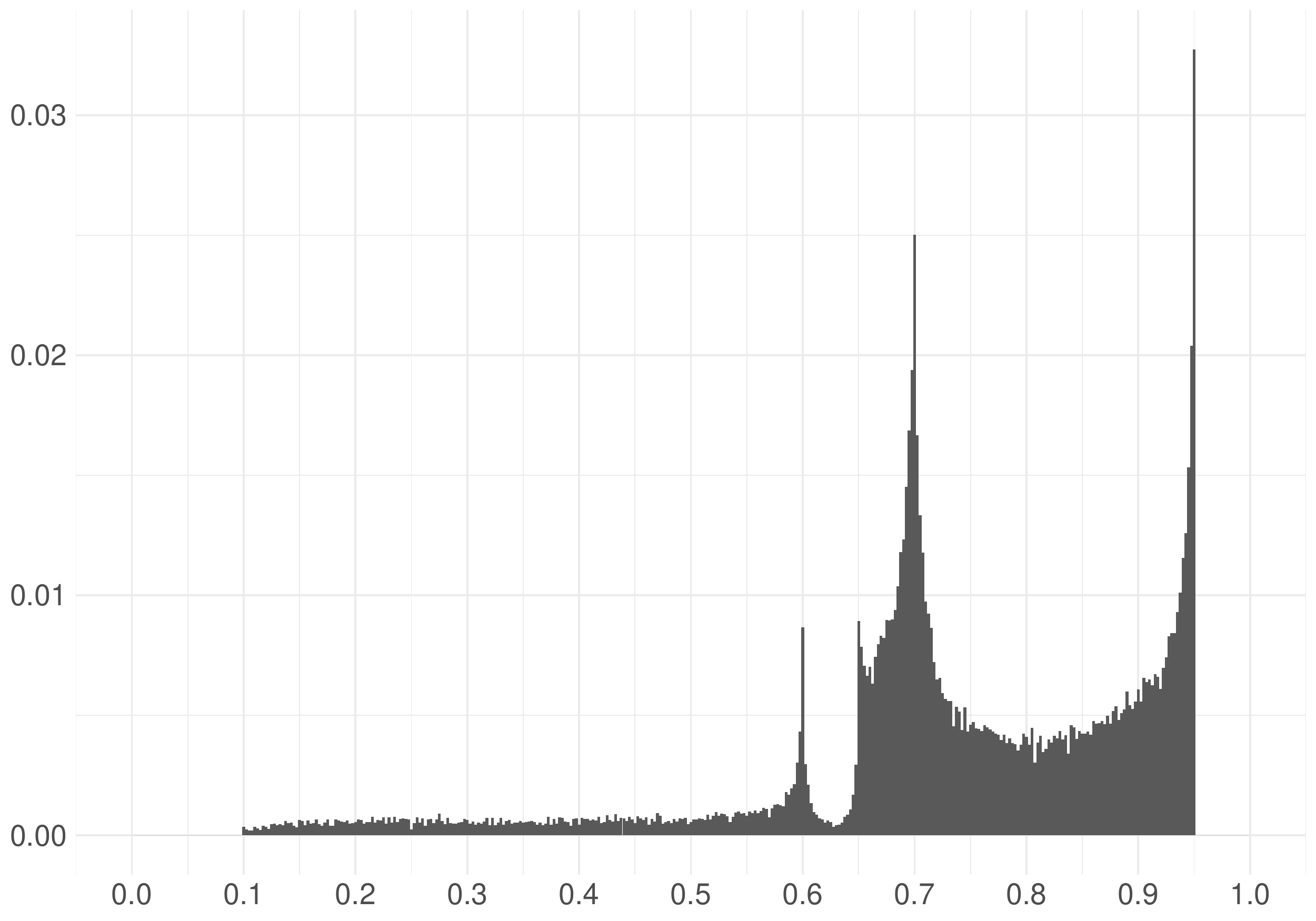}\label{fig:12:1}}
\subfigure[$T=800$, $\phi_a=1.01$, $\phi_b=0.96$]{\includegraphics[width=0.45\linewidth]{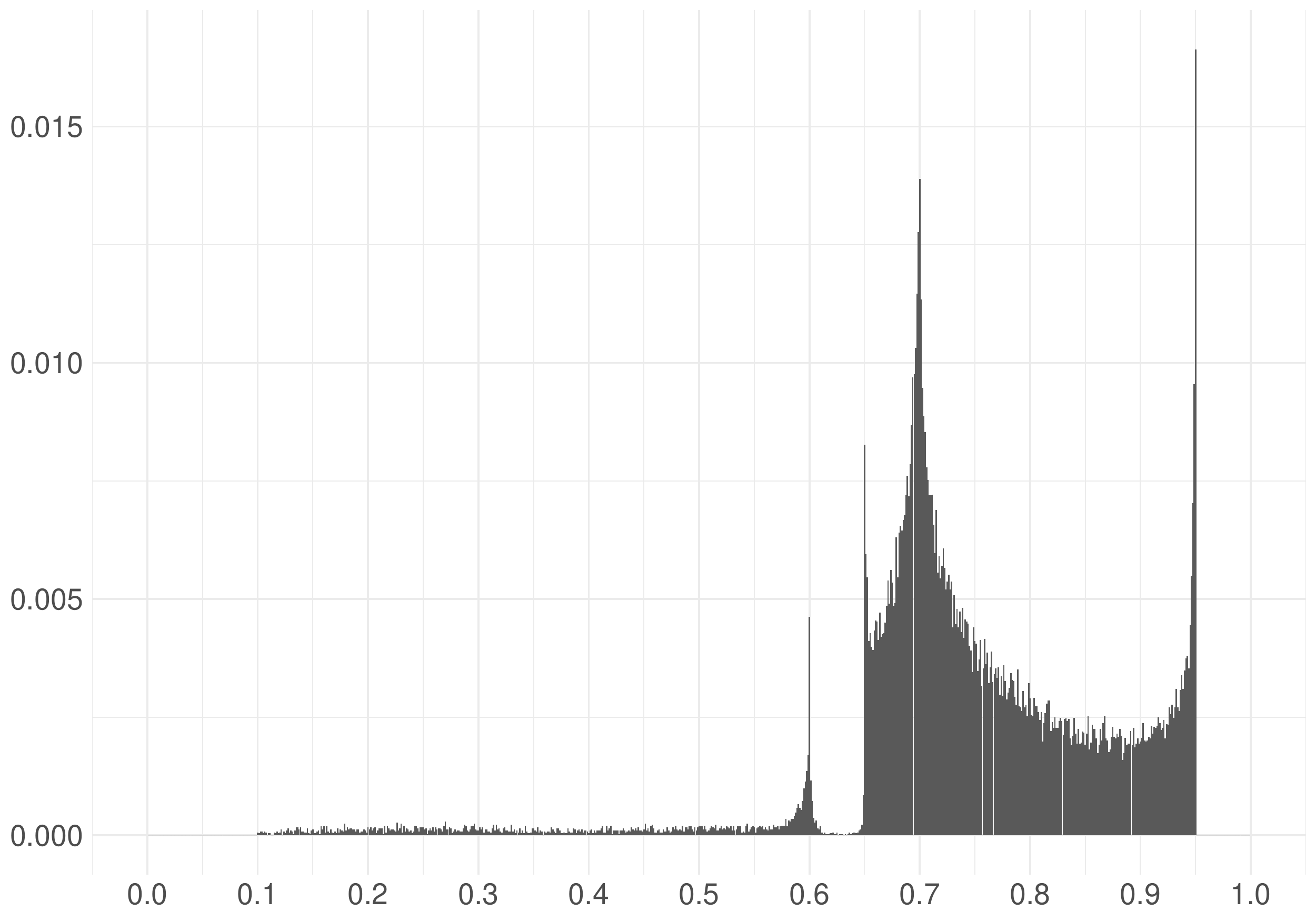}\label{fig:12:2}}\\
\subfigure[$T=400$, $\phi_a=1.05$, $\phi_b=0.96$]{\includegraphics[width=0.45\linewidth]{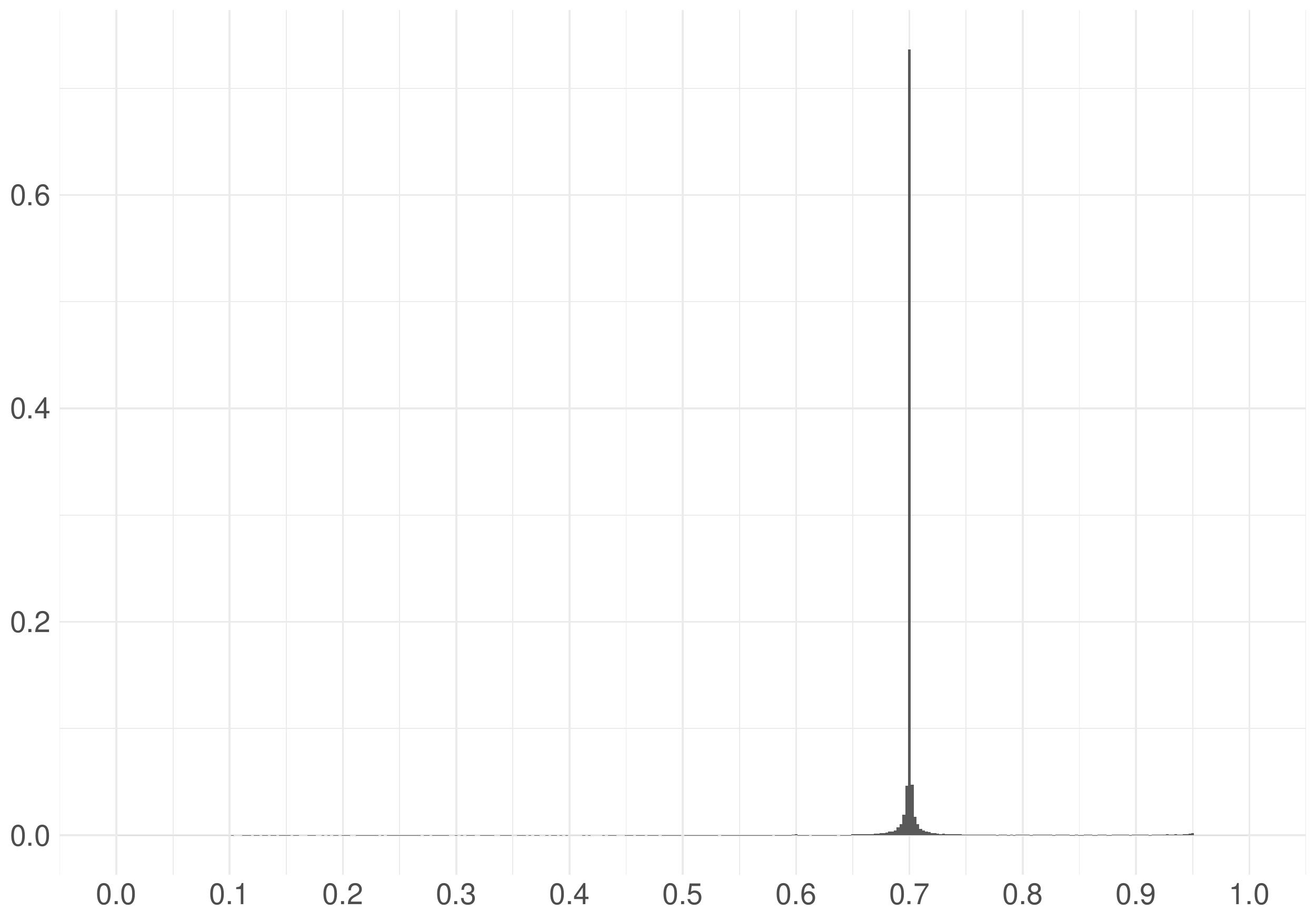}\label{fig:12:3}}
\subfigure[$T=800$, $\phi_a=1.05$, $\phi_b=0.96$]{\includegraphics[width=0.45\linewidth]{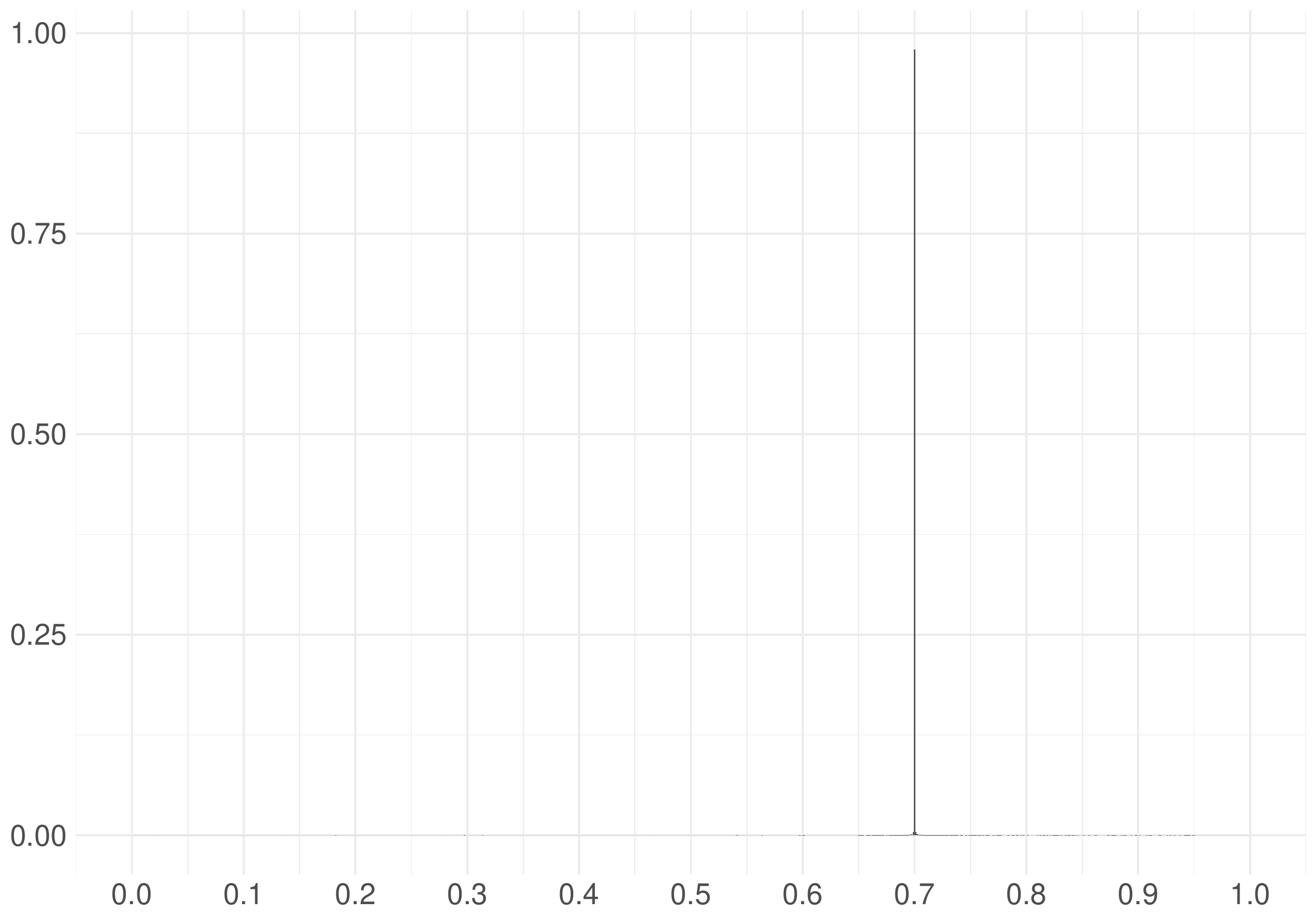}\label{fig:12:4}}\\
\subfigure[$T=400$, $\phi_a=1.09$, $\phi_b=0.96$]{\includegraphics[width=0.45\linewidth]{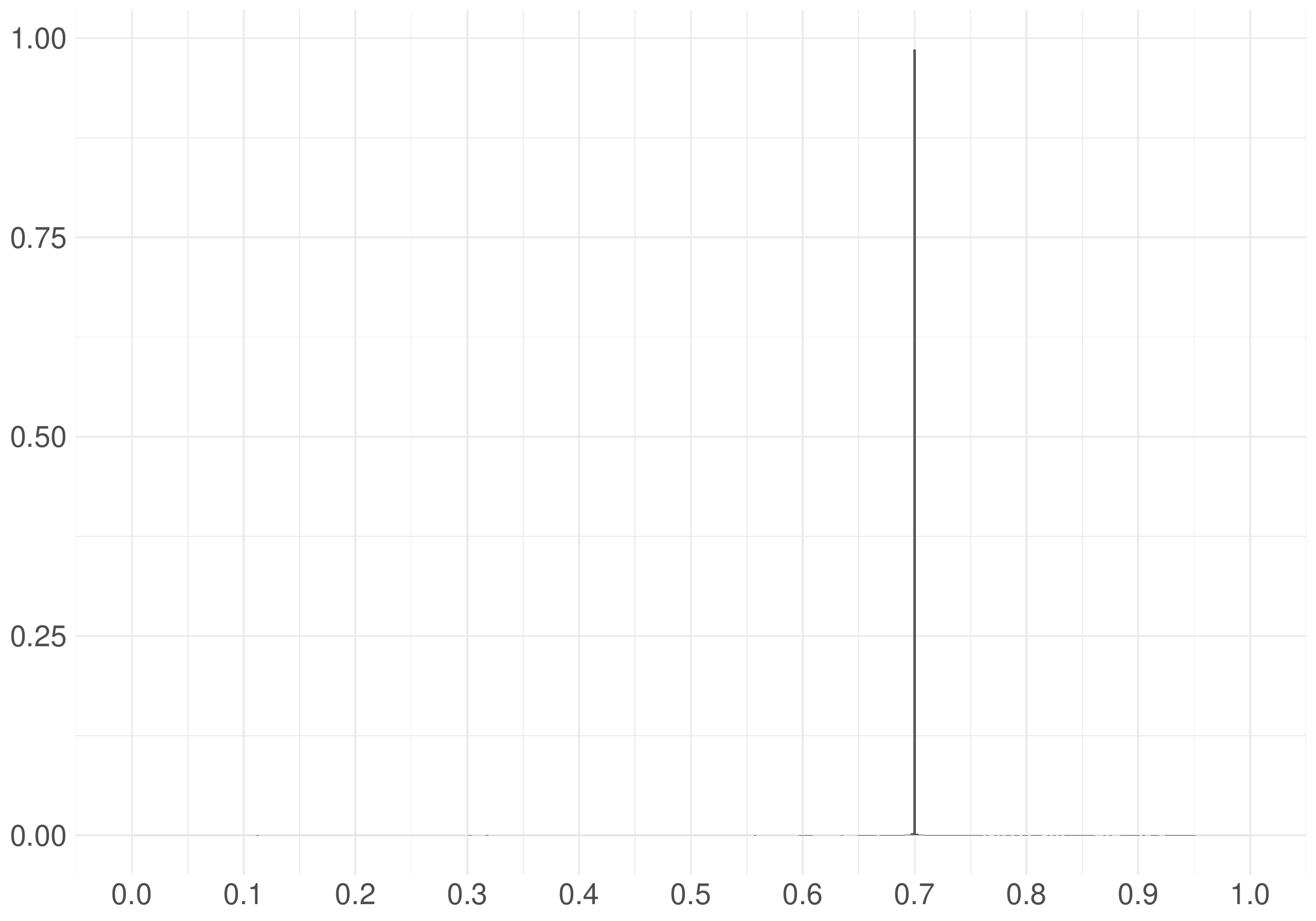}\label{fig:12:5}}
\subfigure[$T=800$, $\phi_a=1.09$, $\phi_b=0.96$]{\includegraphics[width=0.45\linewidth]{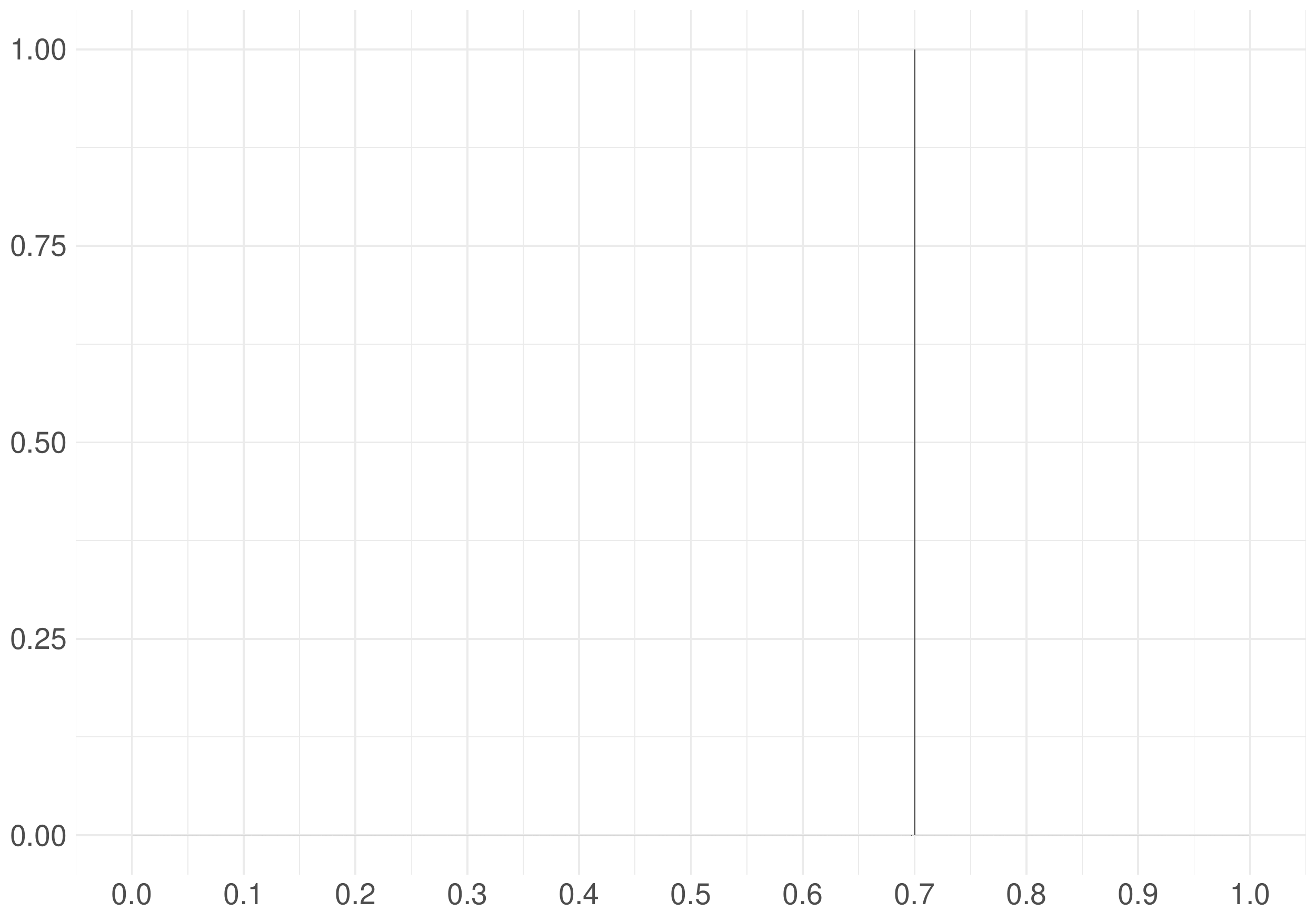}\label{fig:12:6}}
\end{center}%
\caption{Histograms of $\hat{k}_r$ % (left) and $\hat{k}_r$ (right) 
for $(\tau_e,\tau_c,\tau_r)=(0.4,0.6,0.7)$}
\label{fig12}
%\centering
%\footnotesize{OLS}
\end{figure}

\begin{figure}[h]%
\begin{center}%
\subfigure[$T=400$, $\phi_a=1.05$, $\phi_b=0.98$]{\includegraphics[width=0.45\linewidth]{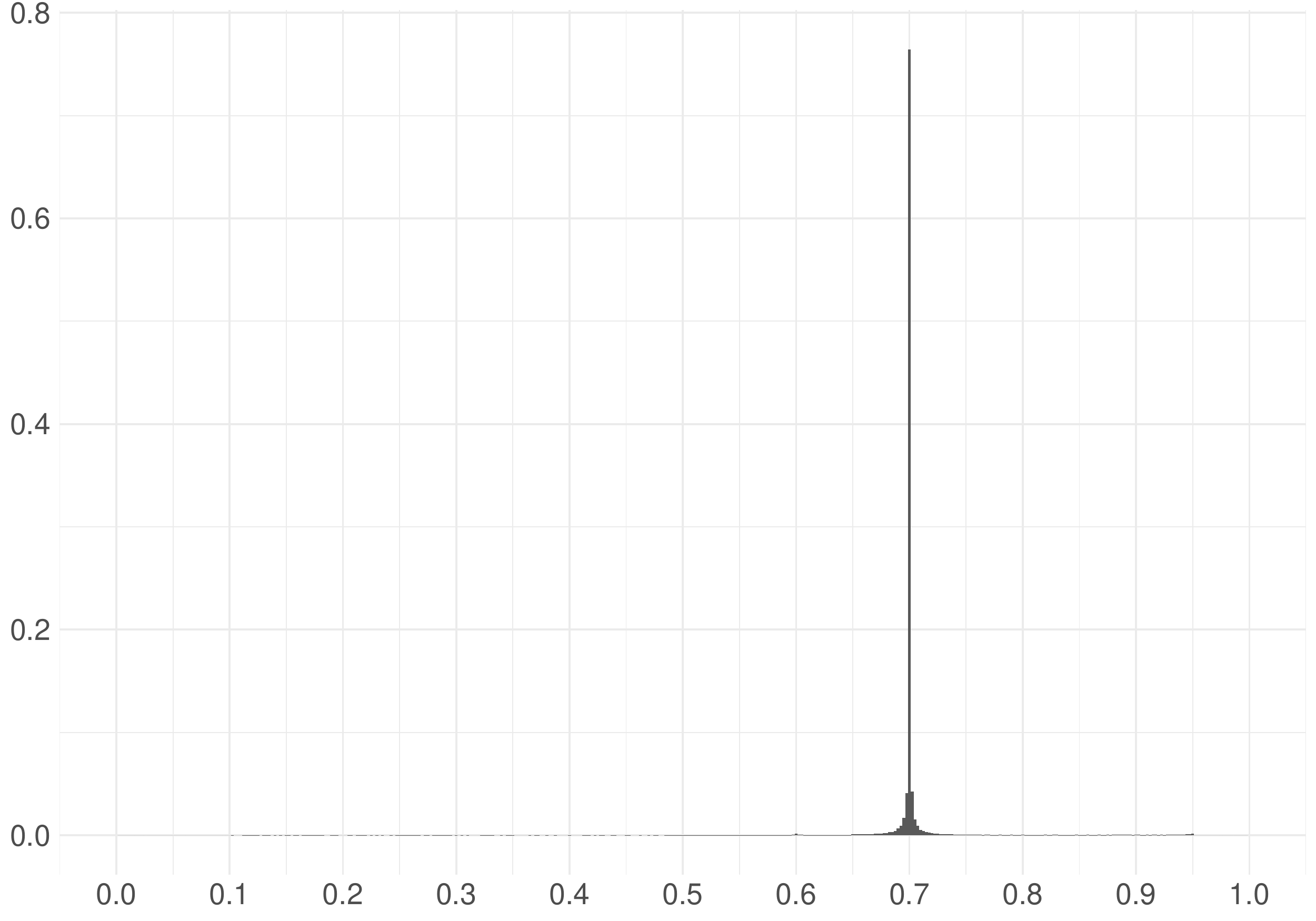}\label{fig:13:1}}
\subfigure[$T=800$, $\phi_a=1.05$, $\phi_b=0.98$]{\includegraphics[width=0.45\linewidth]{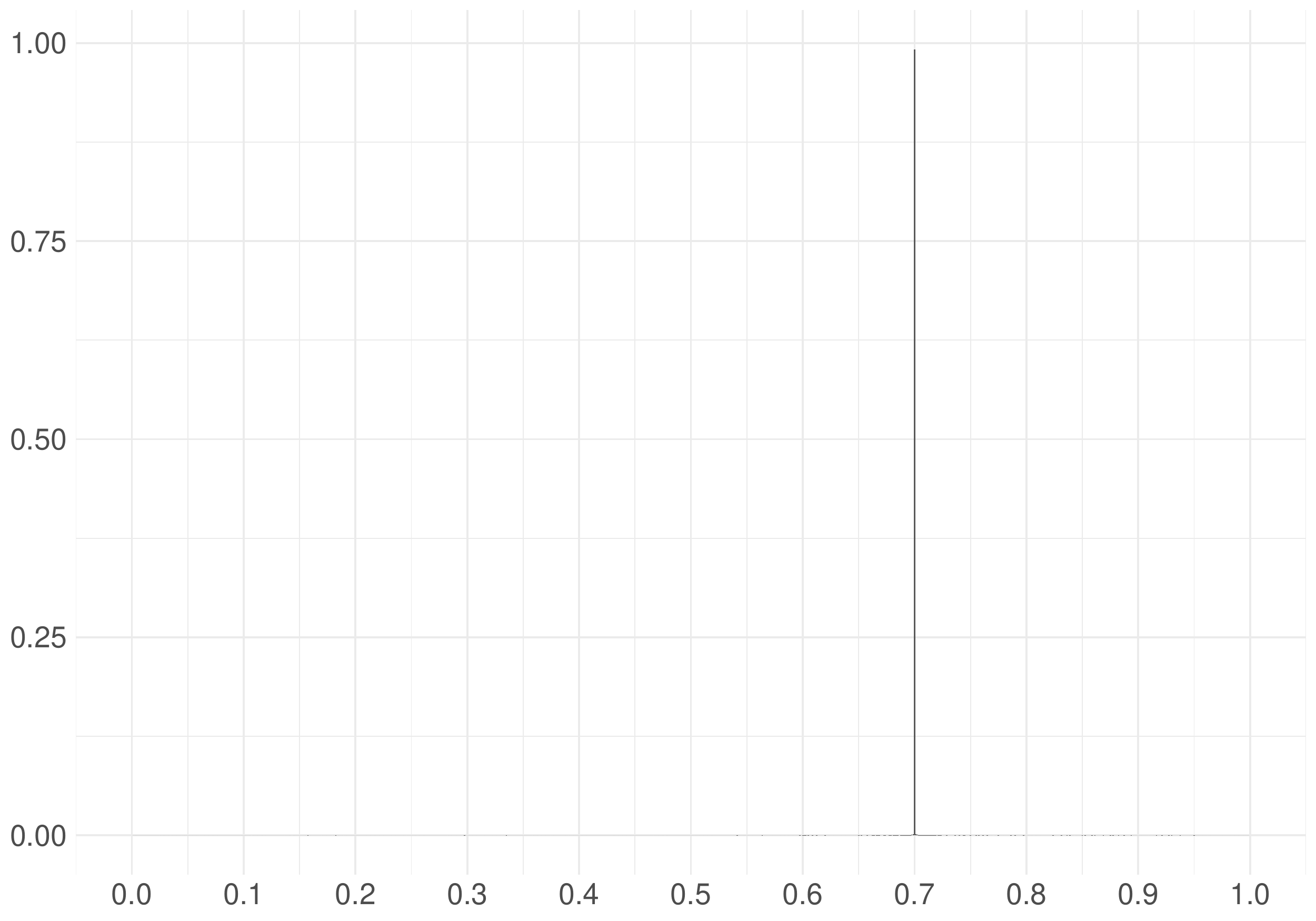}\label{fig:13:2}}\\
\subfigure[$T=400$, $\phi_a=1.05$, $\phi_b=0.96$]{\includegraphics[width=0.45\linewidth]{graph/NV_k_r_T=400_1.05_0.96_Model1s0.s11.pdf}\label{fig:13:3}}
\subfigure[$T=800$, $\phi_a=1.05$, $\phi_b=0.96$]{\includegraphics[width=0.45\linewidth]{graph/NV_k_r_T=800_1.05_0.96_Model1s0.s11.pdf}\label{fig:13:4}}\\
\subfigure[$T=400$, $\phi_a=1.05$, $\phi_b=0.94$]{\includegraphics[width=0.45\linewidth]{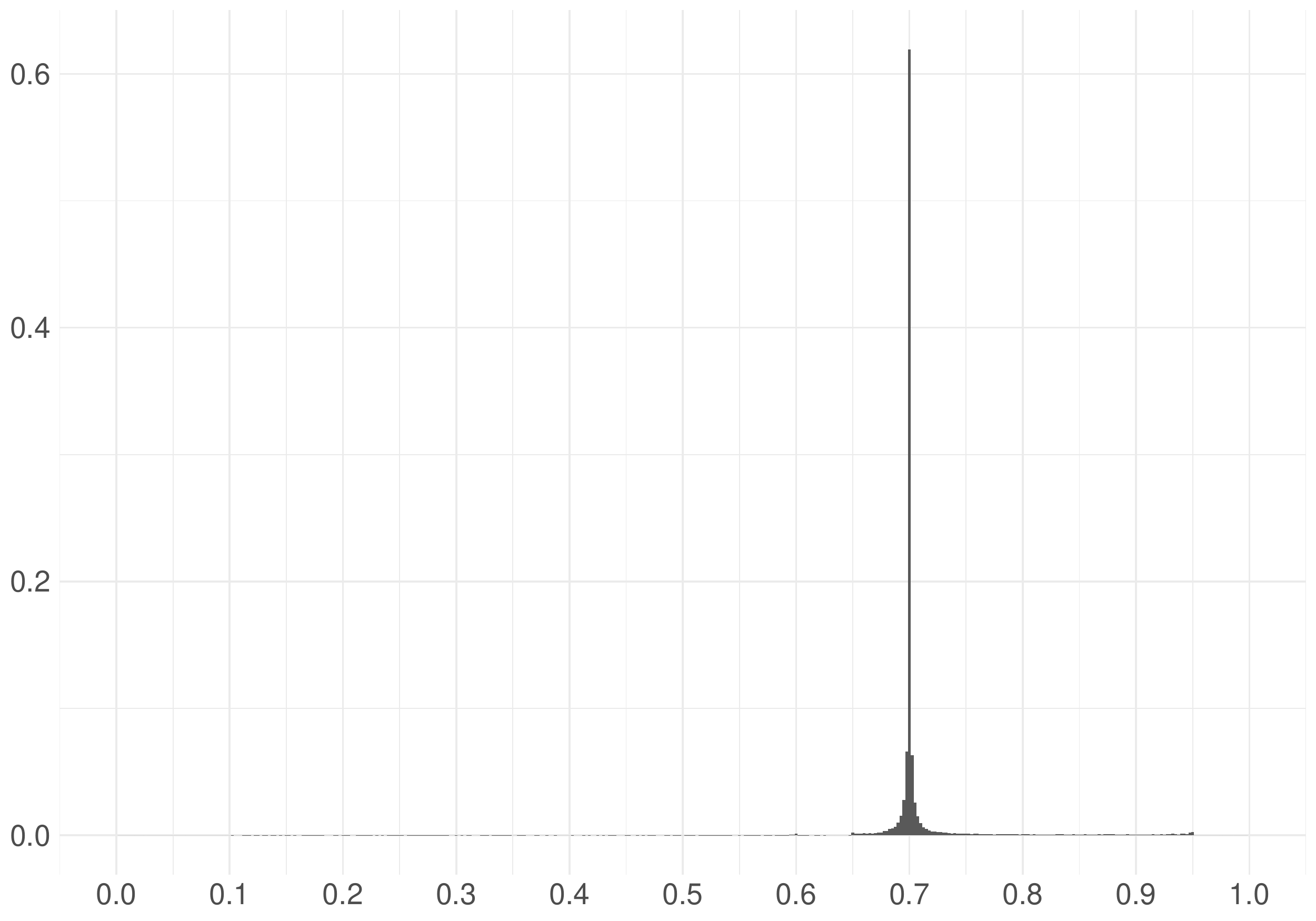}\label{fig:13:5}}
\subfigure[$T=800$, $\phi_a=1.05$, $\phi_b=0.94$]{\includegraphics[width=0.45\linewidth]{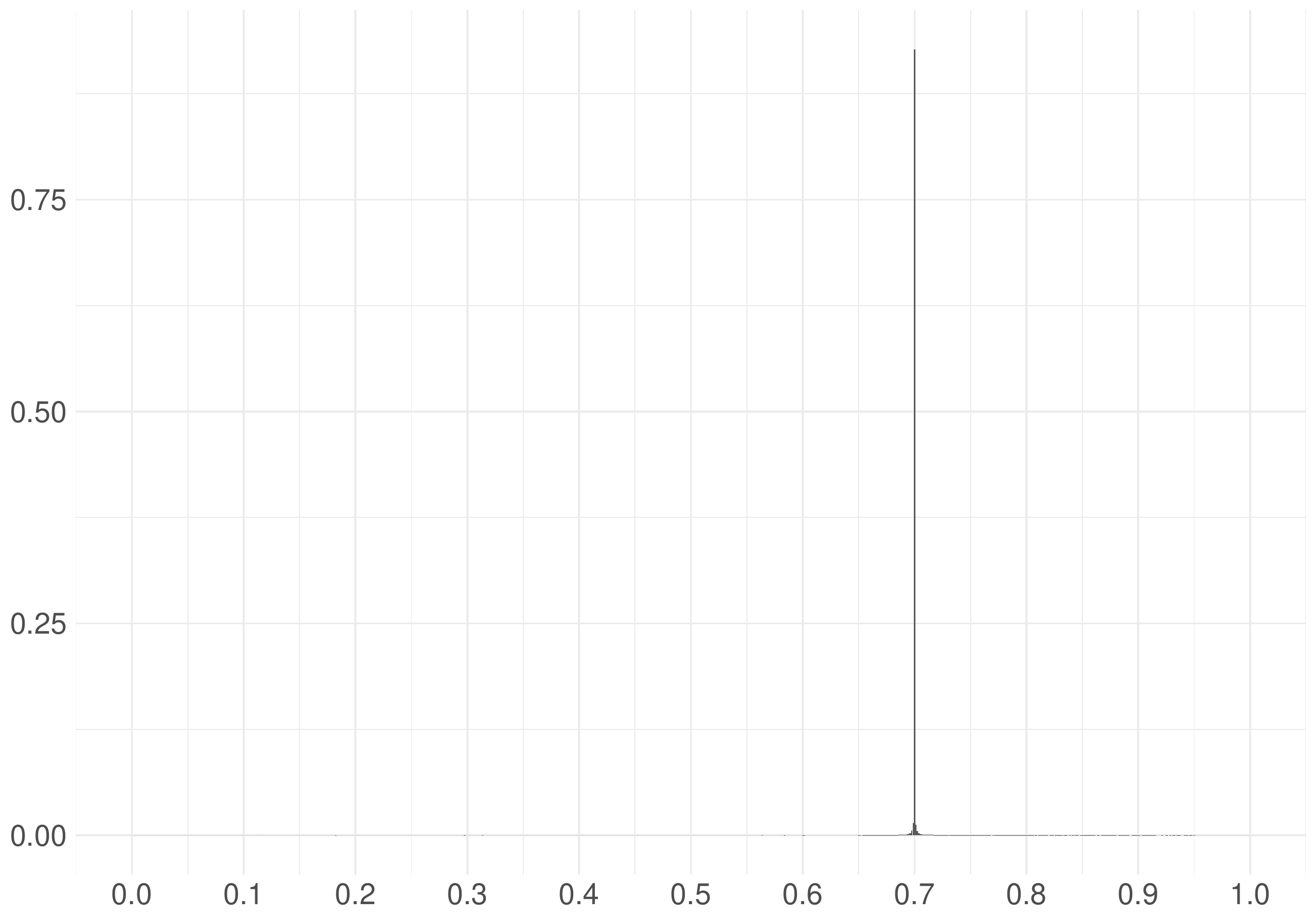}\label{fig:13:6}}
\end{center}%
\caption{Histograms of $\hat{k}_r$ % (left) and $\hat{k}_r$ (right) 
for $(\tau_e,\tau_c,\tau_r)=(0.4,0.6,0.7)$}
\label{fig13}
%\centering
%\footnotesize{OLS}
\end{figure}

\begin{figure}[h]%
\begin{center}%
\includegraphics[width=0.95\linewidth]{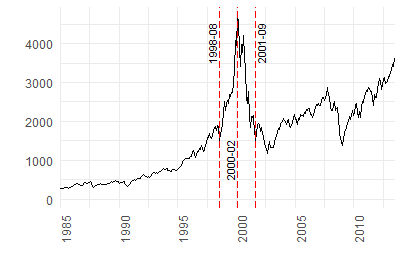}
\end{center}%
\caption{The bubble origination, collapse and recovery dates of the NASDAQ vomposite index}
\label{fig_emp}
%\centering
%\footnotesize{OLS}
\end{figure}

\begin{figure}[h]%
\begin{center}%
\includegraphics[width=0.95\linewidth]{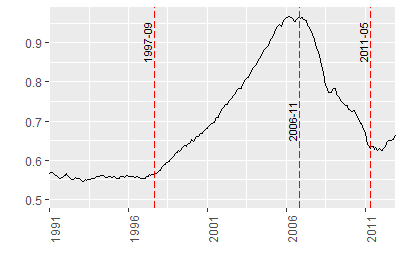}
\end{center}%
\caption{The bubble origination, collapse and recovery dates of U.S.\ house price index}
\label{fig_emp_hpi}
%\centering
%\footnotesize{OLS}
\end{figure}

\end{document}